\documentclass[%
reprint,
superscriptaddress,
preprintnumbers,
amsmath,amssymb,
aps,
prd,
]{revtex4-2}

\usepackage{float}
\usepackage{graphicx}
\usepackage{dcolumn}
\usepackage{bm}
\usepackage{bbold}
\usepackage{braket}
\usepackage{amssymb,amsmath}

\usepackage{color}
\usepackage{amsfonts}
\usepackage{subfigure}
\usepackage{array}
\usepackage{enumerate}

\newcommand{\Tr}{\ensuremath{\operatorname{Tr}}}
\newcommand{\tr}{\ensuremath{\operatorname{tr}}}

\newcolumntype{L}{>{\centering\arraybackslash}m{3cm}}

\definecolor{blue}{rgb}{0,0,1}

\definecolor{green}{rgb}{0,1,0}

\definecolor{red}{rgb}{1,0,0}

\definecolor{gray}{rgb}{.5,.5,.5}

\definecolor{darkgreen}{rgb}{.0,.5,.0}

\usepackage{xspace}
\usepackage{siunitx}
\usepackage{xfrac}
\usepackage{hyperref}
\usepackage[nameinlink]{cleveref}
\usepackage{appendix}
\usepackage{units}
\usepackage{xifthen}
\usepackage{xcolor}
\hypersetup{
	colorlinks,
	linkcolor={red!75!black},
	citecolor={blue!75!black},
	urlcolor={blue!75!black}
}
\def\Fig#1{\Cref{#1}}

\def\Tab#1{\Cref{#1}}
\def\tab#1{\Cref{#1}}

\def\Eq#1{\Cref{#1}}

\def\eq#1{\Cref{#1}}
\def\eqref#1{\Cref{#1}}

\def\sec#1{\Cref{#1}}
\def\app#1{\hyperref[#1]{App.~\ref{#1}}}
\def\app#1{\Cref{#1}}

\def\lA0{{\langle A_0 \rangle}}
\def\bA0{{\bar{A}_0}}

\def\0#1#2{\frac{#1}{#2}}

\graphicspath{{./figures/}{./}}

\begin{document}
	
\title{QCD at finite temperature and density within the fRG approach: An overview}
	
\author{Wei-jie Fu}
\affiliation{School of Physics, Dalian University of Technology, Dalian, 116024,
  P.R. China}

	
\begin{abstract}

In this paper we present an overview on recent progress in studies of QCD at finite temperature and densities within the 
functional renormalization group (fRG) approach. The fRG is a nonperturbative continuum field approach, in which quantum, thermal and density fluctuations are integrated in successively with the evolution of the renormalization group (RG) scale. The fRG results for the QCD phase structure and the location of the critical end point (CEP), the QCD equation of state (EoS), the magnetic EoS, baryon number fluctuations confronted with recent experimental measurements, various critical exponents, spectral functions in the critical region, the dynamical critical exponent, etc., are presented. Recent estimates of the location of the CEP from first-principle QCD calculations within fRG and Dyson-Schwinger Equations, which passes through lattice benchmark tests at small baryon chemical potentials, converge in a rather small region at baryon chemical potentials of about 600 MeV. A region of inhomogeneous instability indicated by a negative wave function renormalization is found with $\mu_B\gtrsim 420$ MeV. It is found that the non-monotonic dependence of the kurtosis of the net-proton number distributions on the beam collision energy observed in experiments, could arise from the increasingly sharp crossover in the regime of low collision energy.

\end{abstract}

\maketitle
\tableofcontents
	
\section{Introduction}
\label{sec:int}

One of the most challenging questions in heavy-ion physics arises from the attempt to understand how the deconfined quarks and gluons, i.e., the quark-gluon plasma (QGP), evolve into the confined hadrons. This evolution involves apparently two different phase transitions: One is the confinement-deconfinement phase transition and the other is the chiral phase transition related to the breaking and restoration of the chiral symmetry of QCD. When the strange quark is in its physical mass and the $u$ and $d$ quarks are massless, i.e., in the chiral limit, the chiral phase transition in the regime of small chemical potential in the QCD phase diagram, see e.g., \Fig{fig:phasedia-Chen2021iuo}, might be of second order, and belongs to the $O(4)$ symmetry universality class \cite{Pisarski:1983ms, Stephanov:2007fk}. With the increase of the baryon chemical potential, the second-order phase transition might be changed into a first-order one at the tricritical point. When the $u$ and $d$ quarks are in their small, but nonvanishing physical masses, due to the explicit breaking of the chiral symmetry, the $O(4)$ second-order chiral phase transition turns into a continuous crossover \cite{Aoki:2006we}, which is also consistent with experimental measurements, cf. e.g., \cite{Andronic:2017pug}. The tricritical point in the phase diagram evolves into a critical end point (CEP), which is the end point of the first-order phase transition line at high baryon chemical potential or densities.

Although the phase transition at the CEP is of second order and belongs to the $Z(2)$ symmetry universality class, the location of CEP and the size of the critical region around the CEP are non-universal. From the paradigm of the QCD phase diagram described above, one can easily find that the CEP plays a pivotal role in understanding the whole QCD phase structure in terms of the temperature and the baryon chemical potential. As a consequence, it becomes an very important task to search for and pin down the location of the CEP in the QCD phase diagram. Lattice simulations provide us with a wealth of knowledge of QCD phase transitions at vanishing and small baryon chemical potential, cf. e.g., \cite{Bazavov:2012vg, Borsanyi:2013hza, Borsanyi:2014ewa, Bellwied:2015rza, Bazavov:2017dus, Bazavov:2017tot, Borsanyi:2018grb, Bazavov:2018mes, Ding:2019prx, Ding:2020xlj, Bazavov:2020bjn, Borsanyi:2020fev}, but due to the sign problem at finite chemical potentials, lattice calculations are usually restricted in the region of $\mu_B/T\lesssim 2 \sim 3$, where no signal of CEP is observed \cite{Karsch:2019mbv}. Notably, recent estimates of the location of CEP from first-principle functional approaches, such as the functional renormalization group (fRG) and Dyson-Schwinger Equations (DSE), which passes through lattice benchmark tests at small baryon chemical potentials, converge in a rather small region at baryon chemical potentials of about 600 MeV \cite{Fu:2019hdw, Gao:2020fbl, Gunkel:2021oya}, and see also, e.g., \cite{Fischer:2018sdj} for related discussions.

Search for the CEP is currently under way or planned at many facilities, see, e.g., \cite{Luo:2017faz, Bzdak:2019pkr,  Andronic:2017pug, Friman:2011zz, Agakishiev:2009am, Abgrall:2014xwa, Sorin:2011zz, NICA-white, Dainese:2019xrz, Yang:2013yeb, Lu:2016htm, Sakaguchi:2017ggo, Sako:2019hzh}. Since the CEP is of second order, the correlation length increases significantly in the critical region in the vicinity of a CEP. Moreover, it is well known that fluctuation observables, e.g., the fluctuations of conserved charges, are very sensitive to the critical dynamics, and the increased correlation length would result in the increase of the fluctuations as well. Therefore, it has been proposed that a non-monotonic dependence of the conserved charge fluctuations on the beam collision energy can be used to search for the CEP in experimental measurements \cite{Stephanov:1999zu, Stephanov:2008qz, Stephanov:2011pb}, cf. also \cite{Luo:2017faz}. In the first phase of the Beam Energy Scan (BES-I) program at the Relativistic Heavy Ion Collider (RHIC) in the last decade, cumulants of net-proton, net-charge and net-kaon multiplicity distributions of different orders, and their correlations have been measured \cite{Adamczyk:2013dal, Adamczyk:2014fia, Luo:2015ewa, Adamczyk:2017wsl, Adam:2019xmk, STAR:2020tga, STAR:2021rls, STAR:2021fge}. Notably, recently a non-monotonic dependence of the kurtosis of the net-proton multiplicity distribution on the collision energy is observed with $3.1\,\sigma$ significance for central gold-on-gold (Au+Au) collisions \cite{STAR:2020tga}.

In this work we would like to present an overview on recent progress in studies of QCD at finite temperature and densities within the fRG approach. The fRG is a nonperturbative continuum field approach, in which quantum, thermal and density fluctuations are integrated in successively with the evolution of the renormalization group (RG) scale \cite{Wetterich:1992yh}, cf. also \cite{Ellwanger:1993mw, Morris:1993qb}. For QCD-related reviews, see, e.g., \cite{Berges:2000ew, Pawlowski:2005xe, Schaefer:2006sr, Gies:2006wv, Rosten:2010vm, Braun:2011pp, Pawlowski:2014aha, Dupuis:2020fhh}. Remarkably, recent years have seen significant progress in first-principle fRG calculations, for example, the state-of-the-art quantitative fRG results for Yang-Mills theory in the vacuum  \cite{Cyrol:2016tym} and at finite temperature \cite{Cyrol:2017qkl}, vacuum QCD results in the quenched approximation \cite{Mitter:2014wpa}, unquenched QCD in the vacuum \cite{Braun:2014ata, Rennecke:2015eba, Cyrol:2017ewj} and at finite temperature and densities \cite{Fu:2019hdw, Braun:2020ada}.

In this paper we try to present a self-contained overview, which include some fundamental derivations. Although new researchers in this field, e.g., students, may find these derivations useful, familiar readers could just skip over them. Furthermore, in this paper we focus on studies of fRG at finite temperature and densities, so we have to give up some topics, which in fact are very important for the developments and applications of the fRG approach, such as the quantitative fRG calculations to QCD in the vacuum \cite{Mitter:2014wpa, Cyrol:2016tym, Cyrol:2017ewj}.

This paper is organized as follows: In \sec{sec:FRGformal} we introduce the formalism of the fRG approach, including the Wetterich equation, the flow equations of correlation functions, the technique of dynamical hadronization, etc. Moreover, we also give a brief discussion about the Wilson's recursion formula and the Polchinski equation, which are closely related to the fRG approach. In \sec{sec:LEFT} we discuss the application of fRG in the low energy effective field theories (LEFTs), including the Nambu--Jona-Lasinio model and the quark-meson model. The relevant results in LEFTs, e.g., the phase structure, the equation of state, baryon number fluctuations, critical exponents, are presented. In \sec{sec:QCD} we turn to the application of fRG to QCD at finite temperature and densities. After a discussion about the flows of the propagators, strong couplings, four-quark couplings and the Yukawa couplings, we present and discuss the relevant results, e.g., the natural emergence of LEFTs from QCD, several different chiral condensates, QCD phase diagram and QCD phase structure, the inhomogeneous instability at large baryon chemical potentials, the magnetic equation of state, etc. In \sec{sec:RealtimefRG} we discuss the real-time fRG. After the derivation of the fRG flows on the Schwinger-Keldysh closed time path, one formulates the real-time effective action in terms of the ``classical'' and ``quantum'' fields in the physical representation. The spectral functions and the dynamical critical exponent of the $O(N)$ scalar theory are discussed. In \sec{sec:summary} a summary with conclusions is given. Moreover, an example for the flow equations of the gluon and ghost self-energies in Yang-Mills theory at finite temperature is given in \app{app:gluGhoSelfEner}. The Fierz-complete basis of four-quark interactions of $N_f=2$ flavors is listed in \app{app:Fierzbasis}, and some useful flow functions are collected in \app{app:flowFunc}.

	
\section{Formalism of the fRG approach}
\label{sec:FRGformal}

We begin the section with a derivation of the Wetterich equation, i.e., the flow equation of the effective action, which is followed by a brief introduction about the Wilson's recursion formula and the Polchinski equation, since they are closely related to the fRG approach. Then, the fRG approach is applied to QCD. A simple method to obtain the flow equations of correlation functions, i.e., \Eq{eq:dtVk}, is presented. An example for the flow equations of the gluon and ghost self-energies in Yang-Mills theory at finite temperature is given in \app{app:gluGhoSelfEner}. Finally, the technique of dynamical hadronization is discussed in \sec{subsec:dynhadron}.

\subsection{Flow equation of the effective action}
\label{subsec:actionflow}

We begin with a generating functional for a classical action $S[\hat\Phi]$ with an infrared (IR) regulator as follows
\begin{align}
  Z_k[J]=&\int (\mathcal{D} \hat\Phi) \exp\Big\{-S[\hat\Phi]-\Delta S_k[\hat\Phi]+J^a\hat\Phi_a\Big\}\,,\label{eq:Zk}
\end{align}
where the field $\hat\Phi$ is a collective symbol for all fields relevant in a specific physical problem, and the hat on the field is used to distinguish from its expected value in the following. It can even include fields which do not appear in the original classical action, and we will come back to this topic in what follows. The suffix of $\hat\Phi$, $a$, denotes not only the discrete degrees of freedom, e.g., the species, inner components of fields, etc., but also the continuous spacetime coordinates or the energy and momenta. The external source $J^a$ is conjugated to $\hat\Phi_a$. A summation or/and integral is assumed for a repeated index as shown in \Eq{eq:Zk}. The $k$-dependent regulator $\Delta S_k[\hat\Phi]$ in \Eq{eq:Zk} is used to suppress quantum fluctuations of momenta $q\lesssim k$, while leave those of $q> k$ untouched. Usually, a bilinear term, convenient in actual computations, is adopted for the regulator, which reads
\begin{align}
  \Delta S_k[\hat\Phi]=&\frac{1}{2}\hat\Phi_a R^{ab}_k\hat\Phi_b\,,\label{eq:DeltaS}
\end{align}
with $R^{ab}_k=R^{ba}_k$ for bosonic indices and $R^{ab}_k=-R^{ba}_k$ for fermionic ones. See e.g., \cite{Pawlowski:2005xe} for discussions about generic regulators. We will see in the following that the IR cutoff $k$ here is essentially the renormalization group (RG) scale.

\begin{figure}[t]
\includegraphics[width=0.4\textwidth]{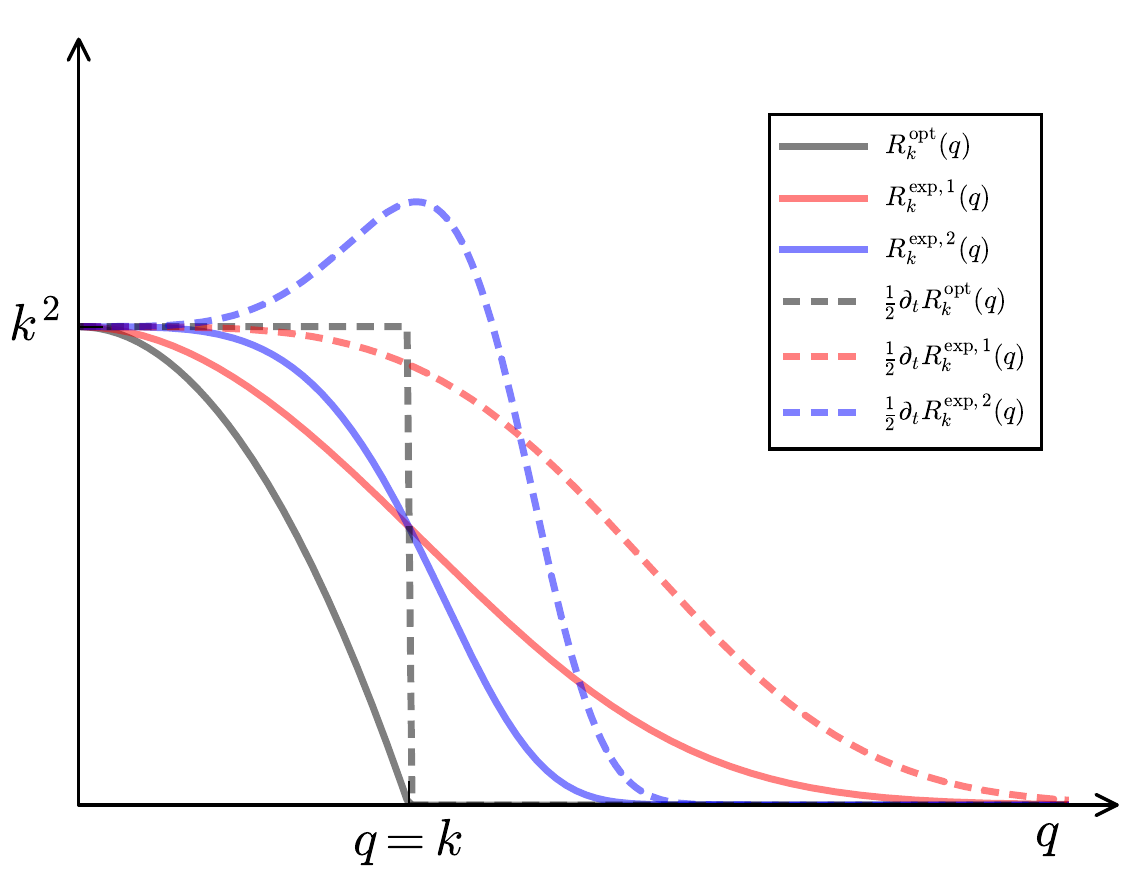}
\caption{Comparison of several different regulators and their derivative with respect to the RG scale $k$ as functions of the momentum $q$ with a fixed $k$.}\label{fig:regulator}
\end{figure}

We proceed to taking a single scalar field $\varphi$ for example, and the relevant regulator reads
\begin{align}
  \Delta S_k[\varphi]=&\frac{1}{2}\int d^4 x d^4 y \varphi(x)R_k(x,y)\varphi(y)\,.\label{eq:DeltaSvarphi}
\end{align}
It is more convenient to work in the momentum space. Employing the Fourier transformation as follows
\begin{align}
    \varphi(x)=&\int\frac{d^{4}q}{(2\pi)^{4}} \varphi(q)e^{iqx}\,,\label{eq:varphip}\\[2ex]
    R_k(x,y)=&\int\frac{d^{4}q}{(2\pi)^{4}}R_k(q)e^{iq(x-y)}\,,\label{eq:Rp}
\end{align}
one is led to
\begin{align}
  \Delta S_k[\varphi]=&\frac{1}{2}\int \frac{d^{4}q}{(2\pi)^{4}} \varphi(-q)R_k(q)\varphi(q)\,.\label{}
\end{align}
Here, the regulator has the following asymptotic properties which read
\begin{align}
  R_{k\rightarrow \infty}(q)\rightarrow \infty, \quad \text{and} \quad R_{k\rightarrow 0}(q)\rightarrow 0\,,\label{}
\end{align}
with a fixed $q$. In order to fulfill the aforementioned requirement, viz., only suppressing quantum fluctuations of momenta $q\lesssim k$ selectively, one could make the choice as follows
\begin{align}
  R_{k}(q)\big\vert_{q<k} \sim k^2, \qquad R_{k}(q)\big\vert_{q>k} \sim 0\,.\label{eq:RkGeneral}
\end{align}
One may have already noticed that there are infinite regulators fulfilling \Eq{eq:RkGeneral}. Here, we present two classes of regulators that are frequently used in literatures: One is the exponential-like regulator as follows
\begin{align}
  R_{k}^{\mathrm{exp}, n}(q)&= q^2 r_{\mathrm{exp}, n}(q^2/k^2)\,,\label{eq:regulatorExp}
\end{align}
with 
\begin{align}
  r_{\mathrm{exp}, n}(x)&= \frac{x^{n-1}}{e^{x^n}-1}\,.\label{eq:regulatorExp2}
\end{align}
The sharpness of regulator in the vicinity of $q=k$ is determined by the parameter $n$, as shown in \Fig{fig:regulator}, where the exponential regulators with $n=1$ and 2 are depicted as functions of $q$ with a fixed $k$. In \Fig{fig:regulator} we also plot another commonly used regulator, i.e., the flat or optimized one \cite{Litim:2000ci,Litim:2001up}, which reads
\begin{align}
  R_{k}^{\mathrm{opt}}(q)&= q^2 r_{\mathrm{opt}}(q^2/k^2)\,,\label{eq:regulatorOpt}
\end{align}
with 
\begin{align}
  r_{\mathrm{opt}}(x)&= \Big(\frac{1}{x}-1\Big)\Theta(1-x)\,,\label{eq:regulatorOpt2}
\end{align}
where $\Theta(x)$ is the Heaviside step function. Moreover, the derivative of regulator with respect to the RG scale $k$, to wit,
\begin{align}
  \partial_t R_{k}(q)&\equiv k \partial_k R_{k}(q)\,,\label{}
\end{align}
is also shown in \Fig{fig:regulator}, where $t$ is usually called as the RG time. For more discussions about regulators, see, e.g., \cite{Fister:2015eca}.

%
%

It is more convenient to use the generating functional for connected correlation functions, viz.,
\begin{align}
  W_k[J]&= \ln  Z_k[J]\,,\label{eq:Wk}
\end{align}
which is also known as the Schwinger function. Then, the expected value of a field is readily obtained as
\begin{align}
  \Phi_a &\equiv \langle \hat\Phi_a \rangle= \frac{\delta W_k[J]}{\delta J^a}\,,  \label{}
\end{align}
and the propagator reads
\begin{align}
  G_{k,ab} &\equiv \langle \hat\Phi_a \hat\Phi_b \rangle_c=\langle \hat\Phi_a \hat\Phi_b \rangle-\langle \hat\Phi_a\rangle\langle \hat\Phi_b\rangle  \nonumber\\[2ex]
  &=\frac{\delta^2 W_k[J]}{\delta J^a J^b}\,,  \label{eq:Gpropa}
\end{align}
where the subscript $c$ stands for ``connected". Moreover, Legendre transformation to the Schwinger function leaves us immediately with the one particle irreducible (1PI) effective action, which reads
\begin{align}
  \Gamma_{k}[\Phi]&=-W_k[J]+J^a\Phi_a-\Delta S_k[\Phi]\,.  \label{eq:Gamk}
\end{align}
In order to take both bosonic and fermionic fields into account all together, we adopt the notation introduced in \cite{Pawlowski:2005xe}, to wit, 
\begin{align}
  J^a\Phi_a&=\gamma^a_{\hspace{0.15cm}b} \Phi_a J^b\,,  \label{eq:JPhi}
\end{align}
with
\begin{align}
  \gamma^a_{\hspace{0.15cm}b}&=\left \{\begin{array}{l}
-\delta^a_{\hspace{0.15cm}b}, \quad \text{a and b are fermionic,}\\
 \delta^a_{\hspace{0.15cm}b}, \quad\hspace{0.25cm} \text{others.}
\end{array} \right.  \label{}
\end{align}
Inserting \Eq{eq:JPhi} into \Eq{eq:Gamk} and differentiating both sides of \Eq{eq:Gamk} with respect to $\Phi_a$, one is led to
\begin{align}
  \frac{\delta(\Gamma_{k}[\Phi]+\Delta S_k[\Phi])}{\delta \Phi_a}&=\gamma^a_{\hspace{0.15cm}b} J^b\,, \label{eq:dGamdPhi}
\end{align}
whereby, the propagator in \Eq{eq:Gpropa} is readily written as the inverse of the second-order derivative of $\Gamma_{k}[\Phi]$ w.r.t. $\Phi$, i.e.,
\begin{align}
  G_{k,ab}&=\gamma^c_{\hspace{0.15cm}a}\Big(\Gamma_{k}^{(2)}[\Phi]+\Delta S_k^{(2)}[\Phi]\Big)^{-1}_{cb}\,,  \label{eq:invgam2}
\end{align}
with
\begin{align}
  \Big(\Gamma_{k}^{(2)}[\Phi]+\Delta S_k^{(2)}[\Phi]\Big)^{ab}&\equiv \frac{\delta^2(\Gamma_{k}[\Phi]+\Delta S_k[\Phi])}{\delta \Phi_a \delta \Phi_b}\,.  \label{eq:d2GadPhi2}
\end{align}

We proceed to considering the evolution of Schwinger function with the RG scale, i.e., the flow equation of $W_k[J]$, which is straightforwardly obtained from Eqs. (\ref{eq:Zk}) and (\ref{eq:Wk}). The resulting flow reads
\begin{align}
  \partial_t W_k[J]&=-\frac{1}{2}\text{STr}\Big[\big(\partial_t R_k\big) G_{k}\Big]-\frac{1}{2}\Phi_a\partial_t R^{ab}_k\Phi_b\,, \label{eq:dtWk}
\end{align}
where we have introduced a notation super trace for compactness, which can also be expressed as  
\begin{align}
  \text{STr}\Big[\big(\partial_t R_k\big) G_{k}\Big]&=\big(\partial_t R^{ab}_k\big) \gamma^c_{\hspace{0.15cm}b}G_{k,ca}\,. \label{}
\end{align}
The factor $\gamma$ in the equation above indicates that the super trace provides an additional minus sign for fermionic degrees of freedom. Note that in deriving \Eq{eq:dtWk} we have used the relation in \Eq{eq:Gpropa}. Applying Legendre transformation in \Eq{eq:Gamk} to the flow equation of Schwinger function in \Eq{eq:dtWk} once more, one immediately arrives at the flow equation of the effective action, as follows
\begin{align}
  \partial_t \Gamma_{k}[\Phi]&=-\partial_t W_k[J]-\partial_t \Delta S_k[\Phi]\nonumber \\[2ex]
  &=\frac{1}{2}\text{STr}\Big[\big(\partial_t R_k\big) G_{k}\Big]\,, \label{eq:WetterichEq}
\end{align}
which is the Wetterich equation \cite{Wetterich:1992yh}. Note that the flow equation of effective action in \Eq{eq:WetterichEq} would be modified when the dynamical hadronization is encoded, see \Eq{eq:FlowQCD}. In \sec{subsec:QCDaction} we would like to give an example for the application of fRG, and postpone discussions of the dynamical hadronization in \sec{subsec:dynhadron}.

\subsection{From Wilson's RG to Polchinski equation}
\label{subsec:WilsonRG}

The idea encoded in the Wetterich equation in \Eq{eq:WetterichEq} is that, integrating out high momentum modes leaves us with a RG rescaled theory, and this theory is invariant at a second-order phase transition. This idea is also reflected in the Wilson's RG and Polchinski equation, to be discussed in this subsection.

\subsubsection{	Wilson's RG and recursion formula}
\label{subsubsec:Wilson}

Here we follow \cite{Ma:2020a} and begin with the Ginzburg-Landau Hamiltonian $\mathcal{K}\equiv H[\sigma]/T$, which reads
\begin{align}
  \mathcal{K}=&\int d^d x \Big[\frac{1}{2}c \big(\nabla \sigma \big)^2+U(\sigma)\Big]\,.\label{eq:mathcalK}
\end{align}
Then one separates the field into two parts as follows
\begin{align}
  \sigma=&\sigma'+\tilde \sigma\,,\label{eq:sigmaSepra}
\end{align}
with 
\begin{align}
  \tilde \sigma (\bm{x})=&L^{-d/2}\sum_{\Lambda/2<q<\Lambda}\sigma_{\bm{q}}e^{i \bm{q}\cdot \bm{x}}\,,\label{eq:tilsigma}
\end{align}
where $L$ denotes the size of the system, $\Lambda$ is a UV cutoff scale and $\Lambda^{-1}$ can be regarded as the lattice spacing. The plane-wave expansion in \Eq{eq:tilsigma} can also be replaced by that in terms of localized wave packets $W_z(\bm{x})$ , i.e., the Wannier functions for the band of plane waves $\Lambda/2<q<\Lambda$, that is,
\begin{align}
  \tilde \sigma (\bm{x})=&\sum_{z}\tilde \sigma_z W_z(\bm{x})\,.\label{}
\end{align}
Obviously, the wave packets have the property $W_z(\bm{x})\sim 0$, if $|\bm{x}-\bm{z}|\gg 2 \Lambda^{-1}$, and one also has
\begin{align}
  \int d^d x  W_z(\bm{x})  W_{z'}(\bm{x})=&\delta_{z z'}\,,\label{}
\end{align}
i.e., they are orthonormal. Substituting \Eq{eq:sigmaSepra} into \Eq{eq:mathcalK} and performing a functional integral over $\tilde \sigma (\bm{x})$ or $\tilde \sigma_z$, one arrives at
\begin{align}
  &\int (\mathcal{D} \tilde \sigma) e^{-\mathcal{K}}=\exp\Big[-\int d^d x \frac{1}{2}c \big(\nabla \sigma' \big)^2\Big]\prod_z I( \sigma')\nonumber\\[2ex]
=&\exp\bigg\{-\int d^d x \Big[\frac{1}{2}c \big(\nabla \sigma' \big)^2+\bar U'(\sigma')\Big]-A L^d\bigg\}\,,\label{eq:Dsigma}
\end{align}
with
\begin{align}
  I(\sigma')\equiv&\Omega^{\frac{1}{2}}\int d y \exp\bigg\{-\frac{c}{2}\bar{q^2}\Omega y^2-\frac{\Omega}{2}\Big[U(\sigma'+y)\nonumber\\[2ex]
&+U(\sigma'-y)\Big]\bigg\}\nonumber\\[2ex]
\equiv& \exp\Big[-\Omega \bar U'(\sigma')-\Omega A\Big]\,,\label{eq:Isigma}
\end{align}
where $\Omega$ is the volume of a block, or the wave packet; the constant $A$ is determined by the condition $\bar U'(0)=0$; The mean square wave vector of the packet reads
\begin{align}
  \bar{q^2}=& \int d^d x  \Big(\nabla W_z(\bm{x})\Big)^2\,.\label{}
\end{align}
In deriving \Eq{eq:Dsigma}, one has neglected the overlap between wave packets, the variation of $\sigma'(\bm{x})$ and the absolute value of $W_z(\bm{x}) $ within a block, and see \cite{Wilson:1971bg, Wilson:1971dh, Ma:2020a} for details. Note that \Eq{eq:Dsigma} is just the first step of the Kadanoff transformation.

The second step of Kadanoff transformation is to make replacement for the remaining field $\sigma'$ and the coordinate in \Eq{eq:Dsigma} as follows
\begin{align}
  \sigma'(\bm{x})\rightarrow& 2^{1-\frac{d}{2}-\frac{\eta}{2}} \sigma(\bm{x}')\,, \qquad \bm{x}'=\frac{\bm{x}}{2}\,,\label{}
\end{align}
and thus one is left with
\begin{align}
  \mathcal{K'}=&\int d^d x' \Big[\frac{1}{2}c' \big(\nabla \sigma \big)^2+U'(\sigma)\Big]\,,\label{eq:mathcalKp}
\end{align}
with
\begin{align}
  c' =&2^{-\eta}c\,,\label{}
\end{align}
and 
\begin{align}
  U'(\sigma) =&-2^{d} \Omega^{-1}\ln\frac{I(2^{1-\frac{d}{2}-\frac{\eta}{2}}\sigma)}{I(0)}\,.\label{}
\end{align}
In order to make $c' =c$ satisfied, one adopts $\eta=0$. Choosing an appropriate value of $c$, such that
\begin{align}
  \frac{c}{2}\bar{q^2}\Omega=&1\,,\label{}
\end{align}
and defining 
\begin{align}
  Q(\sigma)\equiv&\Omega U(\sigma)\,,\label{}
\end{align}
one arrives at
\begin{align}
  Q'(\sigma) =&-2^{d} \ln\frac{I(2^{1-\frac{d}{2}}\sigma)}{I(0)}\,.\label{}
\end{align}
with
\begin{align}
   I(\sigma)=&\int d y \exp\bigg\{-y^2-\frac{1}{2}\Big[Q(\sigma+y)+Q(\sigma-y)\Big]\bigg\}\,,\label{}
\end{align}
which is the Wilson's recursion formula \cite{Wilson:1971bg, Wilson:1971dh}. Note that the constant prefactor $\Omega^{1/2}$ in \Eq{eq:Isigma} is irrelevant.

\subsubsection{	Polchinski equation}
\label{subsubsec:PolchinskiEq}

In \sec{subsubsec:Wilson} we have discussed the viewpoint of Wilson's RG, that is, integrating out modes of high scales successively leaves us with a low energy theory that evolves with the RG scale. This idea is also applied to a generic quantum field theory within the formalism of the functional integral, due to Polchinski \cite{Polchinski:1983gv}. We begin with a generating functional for a scalar field theory in four Euclidean dimensions with a momentum cutoff, i.e.,
\begin{align}
  Z[J]=&\int (\mathcal{D} \phi) \exp\bigg\{\int \frac{d^4 p}{(2\pi)^4}\Big[-\frac{1}{2}\phi(p)\phi(-p)(p^2+m^2)\nonumber\\[2ex]
&\times K^{-1}\big(\frac{p^2}{\Lambda_0^2}\big)+J(p)\phi(-p)\Big]+L_{\mathrm{int}}(\phi)\bigg\}\,,\label{eq:ZJPolch}
\end{align}
with a cutoff function given by
\begin{align}
  K(p^2/\Lambda_0^2)&=\left \{\begin{array}{l}
1, \quad p^2<\Lambda_0^2,\\
0, \quad p^2\gg \Lambda_0^2.
\end{array} \right.  \label{}
\end{align}
Evidently, here $\Lambda_0$ plays a role as a UV cutoff scale, and $L_{\mathrm{int}}$ in \Eq{eq:ZJPolch} is the interaction Lagrangian at the scale $\Lambda_0$, that for example reads
\begin{align}
  L_{\mathrm{int}}(\phi)=&\int d^4 x \Big[-\frac{1}{2}\rho^0_1 \phi^2(x)-\frac{1}{2} \rho^0_2 \big(\partial_{\mu}\phi(x)\big)^2\nonumber\\[2ex]
&-\frac{1}{4!}\rho^0_3 \phi^4(x)\Big] \,,\label{}
\end{align}
where we have used the notation in \cite{Polchinski:1983gv}, and $\rho^0_a$'s stand for bare quantities.

One would like to integrate out the high momentum modes of $\phi$ and reduce the UV cutoff $\Lambda_0$ to a lower value, say $\Lambda$. In the meantime, one chooses $|m^2|<\Lambda$ and $J(p)=0$ for $p^2>\Lambda^2$. As we have discussed in \sec{subsubsec:Wilson}, when high momentum modes are integrated out, new effective interactions, included in the potential $U'$ in \Eq{eq:mathcalKp} or the interaction Lagrangian $L_{\mathrm{int}}$ in \Eq{eq:ZJPolch}, are generated. Thus, one is led to the following functional integral 
\begin{align}
  Z[J, L, \Lambda]=&\int (\mathcal{D} \phi) \exp\bigg\{\int \frac{d^4 p}{(2\pi)^4}\Big[-\frac{1}{2}\phi(p)\phi(-p)(p^2\nonumber\\[2ex]
&+m^2) K^{-1}\big(\frac{p^2}{\Lambda^2}\big)+J(p)\phi(-p)\Big]+L(\phi, \Lambda)\bigg\}\,.\label{eq:ZJPolchLam}
\end{align}
If one wishes to take the generating functional $Z[J, L, \Lambda]$ on the l.h.s. of the equation above to be independent of the scale $\Lambda$, i.e.,
\begin{align}
  \Lambda \frac{d Z[J, L, \Lambda]}{d \Lambda} =&0\,,\label{}
\end{align}
the following evolution equation for the interaction Lagrangian in \Eq{eq:ZJPolchLam} has to be satisfied, to wit,
\begin{align}
  \Lambda \frac{\partial L(\phi, \Lambda)}{\partial \Lambda} =&-\int \frac{d^4 p}{(2\pi)^4}\frac{1}{2}\frac{1}{p^2+m^2}\Lambda \frac{\partial K\big(\frac{p^2}{\Lambda^2}\big)}{\partial \Lambda}\nonumber\\[2ex]
&\times\bigg[\frac{\delta^2 L}{\delta \phi(p)\delta \phi(-p)}+\frac{\delta L}{\delta \phi(p)}\frac{\delta L}{\delta \phi(-p)}\bigg]\,,\label{}
\end{align}
which is the Polchinski equation \cite{Polchinski:1983gv}.

\subsection{Application to QCD}
\label{subsec:QCDaction}

%
\begin{figure}[t]
\includegraphics[width=0.45\textwidth]{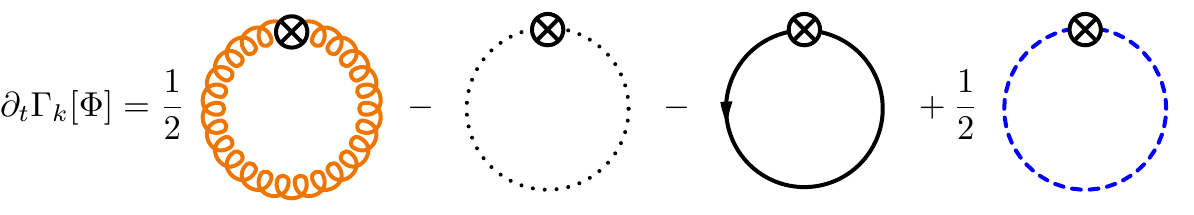}
\caption{Diagrammatic representation of the flow equation for QCD effective action. The lines denote full propagators for the gluon, ghost, quark, and meson, respectively. The crossed circles stand for the infrared regulators.}\label{fig:QCD_equation}
\end{figure}
%

In this section we would like to apply the formalism of fRG discussed above to an effective action of rebosonized QCD  in \cite{Fu:2019hdw}. The truncation for the Euclidean effective action reads
\begin{align}
  &\Gamma_{k}[\Phi]\nonumber\\[2ex]
=&\int_{x} \bigg\{\frac{1}{4}F^a_{\mu\nu}F^a_{\mu\nu}+Z_{c,k} \big(\partial_{\mu}\bar{c}^a\big)D_{\mu} ^{ab}c^b+\frac{1}{2\xi}\big(\partial_{\mu} A^a_{\mu}\big)^2\nonumber \\[2ex]
  &+Z_{q,k}\bar{q}\big(\gamma_{\mu}D_{\mu}-\gamma_0 \hat \mu\big) q+m_s(\sigma_s)\bar{q}_s q_s -\lambda_{q,k}\Big[\big(\bar{q}_l\,T^0 q_l \big)^2\nonumber\\[2ex]
  &+\big(\bar{q}_l\,i\gamma_5 \bm{T} q_l \big)^2\Big]+h_k\bar{q}_l\big(T^0\sigma+i \gamma_5\bm{T}\cdot\bm{\pi} \big)q_l\nonumber  \\[2ex]
  &+\frac{1}{2}Z_{\phi,k}(\partial_{\mu}\phi)^2+V_{k}(\rho,A_0)-c_\sigma\,\sigma -\frac{1}{\sqrt{2}}\, c_{\sigma_s}\, \sigma_s\bigg\}\nonumber\\[2ex]
  &+\Delta\Gamma_{\mathrm{glue}}\,, \label{eq:QCDaction}
\end{align}
with  $\int_{x}=\int_0^{1/T}d x_0 \int d^3 x$, $T$ being the temperature. One can see that field contents in \Eq{eq:QCDaction} include not only the fundamental fields in QCD, i.e., the gluon, Faddeev-Popov ghost, and the quark, but also the composite fields $\phi=(\sigma,\bm{\pi})$, the scalar and pseudo-scalar mesons respectively. Note that here the mesonic fields are not added by hands, but rather dynamically generated and transferred from the fundamental degrees of freedom via the dynamical hadronization technique, described in detail in \sec{subsec:dynhadron}. Consequently, there is no double counting for the degrees of freedom. In short, one is left with $\Phi=(A, c, \bar{c}, q, \bar{q}, \sigma, \pi)$. The $h_k$ and $\lambda_{q,k}$ in \Eq{eq:QCDaction} denote the Yukawa coupling and the four-quark coupling, respectively, 

The first line on the r.h.s. of \Eq{eq:QCDaction} denotes the classical action for the glue sector, while its non-classical contributions are collected in $\Delta\Gamma_{\mathrm{glue}}$. The gauge parameter $\xi=0$, i.e., the Landau gauge, is commonly adopted in the computation of functional approaches. The wave function renormalization $Z_{\Phi,k}$ of field $\Phi$ is defined as
\begin{align}
  \bar \Phi &=Z_{\Phi,k}^{1/2} \Phi\,,  \label{eq:barPhi}
\end{align}
with the renormalized field $\bar \Phi$. The gluonic field strength tensor reads
\begin{align}
  F^a_{\mu\nu}&=Z_{A,k}^{1/2}\big(\partial_{\mu}A^a_{\nu}-\partial_{\nu}A^a_{\mu}+Z_{A,k}^{1/2}\bar{g}_{\text{\tiny glue},k} f^{abc}A^b_{\mu}A^c_{\nu}\big)\,.\label{eq:Fmunua}
\end{align}
Note that although different strong couplings are identical in the perturbative region, they can deviate from each other in the nonperturbative or even semiperturbative regime \cite{Mitter:2014wpa, Cyrol:2016tym, Cyrol:2017ewj, Cyrol:2017qkl}, and see also relevant discussions in \sec{subsec:StrongcouplingQCD}. Therefore, it is necessary to distinguish different strong couplings. The renormalized strong couplings in the glue sector read
\begin{align}
  \bar{g}_{A^3,k} =& \frac{\lambda_{A^3,k}}{Z_{A,k}^{3/2}}\,,\quad \bar{g}_{A^4,k} =\frac{\lambda_{A^4,k}^{1/2}}{Z_{A,k}}\,,\quad \bar{g}_{\bar c c A,k} =\frac{\lambda_{\bar c c A,k}}{Z_{A,k}^{1/2}\,Z_{c,k} }\,, \label{eq:strongcoupGluon}
\end{align}
where $\lambda_{A^3,k}$, $\lambda_{A^4,k}$ and $\lambda_{\bar c c A,k}$ stand for the three-gluon, four-gluon, ghost-gluon dressing functions, respectively, as shown in \Eq{eq:VertexA3}, \Eq{eq:VertexA4}, \Eq{eq:VertexbccA}. In \Eq{eq:Fmunua} the gluonic strong couplings are denoted collectively as $\bar{g}_{\text{\tiny glue},k}$. In the same way, the quark-gluon coupling reads
\begin{align}
  \bar{g}_{\bar q  q A, k} =& \frac{\lambda_{\bar q  q A,k}}{Z_{A,k}^{1/2}\,Z_{q,k}}\,, \label{eq:gbqqa}
\end{align}
with the quark-gluon dressing function $\lambda_{\bar q  q A,k}$. The covariant derivative in the fundamental and adjoint representations of the color $SU(N_c)$ group reads
\begin{align}
  D_{\mu}&=\partial_{\mu}-i Z_{A,k}^{1/2}\bar{g}_{\bar q  q A, k} A^a_{\mu} t^a\,, \label{eq:Dmu}\\[2ex]
  D^{ab}_{\mu}&=\partial_{\mu} \delta^{ab}-Z_{A,k}^{1/2}\bar{g}_{\bar c c A,k} f^{abc}A^c_{\mu}\,, \label{eq:Dmuab}
 \end{align}
 respectively. Here $f^{abc}$ is the antisymmetric structure constant of the $SU(N_c)$ group, determined from its Lie algebra $[t^a,t^b]=i f^{abc} t^c$, where the generators have the normalization $\Tr t^a t^b=(1/2)\delta^{ab}$. 
 
In \Eq{eq:QCDaction} the formalism of $N_f=3$ flavor quark is built upon that of $N_{f}=2$ by means of addition of a dynamical strange quark $q_s$, whose constituent quark mass $m_s(\sigma_s)$ is determined self-consistently from a extended effective potential of $SU(N_f=2)$, and see \cite{Fu:2019hdw} for more details. Therefore, we have the quark field $q=(q_l, q_s)$, where the $u$ and $d$ light quarks are denoted by $q_l=(q_u, q_d)$. The light quarks interact with themself through the four-quark coupling in the $\sigma-\pi$ channel, and they are also coupled with the $\sigma$ and $\pi$ mesons via the Yukawa coupling. Here $T^i$ ($i=1$, 2, 3) are the generators of the group $SU(N_f=2)$ in the flavor space with $\Tr T^i T^j=(1/2)\delta^{ij}$ and $T^{0}=(1/\sqrt{2N_{f}})\mathbb{1}_{N_{f}\times N_{f}}$ with $N_f=2$. In \Eq{eq:QCDaction} $\hat \mu=\mathrm{diag}(\mu_u, \mu_d, \mu_s)$ stands for the matrix of quark chemical potentials.

The effective potential in \Eq{eq:QCDaction} can be decomposed into two parts as follows
\begin{align}
  V_k(\rho,A_0) &=V_{\mathrm{glue},k}(A_0) + V_{\mathrm{mat},k}(\rho,A_0)\,.  \label{eq:EffPot}
\end{align}
The first term on the r.h.s. of equation above is the glue potential, or the Polyakov loop potential. The temporal gluon field $A_0$ is intimately related to the Polyakov loop $L[A_0]$, see e.g., \cite{Herbst:2015ona}; the second term is the mesonic effective potential of $N_f=2$ flavors, which is $O(4)$-invariant with $\rho=\phi^2/2$. Moreover, in the effective action in \Eq{eq:QCDaction} $c_\sigma$ and $c_{\sigma_s}$ are the parameters of explicit chiral symmetry breaking for the light and strange scalars $\sigma$, $\sigma_s$, respectively. 

Applying the fRG flow equation in \Eq{eq:WetterichEq} to the QCD effective action in \Eq{eq:QCDaction}, one immediately arrives at the QCD flow equation, which is depicted in \Fig{fig:QCD_equation}. One can see that the flow of effective potential receives contributions from the gluon, ghost, quark, and composite fields separately, each of which is one-loop structure. It should be emphasized that, although it is a one-loop structure, the flow is composed of full propagators which are in turn dependent on the second derivative of the effective action, see \Eq{eq:invgam2}. Thus, the flow equation in \Fig{fig:QCD_equation} a functional self-consistent differential equation, which will be explored further in the following. 

\subsection{Flow equations of correlation functions}
\label{subsec:correlationflow}

Combining \Eq{eq:invgam2} one can reformulate Wetterich equation in \Eq{eq:WetterichEq} slightly such that
\begin{align}
  \partial_t \Gamma_{k}[\Phi]&=\frac{1}{2}\mathrm{STr}\Big[\tilde{\partial_{t}} \ln\big(\Gamma_{k}^{(2)}[\Phi]+R_{k}\big)\Big]\,, \label{eq:WetterichEq2}
\end{align}
where the differential operator with a tilde, $\tilde{\partial_{t}}$, hits the RG-scale dependence only through the regulator. Note that $\Gamma_{k}^{(2)}[\Phi]$ in \Eq{eq:WetterichEq2} is a bit different from that in \Eq{eq:d2GadPhi2}, and here the factor $\gamma^a_{\hspace{0.15cm}b}$ is absorbed in $\Gamma_{k}^{(2)}[\Phi]$. A convenient way to take this into account is to use the definition as follows
\begin{align}
  \big(\Gamma_{k}^{(2)}[\Phi]\big)^{ab}&\equiv\frac{\overrightarrow{\delta}}{\delta\Phi_a}\Gamma_{k}[\Phi]\frac{\overleftarrow{\delta}}{\delta\Phi_b}\,,
 \label{eq:Gamma2}
\end{align}
where the left and right derivatives have been adopted.

In order to derive the flow equations for various correlation functions of different orders, it is useful to express $\Gamma_{k}^{(2)}$ in \Eq{eq:Gamma2} in terms of a matrix, and the indices of matrix correspond to different fields involved in the theory concerned, e.g., $\Phi=(A, c, \bar{c}, q, \bar{q}, \sigma, \pi)$ in \sec{subsec:QCDaction}. This matrix is also called as the fluctuation matrix \cite{Gies:2001nw}. Then, one can make the division as follows
\begin{align}
  \Gamma_{k}^{(2)}+R_{k}&=\mathcal{P}+\mathcal{F}\,,
 \label{eq:flucmatrdecom}
\end{align}
where $\mathcal{P}$ is the matrix of two-point correlation functions including the regulators, and its inverse gives rise to propagators; $\mathcal{F}$ is the matrix of interaction which includes $n$-point correlation functions with $n>2$, and thus terms in $\mathcal{F}$ have the field dependence. Substituting \Eq{eq:flucmatrdecom} into \eq{eq:WetterichEq2} and making the Taylor expansion in order of $\mathcal{F}/\mathcal{P}$, one arrives at
\begin{align}
  \partial_{t}\Gamma_{k}=&\frac{1}{2}\mathrm{STr}\Big[\tilde{\partial}_{t} \ln(\mathcal{P}+\mathcal{F})\Big]\nonumber \\[2ex]
 =&\frac{1}{2}\mathrm{STr}\tilde{\partial}_{t} \ln\mathcal{P}+\frac{1}{2}\mathrm{STr}\tilde{\partial}_{t}\Big(\frac{1}{\mathcal{P}}\mathcal{F}\Big)-\frac{1}{4}\mathrm{STr}\tilde{\partial}_{t}\Big(\frac{1}{\mathcal{P}}\mathcal{F}\Big)^2\nonumber \\[2ex]
 &+\frac{1}{6}\mathrm{STr}\tilde{\partial}_{t}\Big(\frac{1}{\mathcal{P}}\mathcal{F}\Big)^3-\frac{1}{8}\mathrm{STr}\tilde{\partial}_{t}\Big(\frac{1}{\mathcal{P}}\mathcal{F}\Big)^4+\cdots\,,
 \label{eq:WEqexpand}
\end{align}
from which it is straightforward to obtain the flow equations for various inverse propagators and vertices at some appropriate orders.

Moreover, one can also employ some well-developed Mathematica packages, e.g., DoFun \cite{Huber:2011qr, Huber:2019dkb}, QMeS-Derivation \cite{Pawlowski:2021tkk}, to derive the flow equations for correlation functions. We refrain from elaborating on details about usage of these Mathematica packages in this review, which interested readers can find in the references above, but rather would like to introduce another easy-to-use approach to derive the flow equations described in detail in \sec{subsubsec:flowDeriv}.

\subsubsection{A simple approach to derivation of flow equations of correlation functions}
\label{subsubsec:flowDeriv}

%
\begin{figure}[t]
\includegraphics[width=0.48\textwidth]{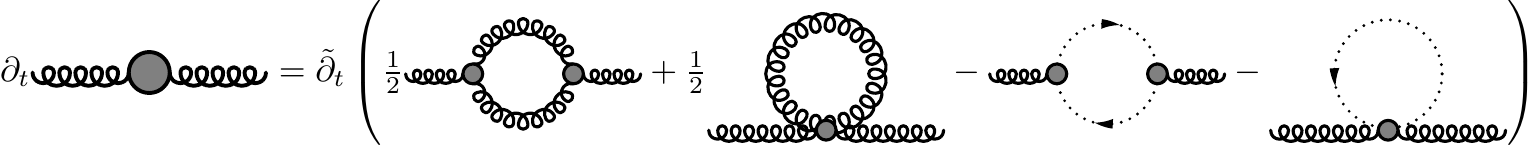}
\caption{Diagrammatic representation of the flow equation for the gluon self-energy in Yang-Mills theory, where the last diagram arises from the non-classical two-ghost--two-gluon vertex.}\label{fig:A2-equ}
\end{figure}
%

We proceed with defining a generic 1PI $n$-point correlation function or vertex, as follows
\begin{align}
  V_{k,\Phi_{a_1}\cdots\Phi_{a_n}}^{(n)}&\equiv -\Gamma_{k,\Phi_{a_1}\cdots\Phi_{a_n}}^{(n)}=-\left. \left(\frac{\delta^{n} \Gamma_{k}[\Phi]}{\delta \Phi_{a_1}\cdots\delta \Phi_{a_n}}\right)\right |_{\Phi=\langle \Phi\rangle}\,,\label{eq:1PIVert}
\end{align}
where $\langle \Phi\rangle$ denotes the value of $\Phi$ on its equation of motion (EoM). The flow equation of vertex $V_{k}^{(n)}$ in \Eq{eq:1PIVert} can be represented schematically as the equation as follows
\begin{align}
  \partial_{t} V_{k,\Phi_{a_1}\cdots\Phi_{a_n}}^{(n)}&=\tilde{\partial}_{t}\left(\begin{array}{l}\mathrm{all\;one\!-\! loop\; correction} \\ \mathrm{diagrams\;of\;}V_{k,\Phi_{a_1}\cdots\Phi_{a_n}}^{(n)} \end{array}\right)\,.\label{eq:dtVk}
\end{align}
Note that the one-loop diagrams on the r.h.s. are comprised of full propagators and vertices. As we have mentioned above, the partial derivative with a tilde $\tilde{\partial_{t}}$ hits the RG-scale dependence only through the regulator, and thus its implementation on diagrams would give rise to the insertion of a regulator for each inner propagator.

As an example, we present the flow equation of the gluon self-energy in Yang-Mills theory in \Fig{fig:A2-equ}. One can see that it receives contributions from the gluon loop, the tadpole of the gluon, the ghost loop, and the tadpole of the ghost. Note that the last diagram on the r.h.s. of equation in \Fig{fig:A2-equ} arises from the non-classical two-ghost--two-gluon vertex. Remarkably, the factors in front of each diagram are in agreement with those in the perturbation theory. It, however, should be emphasized once more that although these diagrams are very similar with the formalism of perturbation theory, they are essentially nonperturbative, since both propagators and vertices in these diagrams are the full ones.

In order to let readers be familiar with the computation in fRG, we present some details about the flow equations of the gluon and ghost self-energies in Yang-Mills theory at finite temperature in \app{app:gluGhoSelfEner}.

\subsection{Dynamical hadronization}
\label{subsec:dynhadron}

In \sec{subsec:QCDaction} we have mentioned that the mesonic fields in \Eq{eq:QCDaction} are not added by hands. On the contrary, these composite degrees of freedom are dynamically generated from fundamental ones with the evolution of the RG scale from the ultraviolet (UV) toward infrared (IR) limit. This is done via a technique called the dynamical hadronization, which was proposed in \cite{Gies:2001nw, Gies:2002hq}, and subsequently the formalism was further developed in \cite{Pawlowski:2005xe}. Notably, recently the explicit chiral symmetry breaking and its role within the dynamical hadronization have been investigated in detail in \cite{Fu:2019hdw}, and a flow of dynamical hadronization with manifest chiral symmetry is put forward therein. In this section, we follow \cite{Fu:2019hdw} and present the derivation of flow equation of the dynamical hadronization.

We denote the original or fundamental degrees of freedom in QCD as $\hat \varphi=(\hat A, \hat c, \hat{\bar{c}}, \hat q, \hat{\bar{q}})$ with the expected value $\varphi=\langle \hat \varphi \rangle$, whereas composite degrees of freedom are introduced via a RG scale $k$-dependent composite field $\hat \phi_k(\hat \varphi)$ \cite{Gies:2001nw, Gies:2002hq, Pawlowski:2005xe}, which is a function of the fundamental field $\hat \varphi$. Then the superfield reads
\begin{align}
  \Phi&=(\varphi, \phi_k)=(A, c, \bar{c}, q, \bar{q}, \phi_k)\,,
 \label{eq:Phi}
\end{align}
with
\begin{align}
  \phi_k&=\langle \hat \phi_k(\hat \varphi) \rangle\,.\label{eq:phik}
\end{align}
The generating functional in \Eq{eq:Zk} is modified a bit such that
\begin{align}
  &Z_k[J]=\exp\big(W_k[J]\big)\nonumber \\[2ex]
 =&\int (\mathcal{D} \hat \varphi) \exp\Big\{-S_{\mathrm{QCD}}[\hat\varphi]-\Delta S_k[\hat \varphi, \hat \phi_k]+J_{\varphi}\cdot \hat \varphi \nonumber \\[2ex]
 &+J_{\phi}\cdot \hat \phi_k \Big\}\,,\label{eq:Zk-2}
\end{align}
where the external source $J=(J_{\varphi}, J_{\phi})$ with $J_{\varphi}=(J_{A}, J_{c}, J_{\bar{c}}, J_{q}, J_{\bar{q}})$ is conjugated to the field $\Phi=(\varphi, \phi_k)$, distinguished with different labels of indices, i.e.,
\begin{align}
  J\cdot \hat \Phi&=J^a \hat \Phi_a, \quad J_{\varphi}\cdot \hat \varphi=J_{\varphi}^{\alpha} \hat \varphi_{\alpha}, \quad J_{\phi}\cdot \hat \phi_k=J_{\phi}^i \hat \phi_{k,i}\,.\label{eq:JaPhia}
\end{align}
The regulator of bilinear fields in \Eq{eq:Zk-2} reads
\begin{align}
  \Delta S_k[\hat \varphi, \hat \phi_k]&=\Delta S_k[\hat \Phi]=\frac{1}{2}\hat\Phi_a R^{ab}_k\hat\Phi_b\,.
 \label{}
\end{align}
The effective action is obtained via a Legendre transformation to the Schwinger function as shown in \Eq{eq:Gamk}. Note that the term of explicit chiral symmetry breaking in the effective action, e.g., $-c_\sigma \sigma$ or $-c_{\sigma_s} \sigma_s/\sqrt{2}$ in \Eq{eq:QCDaction}, does not contribute to the flow of effective action, and thus the effective action can always be decomposed into that in the case of chiral limit plus an explicit chiral symmetry breaking term. We follow \cite{Fu:2019hdw} and separate the explicit breaking out, such that
\begin{align}
  \Gamma_{k}[\Phi]&=\bar\Gamma_{k}[\Phi]-c_\sigma \sigma\,,  \label{eq:GamSymm}
\end{align}
where we have concentrated on the case of $N_f=2$, that can be easily extended to include the strange quark. $\bar\Gamma_{k}$ in \Eq{eq:GamSymm} stands for the effective action without the explicit chiral symmetry breaking. From \Eq{eq:GamSymm}, one arrives at
\begin{align}
  \Gamma_{k}[\Phi]-J\cdot \Phi&=\bar\Gamma_{k}[\Phi]-c_\sigma \sigma-J\cdot \Phi\nonumber \\[2ex]
 &=\bar\Gamma_{k}[\Phi]- \bar J\cdot \Phi\,,  \label{eq:Gam-Jphi}
\end{align}
with
\begin{align}
  \bar J&=(J_\varphi, J_\sigma+c_\sigma, J_\pi)\,,  \label{}
\end{align}
i.e.,
\begin{align}
  \bar J_\sigma&=J_\sigma+c_\sigma\,, \quad \bar J_\varphi=J_\varphi\,, \quad \bar J_\pi=J_\pi\,. \label{}
\end{align}
\Eq{eq:Gam-Jphi} combined with \Eq{eq:Gamk} leaves us with the relation for Schwinger functions as follows
\begin{align}
 W_k[J]&=\bar W_k[\bar J]\,,  \label{}
\end{align}
where one has
\begin{align}
 \bar W_k[J]&=W_k[ J]\Big|_{c_\sigma \rightarrow 0}\,,  \label{}
\end{align}
that is, $\bar W_k$ denotes the Schwinger function in the absence of the explicit chiral symmetry breaking. Similar with \Eq{eq:WetterichEq}, one arrives at
\begin{align}
  \partial_t \bar \Gamma_{k}[\Phi]&=-\partial_t \bar W_k[\bar J]-\partial_t \Delta S_k[\Phi]\,, \label{eq:dtbarGam}
\end{align}
where $\bar W_k[\bar J]=\ln \bar Z_k[\bar J]$ can be obtained in \Eq{eq:Zk-2} with $\bar J$ in lieu of $J$ and a chiral symmetric $S_{\mathrm{QCD}}[\hat\varphi]$. Subsequently, one arrives at
\begin{align}
 &\partial_t \bar W_k[\bar J]\nonumber \\[2ex]
 =&\frac{1}{\bar Z_k[\bar J]}\int (\mathcal{D} \hat \varphi) \big(-\partial_t\Delta S_k[\hat \Phi]+\bar J_{\phi}\cdot  \partial_t \hat \phi_k\big)\nonumber \\[2ex]
 &\times\exp\Big\{-S_{\mathrm{QCD}}[\hat\varphi]-\Delta S_k[\hat \Phi]+\bar J_{\varphi}\cdot \hat \varphi +\bar J_{\phi}\cdot \hat \phi_k \Big\}\,. \label{eq:dtbarWk}
\end{align}
It is straightforward to obtain
\begin{align}
  \partial_t\Delta S_k[\hat \Phi]&=\frac{1}{2}\hat\Phi_a \big(\partial_t R^{ab}_k \big)\hat\Phi_b+\hat \phi_{k,i}R^{ij}_{k,\phi} \big(\partial_t \hat \phi_{k,j} \big)\,. \label{eq:dtDelSkphik}
\end{align}
Inserting \Eq{eq:dtDelSkphik} into \Eq{eq:dtbarWk} and then to \Eq{eq:dtbarGam}, one is led to
\begin{align}
  \partial_t \bar \Gamma_{k}[\Phi]=&\frac{1}{2}G_{k,ab}\partial_t R^{ab}_k+\langle\hat \phi_{k,i}R^{ij}_{k,\phi} \big(\partial_t \hat \phi_{k,j} \big)\rangle\nonumber \\[2ex]
 &-\bar J_{\phi}^i \langle \partial_t \hat \phi_{k,i}\rangle\,, \label{}
\end{align}
where \Eq{eq:Gpropa} has been used. Employing 
\begin{align}
 \langle\hat \phi_{k,i}R^{ij}_{k,\phi} \big(\partial_t \hat \phi_{k,j} \big)\rangle&=\Big(G_{k,ia}\frac{\delta}{\delta \Phi_a}+\phi_{k,i}\Big)R^{ij}_{k,\phi} \langle\partial_t \hat \phi_{k,j} \rangle\,,  \label{}
\end{align}
and
\begin{align}
 \bar J_{\phi}^i&=\frac{\delta \big(\bar\Gamma_{k}[\Phi]+\Delta S_k[\Phi]\big)}{\delta \phi_{k,i}}\,,  \label{}
\end{align}
one arrives at 
\begin{align}
 \partial_t \bar \Gamma_{k}[\Phi]=&\frac{1}{2}G_{k,ab}\partial_t R^{ab}_k+G_{k,ia}\Big(\frac{\delta}{\delta \Phi_a}\langle\partial_t \hat \phi_{k,j} \rangle\Big) R^{ij}_{k,\phi}\nonumber \\[2ex]
 &-\langle \partial_t \hat \phi_{k,i}\rangle\frac{\delta \bar\Gamma_{k}[\Phi]}{\delta \phi_{k,i}} \,.  \label{eq:dtbarGam2}
\end{align}
Given the relation in \Eq{eq:Gam-Jphi} and the fact that $c_\sigma$ is independent of the RG scale $k$, we finally obtain the flow equation of effective action with dynamical hadronization, to wit,
\begin{align} 
&\partial_{t}\Gamma_{k}[\Phi]\nonumber\\[2ex]
=&\frac{1}{2}\mathrm{STr} \big(G_k[\Phi]\,\partial_t R_k\big)+\mathrm{Tr}\bigg(G_{\phi\Phi_a}[\Phi] \frac{\delta
     \langle \partial_t \hat \phi_k\rangle }{\delta \Phi_a} \,R_\phi\bigg)\nonumber\\[2ex]
&-\int \langle   \partial_t \hat \phi_{k,i}\rangle\, \left( \frac{\delta \Gamma_{k}[\Phi]}{\delta \phi_i}+ c_\sigma \delta_{i\, \sigma}\right)\,, \label{eq:FlowQCD}
\end{align}
where some summations for the indices $\{a\}$ and $\{i\}$ as shown in \Eq{eq:JaPhia} have been replaced with the super trace and trace, respectively; the integral over the spacetime coordinate is recovered for the last term on the r.h.s. of \Eq{eq:FlowQCD}. The propagator $G_{k,ia}$ in \Eq{eq:dtbarGam2} is relabeled with $G_{k,\phi_i \Phi_a}$ that has a clearer physical meaning. One can see that in comparison to \Eq{eq:WetterichEq}, there are two additional terms, i.e., the last two on the r.h.s. of \Eq{eq:FlowQCD}, in the flow equation of the effective action. These additional terms arise from the RG scale dependent composite field in \Eq{eq:phik}, and they can be employed to implement the Hubbard-Stratonovich transformation for every value of the RG scale, which eventually transfers the degrees of freedom from quarks to bound states.

	
\section{Low energy effective field theories}
\label{sec:LEFT}

Prior to discussing properties of the QCD matter at finite temperature and densities in \sec{sec:QCD}, in this section we would like to apply the formalism of fRG in \sec{sec:FRGformal} to low energy effective field theories (LEFTs) firstly. We adopt two commonly used formalisms of LEFTs, i.e., the purely fermionic Nambu--Jona-Lasinio (NJL) model in \sec{subsec:NJL} and the quark-meson (QM) model in \sec{subsec:QM}, respectively.

\subsection{Nambu--Jona-Lasinio model}
\label{subsec:NJL}

%
\begin{figure}[t]
\includegraphics[width=0.48\textwidth]{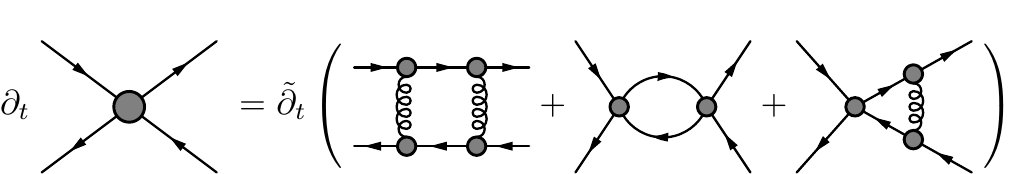}
\caption{Schematic representation of the flow equation for the four-quark coupling. Those on the r.h.s. of equation stand for three classes of diagrams contributing to the flow of four-quark interaction, that is, the two-gluon exchange, purely self-interacting four-quark coupling, and mixture of the quark-gluon and four-quark interactions, respectively. Here, diagrams of different channels of momenta are not distinguished and prefactors for each diagram are not shown.}\label{fig:Gam4-equ}
\end{figure}
%

%
\begin{figure*}[t]
\includegraphics[width=0.35\textwidth]{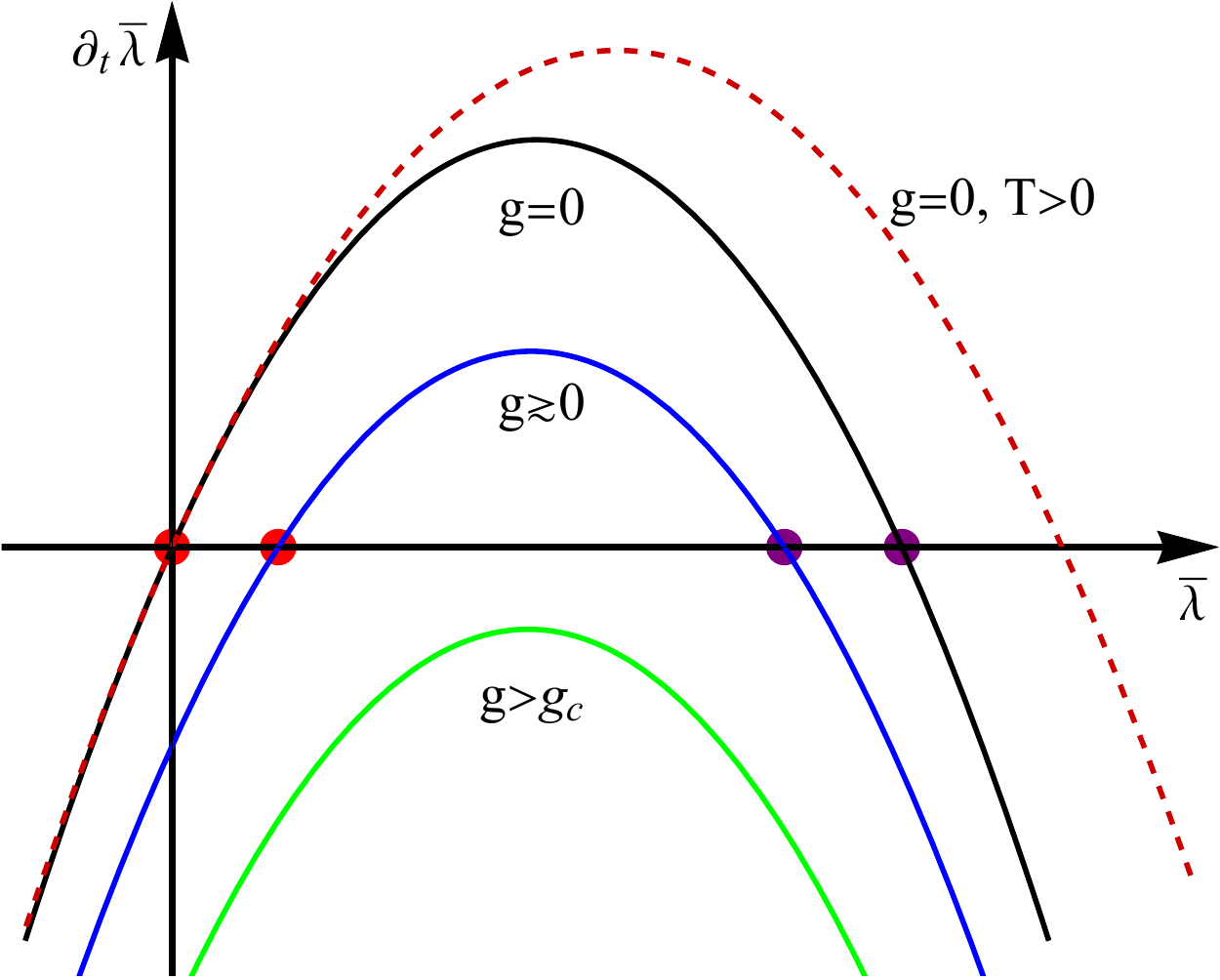}\hspace{2.5cm}
\includegraphics[width=0.35\textwidth]{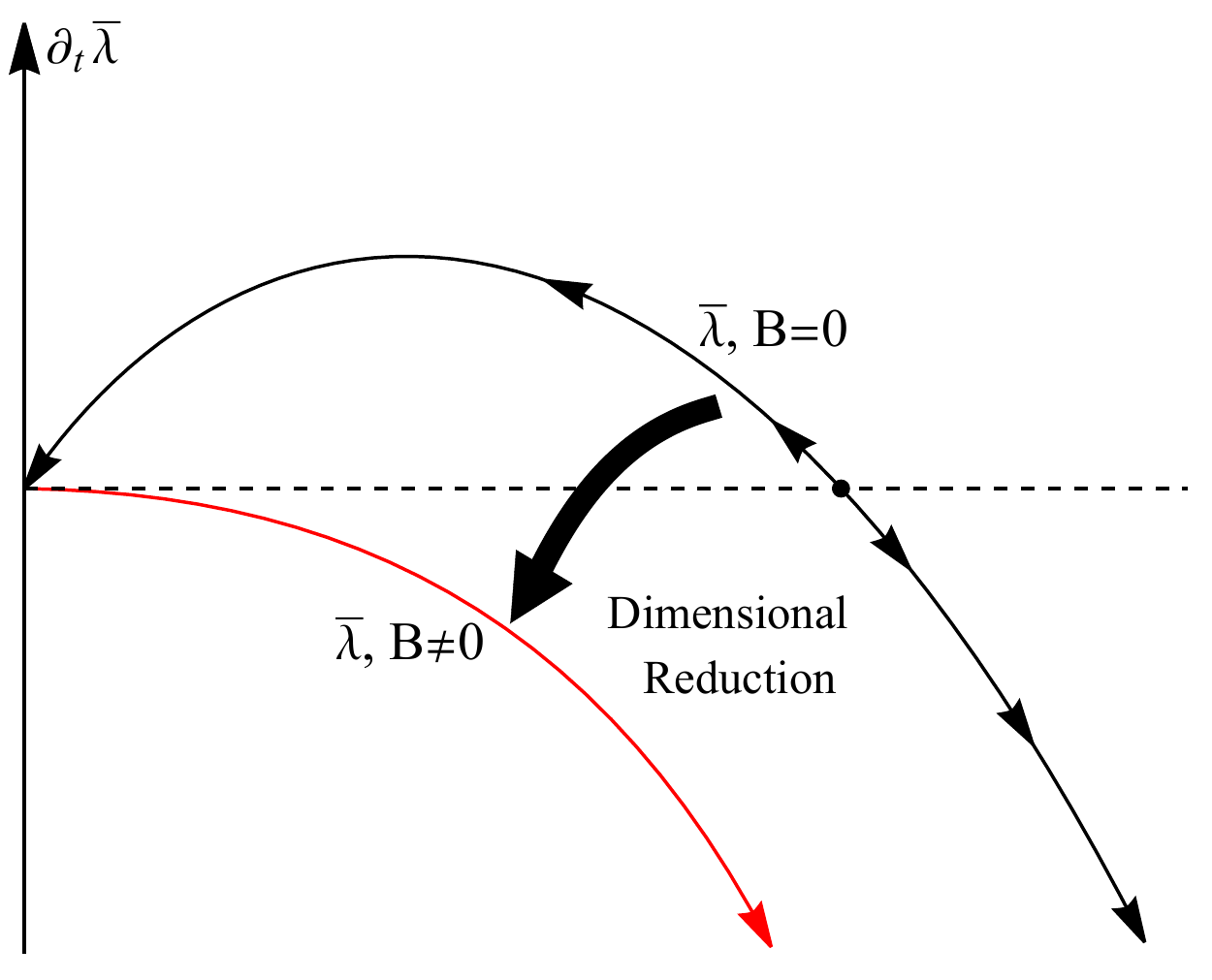}
\caption{Left panel: Sketch of a typical $\beta$ function for the dimensionless four-quark coupling $\bar \lambda\equiv k^2 \lambda$ in different cases, where $g$ is the gauge coupling and $T$ is the temperature. The plot is adopted from \cite{Braun:2006jd}.
Right panel: Comparison between the $\beta$ functions of $\bar \lambda$ with and without an external magnetic field. The inclusion of a magnetic field results in that the chiral symmetry is always broken due to the dimensional reduction. The plot is adopted from \cite{Fukushima:2012xw}.}\label{fig:FourFermiBetaFun}
\end{figure*}
%

One prominent feature characteristic to the nonperturbative QCD is the dynamical chiral symmetry breaking, which is regarded as being responsible for the origin of the $\sim 98\%$ mass of visible matter in the universe \cite{Nambu:1961tp, Nambu:1961fr, Marciano:1977su, Roberts:2021xnz}, in contradistinction to the $\sim 2\%$ electroweak mass. Within the fRG approach, the dynamical breaking or restoration of the chiral symmetry is well encoded in the four-fermion flows, that is illustrated briefly in what follows. For more details, see, e.g., \cite{Fukushima:2012xw, Aoki:2015mqa, Springer:2016cji, Fu:2017vvg, Aoki:2017rjl, Braun:2017srn, Braun:2018bik, Braun:2019aow, Leonhardt:2019fua, Braun:2021uua} and a related review \cite{Braun:2011pp}.

Using the method to derive the flow equation for a generic vertex as shown in \Eq{eq:dtVk}, one is able to obtain the flow equation of four-quark coupling, depicted schematically in \Fig{fig:Gam4-equ}. Here we refrain from going into the details of a realistic calculation, but rather try to infer behaviors of the four-quark flow connected to breaking or restoration of the chiral symmetry. It follows from \Fig{fig:Gam4-equ} that the $\beta$ function for the dimensionless four-quark coupling $\bar \lambda\equiv k^2 \lambda$ reads
\begin{align}
 \beta \equiv&\partial_t \bar \lambda=(d-2)\bar \lambda-a\bar \lambda^2-b\bar \lambda g^2-c g^4\,,  \label{eq:betaFunlam}
\end{align}
with the dimension of spacetime $d=4$ and the strong coupling $g$. Note that apart from the first term on the r.h.s. of \Eq{eq:betaFunlam} arising from the dimension of $\lambda$, the remaining three terms corresponds to the three classes of diagrams in \Fig{fig:FourFermiBetaFun} one by one, and their coefficients are denoted by $a$, $b$, $c$, respectively. Further computation indicates one has $a>0$ and $c>0$ \cite{Braun:2006jd}. 

In the left panel of \Fig{fig:FourFermiBetaFun}, a typical $\beta$ function is plotted as a function of the dimensionless four-quark coupling schematically in different cases. When the gauge coupling $g$ is vanishing and at zero temperature, there are two fixed points: One is the Gaussian IR fixed point $\bar \lambda=0$, the other the UV attractive fixed point at a nonvanishing $\bar \lambda$, and they are shown in the plot by red and purple dots, respectively. The position of the UV fixed point determines a critical value $\bar \lambda_c$, which is necessitated in order to break the chiral symmetry, since only when $\bar \lambda>\bar \lambda_c$, the four-quark coupling grows large and eventually diverges with the decreasing RG scale. When the temperature is turned on, the IR fixed point remains at the origin while the UV fixed point move towards right, as shown by the red dashed line. As a consequence, the broken chiral symmetry in the vacuum is restored at a finite  $T$, if one has a value of $\bar \lambda$ with $\bar \lambda_c(T=0)<\bar \lambda<\bar \lambda_c(T)$. When the strong coupling is nonzero, the parabola of $\beta$ function moves downwards globally as shown by the blue curve in the left panel of \Fig{fig:FourFermiBetaFun}. There is a critical value of the strong coupling, say $g_c$, once one has $g>g_c$, the whole curve of the beta function is below the line $\beta=0$, which implies that the chiral symmetry breaking is bound to occur, no matter how large the initial value of $\bar \lambda$ is.

The four-fermion flow is also well suited for an analysis of the chiral symmetry breaking in an external magnetic field. The inclusion of a magnetic field would modify the four-quark flows as well as the fixed-point structure \cite{Fukushima:2012xw, Fu:2017vvg}. The plot in the right panel of \Fig{fig:FourFermiBetaFun} is obtained in \cite{Fukushima:2012xw}, which demonstrates that in the case with a magnetic field the flow pattern of the four-quark coupling has been changed and the chiral symmetry is always broken. This is in fact due to the dimensional reduction under a finite external magnetic field. Moreover, it is found that once the in-medium effects of temperature and densities are implemented, long-range correlations are screened and the vanishing critical coupling shown by the red line in the right panel of \Fig{fig:FourFermiBetaFun} is not zero anymore  \cite{Fukushima:2012xw}.

Recently, a Fierz-complete four-fermion model is employed to investigate the phase structure at finite temperature and quark chemical potential, and it is found that the inclusion of four-quark channels other than the conventionally used scalar-pseudoscalar ones not only plays an important role in the phase diagram at large chemical potential, but also affects the dynamics at small chemical potential \cite{Braun:2017srn}. The related fixed-point structure is analyzed at finite chemical potential in the Fierz-complete NJL model, and it is found the dynamics is dominated by diquarks at large chemical potential \cite{Braun:2018bik}. Resorting to the fRG flows of four-quark interactions, one is able to observe the natural emergence of the NJL model at intermediate and low energy scales from fundamental quark-gluon interactions \cite{Braun:2019aow}. Equation of state of nuclear matter at supranuclear densities is also studied in the Fierz-complete setting \cite{Leonhardt:2019fua}, and a maximal speed of sound is found at supranuclear densities, that is related to the emergence of color superconductivity in the regime of high densities, i.e., the formation of a diquark gap, see also \cite{Braun:2021uua} for details. Moreover,  the $U_{\mathrm{A}}(1)$ symmetry and its effect on the chiral phase transition are investigated in the Fierz-complete basis \cite{Braun:2020mhk}.

Up to now, we have only discussed the chiral symmetry and its breaking by means of the four-fermion flows. As mentioned in the beginning of \sec{subsec:NJL}, a direct consequence of the dynamical chiral symmetry breaking is the production of mass, that is our central concern in the following. Here, we discuss a recent progress in understanding the quark mass generation and the emergence of bound states in terms of RG flows \cite{Fu:2022a}. Considering only the quark degrees of freedom in \Eq{eq:QCDaction} and extending the scalar-pseudoscalar four-quark interaction to a Fierz-complete set of $N_f=2$ flavors, denoted by $\mathcal{B}$ in the following, one arrives at a RG-scale dependent effective action, given by
\begin{align}
  &\Gamma_k[\Phi]\nonumber\\[2ex]
=&\int_{x,y}\Big[Z_{q,k}(x,y)\bar q(x) \gamma_\mu \partial_\mu q(y)+m_{q,k}(x,y)\bar q(x) q(y)\Big]\nonumber\\[2ex]
-& \int_{x,y,w,z}\sum_{\alpha\in \mathcal{B}}\lambda_{\alpha,k}(x,y,w,z){\mathcal{O}}^\alpha_{ijlm}{\bar{q}}_i(x)q_j(y){\bar{q}}_l(w)q_m(z)\,, \label{eq:NJLaction}
\end{align}
with $\Phi=(q,\bar q)$ and $\int_{x,\cdots}\equiv \int d^4 x \int\cdots$. Here ${\mathcal{O}}^\alpha_{ijlm}$ stands for the four-quark operator of channel $\alpha$, where the indices $i$, $j$, $l$, $m$ run over the Dirac, flavor and color ($N_c=3$) spaces, and the related coupling strength is given by $\lambda_{\alpha,k}$. In the same way, summation is assumed for repeated indices. In \app{app:Fierzbasis} ten independent Fierz-complete channels of four-quark interactions of $N_f=2$ flavors are presented.

The two-quark correlation function reads
\begin{align}
  \parbox[c]{0.15\textwidth}{\includegraphics[width=0.15\textwidth]{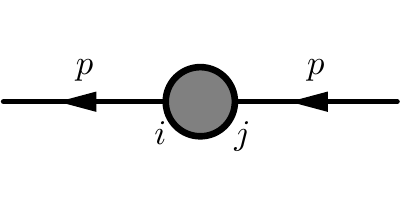}}&\equiv -\Gamma^{(2)\bar q q}_{k, ij}(p^\prime,p)\,,\label{}
\end{align}
with
\begin{align}
 &\Gamma^{(2)\bar q q}_{k, ij}(p^\prime,p)\equiv -\frac{\delta^2 \Gamma_k}{\delta \bar{q}_i(p^\prime)\delta q_j(p)}\bigg|_{\Phi=0}\nonumber\\[2ex]
=&\Big[Z_{q,k}(p)i (\gamma\cdot p)_{ij}+m_{q,k}(p)\delta_{ij}\Big](2\pi)^4\delta^4(p^\prime+p)\,.\label{eq:Gamma2bqq}
\end{align}
Then one arrives at the quark propagator with an IR regulator, viz.,
\begin{align}
  G^{q \bar q}_k(p,p^\prime)&= \big[\Gamma^{(2)\bar q q}_{k}+R^{\bar q q}_k\big]^{-1}\nonumber\\[2ex]
 &=G^{q}_k(p)(2\pi)^4\delta^4(p^\prime+p)\,,\label{}
\end{align}
with a fermionic regulator given by
\begin{align}
  R^{\bar q q}_k&=Z_{q,k}r_F(p^2/k^2)i \gamma\cdot p\,,\label{eq:Rbqq4d}
\end{align}
and
\begin{align}
  G^{q}_k(p)&=\frac{1}{Z_{q,k}(p)i \gamma\cdot p+Z_{q,k}r_F(p^2/k^2)i \gamma\cdot p+m_{q,k}(p)}\,.\label{}
\end{align}
Note that the fermionic regulator in \Eq{eq:Rbqq4d}, in comparison to the bosonic one in Eqs. (\ref{eq:regulatorExp}) and (\ref{eq:regulatorOpt}), is implemented in the vector channel rather than the scalar one, which guarantees that the chiral symmetry is not broken by the regulator. In the same way, one is allowed to make a choice for the specific formalism of the fermionic regulator, e.g., the optimized one,
\begin{align}
  r_{F, \mathrm{opt}}(x)&=\left(\frac{1}{\sqrt{x}}-1\right)\Theta(1-x)\,,\label{eq:rFopt}
\end{align}
or the exponential one in \Eq{eq:regulatorExp2},
\begin{align}
  r_{F}(x)&=r_{\mathrm{exp}, n}(x)= \frac{x^{n-1}}{e^{x^n}-1}\,,\label{}
\end{align}
and even the simplest exponential regulator as follows
\begin{align}
  r_{F, \mathrm{exp}}(x)&= \frac{1}{x}e^{-x}\,.\label{}
\end{align}

The four-quark correlation function reads
\begin{align}
  \parbox[c]{0.12\textwidth}{\includegraphics[width=0.12\textwidth]{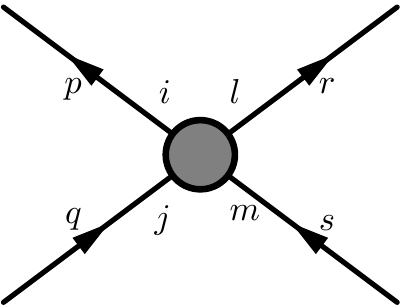}}&\equiv -\Gamma^{(4)\bar q q \bar q q}_{k, ijlm}(p,q,r,s)\,.\label{}
\end{align}
with
\begin{align}
 &\Gamma^{(4)\bar q q \bar q q}_{k, ijlm}(p,q,r,s)\equiv \frac{\delta^4 \Gamma_k}{\delta \bar{q}_i(p)\delta q_j(q)\delta \bar{q}_l(r)\delta q_m(s)}\bigg|_{\Phi=0}\nonumber\\[2ex]
=&2\sum_{\alpha\in \mathcal{B}}\Big(\lambda_{\alpha,k}^{S}(p,q,r,s)({\mathcal{O}}^\alpha_{ijlm}-{\mathcal{O}}^\alpha_{ljim})\nonumber\\[2ex]
&\hspace{0.5cm}+\lambda_{\alpha,k}^{A}(p,q,r,s)({\mathcal{O}}^\alpha_{ijlm}+{\mathcal{O}}^\alpha_{ljim})\Big)\nonumber\\[2ex]
  &\hspace{0.5cm}\times(2\pi)^4\delta^4(p+q+r+s)\,,\label{eq:Gam4}
\end{align}
where the symmetric and antisymmetric four-quark couplings read
\begin{align}
 \lambda_{\alpha,k}^{S}(p,q,r,s)\equiv&\big(\lambda_{\alpha,k}(p,q,r,s)+\lambda_{\alpha,k}(r,q,p,s)\big)/2\,,\label{eq:lambdaS}\\[2ex]
 \lambda_{\alpha,k}^{A}(p,q,r,s)\equiv&\big(\lambda_{\alpha,k}(p,q,r,s)-\lambda_{\alpha,k}(r,q,p,s)\big)/2\,.\label{eq:lambdaA}
\end{align}

%
\begin{figure}[t]
\includegraphics[width=0.48\textwidth]{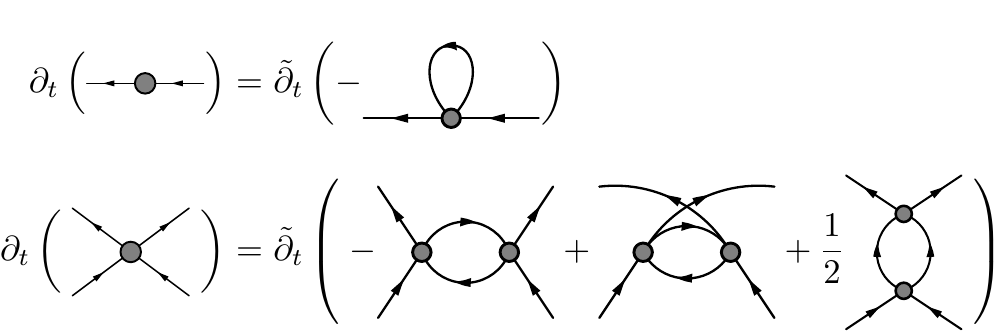}
\caption{Diagrammatic representation of the flow equations for the two- and four-quark correlation functions, where prefactors and signs for each diagram are also included. The three diagrams on the r.h.s. of the flow equation of four-quark coupling stand for the $t$-, $u$-, and $s$-channels, respectively.}\label{fig:Gam2Gam4-equ}
\end{figure}
%

Neglecting the diagrams including the quark-gluon interaction in \Fig{fig:Gam4-equ}  and showing explicitly different channels of momenta, one is able to obtain the flow equation of four-quark coupling within the purely fermionic effective action in \Eq{eq:NJLaction}, shown diagrammatically in the second line of \Fig{fig:Gam2Gam4-equ}. We also depict the flow equation of the two-quark correlation function, i.e., the quark self-energy, in \Fig{fig:Gam2Gam4-equ}.

\subsubsection{	Quark mass production}
\label{subsubsec:quarkmass}

%
\begin{figure}[t]
\includegraphics[width=0.48\textwidth]{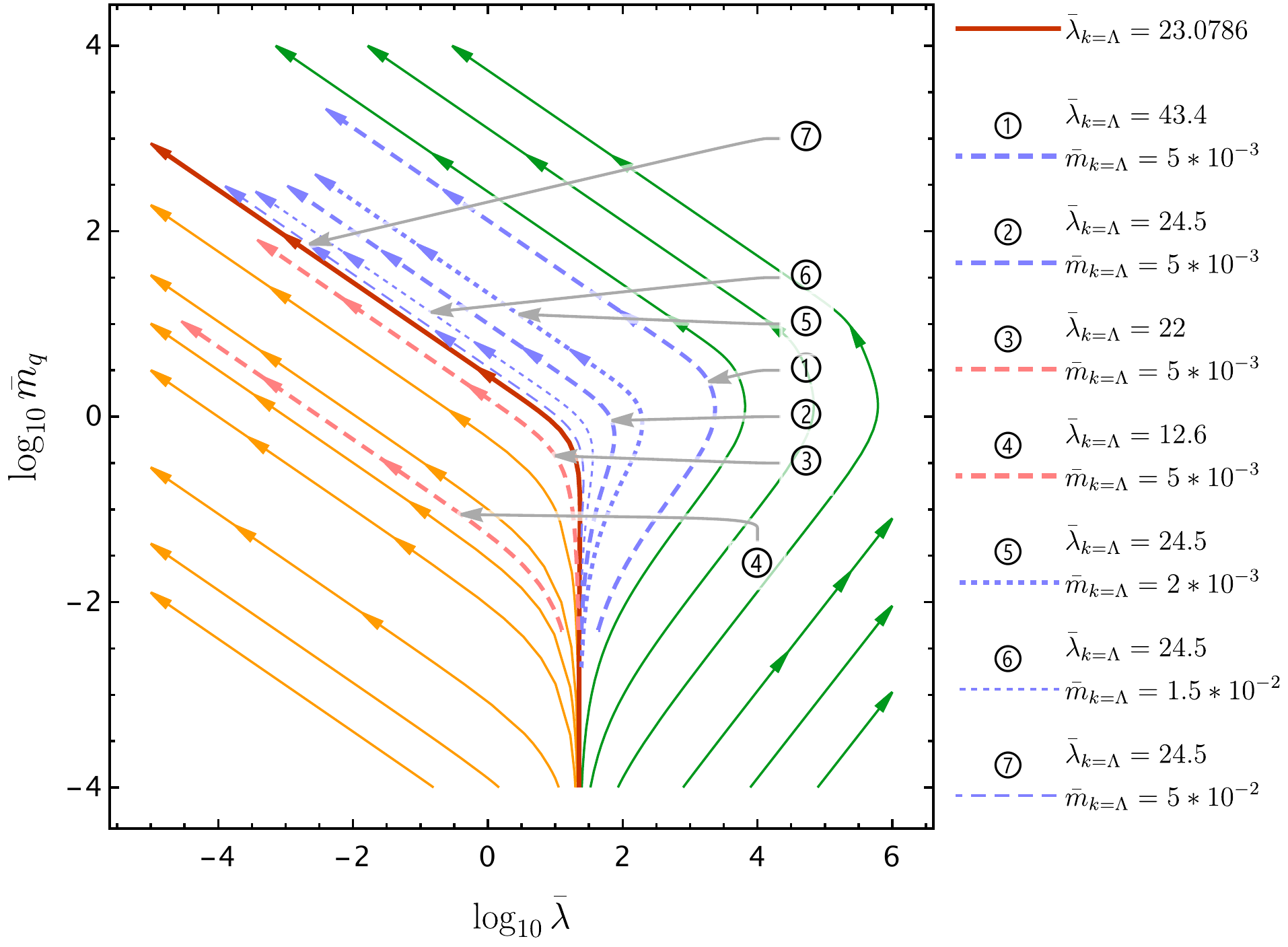}
\caption{RG flows in the plane spanned by the dimensionless quark mass $\bar m_{q}$ and the four-quark coupling of $\sigma$-$\pi$ channel $\bar \lambda_{\sigma\!-\!\pi}$, where logarithmic values of these two variables are used. Several evolutional trajectories with different initial conditions are labeled with numbers in circles, and arrows on trajectories point towards the IR direction. Plot is adopted from \cite{Fu:2022a}.}\label{fig:flowdiagram}
\end{figure}
%

%
\begin{figure*}[t]
\includegraphics[width=0.9\textwidth]{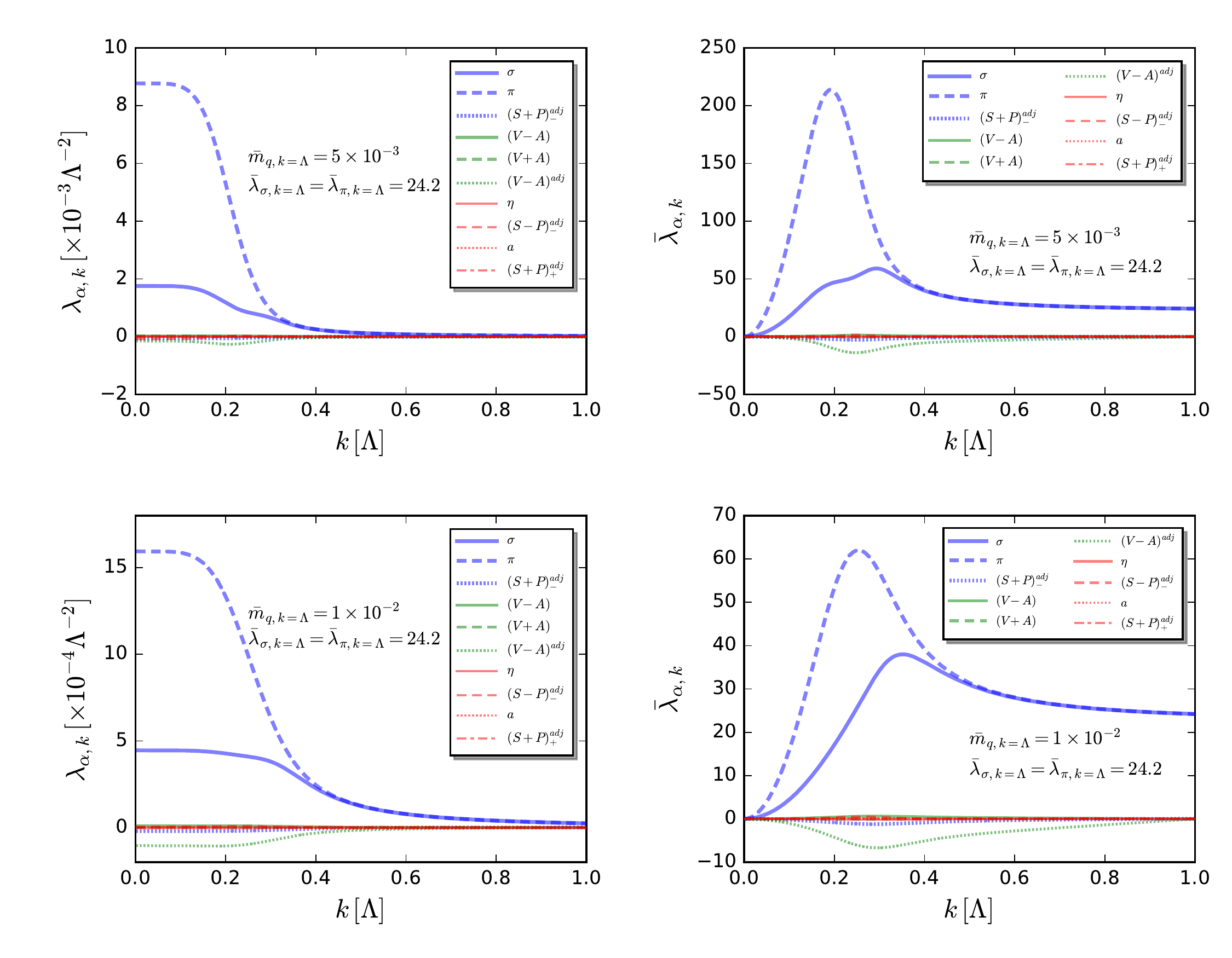}
\caption{Four-quark couplings $\lambda_{\alpha,k}$ and their dimensionless counterparts $\bar \lambda_{\alpha,k}=\lambda_{\alpha,k} k^2$ for the ten Fierz-complete channels as functions of the RG scale $k$. Initial values of couplings are chosen as follows, $\bar \lambda_{\pi,k=\Lambda}=\bar \lambda_{\sigma,k=\Lambda}=24.2$ and $\bar \lambda_{\alpha,k=\Lambda}=0$ ($\alpha \notin \{\sigma, \pi\}$) for other channels. $\bar m_{q,k=\Lambda}=5\times 10^{-3}$ (top panels) and $1\times 10^{-2}$ (bottom panels) are adopted for the initial value of quark mass. Plot is adopted from \cite{Fu:2022a}.}\label{fig:lam-k4d-10ch}
\end{figure*}
%

%
\begin{figure}[t]
\includegraphics[width=0.45\textwidth]{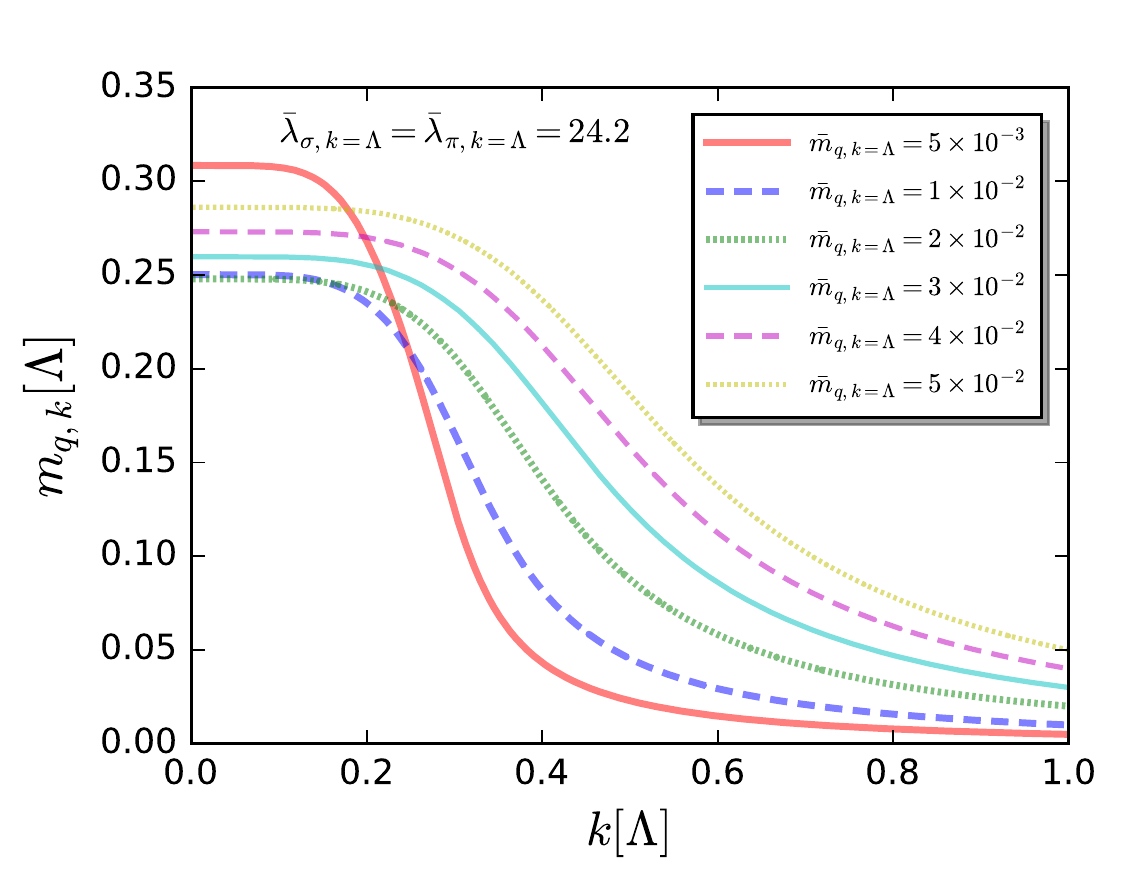}
\caption{Evolution of the quark mass with the RG scale obtained within the Fierz-complete basis of four-quark interactions. Initial values of couplings are chosen as follows, $\bar \lambda_{\pi,k=\Lambda}=\bar \lambda_{\sigma,k=\Lambda}=24.2$ and $\bar \lambda_{\alpha,k=\Lambda}=0$ ($\alpha \notin \{\sigma, \pi\}$) for other channels. Results obtained from several different initial values of $\bar m_{q,k=\Lambda}$ are compared. Plot is adopted from \cite{Fu:2022a}.}\label{fig:mq-k4d-10ch}
\end{figure}
%

Following \cite{Fu:2022a} we assume that the antisymmetric four-quark couplings $\lambda_{\alpha,k}^{A}$'s in \Eq{eq:lambdaA} are vanishing in order to simplify calculations. Then, Eqs. (\ref{eq:lambdaS}) and (\ref{eq:lambdaA}) leaves us with
\begin{align}
  \lambda_{\alpha,k}^{S}(p,q,r,s)&=\lambda_{\alpha,k}(p,q,r,s)=\lambda_{\alpha,k}(r,q,p,s)\,,\label{}
\end{align}
and its flow equation reads
{\allowdisplaybreaks
\begin{align}
  &\partial_t \lambda_{\alpha,k}(p_1,p_2,p_3,p_4)\nonumber\\[2ex]
  =&\sum_{\alpha^{\prime},\alpha^{\prime\prime}\in \mathcal{B}}\int \frac{d^4 q}{(2\pi)^4}\Big[\lambda_{\alpha^{\prime},k}(p_1,p_2,q+p_2-p_1,q)\nonumber\\[2ex]
  &\times\lambda_{\alpha^{\prime\prime},k}(p_3,p_4,q,q+p_2-p_1)\mathcal{F}^{t}_{\alpha^{\prime}\alpha^{\prime\prime},\alpha}\nonumber\\[2ex]
  &+\lambda_{\alpha^{\prime},k}(p_3,p_2,q+p_2-p_3,q)\nonumber\\[2ex]
  &\times\lambda_{\alpha^{\prime\prime},k}(p_1,p_4,q,q+p_2-p_3)\mathcal{F}^{u}_{\alpha^{\prime}\alpha^{\prime\prime},\alpha}\nonumber\\[2ex]
  &+\lambda_{\alpha^{\prime},k}(p_1,q,p_3,-q+p_1+p_3)\nonumber\\[2ex]
  &\times\lambda_{\alpha^{\prime\prime},k}(q,p_2,-q+p_1+p_3,p_4)\mathcal{F}^{s}_{\alpha^{\prime}\alpha^{\prime\prime},\alpha}\Big]\,,\label{eq:dtlambdaS}
\end{align}}
where the superscripts $t$, $u$, and $s$ of coefficient $\mathcal{F}$'s indicate that the relevant terms in \Eq{eq:dtlambdaS} arise from the corresponding loop diagrams in the flow of four-quark coupling in \Fig{fig:Gam2Gam4-equ}. The coefficient $\mathcal{F}$'s depend on the quark propagators and regulators, and see \cite{Fu:2022a} for their explicit expressions. The flow of quark mass is readily obtained by projecting the flow equation of the quark self-energy in the first line in \Fig{fig:Gam2Gam4-equ} onto the scalar channel, which reads
{\allowdisplaybreaks
\begin{align}
  &\partial_t m_{q,k}(p)\nonumber\\[2ex]
  =&\int \frac{d^4 q}{(2\pi)^4}\big(\tilde \partial_t \bar G^q_k(q)\big)m_{q,k}(q)\Big[\frac{3}{2}\lambda_{\pi,k}(p,p,q,q)\nonumber\\[2ex]
  &+\frac{23}{2}\lambda_{\sigma,k}(p,p,q,q)-\frac{3}{2}\lambda_{a,k}(p,p,q,q)\nonumber\\[2ex]
  &+\frac{1}{2}\lambda_{\eta,k}(p,p,q,q)+\frac{8}{3}\lambda_{(S+P)_{-}^{\mathrm{adj}},k}(p,p,q,q)\nonumber\\[2ex]
  &-\frac{16}{3}\lambda_{(S+P)_{+}^{\mathrm{adj}},k}(p,p,q,q)-4\lambda_{(V+A),k}(p,p,q,q)\Big]\,,\label{eq:dtmq}
\end{align}}
with
\begin{align}
 \tilde \partial_t \bar G^q_k(q)&=-2\big(\bar G^q_k(q)\big)^2 Z_{q,k}^{R}(q)q^2 \partial_t R_{F,k}(q)\,.\label{eq:dtbarG}\end{align}
Here one has $Z_{q,k}^{R}(q)=Z_{q,k}(q)+R_{F,k}(q)$ with $R_{F,k}(q)=Z_{q,k}r_F(q^2/k^2)$, and 
\begin{align}
 \bar G^q_k(q)&=\frac{1}{\big(Z_{q,k}^{R}(q)\big)^2 q^2+m_{q,k}^2(q)}\,.\label{}
\end{align}
For the moment, we assume $Z_{q,k}=1$ and use the truncation as follows
\begin{align}
  \lambda_{\alpha,k}&=\lambda_{\alpha,k}(p_i=0)\,,\qquad (i=1,\cdots,4)\,,\label{eq:lamp0}\\[2ex]
  m_{q,k}&=m_{q,k}(p=0)\,,\label{eq:mqp0}
\end{align}
that is, neglecting the momentum dependence of the four-quark coupling and quark mass. The dimensionless four-quark coupling $\bar \lambda_{\alpha,k}=\lambda_{\alpha,k} k^2$ and quark mass $\bar m_{q,k}=m_{q,k}/k$ are also very useful in the following.

In order to focus on the mechanism of quark mass production, we make a further approximation as follows
\begin{align}
  \lambda_{\alpha,k}&=0\,,\qquad (\alpha \notin \{\sigma, \pi\})\,,\label{}\\[2ex]
  \lambda_{\sigma\!-\!\pi,k}&\equiv \lambda_{\sigma,k}=\lambda_{\pi,k} \,,\label{}
\end{align}
i.e., only keeping the scalar-pseudoscalar $\sigma$ and $\pi$ channel. Then the flow equations in Eqs. (\ref{eq:dtlambdaS}) and (\ref{eq:dtmq}) are simplified as
\begin{align}
  \partial_{t} \bar \lambda_{\sigma-\pi}=&2\bar \lambda_{\sigma-\pi}+\frac{\bar \lambda_{\sigma-\pi}^2}{2 \pi^2}\int_0^\infty dx\, x^3 {r_F}^{\prime}(x)\nonumber\\[2ex]
  &\times \Big[-4\bar m_{q}^2+7x \big(1+r_F(x)\big)^2\Big]\nonumber\\[2ex]
  &\times\frac{1+r_F(x)}{\Big[\big(1+r_F(x)\big)^2 x+\bar m_{q}^2\Big]^3}\,,\label{eq:dtlambar}
\end{align}
and
\begin{align}
  \partial_{t}\bar m_{q}=&-\bar m_{q}+\bar m_{q} \bar \lambda_{\sigma-\pi}\frac{13}{4 \pi^2}\int_0^\infty dx\, x^3 {r_F}^{\prime}(x)\nonumber\\[2ex]
  &\times\frac{1+r_F(x)}{\Big[\big(1+r_F(x)\big)^2 x+\bar m_{q}^2\Big]^2}\,,\label{eq:dtmqbar}
\end{align}
respectively. Here we have used the dimensionless variables, which entails that the RG scale $k$-dependence for these equations is removed. RG flows of $ \bar \lambda_{\sigma-\pi}$ and $\bar m_{q}$ in Eqs. (\ref{eq:dtlambar}) and (\ref{eq:dtmqbar}) are depicted in \Fig{fig:flowdiagram}.

The plane in \Fig{fig:flowdiagram} is segmented into two parts by the red solid line. In the chiral limit the red line crosses the $x$-axis  at the UV fixed point, as shown in \Fig{fig:FourFermiBetaFun}, and the critical value here is $\bar \lambda_{\sigma-\pi}^{c}=23.08$. Interestingly, the UV critical point is extended to being an approximate critical line in the flow diagram in \Fig{fig:flowdiagram}, where the word ``approximate'' is used because the exact chiral symmetry is lost once the quark mass is nonzero. On the l.h.s. of the critical line, there is little dynamical chiral symmetry breaking and the quark mass is dominated by the current mass, while on the r.h.s. the dynamical chiral symmetry breaking plays a dominant role. Furthermore, it is observed that in the regime of dynamical chiral symmetry breaking, that is, on the r.h.s. of the red line, the dimensionless four-quark coupling increases firstly and then decreases. This is due to the competition between the flow equations of the quark self-energy and the four-quark coupling shown in \Fig{fig:Gam2Gam4-equ}, where the fish diagrams drive the dynamical breaking of chiral symmetry and result in the increase of the quark mass via the flow of quark self-energy, and in turn the increase of quark mass suppresses fluctuations of the four-quark flow. Finally, a balance is obtained with the decrease of RG scale, where the dimensional $m_{q, k}$ and $\lambda_{\sigma-\pi, k}$ are not dependent on $k$ any more. 

In \Fig{fig:lam-k4d-10ch} the evolution of four-quark couplings $\lambda_{\alpha,k}$ and their dimensionless counterparts $\bar \lambda_{\alpha,k}=\lambda_{\alpha,k} k^2$ with the RG scale for ten Fierz-complete channels is shown. The results are obtained from calculations, in which the initial values of couplings are chosen to be $\bar \lambda_{\pi,k=\Lambda}=\bar \lambda_{\sigma,k=\Lambda}=24.2$ and $\bar \lambda_{\alpha,k=\Lambda}=0$ ($\alpha \notin \{\sigma, \pi\}$) for other channels, i.e., the coupling strength of channels except the $\sigma$ and $\pi$ ones is assumed to be vanishing at the UV cutoff. In the meantime, results obtained from two initial values of quark mass, viz., $\bar m_{q,k=\Lambda}=5\times 10^{-3}$ (top panels) and $1\times 10^{-2}$ (bottom panels), are compared. In both cases one finds that the $\pi$ and $\sigma$ channels play a dominant role in the whole range of RG scale, and they are no longer degenerate with the scale evolving towards the IR limit. The strength of the $\pi$ channel is larger than that of the $\sigma$ channel. Moreover, one observes that the interaction strength of channels $(V-A)^{\mathrm{adj}}$ in \Eq{eq:VmAadjCh} and $(S+P)^{\mathrm{adj}}_{-}$ in \Eq{eq:SpPmadjCh} are also excited to some values, though they are significantly smaller than those of the $\pi$ and $\sigma$ channels. On the contrary, magnitudes of the remaining channels are very small, and they could be neglected in the whole range of RG scale. In \Fig{fig:mq-k4d-10ch} dependence of the quark mass on the RG scale is shown, and in the same way calculations are done with the Fierz-complete basis of four-quark interactions. Same initial values of the four-quark couplings as those in \Fig{fig:lam-k4d-10ch} are employed, and results obtained from different initial values of the quark mass are compared.

\subsubsection{	Natural emergence of bound states}
\label{subsubsec:boundStates}

%
\begin{figure}[t]
\includegraphics[width=0.45\textwidth]{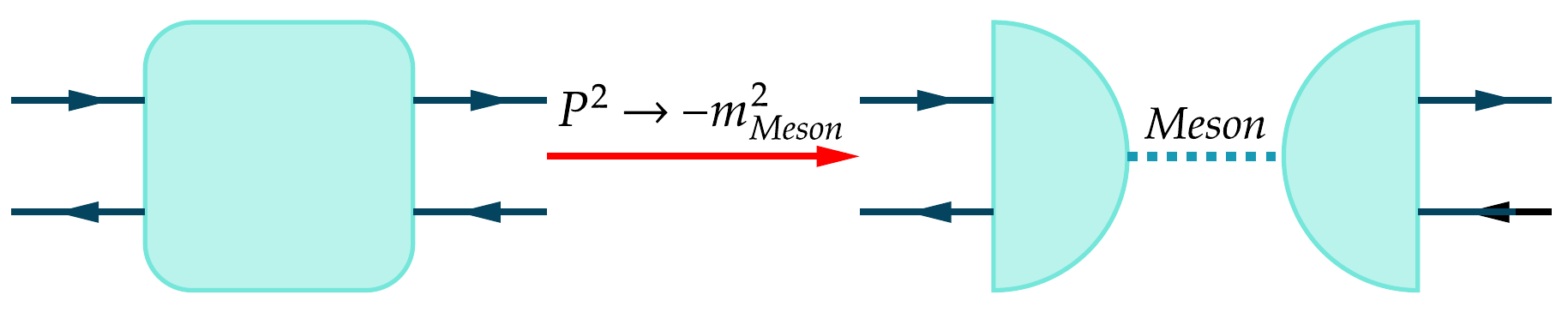}
\caption{Schematic diagram showing emergence of a resonance at the pole of relevant meson mass from the four-quark vertex, where the square and half-circles stand for the full four-quark and quark-meson vertices, respectively. The dashed line denotes the meson propagator.}\label{fig:resonance}
\end{figure}
%

%
\begin{figure*}[t]
\includegraphics[width=0.52\textwidth]{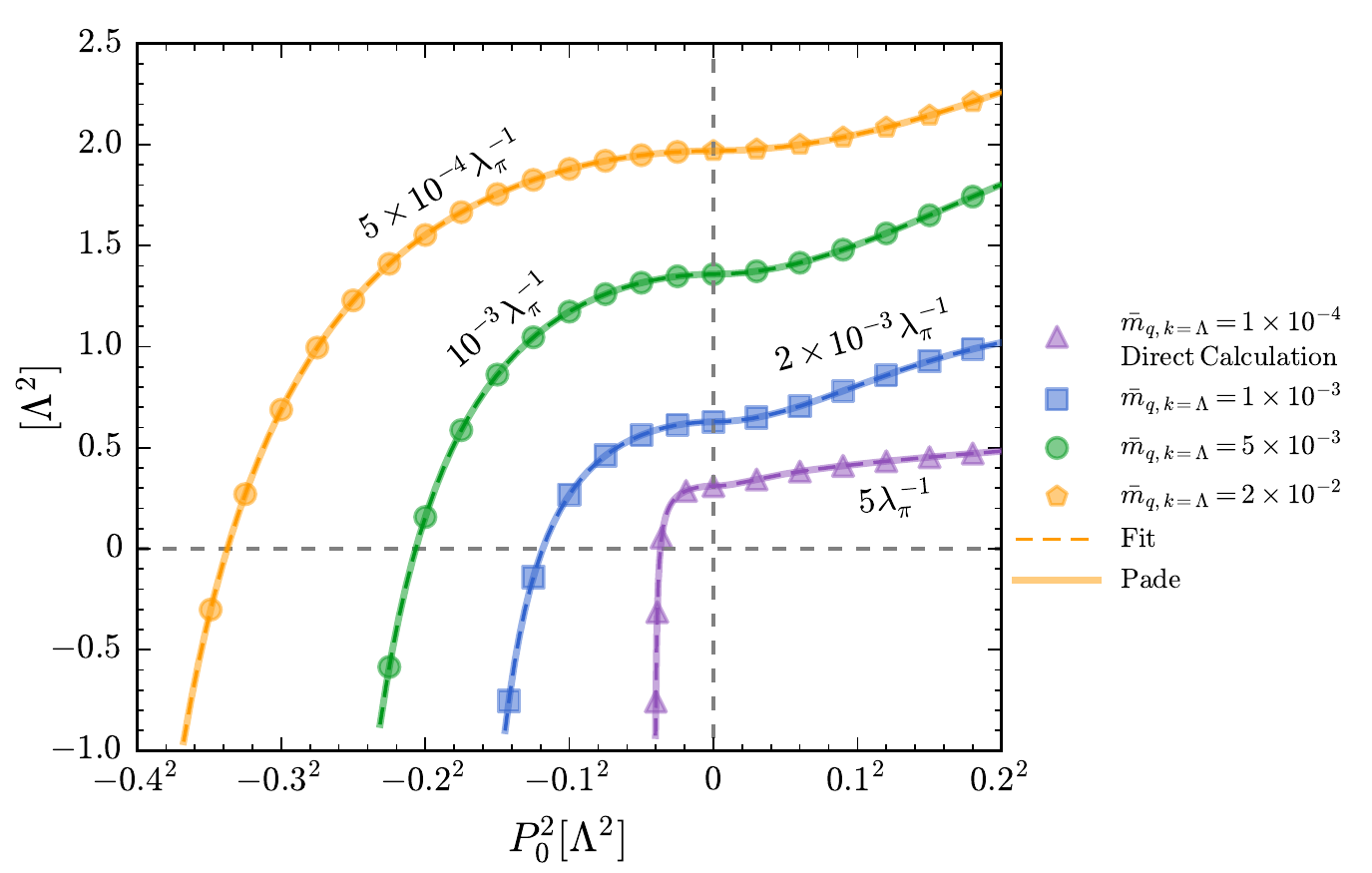}
\includegraphics[width=0.45\textwidth]{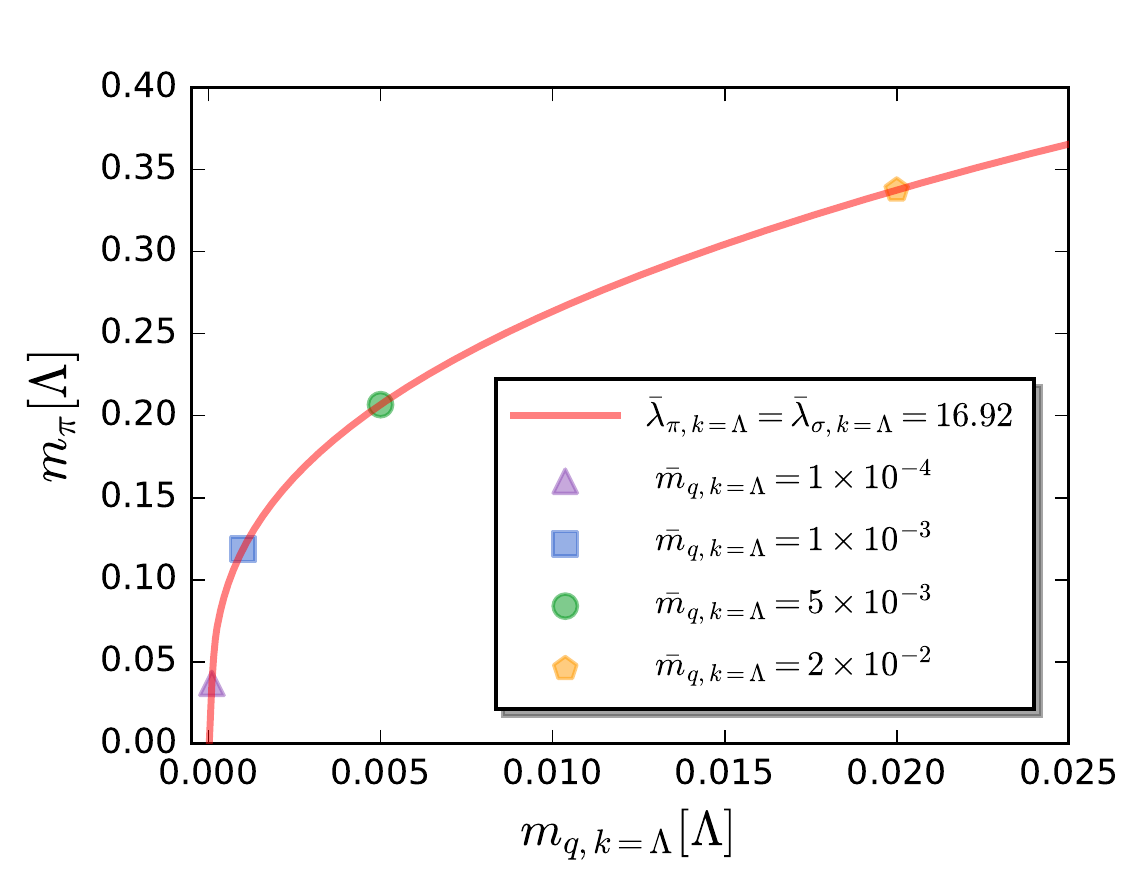}
\caption{Left panel: Dependence of $1/\lambda_{\pi,k=0}$, i.e., the inverse four-quark coupling of the pion channel, on the Mandelstam variable $t=P^2=P_0^2+\vec{P}^2$ with $\vec{P}=0$, where a $3d$ regulator is used. Data points stand for results computed directly from the analytic flow equation in \Eq{eq:dtlampi} both in the Euclidean ($P_0^2>0$) and Minkowski ($P_0^2<0$) regimes. The solid and dashed lines show results of analytic continuation from $P_0^2>0$ to $P_0^2<0$ based on the Pad\'e approximation and the fitting function in \Eq{eq:lampifit}, respectively. Several different values of the quark mass at the UV cutoff $\bar m_{q,k=\Lambda}$ are adopted, while the initial values of four-quark couplings are fixed with $\bar \lambda_{\pi,k=\Lambda}=\bar \lambda_{\sigma,k=\Lambda}=16.92$ and $\bar \lambda_{\alpha,k=\Lambda}=0$ ($\alpha \notin \{\sigma, \pi\}$). Right panel: Pion mass as a function of the quark mass at the UV cutoff $\bar m_{q,k=\Lambda}$, where the flow equation of four-quark coupling in \Eq{eq:dtlampi} is solved directly in the Minkowski spacetime with a $3d$ regulator. Here same initial values of four-quark couplings as those in the left panel are used. The several values of the pion mass extracted in the left panel are also shown on the curve in scattering points. Plots are adopted from \cite{Fu:2022a}.}\label{fig:invlam-p2-3d}
\end{figure*}
%

Properties of bound states of quarks or antiquarks, e.g., the pions and nucleons,  in principle can be inferred from the relevant four-point and six-point vertices of quarks, in some specific channels and regimes of momenta \cite{Eichmann:2016yit}. In \Fig{fig:resonance} a sketch map shows how this happens at the example of mesons. The square denotes a four-point vertex of quark and antiquark. If the total momentum of a quark and an antiquark, denoted by $P$ here, is in the Minkowski spacetime and in the vicinity of the on-shell pole mass of a meson in some channel, i.e., $P^2\sim-m_{\mathrm{meson}}^2$, the full four-quark vertex can be well described by a resonance of the meson, as shown on the r.h.s. of \Fig{fig:resonance}, where two quark-meson vertices are connected with the propagator of meson. Therefore, one has to calculate the full four-quark vertex or quark-meson vertex in some specific regime of momentum, which is usually realized by resuming a four-quark kernel to the order of infinity in the formalism of Bethe-Salpeter equations \cite{Salpeter:1951sz, Nakanishi:1969ph}. Note that the necessary resummation for the four-quark vertex is well included in the flow equation of four-quark couplings in \Eq{eq:dtlambdaS}, and it is, therefore, natural to expect that the RG flows are also well-suited for the description of bound states as same as the quark mass production in \sec{subsubsec:quarkmass}. Moreover, the advantage of RG flows is evident, that is, the self-consistency between 
the bound states encoded in the flow of four-quark vertices in \Eq{eq:dtlambdaS} and that of quark mass gap in \Eq{eq:dtmq} can be well guaranteed, once a truncation is made on the level of the effective action, such as that in \Eq{eq:NJLaction}.

In order to investigate the resonance behavior of four-quark vertices in \Eq{eq:dtlambdaS}, one has to go beyond the truncation in \Eq{eq:lamp0} and include appropriate momentum dependence for the four-quark vertices. The external momenta of couplings in \Eq{eq:dtlambdaS} are parameterized as follows
\begin{align}
  p_1=&p+\frac{P}{2}\,,\qquad p_2=p-\frac{P}{2}\,,\label{eq:p1p2}\\[2ex]
  p_3=&p^{\prime}-\frac{P}{2}\,,\qquad p_4=p^{\prime}+\frac{P}{2}\,.\label{eq:p3p4}
\end{align}
Then, one is left with the relevant Mandelstam variables given by
\begin{align}
  s=&(p_1+p_3)^2=(p+p^{\prime})^2\,,\label{eq:sMand}\\[2ex]
  t=&(p_1-p_2)^2=P^2\,,\label{eq:tMand}\\[2ex]
  u=&(p_1-p_4)^2=(p-p^{\prime})^2\,.\label{eq:uMand}
\end{align}

In the following, we focus on the $\pi$ meson and assume, that the total momentum of quark and antiquark in the $t$-channel is near the regime of on-shell pion mass, viz., one has the $t$-variable $P^2\sim -m_\pi^2$ in \Eq{eq:tMand}. Consequently, the four-quark coupling of the pion channel would be significantly larger than those of other channels, and its dependence on external momenta would be dominated by the $t$-variable. Thus, one is allowed to make the approximation as follows
\begin{align}
  \lambda_{\pi,k}(p_1,p_2,p_3,p_4)&\simeq \lambda_{\pi,k}(P^2) \,,\\[2ex]
  \lambda_{\alpha,k}(p_1,p_2,p_3,p_4)&\simeq \lambda_{\alpha,k}(0) \,,\qquad \alpha \ne \pi. \label{}
\end{align}
Furthermore, insofar as the four-quark vertices on the r.h.s. of the flow of coupling in the second line of \Fig{fig:Gam2Gam4-equ},  a simple analysis of relevant momenta for each vertex indicates, that the $t$-variable dependence is only required to be kept for the vertices in the diagram of $t$ channel, i.e., the first diagram. One is thus 
allowed to simplify the flow equation of $\lambda_{\pi,k}$ in \Eq{eq:dtlambdaS} as
\begin{align}
  \partial_t \lambda_{\pi,k}(P^2)&=\mathcal{C}_{k}(P^2)\lambda_{\pi,k}^2(P^2)+\mathcal{A}_{k}(t,u,s)\,,\label{eq:dtlampi}
\end{align}
with two coefficients given by
\begin{align}
  \mathcal{C}_{k}(P^2)&=\int \frac{d^4 q}{(2\pi)^4}\mathcal{F}^{t}_{\pi \pi,\pi}\,,\label{}
\end{align}
and
\begin{align}
  \mathcal{A}_{k}(t,u,s)&=\int \frac{d^4 q}{(2\pi)^4}\bigg\{\sum_{\alpha^{\prime},\alpha^{\prime\prime}\in \mathcal{B}}\Big[\lambda_{\alpha^{\prime},k}\lambda_{\alpha^{\prime\prime},k}\big(\mathcal{F}^{t}_{\alpha^{\prime}\alpha^{\prime\prime},\pi}\nonumber\\[2ex]
  &+\mathcal{F}^{u}_{\alpha^{\prime}\alpha^{\prime\prime},\pi}+\mathcal{F}^{s}_{\alpha^{\prime}\alpha^{\prime\prime},\pi}\big)\Big]-\lambda_{\pi,k}^2\mathcal{F}^{t}_{\pi \pi,\pi}\bigg\}\,.\label{eq:Ak}
\end{align}
Note that all the four-quark couplings in \Eq{eq:Ak} are momentum independent. If one adopts further $p=p^{\prime}=0$ in Eqs. (\ref{eq:p1p2}) and  (\ref{eq:p3p4}), two Mandelstam variables are vanishing, i.e., $s=u=0$, and one arrives at
\begin{align}
  \mathcal{A}_{k}(t,u,s)&\rightarrow\mathcal{A}_{k}(P^2)\,.\label{}
\end{align}

One is able to observe the natural emergence of a bound state arising from resummation of the four-quark vertex from \Eq{eq:dtlampi}, whose solution is readily obtained once the last term on the r.h.s. is ignored. One has 
\begin{align}
  \lambda_{\pi,k=0}(P^2)&=\frac{\lambda_{\pi,k=\Lambda}}{1-\lambda_{\pi,k=\Lambda}\int^0_{\Lambda}\mathcal{C}_{k}(P^2)\frac{dk}{k}}\,,\label{eq:lampik0}
\end{align}
with $\lambda_{\pi,k=\Lambda}$ the four-quark coupling strength of the pion channel at the UV cutoff, that is independent of external momenta. Evidently, when a value of $P^2$ is chosen appropriately, such that the denominator in \Eq{eq:lampik0} is vanishing, the four-quark coupling in the IR $\lambda_{\pi,k=0}$ is divergent. As a consequence, one can employ this condition to determine the pole mass of the bound state, i.e., the pion mass, which reads
\begin{align}
  1-\lambda_{\pi,k=\Lambda}\int^0_{\Lambda}\mathcal{C}_{k}(P^2=-m_{\pi}^2)\frac{dk}{k}=0\,.\label{eq:polemass}
\end{align}
When the coefficient $\mathcal{A}_{k}(P^2)$ in \Eq{eq:dtlampi} is taken into account, there is no analytic solution anymore. However, as would be shown in the following, direct numerical calculation of \Eq{eq:dtlampi} indicates that the qualitative behavior of pole displayed by \Eq{eq:lampik0} is not changed.

In the left panel of \Fig{fig:invlam-p2-3d} the inverse four-quark coupling of the pion channel in the IR limit, i.e., $1/\lambda_{\pi,k=0}$, is shown as a function of the Mandelstam variable $t=P^2=P_0^2+\vec{P}^2$ with $\vec{P}=0$. In order to solve the flow equation of four-quark coupling in \Eq{eq:dtlampi} directly in the Minkowski spacetime with $P^2<0$, one has employed the $3d$ regulator as follows
\begin{align}
  R^{\bar q q}_k&=Z_{q,k}r_F({\vec{p}}^2/k^2)i \vec{\gamma}\cdot \vec{p}\,,\label{eq:Rbqq3d}
\end{align}
in lieu of the $4d$ one in \Eq{eq:Rbqq4d}. Relevant results are shown in the plot by scattering points, where different symbols correspond to several different values of the quark mass at the UV cutoff scale $\bar m_{q,k=\Lambda}$. The $3d$ regulator allows one to compute the flow equation of the four-quark coupling not only in the Euclidean ($P^2>0$) but also Minkowski ($P^2<0$) regimes, viz., the part on the r.h.s. of the dashed vertical line in the plot and that on the l.h.s., respectively. As shown in Eqs. (\ref{eq:lampik0}) and (\ref{eq:polemass}), the pole mass of pion is determined from position of the zero point of $1/\lambda_{\pi,k=0}$, that is, the crossing point between the horizontal dashed line and those of $1/\lambda_{\pi,k=0}$ in the left panel of \Fig{fig:invlam-p2-3d}. Besides the direct calculations, results of analytic continuation from the Euclidean to Minkowski regimes are also presented in the left panel of \Fig{fig:invlam-p2-3d}. Two methods of analytic continuation are employed. One is to fit a simple function which reads
\begin{align}
  \lambda_{\pi,k=0}(P^2)&\approx \frac{a_0+a_2 P^2+a_4 P^4}{c_0+P^2+c_4 P^4}\,.\label{eq:lampifit}
\end{align}
Here the coefficients $a_0$, $a_2$, $a_4$, $c_0$, and $c_4$ are determined by fitting the numerical results of $P_0^2>0$ from \Eq{eq:dtlampi}, and then results of $P_0^2<0$ are predicted by \Eq{eq:lampifit}. The other is the Pad\'e approximation, where the simple function on the r.h.s. of \Eq{eq:lampifit} is replaced by a Pad\'e fraction, to wit,
\begin{align}
  \lambda_{\pi,k=0}[n,n](P^2)&\approx \lambda_{\pi,k=0}(P^2)\,,\qquad P^2>0\,.\label{eq:lamPade}
\end{align}
Here, a diagonal fraction with the same order of the polynomials $n$ in the numerator and denominator is used. Note that the simple function in \Eq{eq:lampifit} is in fact a Pad\'e fraction of order $n=2$. The order of Pad\'e fraction is varied with $n= 25 \sim 100$ in the calculations. In the left panel of \Fig{fig:invlam-p2-3d} the analytically continued results based on the two methods are in comparison to those of direct computation in the Minkowski region with $P^2<0$. Remarkably, it is found that both the analytically continued results are in excellent agreement with the data points obtained from \Eq{eq:dtlampi}. Moreover, in order to verify Goldstone theorem and the nature of Goldstone boson of the pion in the RG flow, one shows the extracted pion mass as a function of the quark mass at the UV cutoff in the right panel of \Fig{fig:invlam-p2-3d}. Evidently, the pion mass is found to decrease with the decreasing current quark mass, and it is exactly massless in the chiral limit.

%
\begin{figure}[t]
\includegraphics[width=0.5\textwidth]{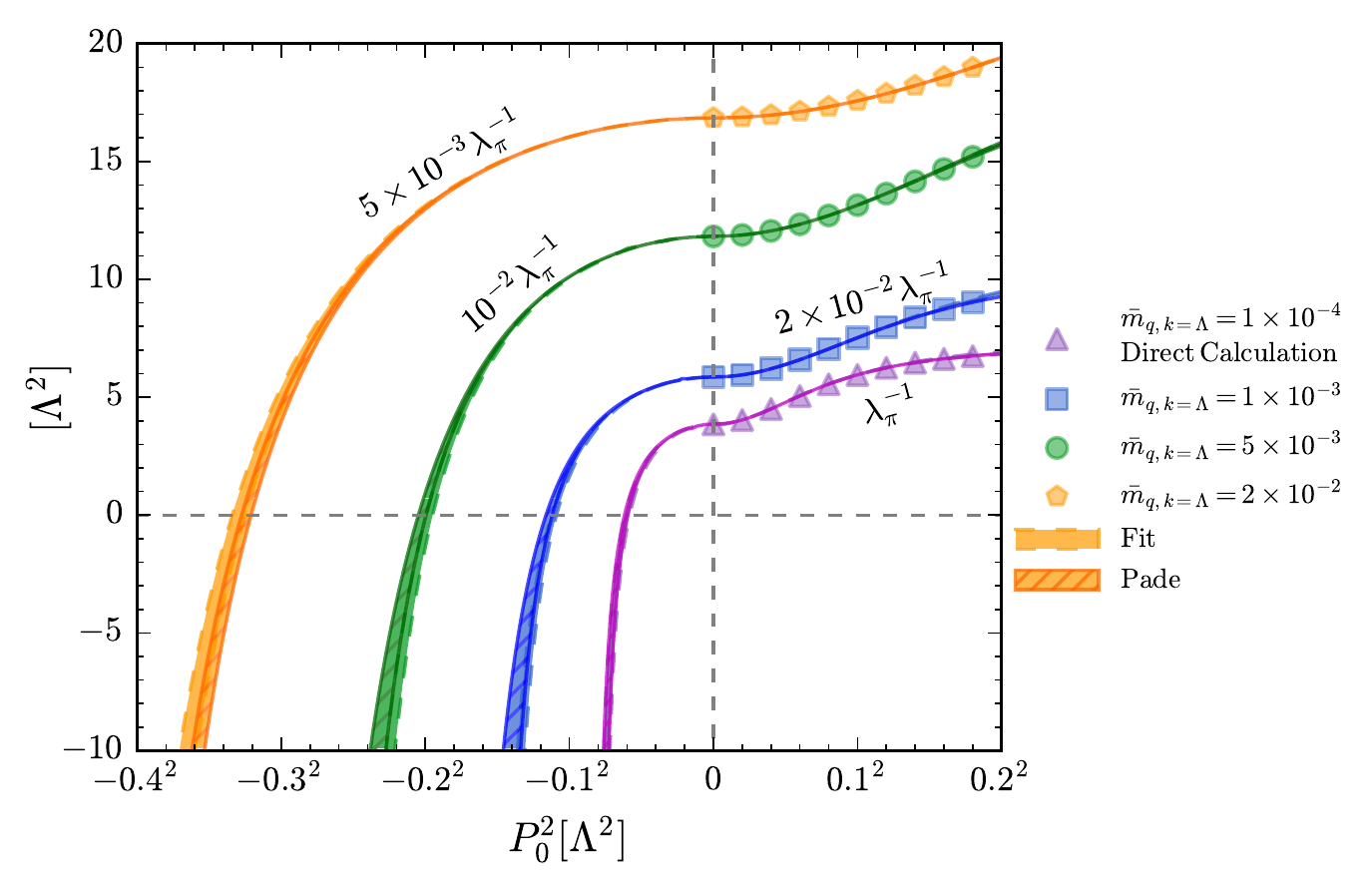}
\caption{Dependence of $1/\lambda_{\pi,k=0}$, i.e., the inverse four-quark coupling of the $\pi$ channel, on the Mandelstam variable $t=P^2=P_0^2+\vec{P}^2$ with $\vec{P}=0$, obtained with the $4d$ regulator. Data points stand for the results calculated in the flow equation in the Euclidean ($P_0^2>0$) region. The solid and dashed lines show analytically continued results from $P_0^2>0$ to $P_0^2<0$ based on the Pad\'e approximation and the fitting function in \Eq{eq:lampifit}, respectively. Several different values of the quark mass at the UV cutoff $\bar m_{q,k=\Lambda}$ are adopted, while the initial values of four-quark couplings are fixed with $\bar \lambda_{\pi,k=\Lambda}=\bar \lambda_{\sigma,k=\Lambda}=22.55$ and $\bar \lambda_{\alpha,k=\Lambda}=0$ ($\alpha \notin \{\sigma, \pi\}$). The plot is adopted from \cite{Fu:2022a}.}\label{fig:invlam-p2-4d}
\end{figure}
%

%
\begin{table}[t]
  \begin{center}
  \begin{tabular}{ccccc}
    \hline\hline & & & &  \\[-2ex]   
    $m_{q,k=\Lambda} [\Lambda]$ & $10^{-4}$ & $10^{-3}$ & $5\times10^{-3}$ & $2\times10^{-2}$ \\[1ex]
    \hline & & & &  \\[-2ex]
    Fit & 0.0597(15) & 0.1107(17) & 0.1971(23) & 0.330(4)  \\[1ex]
    Pad\'e & 0.0607(10) & 0.1140(22) & 0.2021(26) & 0.3242(33)  \\[1ex]    
    \hline\hline
  \end{tabular}
  \caption{Analytically continued results of the pole mass of pion (in unit of $\Lambda$) for different values of the quark mass at the UV cutoff $k=\Lambda$, where a $4d$ regulator is used. ``Fit'' and ``Pad\'e'' stand for the method used to do the analytic continuation, i.e., the fit of a simple rational function in \Eq{eq:lampifit} and the Pad\'e approximation in \Eq{eq:lamPade}, respectively. The initial values of four-quark couplings are fixed with $\bar \lambda_{\pi,k=\Lambda}=\bar \lambda_{\sigma,k=\Lambda}=22.55$ and $\bar \lambda_{\alpha,k=\Lambda}=0$ ($\alpha \notin \{\sigma, \pi\}$). The table is adopted from \cite{Fu:2022a}.} 
  \label{tab:mpi-mqUV-4d}
  \end{center}\vspace{-0.5cm}
\end{table}
%

In \Fig{fig:invlam-p2-4d} one shows the same physical quantities as the left panel of \Fig{fig:invlam-p2-3d}, but obtained with the $4d$ regulator. Quite apparently, in the $4d$ case direct calculation of the flow equation in \Eq{eq:dtlampi} is only accessible in the Euclidean region, as shown in the scattering points in \Fig{fig:invlam-p2-4d}. Thus one has to rely on the analytic continuation to infer the pole mass of pion, and the relevant results are also presented in the plot, where the same two methods of analytic continuation as the case of $3d$ regulator is used. It is found that, in comparison to the analytically continued results of the $3d$ regulator, those of the $4d$ case have significant larger errors, as shown by the bands in \Fig{fig:invlam-p2-4d}. The errors are inferred from varying the range of $P_0$, i.e., $(0, 0.1\Lambda)$, $(0, 0.2\Lambda)$, $(0, 0.3\Lambda)$, that is used to fix the analytically continued functions in \Eq{eq:lampifit} or \Eq{eq:lamPade}. In \Tab{tab:mpi-mqUV-4d} one shows the analytically continued values of the pole mass of pion.

\subsection{Quark-meson model}
\label{subsec:QM}

In this section we discuss another formalism of the low energy effective field theories, i.e., the quark-meson model, which in principle can be obtained from the NJL model via the bosonization with the Hubbard-Stratonovich transformation. The effective interactions between quarks and mesons in the rebosonized  QCD effective action in \Eq{eq:QCDaction}, and their natural emergence resulting from the RG evolution of fundamental degrees of freedom will be discussed in detail in \sec{subsec:LEFTemergQCD}. There is a wealth of studies in the literature relevant to the QM model within the fRG approach, see, e.g., \cite{Papp:1999he, Schaefer:2004en, Schaefer:2006ds,  Schaefer:2006sr, Skokov:2010wb, Herbst:2010rf, Skokov:2010wb, Skokov:2010uh, Braun:2011iz, Strodthoff:2011tz, Schaefer:2011ex, Aoki:2012mj, Kamikado:2012cp, Jiang:2012wm, Haas:2013qwp, Mitter:2013fxa, Herbst:2013ufa, Herbst:2013ail, Tripolt:2013jra, Tripolt:2013zfa, Grahl:2013pba, Morita:2013tu, Pawlowski:2014zaa, Helmboldt:2014iya, Fu:2015naa, Fu:2015amv, Khan:2015puu, Wang:2015bky, Mueller:2015fka, Eser:2015pka, Weyrich:2015hha, Fejos:2015xca, Jiang:2015xqz, Fu:2016tey, Rennecke:2016tkm, Jung:2016yxl, Jung:2016yxl, Fejos:2016hbp, Almasi:2016zqf, Posfay:2016ygf, Yokota:2016tip, Resch:2017vjs, Tripolt:2017zgc, Fejos:2017kpq, Yokota:2017uzu, Almasi:2017bhq, Zhang:2017icm, Fu:2018qsk, Fu:2018swz, Sun:2018ozp, Wen:2018nkn, Fejos:2018dyy, Braun:2018svj, Li:2019nzj, Wen:2019ruz, Yin:2019ebz, Li:2019nzj, Li:2019chs, Fu:2021oaw, Chen:2021iuo}.

For the moment, we only consider the degrees of freedom of quarks and mesons and their interactions via the Yukawa coupling in \Eq{eq:QCDaction}. The $N_f=2$ flavor QM effective action reads
\begin{align}
\Gamma_k=&\int_x \bigg\{Z_{q,k}\bar{q} \Big [\gamma_\mu \partial_\mu -\gamma_0(\hat\mu+igA_0) \Big ]q+\frac{1}{2}Z_{\phi,k}(\partial_\mu \phi)^2\nonumber\\[2ex]
&+h_k\bar{q}\big(T^0\sigma+i\gamma_5\bm{T}\cdot \bm{\pi}\big)q+V_k(\rho,A_0)-c\sigma \bigg\}\,,\label{eq:QMaction}
\end{align}
Note that explanations for most notations in \Eq{eq:QMaction} can be found in \sec{subsec:QCDaction} below \Eq{eq:QCDaction}. The effective potential in \Eq{eq:QMaction} can be decomposed into a sum of the contribution from the glue sector and that from the matter sector, which corresponds to the first and last two loops of the flow in \Fig{fig:QCD_equation}, respectively. Thus, one is led to
\begin{align}
V_k(\rho,A_0)=&V_{\mathrm{glue},k}(A_0)+V_{\mathrm{mat},k}(\rho,A_0)\,,\label{eq:Vtotal}
\end{align}
where the first term on the r.h.s. is the glue potential or the Polyakov loop potential, since it is usually reformulated by means of the Polyakov loop $L(A_0)$, and the latter can be calculated through the flow equation within the QM model. In the following, we still use $V_k(\rho)$ to stand for $V_{\mathrm{mat},k}$ for simplicity in a slight abuse of notation.

\subsubsection{	Flow of the effective potential}
\label{subsubsec:flowVQM}

The flow equation of the effective potential reads
\begin{align}
\partial_t V_{k}(\rho)=&\frac{k^4}{4\pi^2} \bigg [\big(N^2_f-1\big) l^{(B,4)}_{0}(\tilde{m}^{2}_{\pi,k},\eta_{\phi,k};T)\nonumber\\[2ex]
&+l^{(B,4)}_{0}(\tilde{m}^{2}_{\sigma,k},\eta_{\phi,k};T)\nonumber\\[2ex]
&-4N_c N_f l^{(F,4)}_{0}(\tilde{m}^{2}_{q,k},\eta_{q,k};T,\mu)\bigg]\,, \label{eq:flowV}
\end{align}
with the threshold functions given by
\begin{align}
  &l^{(B,4)}_{0}(\tilde{m}^{2}_{\phi,k},\eta_{\phi,k};T)\nonumber\\[2ex]
  =&\frac{2}{3}\left(1-\frac{\eta_{\phi,k}}{5}\right)\frac{1}{\sqrt{1+\tilde{m}^{2}_{\phi,k}}}\left(\frac{1}{2}+n_B(\tilde{m}^{2}_{\phi,k};T)\right)\,, \label{}
\end{align}
and
\begin{align}
  &l_0^{(F,4)}(\tilde{m}^{2}_{q,k},\eta_{q,k};T,\mu)\nonumber\\[2ex]
  =&\frac{2}{3}\left( 1-\frac{\eta_{q,k}}{4} \right)\frac{1}{2\sqrt{1+\tilde{m}^{2}_{q,k}}}\nonumber\\[2ex]
    &\hspace{0.cm}\times\Big(1-n_{F}(\tilde{m}^{2}_{q,k};T,\mu,L,\bar{L})-n_{F}(\tilde{m}^{2}_{q,k};T,-\mu,\bar{L},L)\Big)\,,\label{eq:l0F}
\end{align} 
where the bosonic distribution function reads
\begin{align}
  n_B(\tilde{m}^{2}_{\phi,k};T)&=\frac{1}{\exp\bigg\{\frac{k}{T}\sqrt{1+\tilde{m}^{2}_{\phi,k}}\bigg\}-1}\,,
\end{align} 
and the fermionic one
\begin{align}
n_F(\tilde{m}^{2}_{q,k};T,\mu,L,\bar{L})&=\frac{1+2\bar{L}e^{x/T}+Le^{2x/T}}{1+3\bar{L}e^{x/T}+3Le^{2x/T}+e^{3x/T}}\,,
\end{align} 
with $x=k\sqrt{1+\tilde{m}^{2}_{q,k}}-\mu$. Here, $L$ and $\bar{L}$ are the Polyakov loop and its conjugate. The dimensionless, renormalized meson and quark masses squared in \Eq{eq:flowV} read
\begin{align}
 \tilde{m}^{2}_{\pi,k}=&\frac{V'_{k}(\rho)}{k^2 Z_{\phi,k}}\,,\qquad \tilde{m}^{2}_{\sigma,k}=\frac{V'_{k}(\rho)+2\rho V''_{k}(\rho)}{k^2 Z_{\phi,k}}\,,\\[2ex]
\tilde{m}^{2}_{q,k}=&\frac{h^{2}_{k}\rho}{2k^2Z^{2}_{q,k}}\,.\label{eq:mqk}
\end{align}
The quark and meson anomalous dimensions in \Eq{eq:flowV} are given by
\begin{align}
\eta_{q,k}&=-\frac{\partial_t Z_{q,k}}{Z_{q,k}}\,,\qquad
\eta_{\phi,k}=-\frac{\partial_t Z_{\phi,k}}{Z_{\phi,k}}\,,
\end{align}
computation of which can be found in \sec{subsec:PropQCD}.

There are two classes of methods to solve the flow equation of the effective potential in \Eq{eq:flowV}. The methods of the first class capture local properties of the potential and are very convenient for numerical calculations, but fail to obtain global properties of the potential. A typical method in this class is the Taylor expansion of the effective potential. On the contrary, the methods in the other class are able to provide us with the global properties of the potential, but usually their numerical implements are relatively more difficult. The second class includes the discretization of the potential on a grid \cite{Schaefer:2004en}, the pseudo-spectral methods \cite{Boyd:2000, Borchardt:2015rxa, Borchardt:2016pif, Knorr:2020rpm}, e.g., the Chebyshev expansion of the potential \cite{Chen:2021iuo}, the discontinuous Galerkin method \cite{Grossi:2019urj, Grossi:2021ksl}, etc. In what follows we give a brief discussion about the Taylor expansion and the Chebyshev expansion of the potential.

The Taylor expansion of the effective potential reads
\begin{align}
V_{k}(\rho)&=\sum_{n=0}^{N_v}\frac{\lambda_{n,k}}{n!}(\rho-\kappa_k)^n\,, \label{eq:VTaylor}
\end{align}
with the expansion coefficients $\lambda_{n,k}$ and the expansion point $\kappa_k$ that might be $k$-dependent, where $N_v$ is the maximal expanding order in the calculations. Reformulating \Eq{eq:VTaylor} in terms of the renormalized variables, one is led to 
\begin{align}
\bar V_{k}(\bar \rho)&=\sum_{n=0}^{N_v}\frac{\bar\lambda_{n,k}}{n!}(\bar \rho-\bar \kappa_k)^n\,,\label{eq:VbarTaylor}
\end{align}
with
\begin{align}
\bar V_{k}(\bar \rho)&=V_{\mathrm{mat}, k}(\rho)\,,\qquad \; \bar \rho=Z_{\phi,k} \rho\,,\\[2ex]
    \bar \kappa_k&=Z_{\phi,k}\kappa_k\,, \qquad \bar \lambda_{n,k}=\frac{\lambda_{n,k}}{(Z_{\phi,k})^n}\,,\label{}
\end{align}
Substituting \Eq{eq:VbarTaylor} into the left hand side of \Eq{eq:flowV}, one arrives at
\begin{align}
&\partial^n_{\bar \rho}\left(\partial_t\big|_{\rho} \bar V_{k}(\bar \rho)\right)\Big|_{\bar \rho=\bar \kappa_k}\nonumber\\[2ex]
=&(\partial_t -n\eta_{\phi,k})\bar{\lambda}_{n,k}-(\partial_t \bar \kappa_k+\eta_{\phi,k}\bar \kappa_k)\bar \lambda_{n+1,k}\,.\label{eq:drhoV}
\end{align}
If the expansion point $\kappa_k$ is independent of $k$, i.e., $\partial_t \kappa_k=0$, which is usually called as the fixed point expansion, the expression in the second bracket on the r.h.s. of \Eq{eq:drhoV} is vanishing. One can see that in this case the expansion coefficients of different orders in \Eq{eq:VbarTaylor} decouples from each other in the flow equations, which results in an improved convergence for the Taylor expansion and a better numerical stability \cite{Pawlowski:2014zaa}. Another commonly used choice is the running expansion, whereof one expands the potential around the field on the EoS for every value of $k$, that is,
\begin{align}
\frac{\partial}{\partial \bar \rho}\Big(\bar V_{k}(\bar \rho)-\bar c_k
\bar \sigma \Big)\bigg \vert_{\bar\rho=\bar \kappa_k}&=0\,, \label{eq:Vstat}
\end{align}
with
\begin{align}
\bar \sigma&=Z_{\phi,k}^{1/2} \sigma\,,\qquad \bar c_k=Z_{\phi,k}^{-1/2} c\,.\label{}
\end{align}
Here $c$ is independent of $k$. Then one arrives at the flow of the expansion point, as follows
\begin{align}
\partial_t \bar \kappa_k=&\,-\frac{\bar c_k^2}{\bar{\lambda}_{1,k}^3+\bar c_k^2\bar{\lambda}_{2,k}}\Bigg[\partial_{\bar \rho}\left(\partial_t\big|_{\rho} \bar V_{k}(\bar \rho)\right)\Big|_{\bar \rho=\bar \kappa_k}\nonumber \\[1ex]
&\hspace{0.3cm}+\eta_{\phi,k}\left(\frac{\bar{\lambda}_{1,k}}{2}+\bar\kappa_k\bar{\lambda}_{2,k}\right)\Bigg]\,.\label{eq:flowkappa}
\end{align}
For more discussions about the fixed and running expansions, see, e.g., \cite{Pawlowski:2014zaa, Braun:2014ata, Rennecke:2015lur, Fu:2015naa, Rennecke:2016tkm, Yin:2019ebz}. In the same way, the field dependence of the Yukawa coupling, i.e., $h_k(\rho)$ can also be studied within a similar Taylor expansion \cite{Pawlowski:2014zaa, Yin:2019ebz}, which encodes higher-order quark-meson interactions. Moreover, effects of the thermal splitting for the mesonic wave function renormalization in the heat bath are also investigated in \cite{Yin:2019ebz}.

As for the Chebyshev expansion, the effective potential reads
\begin{align}
  \bar V_k(\bar \rho)&=\sum^{N_v}_{n=1} c_{n,k}T_n(\bar \rho)+\frac{1}{2}c_{0,k}\,,\label{eq:barVk}
\end{align}
with the Chebyshev polynomials $T_n(\bar \rho)$ and the respective expansion coefficients $c_{n,k}$. Then, one arrives at the flow of the effective potential as follows
\begin{align}
  \partial_t\big|_{\rho} \bar V_k(\bar \rho)=&\sum^{N_v}_{n=1} \Big(\partial_t c_{n,k}-d_{n,k}\eta_{\phi,k}(\bar \rho)\bar \rho \Big)T_n(\bar \rho)\nonumber\\[2ex]
&+\frac{1}{2}\Big(\partial_t c_{0,k}-d_{0,k}\eta_{\phi,k}(\bar \rho)\bar \rho \Big)\,,\label{}
\end{align}
where the field dependence of the meson wave function renormalization $Z_{\phi,k}(\rho)$ as well as the meson anomalous dimension $\eta_{\phi,k}(\bar \rho)$ is taken into account. Note that the coefficients $d_{n,k}$ are related to $c_{n,k}$ through a recursion relation as shown in \cite{Chen:2021iuo}. The flow equations of the coefficients in \Eq{eq:barVk} are given by 
\begin{align}
  \partial_t c_{m,k}=&\frac{2}{N+1}\sum^{N}_{i=0}\Big(\partial_t\big|_{\rho} \bar V_k(\bar \rho_i)\Big)T_m(\bar \rho_i) \nonumber\\[2ex]
&+\frac{2}{N+1}\sum^{N_v}_{n=1}\sum^{N}_{i=0}d_{n,k}T_m(\bar \rho_i) T_n(\bar \rho_i)\eta_{\phi,k}(\bar \rho_i)\bar \rho_i \nonumber\\[2ex]
&+\frac{1}{N+1}d_{0,k}\sum^{N}_{i=0}T_m(\bar \rho_i)\eta_{\phi,k}(\bar \rho_i)\bar \rho_i \,,\label{eq:flowCoeff}
\end{align}
where the summation for $i$ is performed on $N+1$ zeros of the polynomial $T_{N+1}(\bar \rho)$.

\subsubsection{Quark-meson model of $N_f=2+1$ flavors}
\label{subsubsec:QM-Nf2plus1}

The effective action for the $N_f=2+1$ flavor QM model, see, e.g., \cite{Mitter:2013fxa, Herbst:2013ufa, Rennecke:2016tkm, Fu:2018qsk, Wen:2018nkn, Wen:2019ruz}, reads
\begin{align}
  \Gamma_{k}=&\int_x \bigg\{ Z_{q,k}\bar{q} \Big [\gamma_\mu \partial_\mu -\gamma_0(\hat\mu+igA_0) \Big ]q+h_k\,\bar{q} \,\Sigma_5 q\nonumber\\[2ex]
&+Z_{\phi,k}\text{tr}(\bar D_\mu \Sigma \cdot \bar D_\mu\Sigma^\dagger)+V_{\text{\tiny{glue}}}(L, \bar L)+V_k(\rho_1,\rho_2)\nonumber\\[2ex]
&-c_A \xi-c_l \sigma_l-\frac{1}{\sqrt{2}}\, c_{s} \sigma_s\bigg\}\,,\label{eq:QM2p1action}
\end{align}
where the scalar and pseudoscalar mesonic fields of nonets are in the adjoint representation of the $U(N_f=3)$ group, i.e.,
\begin{align}
  \Sigma&=T^a(\sigma^a+i \pi^a)\,, \quad a=0,\,1,...,8\,. \label{eq:SigmaMeson}
\end{align}
with $T^{0}=1/\sqrt{2N_{f}}\mathbb{1}_{N_{f}\times N_{f}}$ and $T^a=\lambda^a/2$ for $a=1,...,8$. Here $\lambda^a$ are the Gell-Mann matrices. The covariant derivative on the mesonic fields is given by
\begin{align}
\bar D_\mu \Sigma&=\partial_\mu +\delta_{\mu 0}\big[\hat \mu ,\Sigma \big]\,.\label{}
\end{align}
where $\big[\hat \mu ,\Sigma \big]$ denotes the commutator between the matrix of chemical potentials and the meson matrix in \Eq{eq:SigmaMeson}. Note that although mesons do not carry the baryon chemical potential, they might have chemical potentials for the electric charge and the strangeness. The quark chemical potentials are related to those of conserved charges through the relations as follows
\begin{align}
\mu_u=&\frac{1}{3}\mu_B+\frac{2}{3}\mu_Q\,,\label{eq:muu}\\[2ex]
\mu_d=&\frac{1}{3}\mu_B-\frac{1}{3}\mu_Q\,,\label{eq:mud}\\[2ex]
\mu_s=&\frac{1}{3}\mu_B-\frac{1}{3}\mu_Q-\mu_S\,.\label{eq:mus}
\end{align}
For the Yukawa coupling, one has
\begin{align}
  \Sigma_5=&T^a(\sigma^a+i \gamma_5\pi^a)\,. \label{}
\end{align}
In contradistinction to the two-flavor case,  the effective potential in \Eq{eq:QM2p1action} is a function of two chiral invariants, to wit,
\begin{align}
  \rho_1&=\text{tr}(\Sigma \cdot \Sigma^\dagger)\,, \label{eq:rho1}\\[2ex]
  \rho_2&=\text{tr}\Big(\Sigma \cdot \Sigma^\dagger-\frac{1}{3}\,\rho_1\,\mathbb{1}_{3\times 3}\Big)^2 \,.\label{eq:rho2}
\end{align}
which are both invariant not only under $SU_{\text{A}}(3)$ but also $U_{\text{A}}(1)$. On the EoM these two invariants read
\begin{align}
   \rho_1\big|_{\mathrm{EoM}}&=\frac{1}{2}(\sigma_l^2+\sigma_s^2)\,,\label{eq:rho1EoM}\\[2ex]
   \rho_2\big|_{\mathrm{EoM}}&=\frac{1}{24}(\sigma_l^2-2\sigma_s^2)^2\,,\label{eq:rho2EoM}
\end{align}
where the light and strange fields are related to the zeroth and eighth components in \Eq{eq:SigmaMeson} through the relation as follows
\begin{align}
  \begin{pmatrix} \phi_l \\ \phi_s \end{pmatrix}
  &=\frac{1}{\sqrt{3}}\begin{pmatrix} 1 & \sqrt{2}\\ -\sqrt{2} & 1 \end{pmatrix}
  \begin{pmatrix}\phi_8\\ \phi_0 \end{pmatrix}\,. \label{}
\end{align}
The term of Kobayashi-Maskawa-'t Hooft determinant in \Eq{eq:QM2p1action} preserves $SU_{\text{A}}(3)$ while breaks $U_{\text{A}}(1)$ with
\begin{align}
  \xi&=\det(\Sigma)+\det(\Sigma^\dagger)\,, \label{}
\end{align}
and the breaking strength is described by the coefficient $c_A$. The light and strange quark masses read
\begin{align}
  m_{l,k}&=\frac{h_k}{2}\sigma_l, \qquad m_{s,k}=\frac{h_k}{\sqrt{2}} \sigma_s\,,\label{}
\end{align}
and the pion and kaon decay constants are given by 
\begin{align}
  f_\pi=\sigma_l, \quad f_K=\frac{\sigma_l+\sqrt{2}\,\sigma_s}{2}\,,\label{eq:fpiK}
\end{align}
where $\sigma_l$ and $\sigma_s$ are on their respective equations of motion. For more discussions about the flow equations in the $N_f=2+1$ flavor QM model, see the aforementioned references in this section.

\subsubsection{	Phase structure}
\label{subsubsec:phasediagramQM}

%
\begin{figure}[t]
\includegraphics[width=0.48\textwidth]{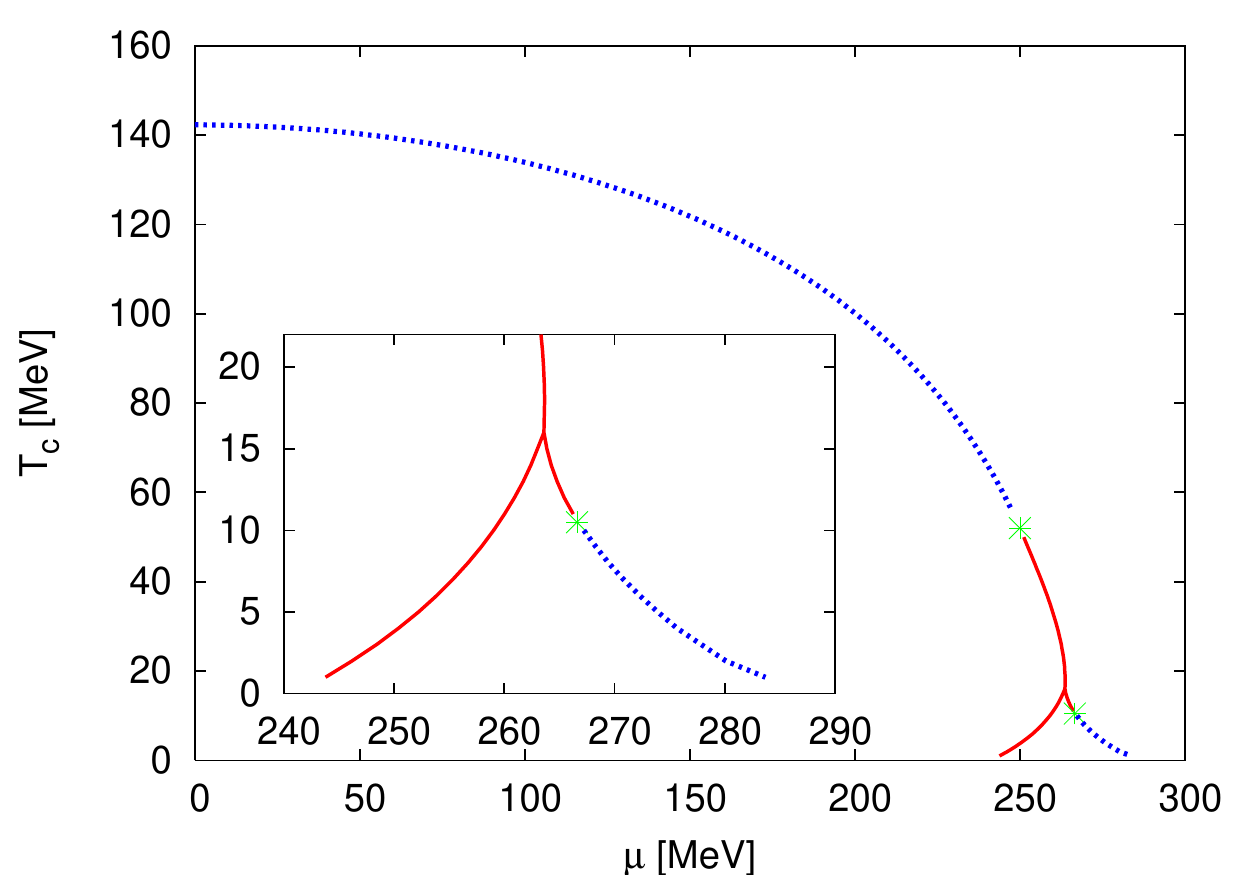}
\caption{Phase diagram of the $N_f=2$ flavor QM model in the chiral limit in the plane of the temperature and quark chemical potential obtained with discretization of the potential on a grid in \cite{Schaefer:2004en}. The plot is adopted from \cite{Schaefer:2004en}.}\label{fig:phase_diagram-Schaefer2004en}
\end{figure}
%

%
\begin{figure}[t]
\includegraphics[width=0.45\textwidth]{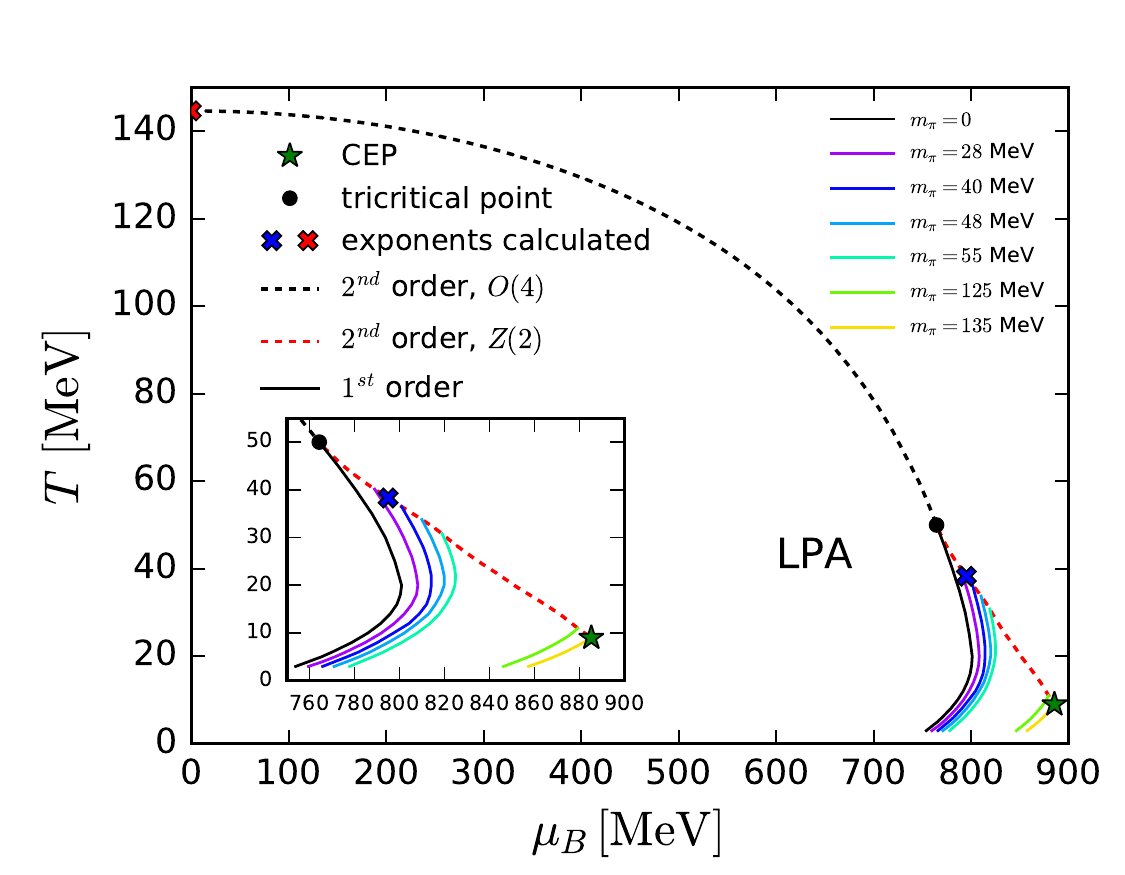}
\caption{Phase diagram of the $N_f=2$ flavor QM model in the $T-\mu_B$ plane obtained with the Chebyshev expansion of the potential in \cite{Chen:2021iuo}. The plot is adopted from \cite{Chen:2021iuo}.}\label{fig:phasedia-Chen2021iuo}
\end{figure}
%

%
\begin{figure}[t]
\includegraphics[width=0.48\textwidth]{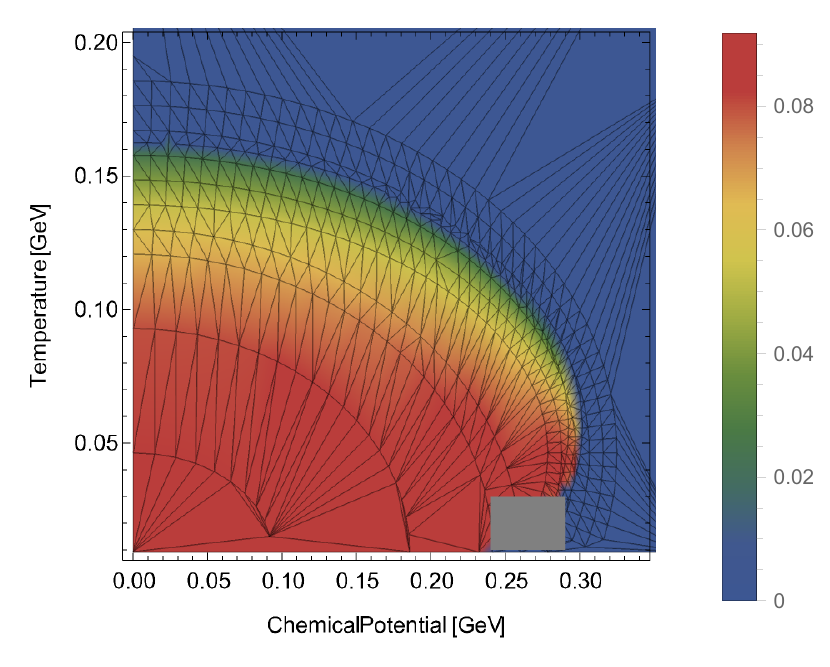}
\caption{Phase diagram of the $N_f=2$ flavor QM model in the $T-\mu_q$ plane obtained with a discontinuous Galerkin method in \cite{Grossi:2021ksl}, where the color stands for the value of the order parameter. The plot is adopted from \cite{Grossi:2021ksl}.}\label{fig:rescale-phasediagm-qm-Grossi2021ksl}
\end{figure}
%

%
\begin{figure*}[t]
\centering
\includegraphics[width=0.45\textwidth]{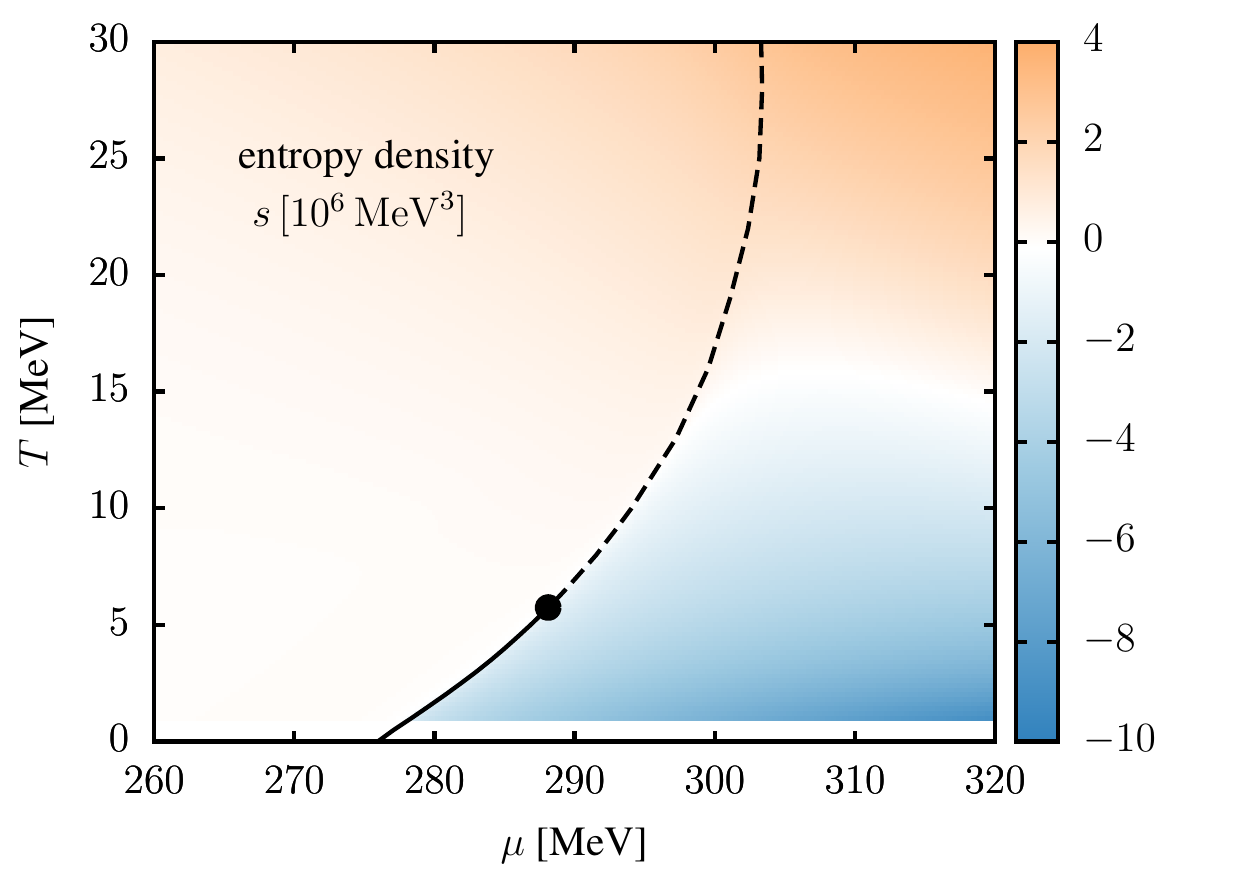}
\includegraphics[width=0.45\textwidth]{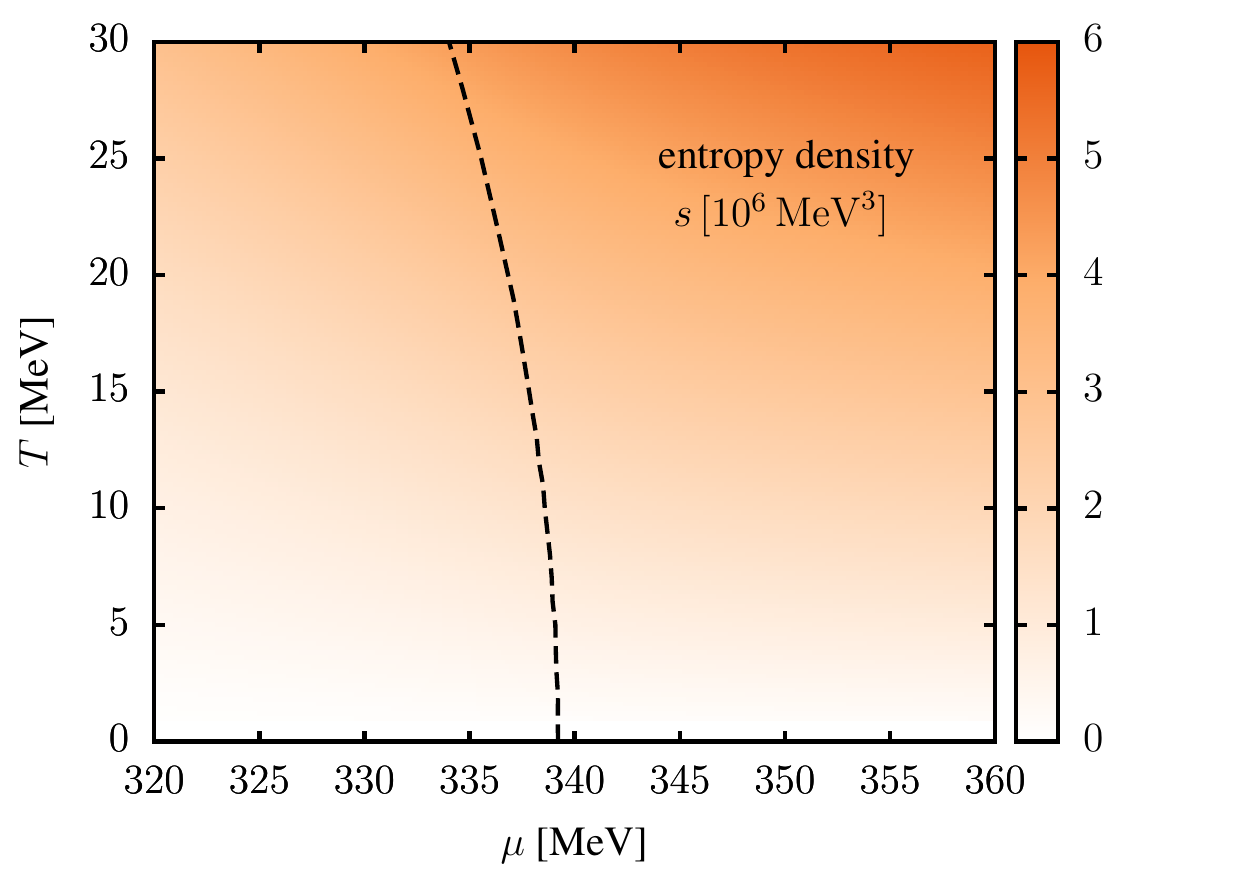}
\caption{Entropy density for the 3$d$ flat regulator (left panel) and 3$d$ mass-like regulator (right panel) close to the zero temperature chiral phase boundary obtained in the QM model in LPA \cite{Otto:2022}, where the solid and dashed lines denote the first-order phase transition and the crossover, respectively, and the black dot stands for the CEP. The blue region shows a negative entropy density. The plots are adopted from \cite{Otto:2022}.}\label{fig:phaseline-Otto2022}
\end{figure*}
%

%
\begin{figure}[t]
\includegraphics[width=0.45\textwidth]{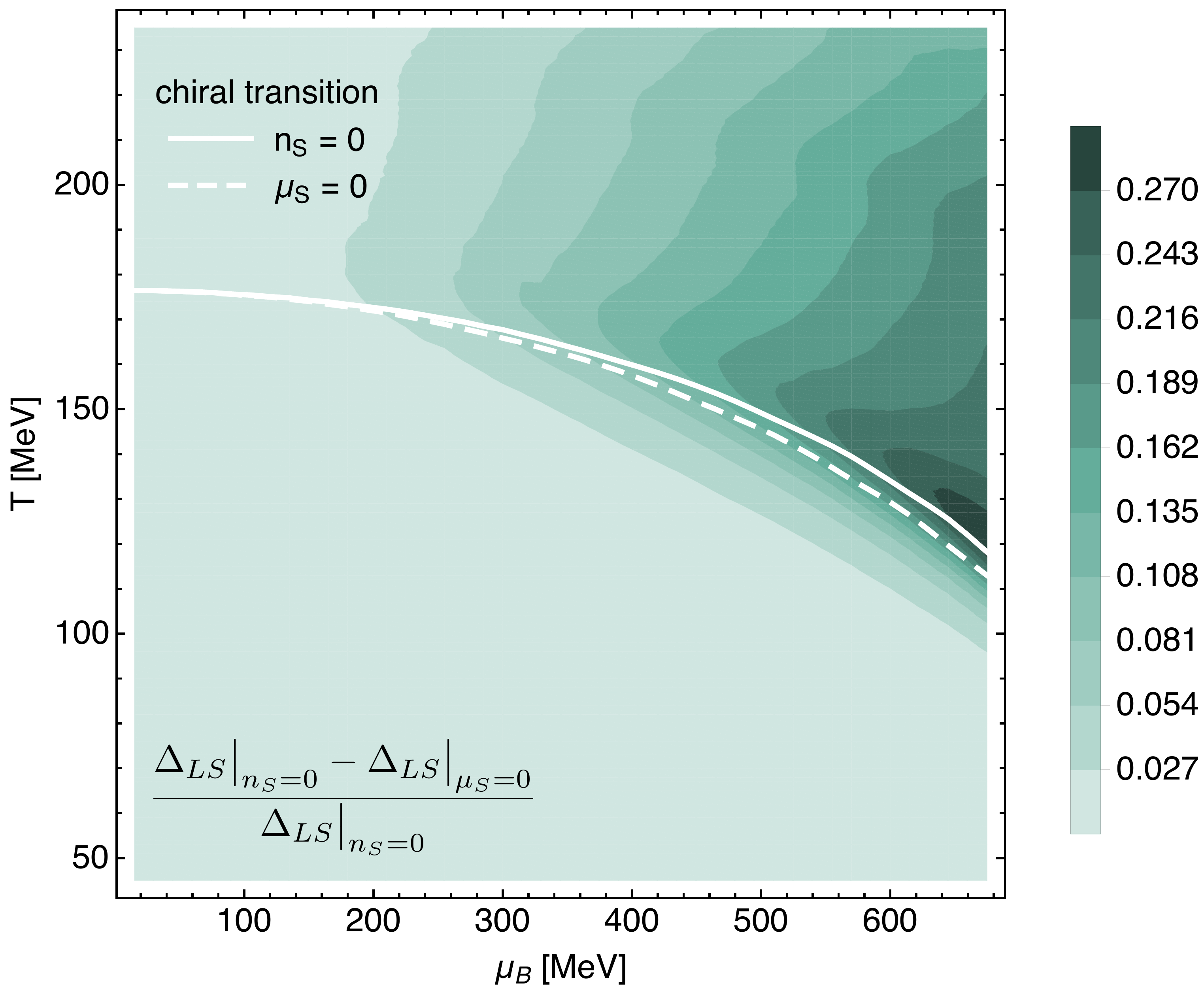}
\caption{Phase diagram of the $N_f=2+1$ flavor QM model in the $T-\mu_B$ plane obtained in \cite{Fu:2018qsk}, where the phase boundaries in the regime of crossover for the two cases: $\mu_S=0$ and the strangeness neutrality $n_S=0$ are compared. The color stands for the relative error of the reduced condensate (see \sec{subsec:condensateQCD}) for the two cases. The plot is adopted from \cite{Fu:2018qsk}.}\label{fig:PBerrCHItex-Fu2018qsk}
\end{figure}
%

In \Fig{fig:phase_diagram-Schaefer2004en} the phase diagram of the $N_f=2$ flavor QM model in the chiral limit obtained in \cite{Schaefer:2004en} is shown. In order to investigate the phase structure, especially in the regime of large chemical potential, the global information of the effective potential in \Eq{eq:flowV} is indispensable. Thus, the potential is discretized on a grid to resolve the flow equation, and see \cite{Schaefer:2004en} for more details. One can see that in the region of small chemical potentials, there is a second-order chiral phase transition denoted by the blue dashed line. With the decrease of the temperature, the second-order phase transition is changed into the first-order one at the tricritical point denoted by a green asterisk. The red solid lines stand for the first-order phase transition. Moreover, with the further decrease of the temperature, one observes that the first-order phase transition splits into two phase transition lines, one of which even evolves again into the second-order phase transition at high chemical potentials. Note that the first-order phase transition line backbends at high chemical potentials \cite{Tripolt:2017zgc}.

In \Fig{fig:phasedia-Chen2021iuo} the phase diagram of the $N_f=2$ flavor QM model obtained with the Chebyshev expansion in \cite{Chen:2021iuo} is shown. The black dashed line denotes the $O(4)$ second-order phase transition in the chiral limit, and the black dot stands for the tricritical point. The back solid line is the first-order phase transition in the chiral limit, where the splitting at large baryon chemical potential as shown in \Fig{fig:phase_diagram-Schaefer2004en} is not shown explicitly here, and only the left branch is depicted. The solid lines of different colors in \Fig{fig:phasedia-Chen2021iuo} correspond to different strengths of the explicit chiral symmetry breaking, i.e., different pion masses, and their end points, that is, the critical end points (CEP) comprise a $Z(2)$ second-order phase transition line, which is denoted by the red dashed line.

As we have mentioned above, the first-order phase transition line backbends at large chemical potential. It is conjectured that this artifact might arise from the defect that the discontinuity for the potential at high baryon chemical potential is not well dealt with within the grid or pseudo-spectral methods. Recently, a discontinuous Galerkin method has been developed to resolve the flow equation of the effective potential \cite{Grossi:2019urj, Grossi:2021ksl}, and the relevant result for the two-flavor phase diagram is shown in \Fig{fig:rescale-phasediagm-qm-Grossi2021ksl} \cite{Grossi:2021ksl}. Within this approach, the formation and propagation of shocks are allowed, and see \cite{Grossi:2019urj, Grossi:2021ksl} for a detailed discussion. 

However, very recently another research in \cite{Otto:2022} finds that the back-bending behavior of the chiral phase boundary at large chemical potential has another different origin. It is found that the appearance of the back-bending behavior depends on the employed regulator \cite{Otto:2022}. In \Fig{fig:phaseline-Otto2022} results of the chiral phase boundary and the entropy density obtained with two different regulators are compared. The left panel corresponds to the usually used 3$d$ flat or optimized regulator, cf. \Eq{eq:regulatorOpt2} and \Eq{eq:rFopt}, and the right is obtained with a 3$d$ mass-like regulator, see \cite{Otto:2022} for more details. The same truncation, i.e., the LPA approximation, is used in both calculations. It is observed that the usually used flat (Litim) regulator results in a back-bending of the chiral phase line at large chemical potential, which is also accompanied by a region of negative entropy density in the chirally symmetric phase, as shown by the blue area in the left panel of \Fig{fig:phaseline-Otto2022}. On the contrary, in the right panel one finds that the chiral phase line shows no back-bending behavior and the entropy density stays positive if a mass-like regulator is used. See \cite{Otto:2022} for a more detailed discussion.

In heavy-ion collisions since the incident nuclei do not carry the strangeness, the produced QCD matter after collisions is of strangeness neutrality, that is, the strangeness density $n_S$ is vanishing. Usually, the condition of strangeness neutrality requires that the chemical potential of strangeness $\mu_S$ in \Eq{eq:mus} is nonvanishing with $\mu_B\ne0$ \cite{Fu:2018qsk, Fu:2018swz}. The influence of the strangeness neutrality on the phase boundary is investigated in \cite{Fu:2018qsk}, which is presented in \Fig{fig:PBerrCHItex-Fu2018qsk}. One can see in comparison to the case of $\mu_S=0$, the phase boundary moves up slightly due to the strangeness neutrality. In another words, the curvature of the phase boundary is decreased a bit by the condition of strangeness neutrality, and see \sec{subsec:phaseStruQCD} for more discussions about the curvature of phase boundary.

\subsubsection{	Equation of state}
\label{subsubsec:EoSQM}

%
\begin{figure}[t]
\includegraphics[width=0.5\textwidth]{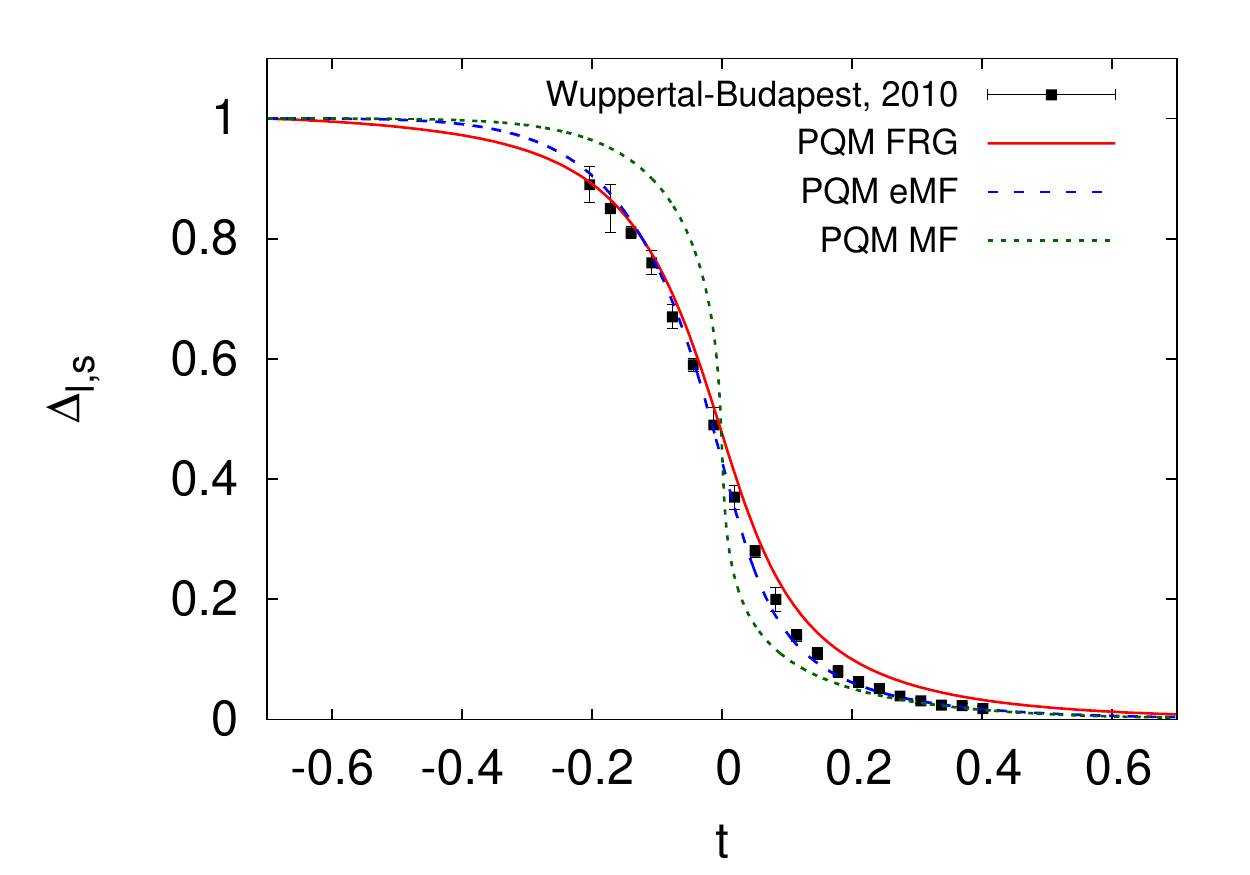}
\caption{Reduced chiral condensate as a function of the reduced temperature in the $N_f=2+1$ flavor QM model obtained in \cite{Herbst:2013ufa}. The fRG result (red solid line) is compared with the lattice result \cite{Borsanyi:2010bp} as well as the mean-field ones. The plot is adopted from \cite{Herbst:2013ufa}.}\label{fig:Deltals-Herbst2013ufa}
\end{figure}
%

%
\begin{figure*}[t]
\centering
\includegraphics[width=0.45\textwidth]{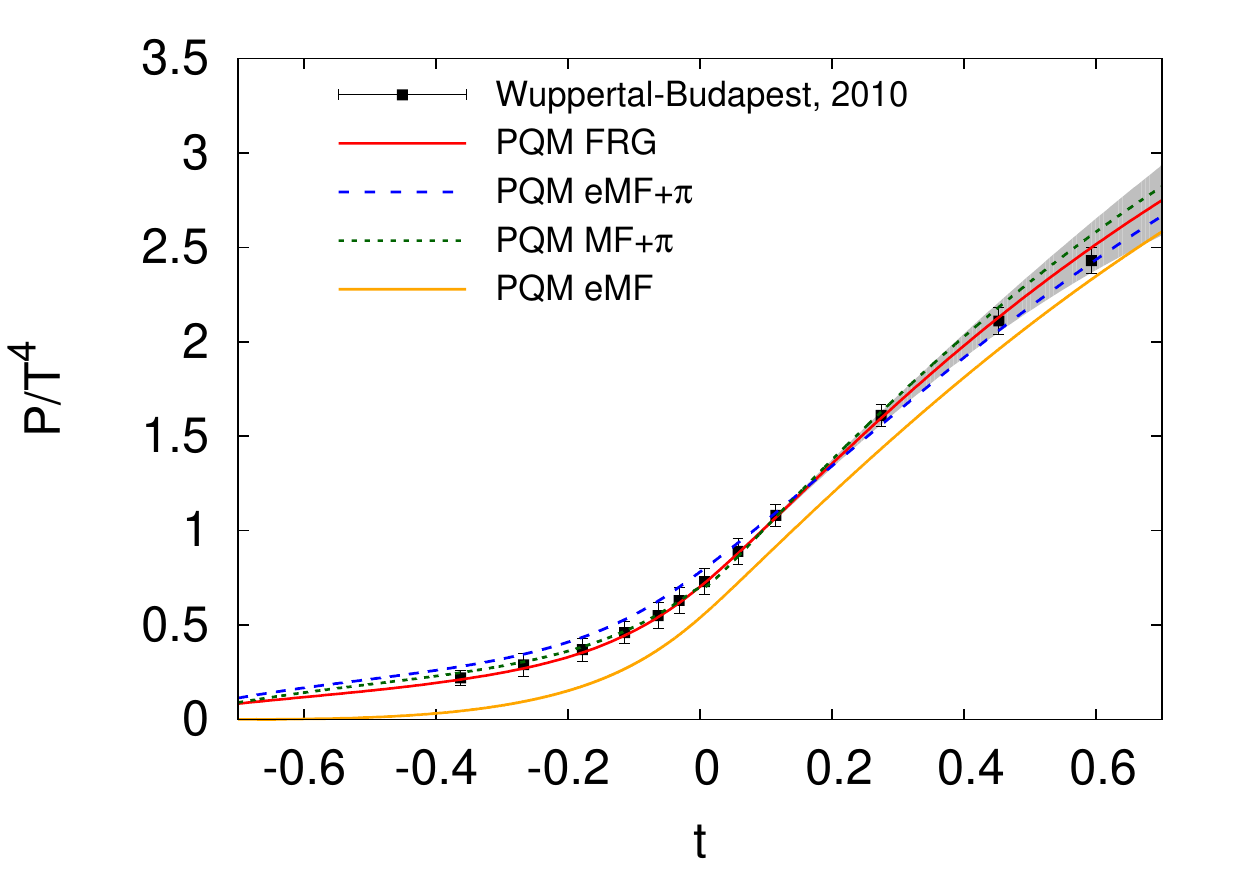}
\includegraphics[width=0.45\textwidth]{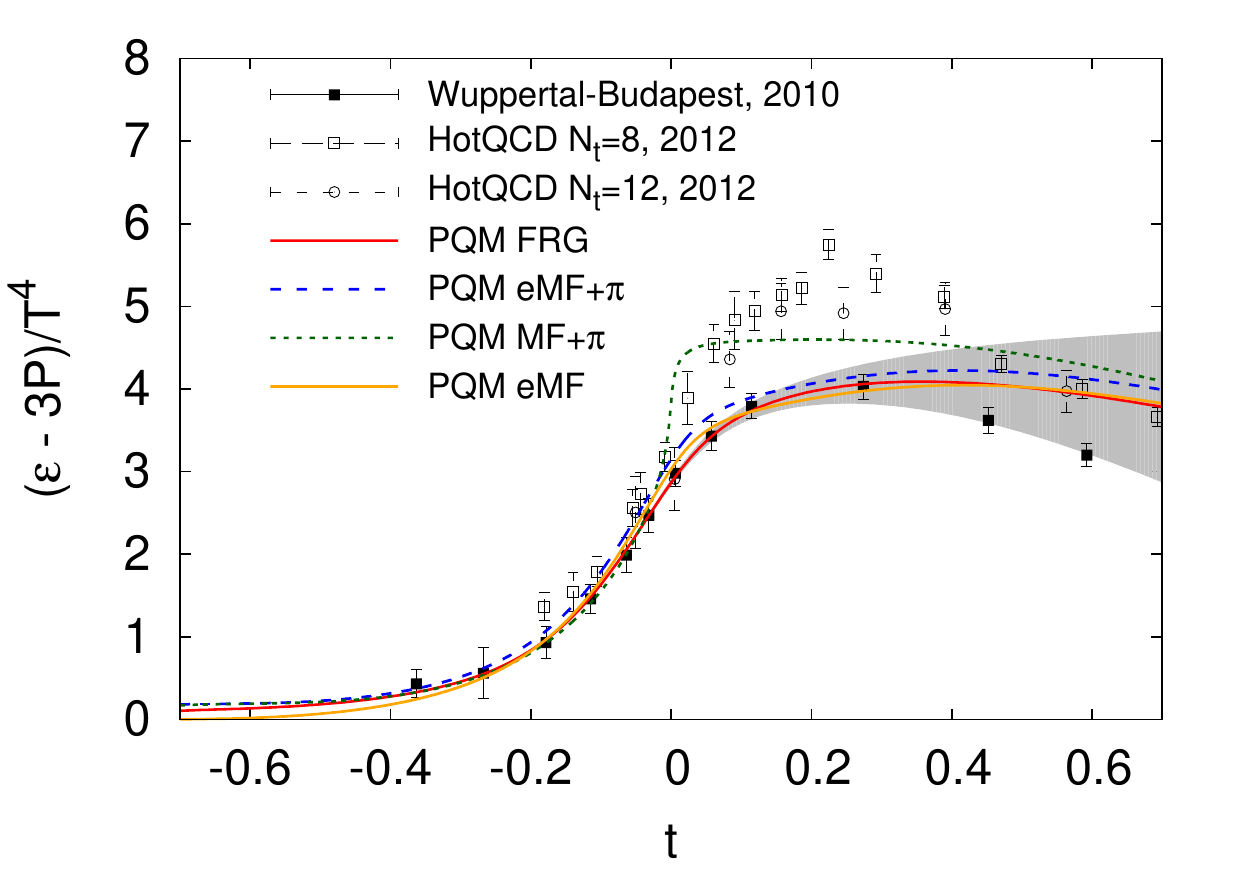}
\caption{Pressure (left panel) and trace anomaly (right panel) as functions of the reduced temperature in the $N_f=2+1$ flavor QM model obtained in \cite{Herbst:2013ufa}. The fRG results are in comparison to the lattice \cite{Bazavov:2012bp, Borsanyi:2010cj} and mean-field results. The plots are adopted from \cite{Herbst:2013ufa}.}\label{fig:Pres-Herbst2013ufa}
\end{figure*}
%

%
\begin{figure*}[t]
\includegraphics[width=0.8\textwidth]{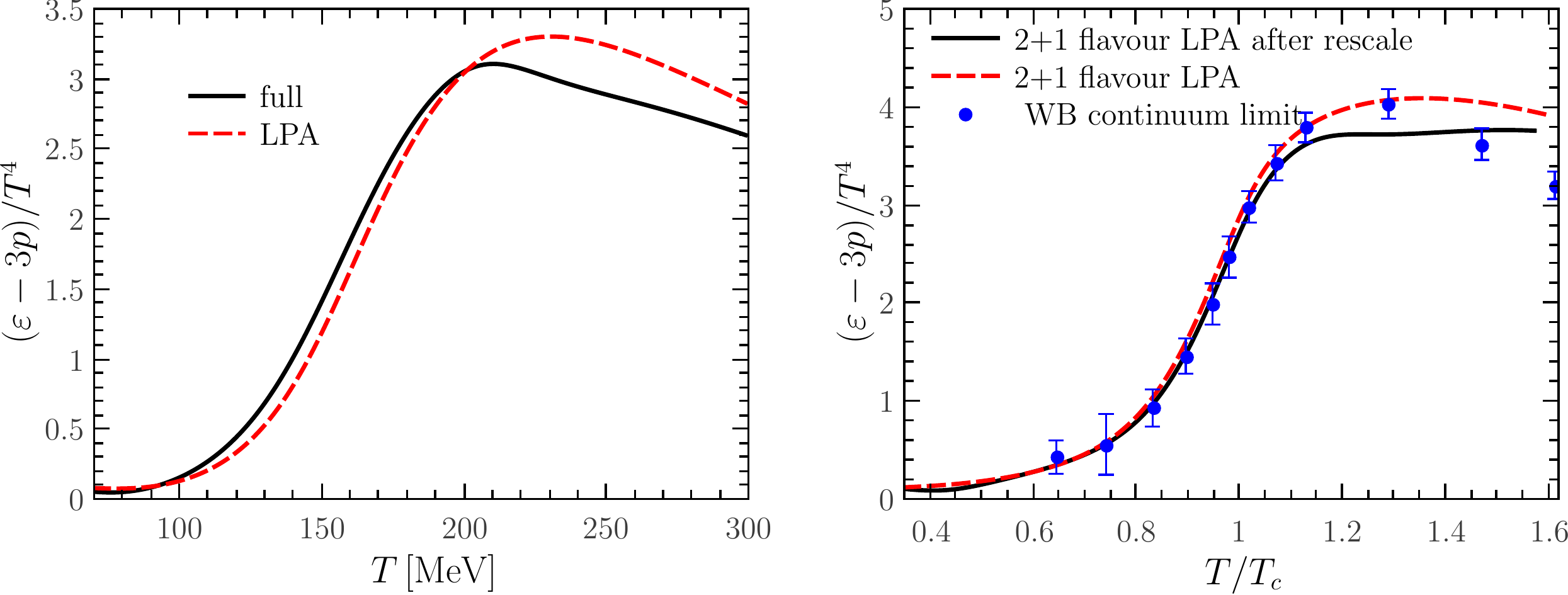}
\caption{Left panel: Trace anomaly of the $N_f=2$ QM model as a function of the temperature. Here the LPA result is compared with that beyond LPA, denoted by ``full'', in which the wave function renormalizations for the mesons and quarks, the RG scale dependence of the Yukawa coupling are taken into account. Right panel: Trace anomaly of the $N_f=2+1$ QM model obtained beyond LPA (labelled with``LPA after rescale'') as a function of the temperature, in comparison to the LPA result \cite{Herbst:2013ufa} and the lattice result from Wuppertal-Budapest collaboration \cite{Borsanyi:2010cj}. Plots are adopted from \cite{Fu:2015naa}.}\label{fig:traceano}
\end{figure*}
%

%
\begin{figure}[t]
\includegraphics[width=0.45\textwidth]{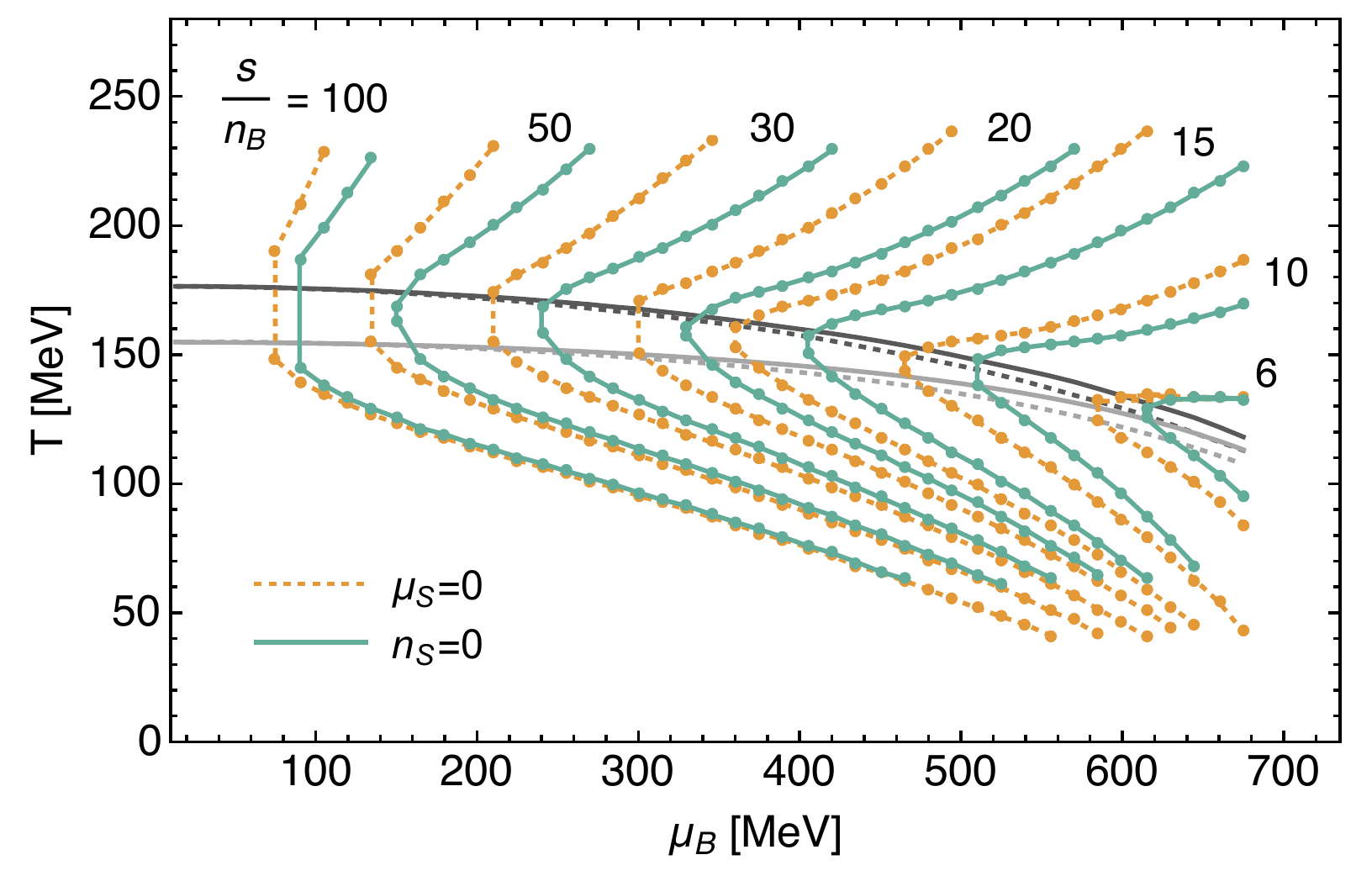}
\caption{Isentropes for different ratios of $s/n_B$ in the phase diagram in the $N_f=2+1$ flavor QM model obtained in \cite{Fu:2018qsk}. Two cases with $\mu_S=0$ and the strangeness neutrality $n_S=0$ are compared. The black and gray lines stand for the chiral and deconfinement crossover lines, respectively. The plot is adopted from \cite{Fu:2018qsk}.}\label{fig:IsenComp-Fu2018qsk}
\end{figure}
%

From the effective action in \Eq{eq:QMaction} or \Eq{eq:QM2p1action}, one is able to obtain the thermodynamic potential density, as follows
\begin{align}
  \Omega[T, \mu]=\frac{T}{V}\left(\Gamma_{k=0}[\Phi_{\mathrm{EoM}}]\Big\vert_{T,\mu}-\Gamma_{k=0}[\Phi_{\mathrm{EoM}}]\Big\vert_{T=\mu=0}\right)\,,\label{eq:Omega}
\end{align}
where $\Phi_{\mathrm{EoM}}$ denotes the fields on the equations of motion. Then the pressure and the entropy density read
\begin{align}
  p&=-\Omega[T, \mu]\,, \label{eq:pressure}
\end{align}
and 
\begin{align}
  s&=\frac{\partial p}{\partial T}\,. \label{}
\end{align}
Moreover, the trace anomaly that is also called as the interaction measure is given by
\begin{align}
  \Delta&=\epsilon-3p\,,\label{}
\end{align}
where the energy density reads
\begin{align}
  \epsilon&=-p+T s+\sum_{f=u,d,s} \mu_f n_f\,,\label{}
\end{align}
with the number density for quark of flavor $f$
\begin{align}
  n_f&=\frac{\partial p}{\partial \mu_f}\,.\label{}
\end{align}

In \Fig{fig:Deltals-Herbst2013ufa} the reduced condensate, cf. \Eq{eq:RedCondSigma} and the relevant discussions in \sec{subsec:condensateQCD}, is shown as a function of the reduced temperature $t=(T-T_{\mathrm{pc}})/T_{\mathrm{pc}}$ obtained in \cite{Herbst:2013ufa}, where $T_{\mathrm{pc}}$ denotes the pseudocritical temperature for the chiral crossover. The fRG calculations are in agreement with the lattice simulations \cite{Borsanyi:2010bp}. Furthermore, the mean-field results are also presented for comparison, where those labelled with ``eMF'' and ``MF''
are obtained by taking into account or not the fermionic vacuum loop contribution in the calculations, respectively, and see, e.g., \cite{Schaefer:2011ex} for more discussions. In \Fig{fig:Pres-Herbst2013ufa} the pressure and the trace anomaly in the $N_f=2+1$ flavor QM model obtained in \cite{Herbst:2013ufa} are shown as functions of the reduced temperature. The fRG results are compared with the lattice results \cite{Bazavov:2012bp, Borsanyi:2010cj}, and it is observed that the fRG results are consistent with the lattice results from the Wuppertal-Budapest collaboration (WB), except that the trace anomaly from fRG is a bit larger than that from WB in the regime of high temperature with $t \gtrsim 0.3$. Moreover, the mean-field results as well as those including a contribution of the thermal pion gas, are also presented for comparison.

The equations of state in \Fig{fig:Pres-Herbst2013ufa} are obtained in fRG within the local potential approximation (LPA), where the RG-scale dependence only enters the effective potential in \Eq{eq:QMaction}. The equations of state are also calculated beyond the LPA approximation in \cite{Fu:2015naa}. The relevant results are presented in \Fig{fig:traceano}. In the calculations the wave function renormalizations for the mesons and quarks, and the RG scale dependence of the Yukawa coupling are taken into account. In \Fig{fig:traceano} the results beyond LPA are in comparison to the lattice and LPA results. One can see that the trace anomaly is decreased a bit beyond LPA in the regime of $T\gtrsim 1.1T_c$. In \Fig{fig:IsenComp-Fu2018qsk} the influence of the strangeness neutrality on the isentropes is presented obtained in \cite{Fu:2018qsk}. It is found that, in comparison to the case of $\mu_S=0$, the isentropes with $n_S=0$ move towards the r.h.s. a bit, and their turning around in the regime of crossover becomes more abruptly. Moreover, isentropes obtained from the Taylor expansion and the grid method for the effective potential are compared, and consistent results are found \cite{Nakano:2009ps}. 

Note that calculations of the equations of state at high baryon densities within the fRG approach have also made progress recently, see, e.g., \cite{Leonhardt:2019fua, Otto:2019zjy, Otto:2020hoz}, and the obtained EoS has been employed to study the phenomenology of compact stars \cite{Otto:2019zjy, Otto:2020hoz}.

\subsubsection{Baryon number fluctuations}
\label{subsubsec:fluctuationsQM}

%
\begin{figure}[t]
\includegraphics[width=0.45\textwidth]{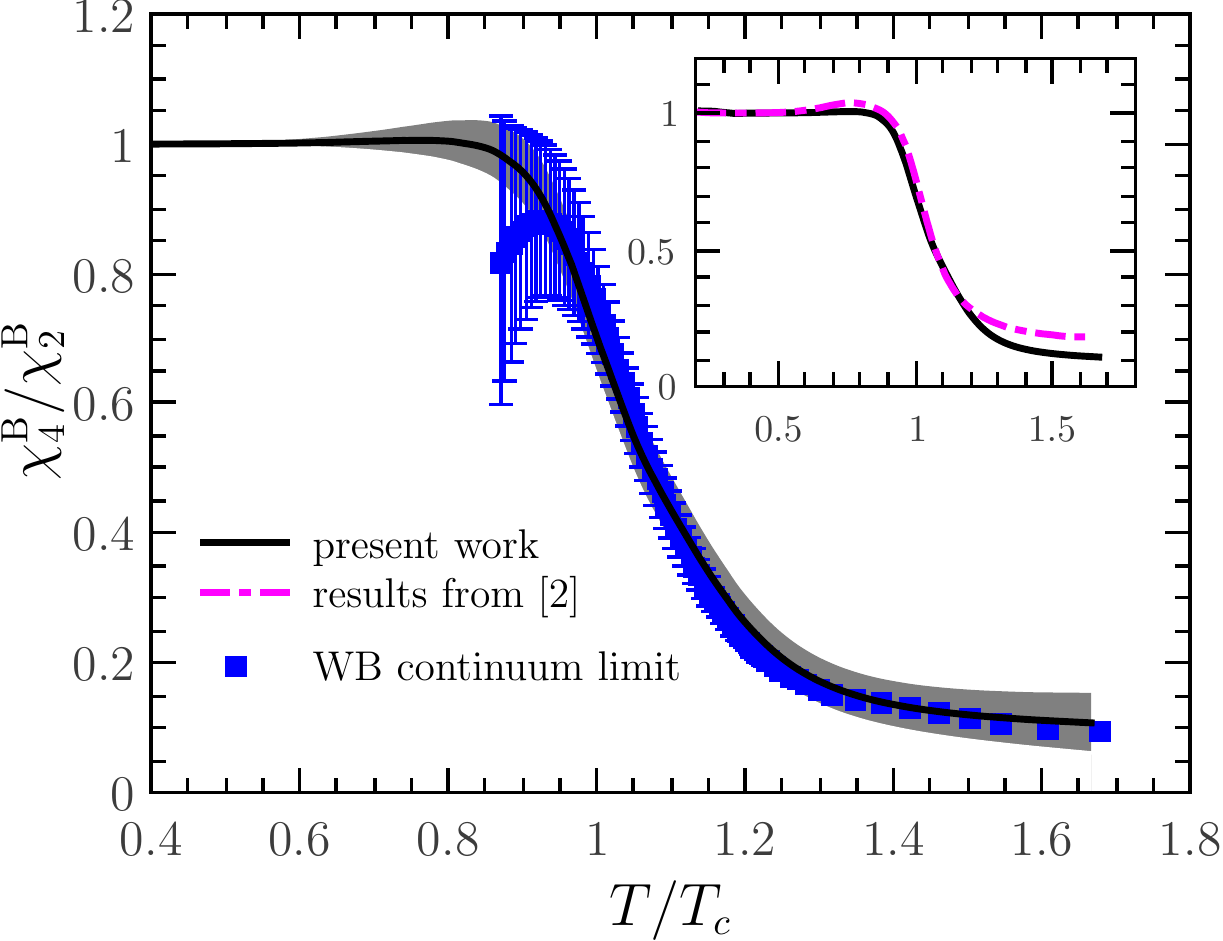}
\caption{Kurtosis of the baryon number distributions as a function of the temperature obtained within the fRG approach to LEFTs \cite{Fu:2016tey}. In the calculations the frequency dependence for the quark wave function renormalization is taken into account, and the relevant results (black solid line) are compared with those without the frequency dependence (pink dashed line) in \cite{Fu:2015amv}. The continuum-extrapolated lattice results from the Wuppertal-Budapest collaboration \cite{Borsanyi:2013hza} are also presented for comparison. The plot is adopted from \cite{Fu:2016tey}.}\label{fig:kurtosis}
\end{figure}
%

%
\begin{figure}[t]
\includegraphics[scale=0.22]{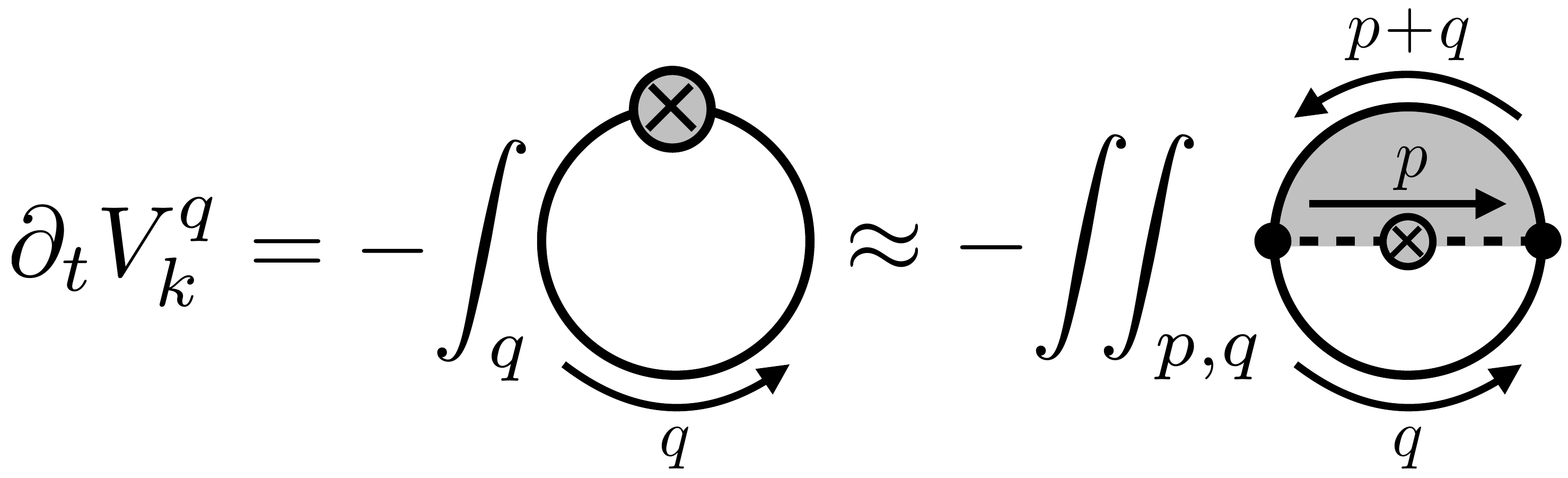}
\caption{Schematic diagram of the frequency dependence of the quark anomalous dimension to the flow equation of the effective potential. The gray area stands for the contribution of the quark anomalous dimension, where a summation for the quark external frequency has been done. The plot is adopted from \cite{Fu:2016tey}.
}\label{fig:qloop}
\end{figure}
%

%
\begin{figure*}[t]
\includegraphics[width=0.9\textwidth]{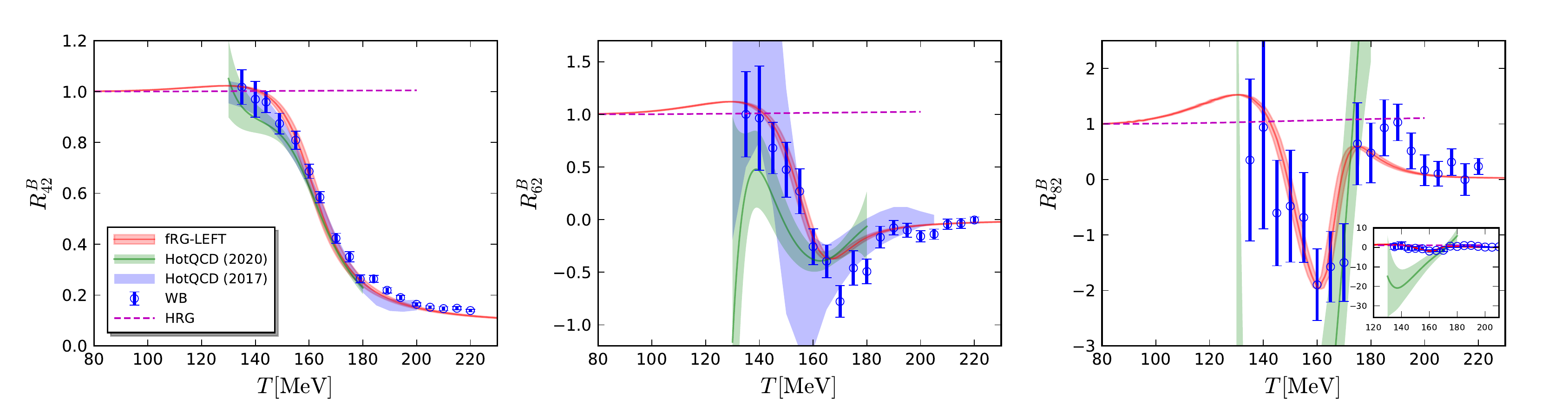}
\caption{Baryon number fluctuations $R^{B}_{42}$ (left panel), $R^{B}_{62}$ (middle panel), and $R^{B}_{82}$ (right panel) as functions of the temperature at $\mu_B=0$, obtained within the fRG approach to a QCD-assisted LEFT in \cite{Fu:2021oaw}. The fRG results are compared with lattice results from the HotQCD collaboration \cite{Bazavov:2017dus, Bazavov:2017tot, Bazavov:2020bjn} and the Wuppertal-Budapest collaboration \cite{Borsanyi:2018grb}. The inlay in the right panel shows the zoomed-out view of $R^{B}_{82}$. The plots are adopted from \cite{Fu:2021oaw}.}\label{fig:R42R62R82-T-muB0}
\end{figure*}
%

%
\begin{figure*}[t]
\centering
\includegraphics[width=0.9\textwidth]{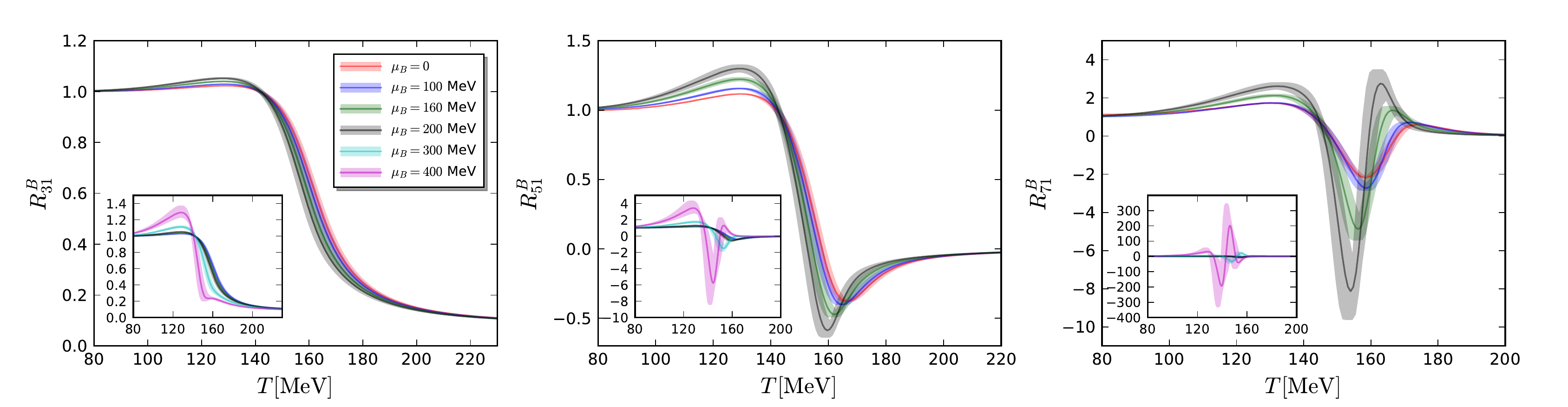}
\includegraphics[width=0.9\textwidth]{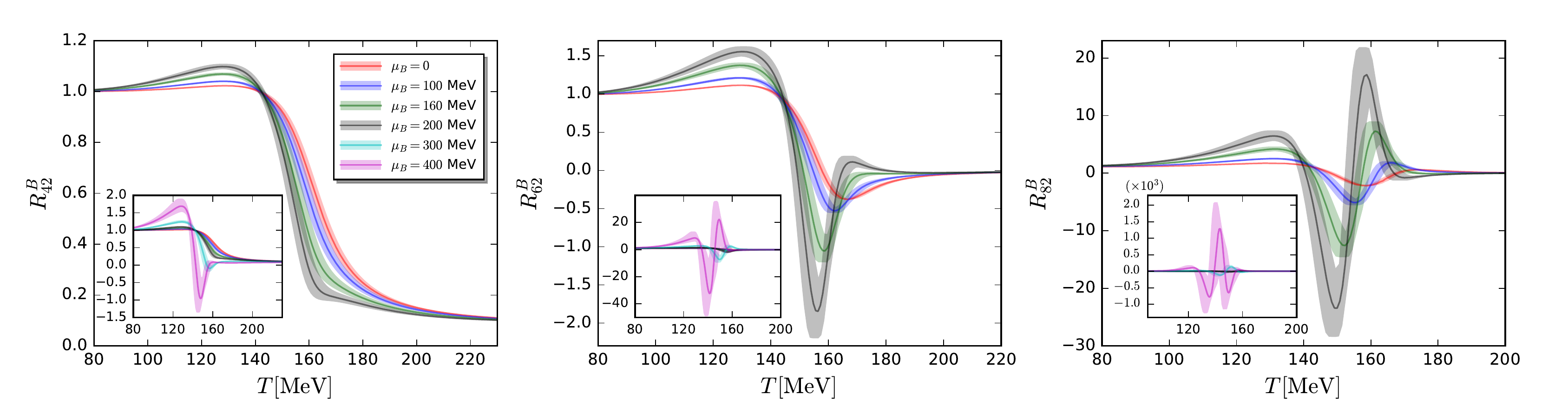}
\caption{Baryon number fluctuations $R^{B}_{31}$ (top-left), $R^{B}_{42}$ (bottom-left), $R^{B}_{51}$ (top-middle), $R^{B}_{62}$ (bottom-middle), $R^{B}_{71}$ (top-right) and $R^{B}_{82}$ (bottom-right) as functions of the temperature with several different values of the baryon chemical potential, obtained within the fRG approach to a QCD-assisted LEFT in \cite{Fu:2021oaw}. The inlays in each plot show the zoomed-out views. The plots are adopted from \cite{Fu:2021oaw}.}\label{fig:Rnm-T-muB0to400}
\end{figure*}
%

%
\begin{figure*}[t]
\centering
\includegraphics[width=0.44\textwidth]{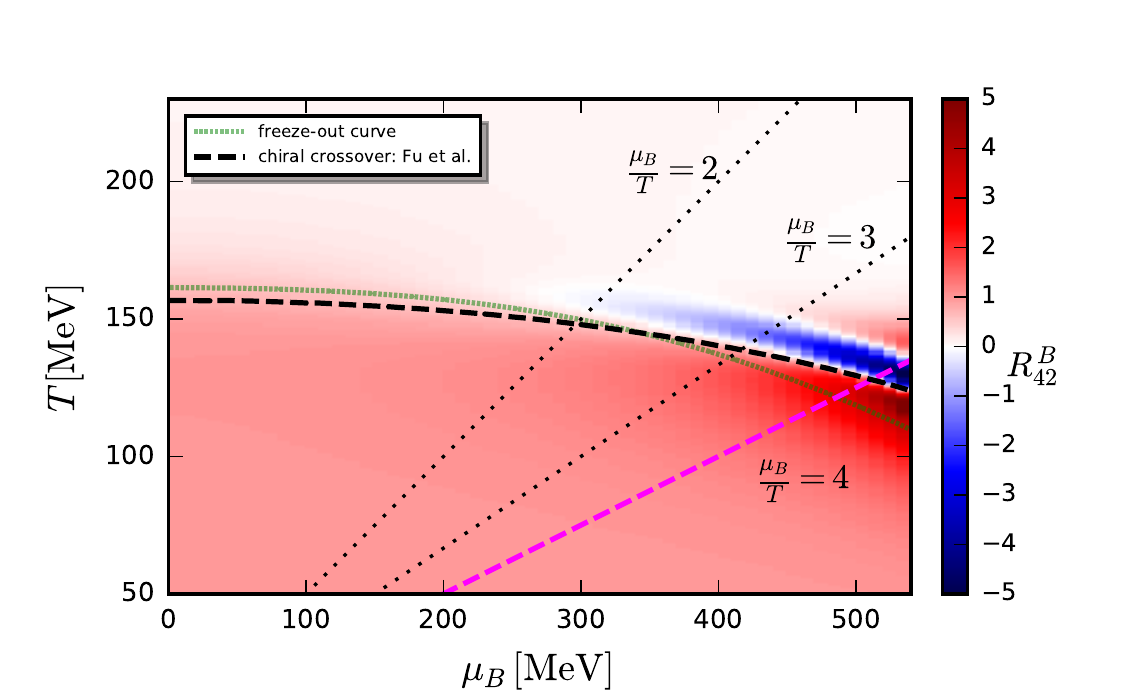}
\includegraphics[width=0.35\textwidth]{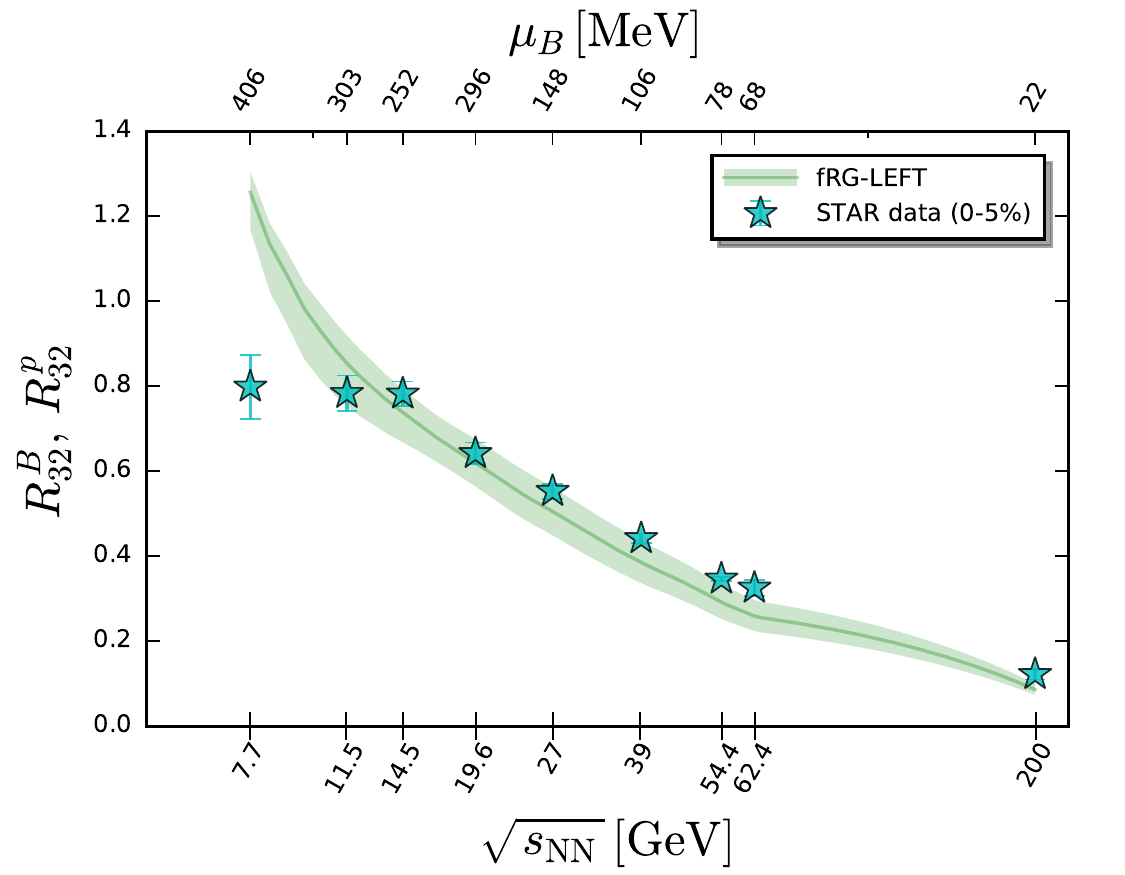}
\caption{Left panel: kurtosis of the baryon number distributions $R^{B}_{42}(T,\mu_B)$ in the phase diagram, obtained within the fRG approach to a QCD-assisted LEFT in \cite{Fu:2021oaw}. The black dashed line stands for the chiral phase boundary in the crossover regime obtained from QCD calculations \cite{Fu:2019hdw}. The green dotted line denotes the chemical freeze-out curve. Right panel: Skewness of the baryon number distributions as a function of the collision energy, obtained within the fRG approach to a QCD-assisted LEFT in \cite{Fu:2021oaw}. The fRG results are compared with the experimental measurements of the skewness of the net-proton number distributions $R^{p}_{32}$ with centrality 0-5\% \cite{STAR:2020tga}. The plots are adopted from \cite{Fu:2021oaw}.}\label{fig:R42phasediagram}
\end{figure*}
%

%
\begin{figure*}[t]
\centering
\includegraphics[width=0.6\textwidth]{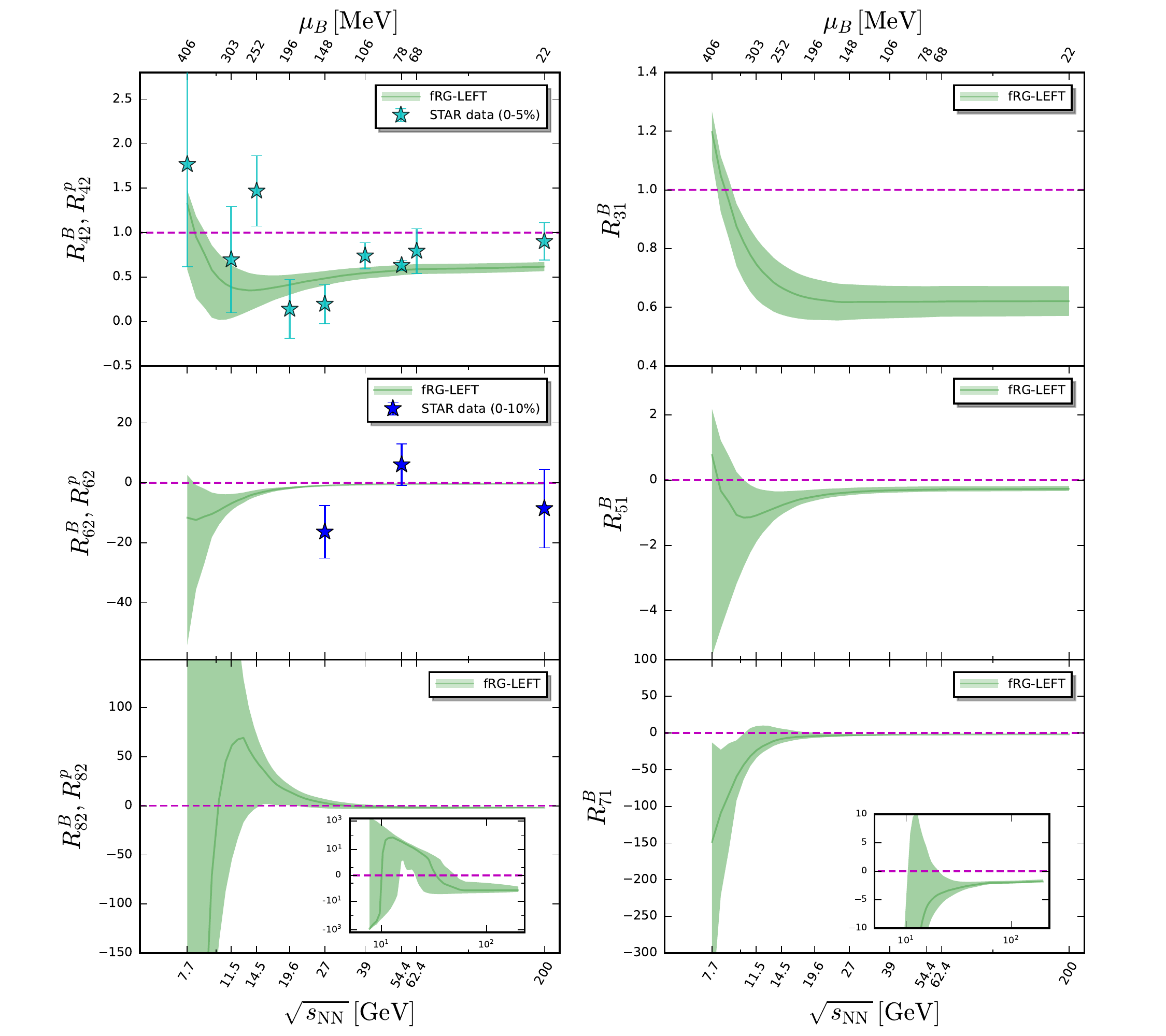}
\caption{Baryon number fluctuations $R^{B}_{42}$ (top-left), $R^{B}_{62}$ (middle-left), $R^{B}_{82}$ (bottom-left), $R^{B}_{31}$ (top-right), $R^{B}_{51}$ (middle-right), $R^{B}_{71}$ (bottom-right) as functions of the collision energy, obtained within the fRG approach to a QCD-assisted LEFT in \cite{Fu:2021oaw}. Experimental data of the kurtosis of the net-proton number distributions $R^{p}_{42}$ with centrality 0-5\% \cite{STAR:2020tga}, and the sixth-order cumulant $R^{p}_{62}$ with centrality 0-10\% \cite{STAR:2021rls} are presented for comparison. The plots are adopted from \cite{Fu:2021oaw}.
}\label{fig:Rm2Rm1-sqrtS}
\end{figure*}
%

%
\begin{figure*}[t]
\includegraphics[width=0.8\textwidth]{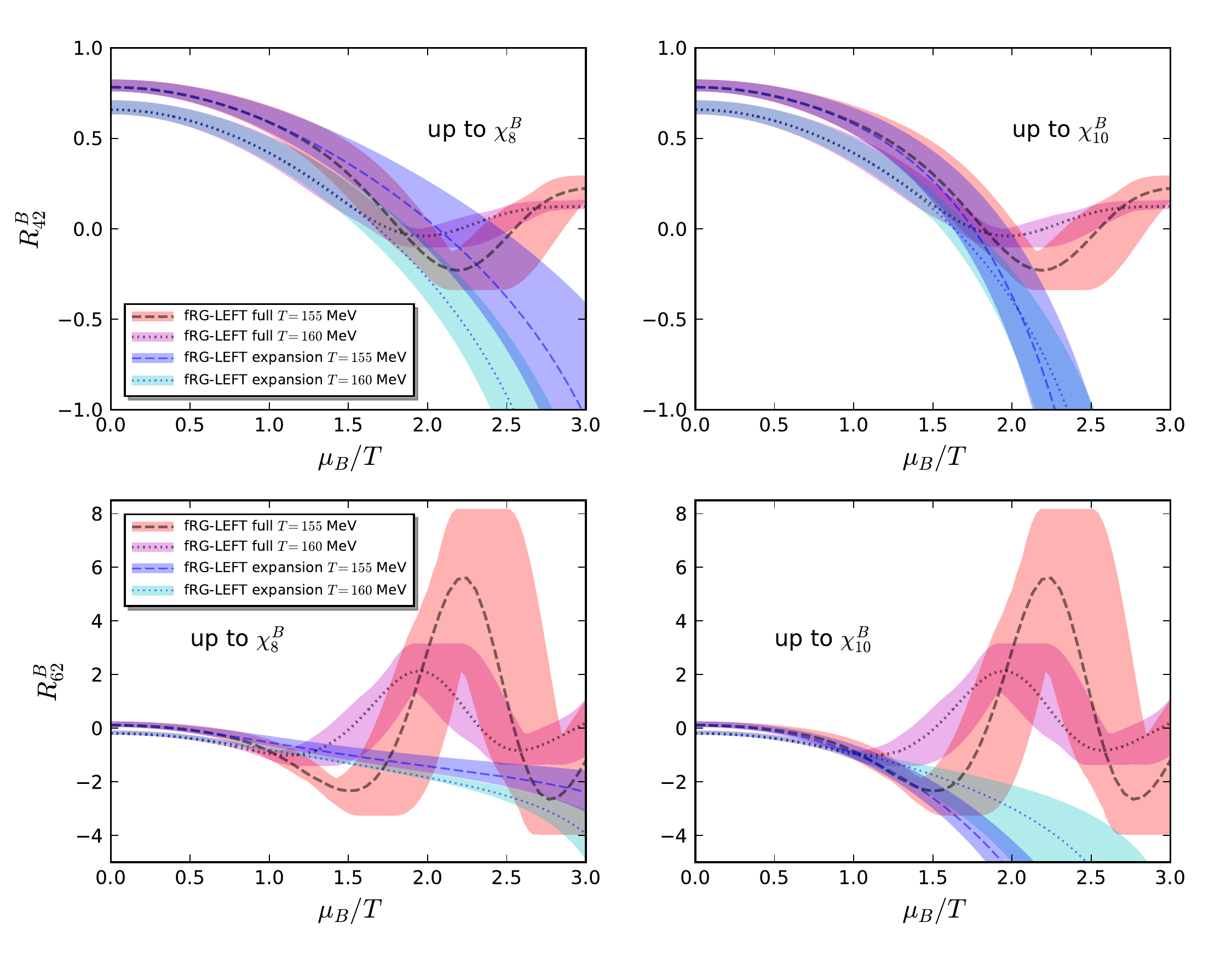}
\caption{Full results of the baryon number fluctuations $R^{B}_{42}$ (upper panels) and $R^{B}_{62}$ (lower panels) as functions of $\mu_B/T$ with two fixed values of the temperature, in comparison to the results of Taylor expansion up to order of $\chi^B_8(0)$ (left panels) and $\chi^B_{10}(0)$ (right panels). Both results are obtained within the fRG approach to a QCD-assisted LEFT in \cite{Fu:2021oaw}. The plots are adopted from \cite{Fu:2021oaw}.}\label{fig:R42R62expansion-muBoT}
\end{figure*}
%

%
\begin{figure}[t]
\includegraphics[width=0.5\textwidth]{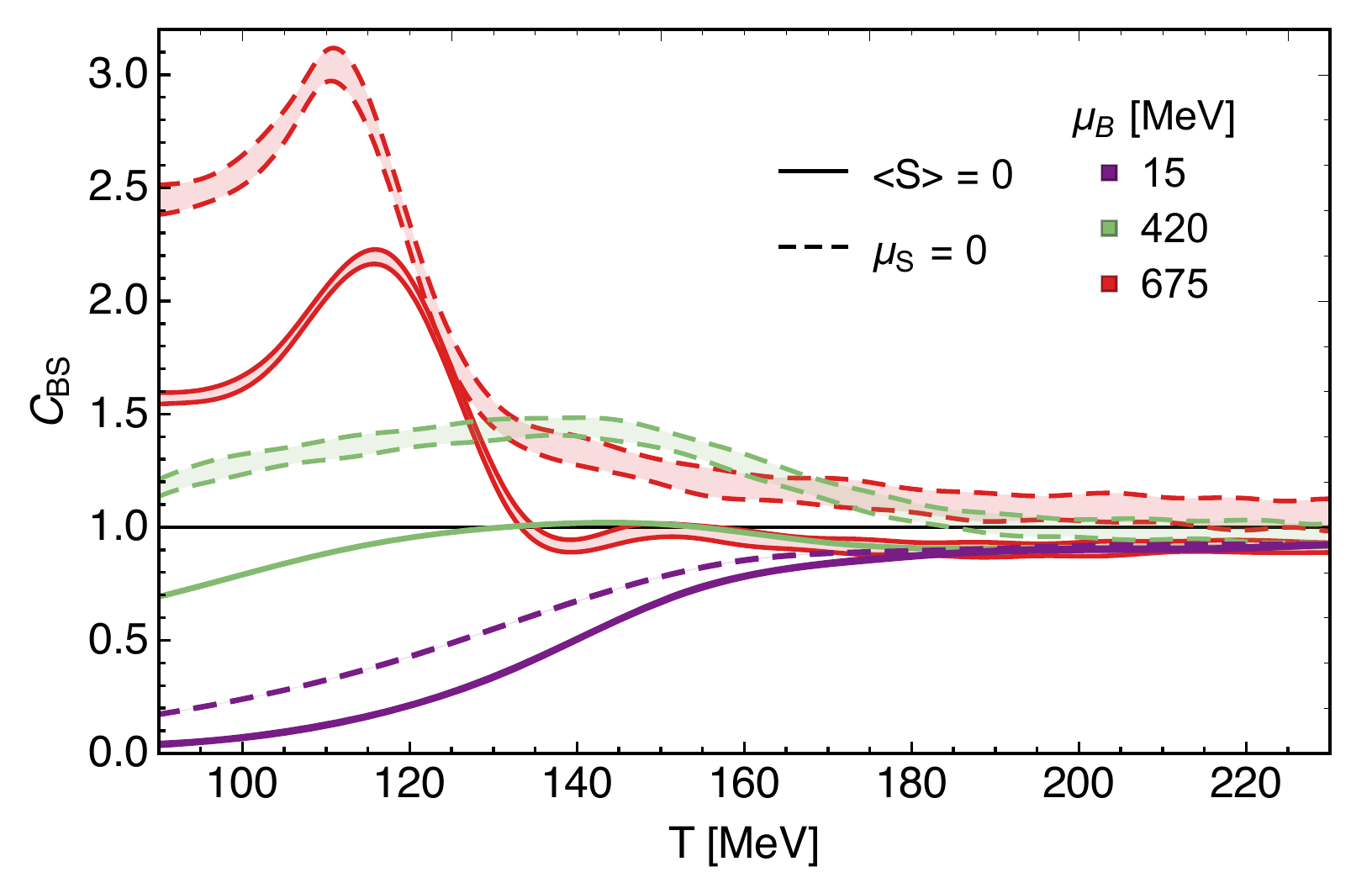}
\caption{Correlation between the baryon number and the strangeness $C_{BS}$ as a function of the temperature with several different values of the baryon chemical potential, obtained within the fRG approach in \cite{Fu:2018swz}. Two cases with $\mu_S=0$ (dashed lines) and the strangeness neutrality $n_S=0$ (solid lines) are compared. The plot is adopted from \cite{Fu:2018swz}.}\label{fig:CBSofTcomp-Fu2018swz}
\end{figure}
%

Differentiating the pressure in \Eq{eq:pressure} with respect to the baryon chemical potential $n$ times, one is led to the $n$-th order generalized susceptibility of the baryon number, i.e.,
\begin{align}
\chi_n^{B}=&\frac{\partial^n}{\partial (\mu_B/T)^n}\frac{p}{T^4}\,,\label{eq:suscept}
\end{align}
which is also related to the $n$-th order cumulant of the net baryon number distributions, that is, 
\begin{align}
  \langle (\delta N_B)^n \rangle=&\sum_{N_B=-\infty}^{\infty}(\delta N_B)^n P(N_B)\,.\label{eq:dNBaver}
\end{align}
with $\delta N_B=N_B-\langle N_B\rangle$. Here $P(N_B)$ denotes the probability distribution of the net baryon number, which can be calculated theoretically from the canonical ensemble with an imaginary chemical potential, and see, e.g., \cite{Morita:2013tu,Sun:2018ozp} for a detailed discussion. In experimental measurements, since neutrons are difficult to detect, the net proton number distribution is measured, as a proxy for the net baryon number distribution, and see, e.g., \cite{Luo:2017faz} for more details. Evidently, $\langle N_B\rangle$ is the ensemble average of the net baryon number.

The relations between the generalized susceptibilities in \Eq{eq:suscept} and the cumulants in \Eq{eq:dNBaver} for the lowest four orders read
{\allowdisplaybreaks
\begin{align}
\chi^B_1=&\frac{1}{VT^3}\braket{N_B}\,,\label{eq:chiB1}\\[2ex]
\chi^B_2=& \frac{1}{VT^3}\braket{(\delta N_B)^2}\,,\label{eq:chiB2}\\[2ex]
\chi^B_3=&\frac{1}{VT^3}\braket{(\delta N_B)^3}\,,\\[2ex]
\chi^B_4=&\frac{1}{VT^3}\Big(\braket{(\delta N_B)^4}-3\braket{(\delta N_B)^2}^2\Big)\,,\label{eq:chiB4}
\end{align}}
which leaves us with the experimental observables as follows
\begin{align}
  M=&VT^3\chi_1^{\text{B}}\,,\qquad \sigma^2=VT^3\chi_2^{\text{B}}\,,\nonumber\\[2ex]
  S=&\frac{\chi_3^{\text{B}}}{\chi_2^{\text{B}}\sigma}\,,\qquad \hspace{0.55cm}\kappa=\frac{\chi_4^{\text{B}}}{\chi_2^{\text{B}}\sigma^2}\,.
\end{align}
Here $M$, $\sigma^2$, $S$, $\kappa$ stand for the mean value, the variance, the skewness, and the kurtosis of the net baryon or proton number distributions. In the same way, calculations can also be extended to the hyper-order baryon number fluctuations, i.e., $\chi_{n>4}^{B}$ \cite{Wagner:2009pm, Karsch:2010hm, Schaefer:2011ex, Friman:2011pf, Fu:2021oaw}. The relations between the hyper-order susceptibilities and the cumulants are given by \cite{Fu:2021oaw}
{\allowdisplaybreaks
\begin{align}
\chi^B_5=&\frac{1}{VT^3}\Big(\braket{(\delta N_B)^5}-10\braket{(\delta N_B)^2}\braket{(\delta N_B)^3}\Big)\,,\\[2ex]
\chi^B_6=&\frac{1}{VT^3}\Big(\braket{(\delta N_B)^6}-15\braket{(\delta N_B)^4}\braket{(\delta N_B)^2}\nonumber\\[2ex]
&-10\braket{(\delta N_B)^3}^2+30\braket{(\delta N_B)^2}^3\Big)\,,\\[2ex]
\chi^B_7=&\frac{1}{VT^3}\Big(\braket{(\delta N_B)^7}-21\braket{(\delta N_B)^5}\braket{(\delta N_B)^2}\nonumber\\[2ex]
&-35\braket{(\delta N_B)^4}\braket{(\delta N_B)^3}\nonumber\\[2ex]
&+210\braket{(\delta N_B)^3}\braket{(\delta N_B)^2}^2\Big)\,,\\[2ex]
\chi^B_8=&\frac{1}{VT^3}\Big(\braket{(\delta N_B)^8}-28\braket{(\delta N_B)^6}\braket{(\delta N_B)^2}\nonumber\\[2ex]
&-56\braket{(\delta N_B)^5}\braket{(\delta N_B)^3}-35\braket{(\delta N_B)^4}^2\nonumber\\[2ex]
&+420\braket{(\delta N_B)^4}\braket{(\delta N_B)^2}^2\nonumber\\[2ex]
&+560\braket{(\delta N_B)^3}^2\braket{(\delta N_B)^2}-630\braket{(\delta N_B)^2}^4\Big)\,.
\end{align}}
It is more convenient to adopt the ratio between two susceptibilities of different orders, say
\begin{align}
R_{nm}^{B}=&\frac{\chi_n^{B}}{\chi_m^{B}}\,,\label{eq:Rnm}
\end{align}
where the explicit volume dependence is removed. The baryon number fluctuations as well as fluctuations of other conserved charges, e.g., the electric charge and the strangeness, and correlations among these conserved charges, have been widely studied in literatures. For lattice simulations, see, e.g.,  \cite{Bazavov:2012vg, Borsanyi:2013hza, Borsanyi:2014ewa, Bazavov:2017dus, Bazavov:2017tot, Borsanyi:2018grb, Bazavov:2020bjn}. Investigations of fluctuations and correlations within the fRG approach to LEFTs can be found in, e.g., \cite{Skokov:2010wb, Skokov:2010uh, Friman:2011pf, Morita:2014fda, Fu:2015naa, Fu:2015amv, Fu:2016tey, Almasi:2017bhq, Fu:2018qsk, Fu:2018swz, Wen:2018nkn, Wen:2019ruz, Fu:2021oaw}, and within the mean-field approximations in, e.g., \cite{Fu:2009wy, Fu:2010ay, Karsch:2010hm, Schaefer:2011ex, Li:2018ygx}. For the relevant studies from Dyson-Schwinger Equations, see, e.g., \cite{Xin:2014ela, Isserstedt:2019pgx}. Remarkably, recently QCD-assisted LEFTs with the fRG approach have been developed and used to study the skewness and kurtosis of the baryon number distributions \cite{Fu:2015naa, Fu:2015amv, Fu:2016tey}, the baryon-strangeness correlations \cite{Fu:2018qsk, Fu:2018swz}, and the hyper-order baryon number fluctuations \cite{Fu:2021oaw}.

In \Fig{fig:kurtosis} the kurtosis of the baryon number distributions is shown as a function of the temperature at vanishing chemical potential. In the calculations the frequency dependence of the quark wave function renormalization is taken into account \cite{Fu:2016tey}, whose contribution to the flow of the effective potential is shown schematically in \Fig{fig:qloop}. It is found that a summation for the quark external frequency is indispensable to the Silver Blaze property at finite chemical potential, see \cite{Fu:2016tey} for a more detailed discussion. From \Fig{fig:kurtosis}, one can see that the summation for the external frequency of the quark wave function renormalization improves the agreement between the fRG results and the lattice ones. 

In \Fig{fig:R42R62R82-T-muB0} the fourth-, sixth-, and eighth-order baryon number fluctuations divided by the quadratic one are shown as functions of the temperature at vanishing baryon chemical potential. The fRG results \cite{Fu:2021oaw} are compared with lattice results from the HotQCD collaboration \cite{Bazavov:2017dus, Bazavov:2017tot, Bazavov:2020bjn} and the Wuppertal-Budapest collaboration (WB) \cite{Borsanyi:2018grb}. It is observed that the fRG results are in quantitative, qualitative agreement with the WB and HotQCD results, respectively. In \Fig{fig:Rnm-T-muB0to400} the baryon number fluctuations of different orders, $R^{B}_{31}$, $R^{B}_{42}$, $R^{B}_{51}$, $R^{B}_{62}$, $R^{B}_{71}$ and $R^{B}_{82}$, are shown as functions of the temperature with several different values of the baryon chemical potential from 0 to 400 MeV. One finds that the dependence of fluctuations on the temperature oscillates more pronouncedly with the increasing order of cumulants. Moreover, with the increase of the baryon chemical potential the chiral crossover grows sharper, which leads to the increase of the magnitude of fluctuations significantly.

In the left panel of \Fig{fig:R42phasediagram} the kurtosis of the baryon number distributions is depicted in the phase diagram \cite{Fu:2021oaw}. The black dashed line denotes the chiral phase boundary in the crossover regime. The green dotted line stands for the chemical freeze-out curve obtained from the STAR freeze-out data \cite{Adamczyk:2017iwn}, see \cite{Fu:2021oaw} for a more detailed discussion. One can see that there is narrow blue area near the phase boundary with $\mu_B \gtrsim 300$ MeV, which indicates that the kurtosis becomes negative in this area. Moreover, it is found that with the increase of the baryon chemical potential, the freeze-out curve moves towards the blue area firstly, and then deviates from it a bit at large baryon chemical potential. The skewness of the baryon number distributions as a function of the collision energy calculated in the fRG approach is presented in the right panel of \Fig{fig:R42phasediagram} \cite{Fu:2021oaw}, which is in comparison to the STAR data of the skewness of the net-proton number distributions $R^{p}_{32}$ with centrality 0-5\% \cite{Adam:2020unf}. It is found that the fRG results are in good agreement with the experimental data except the two lowest energy points, which is attributed to the fact that other effects, e.g., volume fluctuations \cite{Luo:2013bmi, Chatterjee:2019fey, Chatterjee:2020nnn}, the global baryon number conservation \cite{He:2016uei, Braun-Munzinger:2016yjz, Vovchenko:2020tsr}, become important when the beam collision energy is low.

In \Fig{fig:Rm2Rm1-sqrtS} baryon number fluctuations $R^{B}_{42}$, $R^{B}_{62}$, $R^{B}_{82}$, $R^{B}_{31}$, $R^{B}_{51}$, and $R^{B}_{71}$ are shown as functions of the collision energy obtained within the fRG approach \cite{Fu:2021oaw}, where the chemical freeze-out curve obtained from STAR data as shown in the left panel of \Fig{fig:R42phasediagram} is used. The fRG results are compared with experimental data of the kurtosis of the net-proton number distributions $R^{p}_{42}$ with centrality 0-5\% \cite{STAR:2020tga}, and the sixth-order cumulant $R^{p}_{62}$ with centrality 0-10\% \cite{STAR:2021rls}. A non-monotonic dependence of the kurtosis $R^{B}_{42}$ on the collision energy is also observed in the fRG calculations, which is consistent with the experimental measurements of $R^{p}_{42}$. Moreover, it is found that this non-monotonicity arises from the increasingly sharp crossover with the decrease of the collision energy, which is also reflected in the heat map of the kurtosis in the phase diagram in the left panel of \Fig{fig:R42phasediagram}. The fRG results of $R^{B}_{62}$ are also qualitatively consistent with the experimental data of $R^{p}_{62}$.

In \Fig{fig:R42R62expansion-muBoT} the full results of the baryon number fluctuations $R^{B}_{42}$ and $R^{B}_{62}$ are compared with those of Taylor expansion for the baryon chemical potential up to the eighth and tenth orders, which can be used to investigate the convergence properties of the Taylor expansion for the baryon chemical potential. For the two values of the temperature, $T=155$ MeV and 160 MeV, it is found that the full results deviate from those of Taylor expansion significantly with $\mu_B/T \gtrsim 1.2\sim1.5$. The oscillating behavior of the full results at large baryon chemical potential is not captured by the Taylor expansion, which hints that the convergence radius of the Taylor expansion for the baryon chemical potential might be restricted by some singularity, e.g., the Yang--Lee edge singularity, and see, e.g., \cite{Stephanov:2006dn, Mukherjee:2019eou, Connelly:2020gwa, Rennecke:2022ohx} for more discussions.

We close this subsection with a discussion about the correlation between the baryon number and the strangeness, i.e.,
\begin{align}
C_{BS}=&-3\frac{\chi_{11}^{BS}}{\chi_2^{S}}\,,\label{}
\end{align}
which is calculated within the fRG approach in \cite{Fu:2018swz}. There, the influence of the strangeness neutrality on  $C_{BS}$ is investigated, and the relevant results are presented in \Fig{fig:CBSofTcomp-Fu2018swz}. One can see that the correlation between the baryon number and the strangeness is suppressed by the condition of the strangeness neutrality.

\subsubsection{	Critical exponents}
\label{subsubsec:exponentsQM}

%
\begin{table*}[t]
  \begin{center}
  \begin{tabular}{cccccccc}
    \hline\hline & & & & & & &\\[-2ex]
    &Method& $\beta$ & $\delta$ & $\gamma$ & $\nu$ & $\nu_c$ & $\eta$ \\[1ex]
    \hline & & & & & & & \\[-2ex]
    $O(4)$ QM LPA \cite{Chen:2021iuo} & fRG Chebyshev &0.3989(41) &4.975(57) & 1.5458(68) &0.7878(25) &0.3982(17) &0 \\[1ex]
    $O(4)$ QM LPA$'$ \cite{Chen:2021iuo} & fRG Chebyshev &0.3832(31) &4.859(37) &1.4765(76) &0.7475(27)  & 0.4056(19)&0.0252(91)*\\[1ex]
    $Z(2)$ QM LPA \cite{Chen:2021iuo} & fRG Chebyshev &0.3352(12) &4.941(22) &1.3313(96)  &0.6635(17) &0.4007(45)&0\\[1ex]
    $Z(2)$ QM LPA$'$ \cite{Chen:2021iuo} & fRG Chebyshev &0.3259(01) &4.808(14) &1.2362(77) &0.6305(23) &0.4021(43)&0.0337(38)*\\[1ex]
    $O(4)$ scalar theories \cite{Tetradis:1993ts} &fRG Taylor&0.409& 4.80* & 1.556 & 0.791 &        & 0.034 \\[1ex]
    $O(4)$ KT phase transition \cite{VonGersdorff:2000kp} &fRG Taylor &0.387*& 4.73* &        & 0.739 &        & 0.047 \\[1ex]
    $Z(2)$ KT phase transition \cite{VonGersdorff:2000kp} &fRG Taylor &        &         &        & 0.6307  &        & 0.0467         \\[1ex]
    $O(4)$ scalar theories \cite{Litim:2001hk} & fRG Taylor & 0.4022*& 5.00* &    & 0.8043 &     &      \\[1ex]
    $O(4)$ scalar theories LPA\cite{Braun:2007td} &fRG Taylor&0.4030(30)& 4.973(30) &        & 0.8053(60) &       &        \\[1ex]    
    $O(4)$ QM LPA \cite{Stokic:2009uv} & fRG Taylor &  0.402 & 4.818 &  1.575 &  0.787  & 0.396 &  \\[1ex]
    $O(4)$ scalar theories \cite{Bohr:2000gp} & fRG Grid & 0.40 & 4.79 &     & 0.78 &     & 0.037 \\[1ex]
    $Z(2)$ scalar theories \cite{Bohr:2000gp} & fRG Grid & 0.32 & 4.75 &     & 0.64 &     & 0.044 \\[1ex]
    $O(4)$ scalar theories \cite{DePolsi:2020pjk} & fRG DE $\mathcal{O}(\partial^4)$ &   &    &     & 0.7478(9) &     & 0.0360(12) \\[1ex]
    $Z(2)$ scalar theories \cite{Balog:2019rrg,DePolsi:2020pjk} & fRG DE $\mathcal{O}(\partial^6)$ &   &    &     & 0.63012(5) &     & 0.0361(3) \\[1ex]
    $O(4)$ CFTs \cite{Kos:2015mba} & conformal bootstrap &   &    &     & 0.7472(87) &     & 0.0378(32) \\[1ex]
    $Z(2)$ CFTs \cite{Kos:2014bka} & conformal bootstrap &   &    &     & 0.629971(4) &     & 0.0362978(20) \\[1ex]
    $O(4)$ spin model \cite{Kanaya:1994qe}  &Monte Carlo&0.3836(46) & 4.851(22) & 1.477(18) & 0.7479(90) & 0.4019(71)* & 0.025(24)*  \\[1ex]
    $Z(2)$ $d=3$ expansion \cite{ZinnJustin:1999bf} & summed perturbation&0.3258(14) &4.805(17)*  &1.2396(13) &0.6304(13) &0.4027(23) &0.0335(25)      \\[1ex]
    Mean Field  && 1/2 &3&1&1/2&1/3 &0         \\[1ex]
    \hline\hline
  \end{tabular}
  \caption{Comparison of the critical exponents for the 3$d$ $O(4)$ and $Z(2)$ symmetry universality classes from different approaches. The values with an asterisk are derived from the scaling relations in \Eq{eq:relationExpo}. The table is adopted from \cite{Chen:2021iuo}.} 
  \label{tab:exponent}
  \end{center}\vspace{-0.5cm}
\end{table*}
%

%
\begin{figure}[t]
\includegraphics[width=0.48\textwidth]{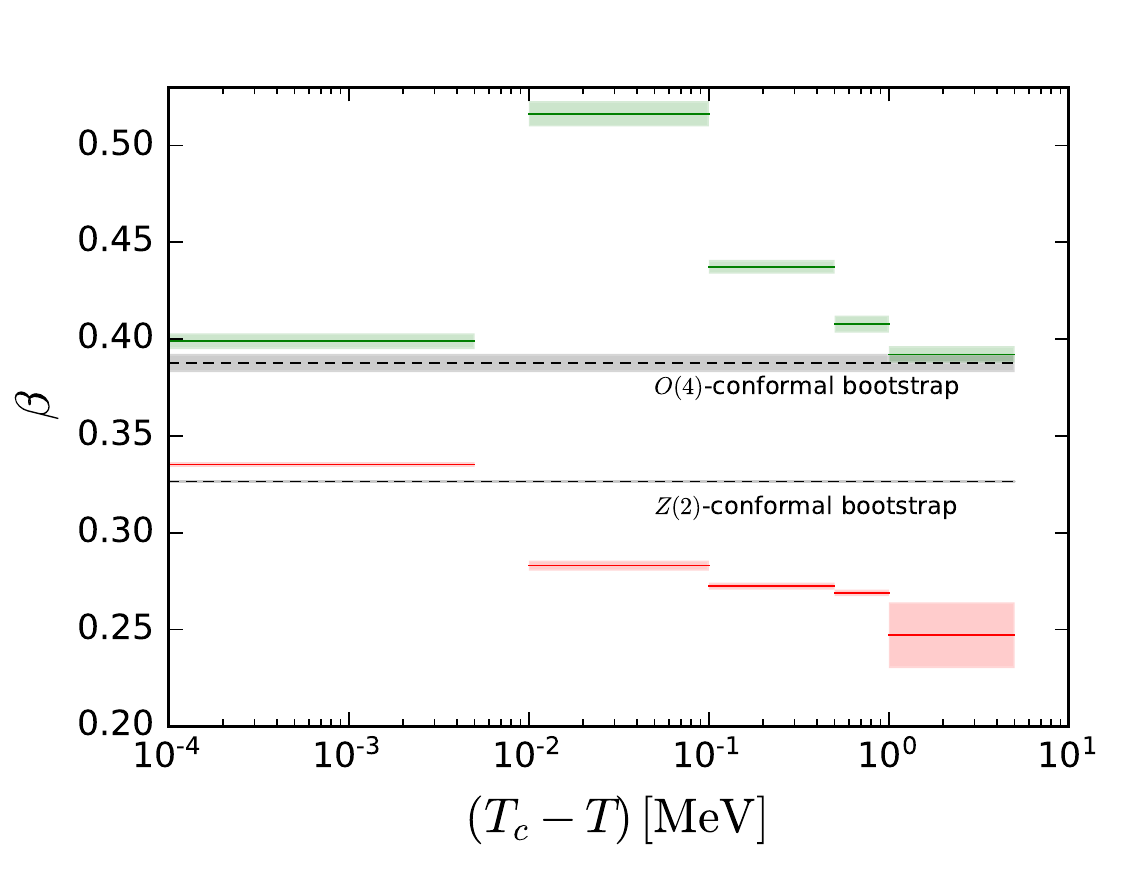}
\caption{Critical exponent $\beta$ for the 3$d$ $O(4)$ (green lines) and $Z(2)$ (red lines) symmetry universality classes extracted from different ranges of the temperature, obtained within the fRG approach in \cite{Chen:2021iuo}. The conformal bootstrap results (gray dashed lines) \cite{Kos:2014bka,Kos:2015mba} are also presented for comparison. The plot is adopted from \cite{Chen:2021iuo}.}\label{fig:beta-scalreg}
\end{figure}
%

According the scaling argument, when a system is in the critical regime, its thermodynamic potential density can be decomposed into a sum of a singular and a regular part \cite{Ma:2020a}, viz.,
\begin{align}
  \Omega\big(t,h\big)&=f_s(t,h)+f_{\mathrm{reg}}(t,h)\,,\label{eq:scalingfunc}
\end{align}
whereof, the singular part satisfies the scaling relation to the leading order as follows
\begin{align}
  f_s(t,h)&=\ell^{-d}f_s(t\,\ell^{y_t},\,h\,\ell^{y_h})\,,\label{eq:scalingrelation}
\end{align}
with a dimensionless rescaling factor $\ell$ and the spacial dimension $d$. In Eqs. (\ref{eq:scalingfunc}) and (\ref{eq:scalingrelation}) $t$ and $h$ stand for the reduced temperature and magnetic field, respectively. They are given by
\begin{align}
  t&=\frac{T-T_c}{T_0}\,,\qquad h=\frac{c}{c_0}\,, \label{eq:th}
\end{align}
where $T_c$ is the critical temperature in the chiral limit, and $T_0$ and $c_0$ are some normalization values for the temperature $T$ and the strength of the explicit chiral symmetry breaking $c$ as shown in \Eq{eq:QMaction}, respectively. In the following, the order parameter will be denoted as the magnetization density, and $c$ as the magnetic field strength, i.e.,
\begin{align}
  M&\equiv \sigma\,,\qquad H\equiv c\,. \label{}
\end{align}
The scaling relation in \Eq{eq:scalingrelation} leaves us with a set of scaling relations for various critical exponents, cf.\cite{Ma:2020a, Braun:2007td}, such as,
\begin{align}
  y_t&=\frac{1}{\nu}\,,\quad\!\! y_h=\frac {\beta\delta}{\nu}\,,\quad\!\! \beta=\frac{\nu}{2} (d-2+\eta)\,,\quad\!\! \gamma=\beta(\delta-1)\,, \nonumber\\[2ex]
\gamma&=(2-\eta)\nu\,,\quad\delta=\frac {d+2-\eta}{d-2+\eta}\,,\quad \nu d=\beta(1+\delta)\,.\label{eq:relationExpo}
\end{align}
The critical behavior of the order parameter is described by the critical exponents $\beta$ and $\delta$, which read
\begin{align}
  M(t,h=0)&\sim (-t)^{\beta}\,,\label{eq:beta} 
\end{align}
with $t<0$, and 
\begin{align}
  M(t=0,h)&\sim h^{1/\delta}\,.\label{eq:delta}
\end{align}
Moreover, one has the susceptibility of the order parameter $\chi$ and the correlation length $\xi$, which scale as
\begin{align}
  \chi&\sim |t|^{-\gamma}\,,\qquad \mathrm{and}\qquad \xi\sim |t|^{-\nu}\,.\label{eq:defiGamNu}
\end{align}

Recently, the pseudo-spectral method of the Chebyshev expansion for the effective potential has been used to calculate the critical exponents \cite{Chen:2021iuo}. In the $N_f=2$ QM model within the fRG approach, the critical exponents for the 3$d$ $O(4)$ and $Z(2)$ symmetry universality classes, which correspond to the black dashed $O(4)$ phase transition line and the red dashed $Z(2)$ phase transition line as shown in \Fig{fig:phasedia-Chen2021iuo} respectively, are calculated with the Chebyshev expansion of the effective potential. Both the LPA and LPA$'$ approximations are used in the calculations. In the LPA$'$ approximation a field-dependent mesonic wave function renormalization is taken into account. The relevant results are show in \tab{tab:exponent}, which are also compared with results of critical exponents from other approaches, e.g., Taylor expansion of the effective potential \cite{Tetradis:1993ts, VonGersdorff:2000kp, Litim:2001hk, Braun:2007td, Stokic:2009uv} and the grid method \cite{Bohr:2000gp} within the fRG approach, the derivative expansions (DE) up to orders of $\mathcal{O}(\partial^4)$ and $\mathcal{O}(\partial^6)$ with the fRG approach \cite{Balog:2019rrg, DePolsi:2020pjk}, the conformal bootstrap for the 3$d$ conformal field theories (CFTs) \cite{Kos:2014bka, Kos:2015mba}, Monte Carlo simulations \cite{Kanaya:1994qe}, and the $d=3$ perturbation expansion \cite{ZinnJustin:1999bf}. In the mean time, the mean-field values of the critical exponents are also shown in the last line of \tab{tab:exponent}.

In \Fig{fig:beta-scalreg} the critical exponent $\beta$ for the 3$d$ $O(4)$ and $Z(2)$ symmetry universality classes extracted from different ranges of the temperature is shown, which is obtained with the Chebyshev expansion for the effective potential in the fRG \cite{Chen:2021iuo}. It is found that only when the temperature is very close to the critical temperature, say $|T-T_c|\lesssim 0.01$ MeV, the fRG results of both the $O(4)$ and $Z(2)$ symmetry universality classes are consistent with the conformal bootstrap results. This indicates that the size of the critical region is extremely small, far smaller than $\sim 1$ MeV. Similar estimates for the size of the critical region are also found in, e.g., \cite{Schaefer:2006ds, Braun:2007td, Braun:2010vd, Klein:2017shl}.

	
\section{QCD at finite temperature and density}
\label{sec:QCD}

In this section we would like to present a brief review on recent progresses in studies of QCD at finite temperature and densities within the fRG approach to the first-principle QCD, which are mainly based on the work in \cite{Fu:2019hdw, Braun:2020ada}. The relevant fRG flows at finite temperature and densities there have been developed from the counterparts in the vacuum in \cite{Braun:2014ata, Rennecke:2015eba}. Furthermore, these flows are also built upon the state-of-the-art quantitative fRG results for Yang-Mills theory in the vacuum  \cite{Cyrol:2016tym} and at finite temperature \cite{Cyrol:2017qkl}, QCD in the vacuum \cite{Mitter:2014wpa, Cyrol:2017ewj}.

In the following we focus on the chiral phase transition. As mentioned above, the QCD phase transitions involve both the chiral phase transition and the confinement-deconfinement phase transition. In fact, related subjects of the deconfinement phase transition, such as the Wilson loop, the Polyakov loop, the flows of the effective action of a background gauge field, the relation between the confinement and correlation functions, etc., have been widely studied within the fRG approach to Yang-Mills theory and QCD. Significant progresses have been made in relevant studies, see \cite{Dupuis:2020fhh} for a recent overview, and also e.g. \cite{Schaefer:2001cn, Pawlowski:2003sk, Fischer:2004uk, Fischer:2006vf, Braun:2007bx, Fischer:2008uz, Fischer:2009tn, Braun:2009gm, Braun:2010cy, Eichhorn:2010zc, Fister:2011uw, Fister:2013bh, Fischer:2013eca, Haas:2013qwp, Herbst:2015ona, Corell:2018yil, Horak:2022aqx, Pawlowski:2022oyq}.

\subsection{Propagators and anomalous dimensions}
\label{subsec:PropQCD}

%
\begin{figure}[t]
\includegraphics[width=0.48\textwidth]{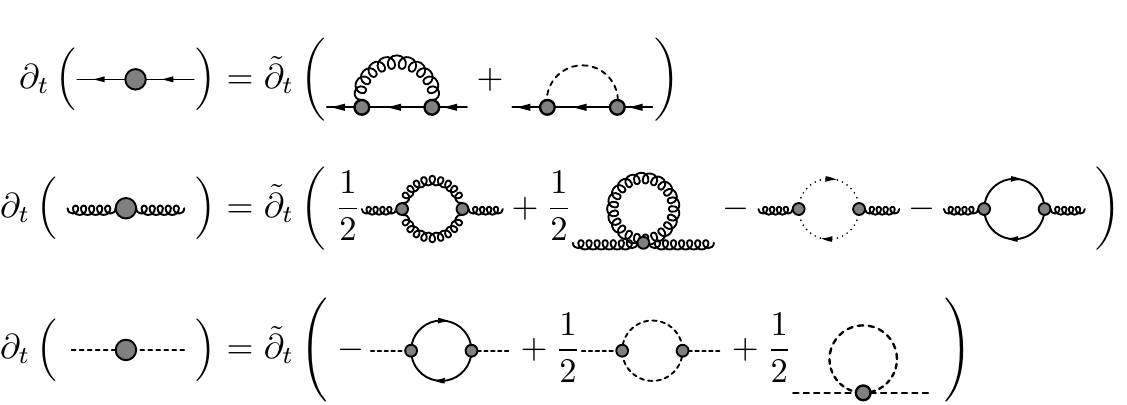}
\caption{Diagrammatic representation of the flow equations for the inverse quark, gluon and meson propagators, respectively, where the dashed lines stand for the mesons and the dotted lines denote the ghost.}\label{fig:prop-equ}
\end{figure}
%

%
\begin{figure*}[t]
	\includegraphics[width=0.98\textwidth]{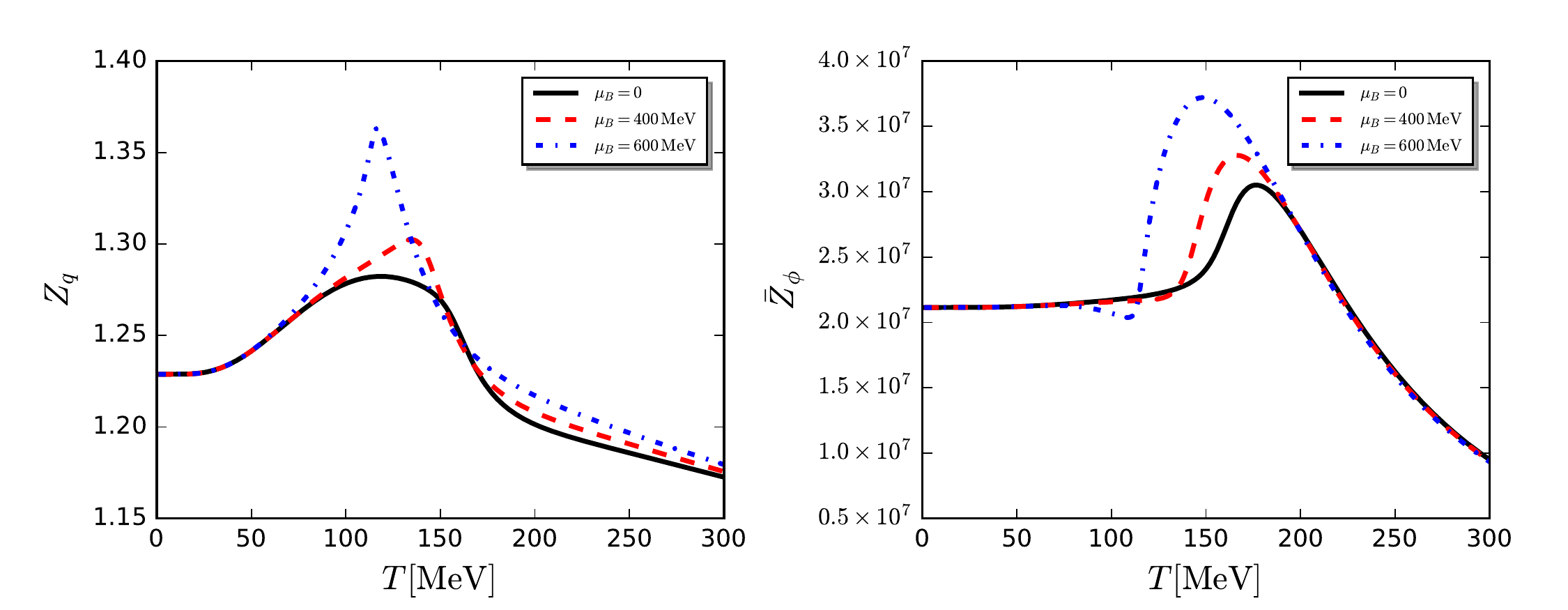}
	\caption{Quark (left panel, $Z_q$) and mesonic (right panel, $\bar Z_\phi$) wave function renormalizations of $N_f=2+1$ flavor QCD obtained in the fRG with RG scale $k=0$, as functions of the temperature with different values of the baryon chemical potential. The plot is adopted from \cite{Fu:2019hdw}.}\label{fig:Zpsi}
\end{figure*}
%

%
\begin{figure}[t]
	\includegraphics[width=0.98\columnwidth]{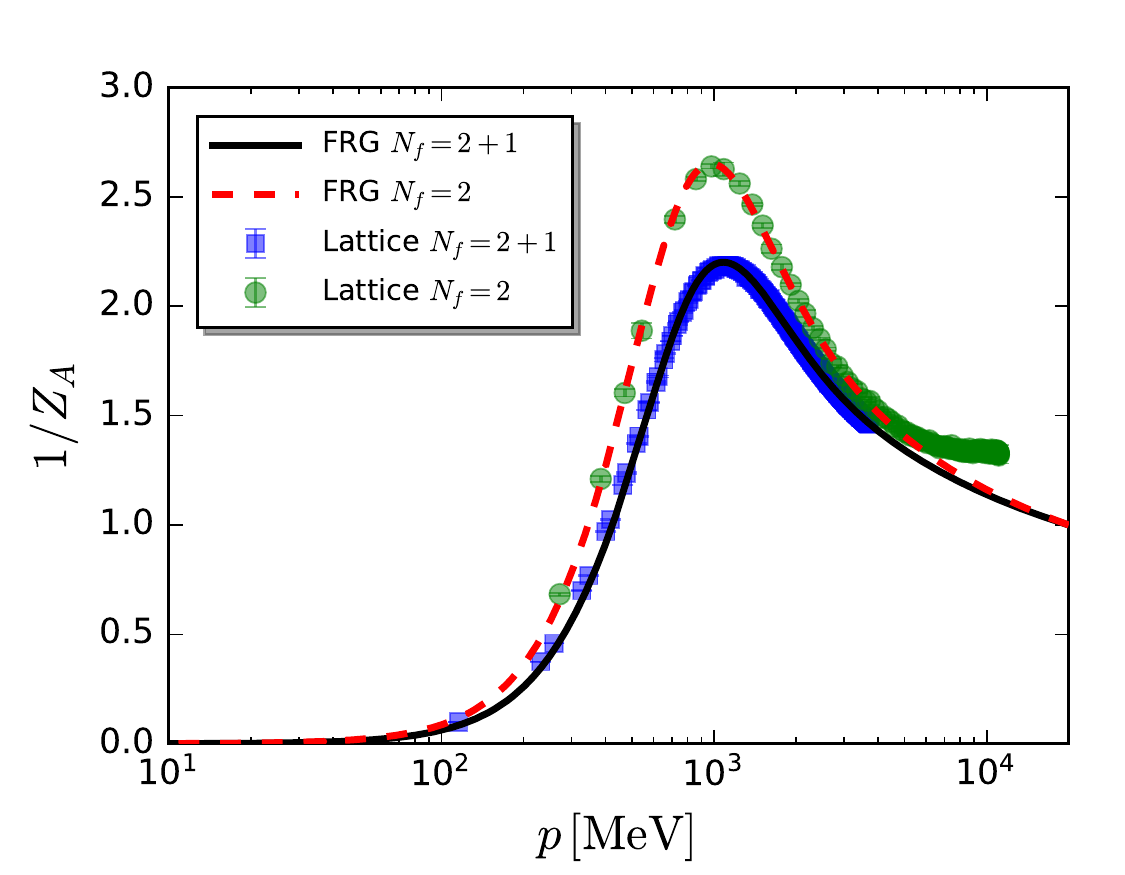}
	\caption{Gluon dressing functions $1/Z_{A}(p)$ of $N_f=2$ and $N_f=2+1$ flavor QCD in the vacuum as functions of the momentum, where the fRG results are in comparison to lattice calculations of $N_f=2$ flavors \cite{Sternbeck:2012qs} and $N_f=2+1$ flavors \cite{Zafeiropoulos:2019flq, Boucaud:2018xup}.The plot is adopted from \cite{Fu:2019hdw}.}\label{fig:inZA_lattice2-2+1}
\end{figure}
%

%
 \begin{figure}[t]
 	\includegraphics[width=0.98\columnwidth]{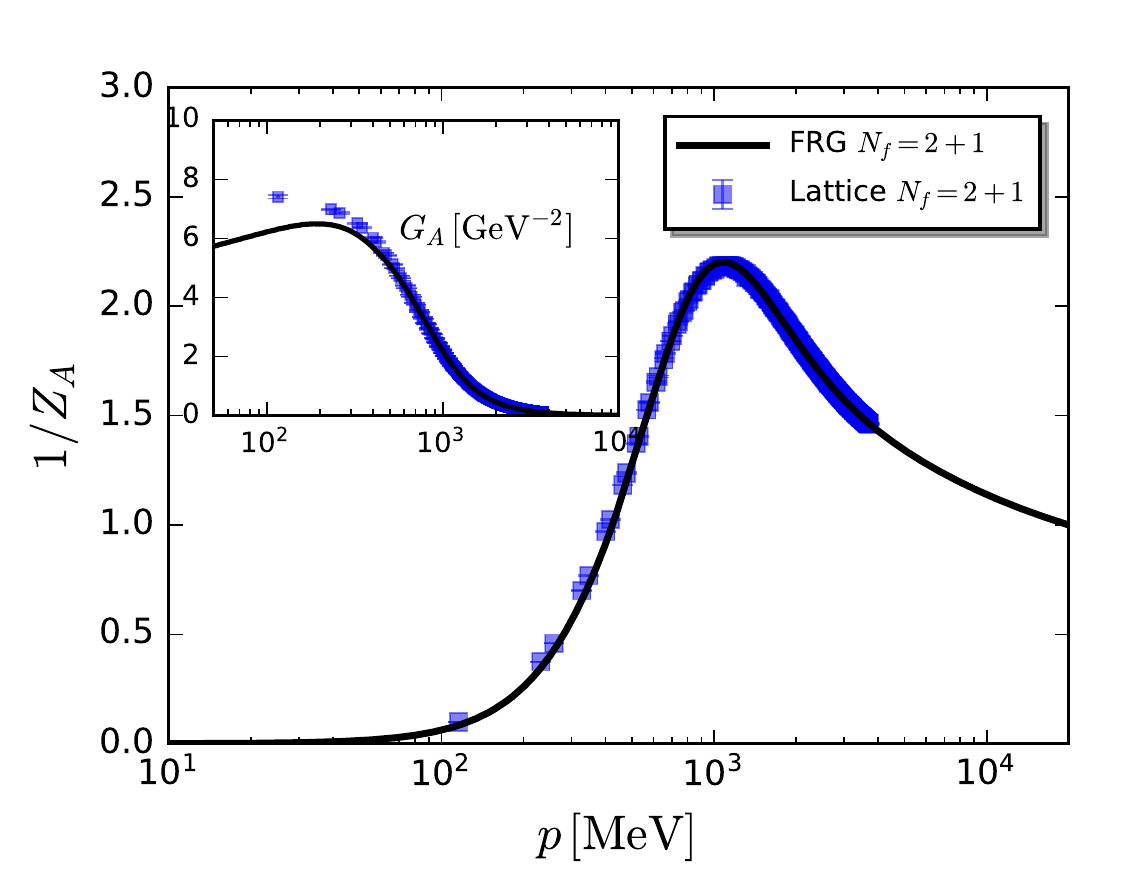}
 	\caption{Gluon dressing function $1/Z_{A}(p)$ and the gluon propagator $G_A=1/[Z_{A}(p)\, p^2]$ (inlay) as functions of momenta for QCD of $N_f=2+1$ flavors in the vacuum, where the fRG results denoted by the black lines are compared with the continuum extrapolated lattice results by the RBC/UKQCD collaboration, see e.g., \cite{Blum:2014tka,Zafeiropoulos:2019flq,Boucaud:2018xup}. For the fRG results the momentum dependence is given by $p^2=k^2$. The plot is adopted from \cite{Fu:2019hdw}.}\label{fig:inZA_lattice}
 \end{figure}
%

%
\begin{figure*}[t]
	\includegraphics[width=0.98\textwidth]{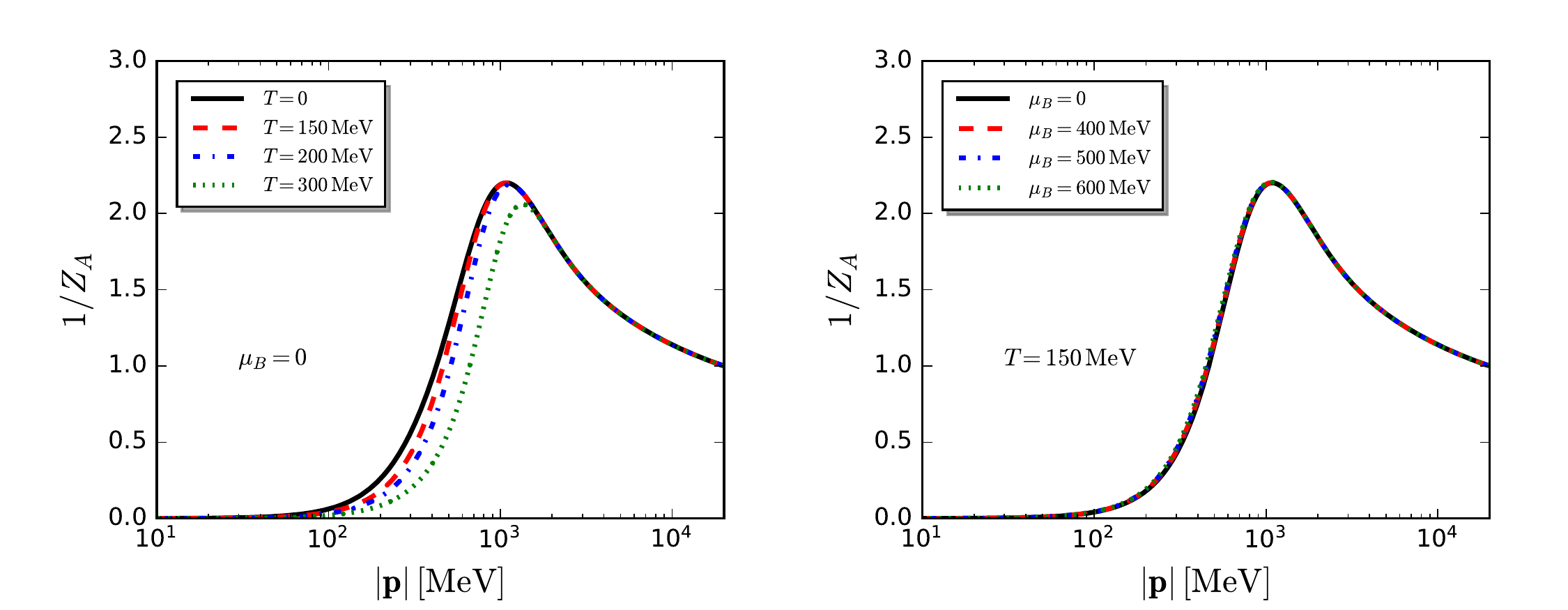}
	\caption{Gluon dressing function $1/Z_{A}$ of the $N_f=2+1$ QCD as a function of spatial momenta $|{\bm p}|$ with several different values of temperature (left panel) and baryon chemical potential (right panel), where the Matsubara frequency $p_0$ is chosen to be vanishing. The identification $\bm {p}^2=k^2$ is used. The plot is adopted from \cite{Fu:2019hdw}.}\label{fig:inZA}
\end{figure*}
%

%
\begin{figure}[t]
	\includegraphics[width=0.98\columnwidth]{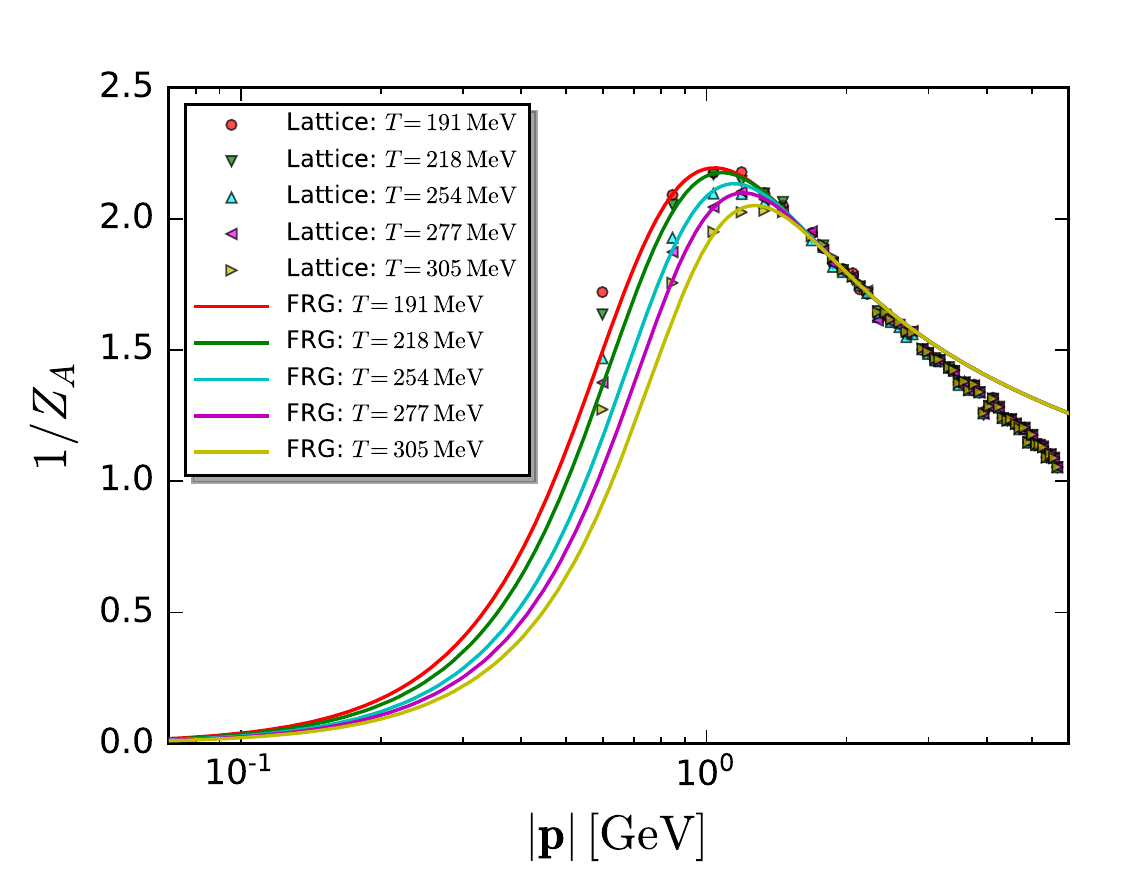}
	\caption{Gluon dressing function $1/Z_{A}$ of $N_f =2 +1$ flavors as a function of spatial momenta $|{\bm p}|$ at several values of temperature obtained in fRG, which is also compared with the lattice results of $N_f=2+1+1$ in \cite{Ilgenfritz:2017kkp}. The plot is adopted from \cite{Fu:2019hdw}.}\label{fig:inZAlattT}
\end{figure}
%

The flow equations of inverse propagators are obtained by taking the second derivative of the Wetterich equation in \Eq{eq:FlowQCD} with respect to respective fields. The resulting flow equations of the inverse quark, gluon and meson propagators are shown diagrammatically in \Fig{fig:prop-equ}. Making appropriate projections for these flow equations, one is able to arrive at the flow of the wave function renormalization $Z_{\Phi,k}$ as shown in \Eq{eq:barPhi} for a given field $\Phi$, which includes nontrivial dispersion relation for the field, and is more conveniently reformulated as the anomalous dimension as follows
\begin{align}
  \eta_{\Phi,k} &=-\frac{\partial_t Z_{\Phi,k}}{Z_{\Phi,k}}\,.  \label{}
\end{align}
The quark two-point correlation function defined in \Eq{eq:Gamma2bqq}, see also \Eq{eq:Gamma2bqq2}, reads
\begin{align}
 \Gamma^{(2)\bar q q}_{k}(p)=& -\frac{\delta^2 \Gamma_k[\Phi]}{\delta \bar{q}(-p)\delta q(p)}\bigg|_{\Phi=\Phi_{\mathrm{EoM}}}\nonumber\\[2ex]
=&Z_{q,k}(p)i \gamma\cdot p+m_{q,k}(p)\,,\label{eq:Gamma2bqq3}
\end{align}
where $\Phi_{\mathrm{EoM}}$ denotes the fields in \Eq{eq:Phi} on their respective equations of motion, that are vanishing except the $\sigma$ field and the temporal gluon field $A_0$. Note that the minus sign on the r.h.s. of \Eq{eq:Gamma2bqq3}, while not appearing in \Eq{eq:Gamma2bqq2}, is due to the fact that the right derivative is used in \Eq{eq:Gamma2bqq2}. Moreover, the quark chemical potentials in \Eq{eq:QCDaction} have been assumed to be vanishing in \Eq{eq:Gamma2bqq3}. Apparently, the quark wave function renormalization and mass are readily obtained by projecting \Eq{eq:Gamma2bqq3} onto the vector and scalar channels, respectively, which read
\begin{align}
 Z_{q,k}(p)=& \frac{1}{4 i } \frac{1}{p^2 }\tr \Big(\gamma\cdot p \, \Gamma^{(2)\bar q q}_{k}(p)\Big)\nonumber\\[2ex]
 m_{q,k}(p)=&\frac{1}{4  } \tr \Big(\Gamma^{(2)\bar q q}_{k}(p)\Big)\,,\label{}
\end{align}
where the trace runs only in the Dirac space. Neglecting the dependence of the quark wave function renormalization on the spacial momentum $\bm{p}$, one arrives at the quark anomalous dimension as follows
\begin{align}
 \eta_{q, k}(p_0)=&\frac{1}{4 Z_{q,k}(p_0)}\nonumber \\[2ex]
          &\times\mathrm{Re}\left[\frac{\partial}{\partial (|\bm{p}|^2)}\mathrm{tr}
            \left(i \bm{\gamma}\cdot\bm{p}\big(\partial_t \Gamma^{(2)\bar q q}_{k}(p)\big)\right)\right]_{\bm{p}=0}\,,\label{eq:etapsi}
\end{align}
where the computation is done at $\bm{p}=0$, and $p_0$ is a small, but nonvanishing frequency. The nontrivial choice of $p_0$ and the fact that the expression in the square bracket in \Eq{eq:etapsi} is complex-valued at nonzero chemical potentials, are both related to constraints from the Silver-Blaze property for the correlation functions at finite chemical potentials, and see, e.g., \cite{Marko:2014hea, Khan:2015puu, Fu:2015amv, Fu:2016tey, Fu:2019hdw} for more details. In \Eq{eq:etapsi} Re denotes that the real part of the expression in the square bracket is adopted. Note that in \Eq{eq:etapsi} the difference between the parts of the quark wave function renormalization longitudinal and transversal to the heat bath is ignored, where the projection is done on the spacial component, and thus the transversal anomalous dimension is used. The explicit expression of $\eta_{q, k}$ is given in \Eq{eq:etapsiexp}. In the left panel of \Fig{fig:Zpsi}, the quark wave function renormalization $Z_{q,k=0}(p_0,\bm{p}=0)$ is depicted as a function of the temperature with different values of the baryon chemical potential.

The mesonic two-point correlation functions, i.e., those of the $\pi$ and $\sigma$ fields, are given by
\begin{align}
 \Gamma^{(2)\pi \pi}_{k}(p)=& \frac{\delta^2 \Gamma_k[\Phi]}{\delta \pi_i(-p)\delta \pi_j(p)}\bigg|_{\Phi=\Phi_{\mathrm{EoM}}}\nonumber\\[2ex]
=&\big(Z_{\pi,k}(p) p^2+m_{\pi,k}^2\big)\delta_{ij}\,,\label{eq:Gam2pi2}
\end{align}
and 
\begin{align}
 \Gamma^{(2)\sigma \sigma}_{k}(p)=& \frac{\delta^2 \Gamma_k[\Phi]}{\delta \sigma(-p)\delta \sigma(p)}\bigg|_{\Phi=\Phi_{\mathrm{EoM}}}\nonumber\\[2ex]
=&Z_{\sigma,k}(p) p^2+m_{\sigma,k}^2\,,\label{eq:Gam2sigma2}
\end{align}
respectively, where the curvature masses of the mesons read
\begin{align}
 m_{\pi,k}^2=& V_k^{\prime}(\rho)\,,\qquad m_{\sigma,k}^2=& V_k^{\prime}(\rho)+2 \rho V_k^{(2)}(\rho)\,.\label{}
\end{align}
As same as the quark wave function renormalization, the thermal splitting of the mesonic wave function renormalization in parts longitudinal and transversal to the heat bath is neglected. Moreover, one has assumed a unique renormalization for both the pion and sigma fields, viz., $Z_{\phi,k}=Z_{\pi,k}=Z_{\sigma,k}$. The validity of these approximations have been verified seriously in \cite{Yin:2019ebz}. The anomalous dimension of mesons at vanishing frequency $p_0=0$ and a finite spacial momentum $\bm{p}$ is given by
\begin{align}
 \eta_{\phi}(0,\bm{p})=&-\frac{\delta_{ij}}{3Z_{\phi}(0,\bm{p})}\frac{\partial_t \Gamma_{\pi_i\pi_j}^{(2)}(0,\bm{p})
  -\partial_t \Gamma_{\pi_i\pi_j}^{(2)}(0,0)}{\bm{p}^2} \,,\label{eq:etaphipk}
\end{align}
Two cases are of special interest: One is the choice $\bm{p}^2=k^2$, i.e., $\eta_{\phi}(0,k)$, which captures the most momentum dependence of the mesonic two-point correlation functions, and thus is usually used in the r.h.s. of flow equations involving mesonic degrees of freedom, for more detailed discussions see \cite{Fu:2019hdw}. The anomalous dimension $\eta_{\phi}(0,k)$ leaves us with the renormalization constant defined as follows
\begin{align}
  \frac{1}{\bar Z_{\phi,k}} \partial_t \bar Z_{\phi,k} \equiv - \eta_\phi(0,k)\,,
  \quad \textrm{with}\quad
  \bar Z_{\phi,k=\Lambda}=1\,. \label{eq:dtbarZphi}
\end{align}
 In the right panel of \Fig{fig:Zpsi}, $\bar Z_{\phi,k=0}$ is depicted as a function of the temperature with different values of the baryon chemical potentials. The other is the case $\bm{p}=0$, and \Eq{eq:etaphipk} is reduced the equation as follows
\begin{align}
  \eta_{\phi}(0)&=-\frac{1}{3Z_{\phi}(0)}\delta_{ij}\left[\frac{\partial}{\partial  \bm{p}^2}\partial_t \Gamma^{(2)}_{\pi_i\pi_j}\right]_{p=0}\,. \label{eq:etaphi}
\end{align}
The related wave function renormalization $Z_{\phi,k}(0)$ is used to extract the renormalized meson mass as
\begin{align}
  \bar m_{\pi}&=\frac{m_{\pi,k=0}}{\sqrt{Z_{\phi,k=0}(0)}}\,,\qquad \bar m_{\sigma}=\frac{m_{\sigma,k=0}}{\sqrt{Z_{\phi,k=0}(0)}}\,, \label{}
\end{align}
where $\bar m_{\pi}$ and $\bar m_{\sigma}$ are approximately equal to their pole masses, respectively. The explicit expressions for \Eq{eq:etaphi} and \Eq{eq:etaphipk} are presented in \Eq{eq:etaphiexp} and \Eq{eq:etaphipkexp}, respectively.

For the ghost anomalous dimension, one extracts it from the relation as follows
\begin{align}
  \eta_{c}=&-\left. \frac{p \partial_p Z^{\mathrm{QCD}}_{c, k=0}(p)}{Z^{\mathrm{QCD}}_{c, k=0}(p)}\right|_{p=k}\,,\label{eq:etaCT}
\end{align}
where $Z^{\mathrm{QCD}}_{c, k=0}(p)$ denotes the momentum-dependent ghost wave function renormalization in QCD with $N_f=2$ flavors in the vacuum obtained in \cite{Cyrol:2017ewj}. Note that the ghost propagator is found to be very insensitive to the effects of finite temperature as well as the quark contributions for $N_f=2$ and $N_f=2+1$ flavors, see e.g. \cite{Cyrol:2017qkl}.

The gluon anomalous dimension is decomposed into a sum of three parts, as follows
\begin{align}
  \eta_{A}=&\eta_{A,\mathrm{vac}}^{\mathrm{QCD}}+\Delta\eta_{A}^{\mathrm{glue}}+\Delta\eta_{A}^{q}\,,\label{eq:etaAT}
\end{align}
where the first term on the r.h.s. denotes the gluon anomalous dimension in the vacuum, and the other two terms stand for the medium contributions to the gluon anomalous dimension from the glue (gluons and ghosts) and quark loops, respectively. The vacuum contribution in \Eq{eq:etaAT} is further expressed as
 \begin{align}
  \eta_{A,\mathrm{vac}}^{\mathrm{QCD}}=&\left. \eta_{A,\mathrm{vac}}^{\mathrm{QCD}}\right|_{N_f=2}+\eta^s_{A,\mathrm{vac}}\,,\label{eq:etaAQCD2+1}
\end{align}
where the two terms on the r.h.s. denote the contributions from the light quarks of $N_f=2$ flavors and the strange quark. The former one is inferred from the gluon dressing function $Z^\textrm{QCD}_{A,k=0}(p)$ for the $N_f=2$ flavor QCD in \cite{Cyrol:2017ewj}, which reads
\begin{align}
  \left.\eta_{A,\mathrm{vac}}^{\mathrm{QCD}}\right|_{N_f=2}=&-\left. \frac{p \partial_p Z^{\mathrm{QCD}}_{A, k=0}(p)}{Z^{\mathrm{QCD}}_{A, k=0}(p)}\right|_{p=k}\,. \label{eq:etaAQCD2}
\end{align}
The explicit expressions of $\Delta\eta_{A}^{q}$ in \Eq{eq:etaAT} and $\eta^s_{A,\mathrm{vac}}$ in \Eq{eq:etaAQCD2+1} can be found in \Eq{eq:DeltaAqexpl}, \Eq{eq:DeltaAqexplMed}, \Eq{eq:etaAs}. Moreover, the in-medium contribution to the gluon anomalous dimension resulting from the glue sector, i.e., the second term on the r.h.s. of \Eq{eq:etaAT}, is taken into account in \Eq{eq:barEtaA}.

In \Fig{fig:inZA_lattice2-2+1} the gluon dressing functions $1/Z_{A}(p)$ of $N_f=2$ and $N_f=2+1$ flavor QCD in the vacuum are shown. The fRG and lattice results are presented. Here the fRG gluon dressing of $N_f=2$ flavors is inputted from \cite{Cyrol:2017ewj}, as shown in \Eq{eq:etaAQCD2}, which is also in quantitative agreement with the lattice result in \cite{Sternbeck:2012qs}. Note that the gluon dressing of $N_f=2+1$ flavors here is a genuine prediction, which is in good agreement with the respective lattice results \cite{Zafeiropoulos:2019flq, Boucaud:2018xup}. In \Fig{fig:inZA_lattice} both the gluon dressing function and the gluon propagator of $N_f=2+1$ flavors are presented. The calculated gluon dressing functions at finite temperature and baryon chemical potential in fRG are shown in \Fig{fig:inZA}. In the left panel of \Fig{fig:inZA} several different values of temperature are chosen with $\mu_B=0$, and it is found that the gluon dressing function $1/Z_{A}$ decreases with the increase of temperature. In the right panel, several different values of $\mu_B$ are adopted while with $T=150$ MeV fixed. It is observed that the dependence of the gluon dressing function on the baryon chemical potential is very small. The gluon dressing functions at finite temperature obtained in fRG are compared with the relevant lattice results from \cite{Ilgenfritz:2017kkp} in \Fig{fig:inZAlattT}, where several different values of temperature are chosen. One can see that the gluon dressing at finite temperature in fRG is comparable to that of lattice QCD.

\subsection{Strong couplings}
\label{subsec:StrongcouplingQCD}

%
\begin{figure}[t]
	\includegraphics[width=0.97\columnwidth]{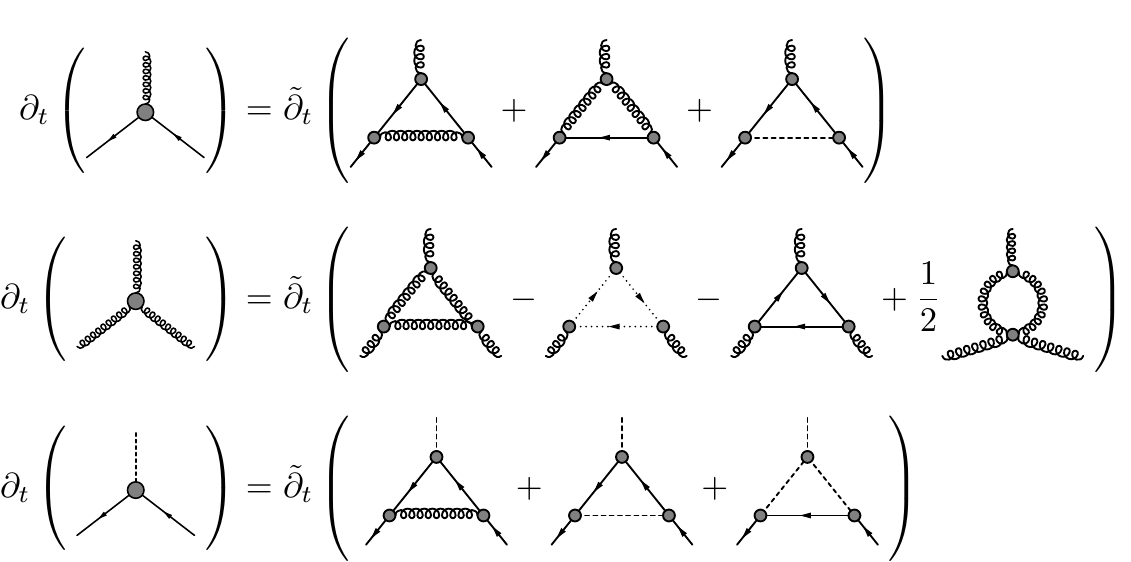}
	\caption{Diagrammatic representation of the flow equations for the quark-gluon, three-gluon, and the quark-meson vertices, respectively.}\label{fig:vertex3point-equ}
\end{figure}
%

%
\begin{figure}[t]
\includegraphics[width=0.98\columnwidth]{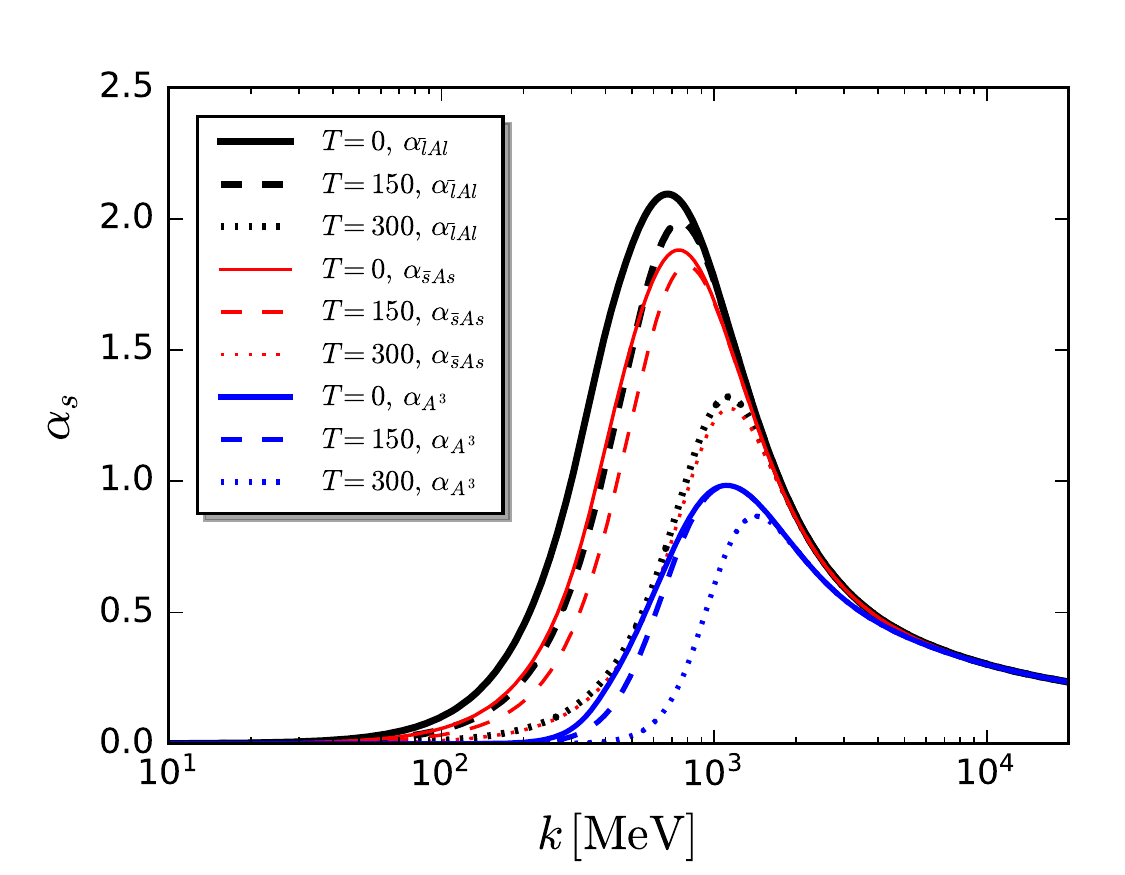}
\caption{Quark-gluon couplings for light quarks ($\alpha_{\bar l l A}$) and strange quarks ($\alpha_{\bar s s A }$), and the three-gluon coupling ($\alpha_{A^3}$) of $N_f=2+1$ flavor QCD as functions of the RG scale $k$ at several values of the temperature and vanishing baryon chemical potential. The plot is adopted from \cite{Fu:2019hdw}.}\label{fig:alphas}
\end{figure}
%

The quark-gluon coupling in \Eq{eq:Dmu}, the ghost-gluon coupling in \Eq{eq:Dmuab}, and the three-gluon or four-gluon coupling in \Eq{eq:Fmunua}, etc., are consistent with one another in the perturbative regime of high energy, due to the Slavnov-Taylor identities (STIs) resulting from the gauge symmetry. However, when the momenta or the RG scale are decreased to $k\lesssim 1\sim 3\,$ GeV, the gluon mass gap begins to affect the running of strong couplings, and also the transversal strong couplings can not be identified with the longitudinal ones via the modified STIs, see, e.g., \cite{Cornwall:1981zr, Aguilar:2011ux, Cyrol:2016tym, Pawlowski:2022oyq} for more details. Consequently, different strong couplings deviate from one another in the low energy regime of $k\lesssim 1\sim 3\,$ GeV \cite{Mitter:2014wpa, Cyrol:2016tym, Cyrol:2017ewj, Cyrol:2017qkl}, and thus it is necessary to distinguish them by adding appropriate suffixes. The couplings of the purely gluonic sector read
\begin{align}
  \alpha_{A^3,k} =& \frac{1}{4\pi} \frac{\lambda_{A^3,k}^2}{Z_{A,k}^{3}}\,,\qquad \alpha_{A^4,k} =\frac{1}{4\pi}  \frac{\lambda_{A^4,k}}{Z_{A,k}^{2}}\,,\\[2ex]
  \alpha_{\bar c c A,k} =& \frac{1}{4\pi} \frac{\lambda_{\bar c c A,k}^2}{Z_{A,k}\,Z_{c,k}^2 }\,, \label{eq:alphasGluon}
\end{align}
where $\lambda_{A^3,k}$, $\lambda_{A^4,k}$ and $\lambda_{\bar c c A,k}$ denote the three-gluon, four-gluon, ghost-gluon dressing functions, respectively, as shown in \Eq{eq:VertexA3}, \Eq{eq:VertexA4}, \Eq{eq:VertexbccA}. The couplings of the matter sector read
\begin{align}
  \alpha_{\bar l  l A,k} =& \frac{1}{4\pi}\frac{\lambda_{\bar l  l A,k}^2}{Z_{A,k}\,Z_{q,k}^2 }\,, \qquad
  \alpha_{\bar s  s A,k} = \frac{1}{4\pi}\frac{\lambda_{\bar s  s A,k}^2}{Z_{A,k}\,Z_{q,k}^2 }\,,\label{eq:alphasMatter}
\end{align}
where the light-quark-gluon coupling and the strange-quark-gluon coupling have been distinguished. From the calculated results in, e.g., \cite{Braun:2014ata, Cyrol:2017ewj}, it is reasonable to adopt the approximation as follows
\begin{align}
  \alpha_{A^4,k} = \alpha_{A^3,k}\,,\qquad  \alpha_{\bar c c A,k} \simeq\alpha_{\bar l  l A,k} \,.\label{eq:IDalphaGlue}
\end{align}

The flow equation of the quark-gluon vertex is presented in the first line of \Fig{fig:vertex3point-equ}. Projecting onto the classical tensor structure of the quark-gluon vertex, denoted by
\begin{align}
  \big(S^{(3)}_{\bar{q}qA}\big)^{a}_{\mu}&\equiv -i \gamma_{\mu}t^{a} \,,\label{eq:S3bqqA}
\end{align}
one is led to the flow of the quark-gluon coupling as follows
\begin{align}
  \partial_t \bar g_{\bar q  q A,k}=&\left( \frac{1}{2}\eta_A+\eta_q\right) \bar g_{\bar q  q A,k}+\frac{1}{8(N_c^2-1)}\nonumber\\[2ex]
  &\times \tr\,\left[\left(\overline{{\mathrm{Flow}}}^{(3)}_{\bar{q} q A}\right)^{a}_{\mu} 
    \left( S^{{(3)}}_{\bar{q} q A}\right)^a_{\mu}\right]\big(\{p\})\,,
 \label{eq:dtg}
\end{align}
where the trace sums over the Dirac and color spaces, and $\{p\}$ stands for the set of external momenta for the vertex. Here $g_{\bar q  q A,k}\equiv \lambda_{\bar q  q A,k}$ is used. In \Eq{eq:dtg} the flow of quark-gluon vertex, i.e., the r.h.s. of the first line in \Fig{fig:vertex3point-equ}, is denoted by
\begin{align}
  \left(\overline{{\mathrm{Flow}}}^{(3)}_{\bar{q} q A}\right)^{a}_{\mu} =&-\frac{1}{Z_{A,k}^{1/2}Z_{q,k}}\left(\partial_t \Gamma_{k}^{(3)\bar{q}qA}\right)^{a}_{\mu}\,,
 \label{}
\end{align}
with
\begin{align}
  &\left(\Gamma_{k}^{(3)\bar{q}qA}\right)^{a}_{\mu}\equiv \frac{\delta}{\delta A^c_{\mu}}\frac{\overrightarrow{\delta}}{\delta\bar{q}}\Gamma_{k}\frac{\overleftarrow{\delta}}{\delta q}\,.\label{}
\end{align}
Note that besides the classical tensor structure of the quark-gluon vertex in \Eq{eq:S3bqqA}, some nonclassical tensor structures also play a sizable role in the dynamical breaking of the chiral symmetry \cite{Mitter:2014wpa,Cyrol:2017ewj, Vujinovic:2018nko}, see, e.g., \cite{Fu:2019hdw} for more detailed discussions. From \Eq{eq:dtg} one formulates the flow of the light-quark--gluon coupling as
\begin{align}
  \partial_t \bar g_{\bar l  l A,k}=&\left( \frac{1}{2}\eta_A+\eta_q\right) \bar g_{\bar l  l A,k}+\overline{\textrm{Flow}}^{(3),A}_{(\bar l  l A)}+\overline{\textrm{Flow}}^{(3),\phi}_{(\bar l  l A)}\,,\label{eq:dtgbllA}
\end{align}
where the second term on the r.h.s. corresponds to the two diagrams on the r.h.s. of the flow equation for the quark-gluon vertex as shown in the first line of \Fig{fig:vertex3point-equ}, and the last term to the last diagram, which arise from the quark-gluon, quark-meson fluctuations, respectively. For the strange-quark--gluon coupling, one has
\begin{align}
  \partial_t \bar g_{\bar s  s A,k}=&\left( \frac{1}{2}\eta_A+\eta_q\right) \bar g_{\bar s  s A,k}+\overline{\textrm{Flow}}^{(3),A}_{(\bar s  s A)}\,,\label{eq:dtgbssA}
\end{align}
where the contribution from the strange-quark--meson interactions are neglected, that is reasonable due to relatively larger masses of the strange quark and strange mesons. The explicit expressions in Eqs. (\ref{eq:dtgbllA}) and  (\ref{eq:dtgbssA}) can be found in \Eq{eq:dtgA} and \Eq{eq:dtgphi}.

In the same way, the flow equation of the three-gluon coupling, $g_{A^3,k}\equiv \lambda_{A^3,k}$, is readily obtained from the flow of the three-gluon vertex in the second line of \Fig{fig:vertex3point-equ}. The flow of the three-gluon coupling is decomposed into a sum of the vacuum part and the in-medium contribution, as follows
\begin{align}
  \partial_t \bar g_{A^3,k}=& \partial_t \bar g_{A^3,k}^{\mathrm{vac}}+\partial_t \Delta \bar g_{A^3,k}\,,\label{}
\end{align}
where the vacuum part, i.e., the first term on the r.h.s. is computed in \cite{Braun:2014ata}, and the second term is identified with the in-medium contribution of the quark-gluon coupling, to wit,
\begin{align}
  \partial_t \Delta \bar g_{A^3,k}=& \partial_t \Delta \bar g_{\bar l  l A,k}\,.\label{}
\end{align}

In \Fig{fig:alphas} the light-quark--gluon coupling, the strange-quark--gluon coupling, and the three-gluon coupling are depicted as functions of the RG scale for several different values of the temperature. It is observed that different couplings are consistent with one another in the regime of $k \gtrsim 3$ GeV, while deviations develop in the nonperturbative or even semiperturbative scale. Moreover, one can see that the strong couplings decrease with the increasing temperature.

\subsection{Dynamical hadronization, four-quark couplings and Yukawa couplings}
\label{subsec:4quarkcouplingQCD}

%
\begin{figure}[t]
\includegraphics[width=0.98\columnwidth]{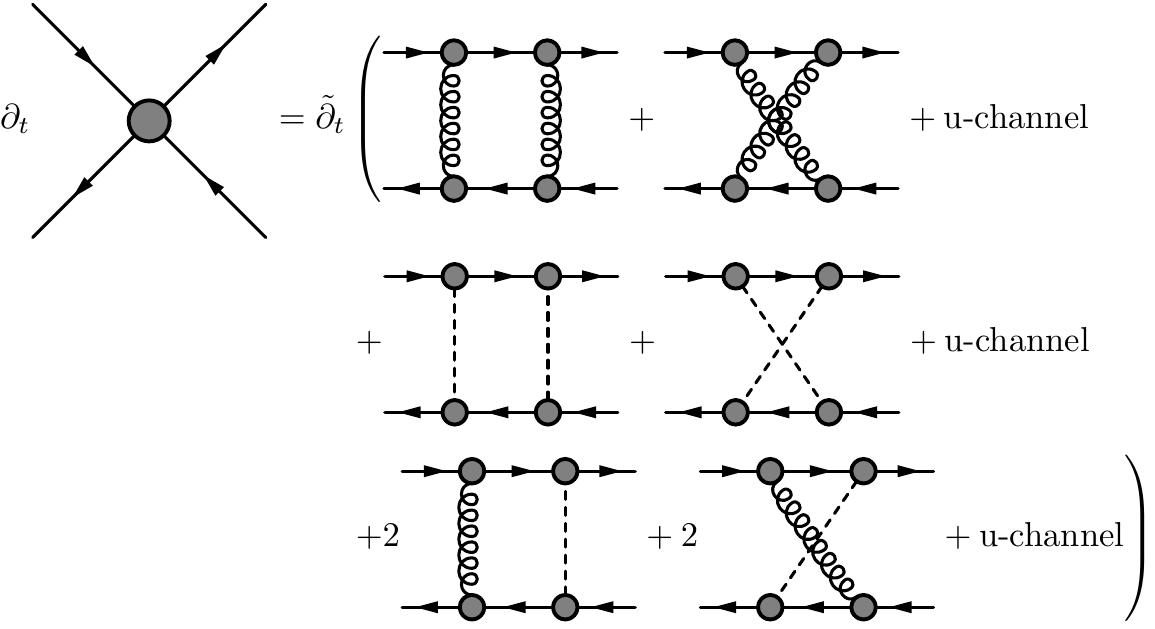}
\caption{Diagrammatic representation of the flow equation for the four-quark vertex. The first line on the r.h.s. denotes the contributions from two gluon exchanges and the second line those from two meson exchanges. The last line stands for the contributions from the mixed diagrams with one gluon exchange and one meson exchange.}\label{fig:v4quark-equ}
\end{figure}
%

%
\begin{figure*}[t]
\includegraphics[width=0.98\textwidth]{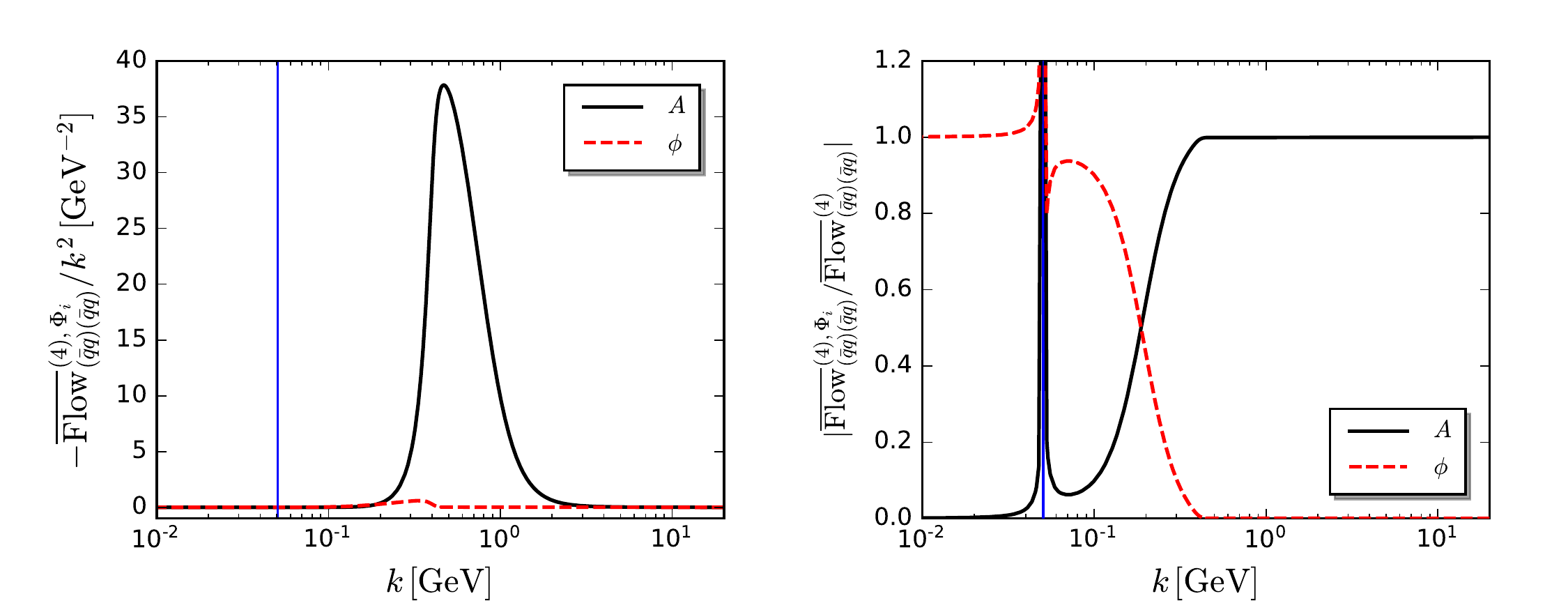}
\caption{Left panel: Comparison between the flow of the four-quark coupling in the $\sigma-\pi$ channel from two gluon exchanges, $\overline{\mathrm{Flow}}^{(4),A}_{(\bar q\tau q)^2}$ in \Eq{eq:Flowbqqbqq2}, as shown in the first line in \Fig{fig:v4quark-equ}, and that from two meson exchanges, $\overline{\mathrm{Flow}}^{(4),\phi}_{(\bar q\tau q)^2}$, in the second line. The flows are depicted as functions of the RG scale with $T=0$ and $\mu_B=0$. Right panel: Absolute values of the ratios $\overline{\mathrm{Flow}}^{(4),A}_{(\bar q\tau q)^2}/\overline{\mathrm{Flow}}^{(4)}_{(\bar q\tau q)^2}$ and $\overline{\mathrm{Flow}}^{(4),\phi}_{(\bar q\tau q)^2}/\overline{\mathrm{Flow}}^{(4)}_{(\bar q\tau q)^2}$ as functions of the RG scale. The position where $\overline{\mathrm{Flow}}^{(4),\phi}_{(\bar q\tau q)^2}$ changes sign is labeled by the blue vertical line in both plots. The plots are adopted from \cite{Fu:2019hdw}.}\label{fig:flow4fermi}
\end{figure*}
%

%
\begin{figure*}[t]
	\includegraphics[width=0.98\textwidth]{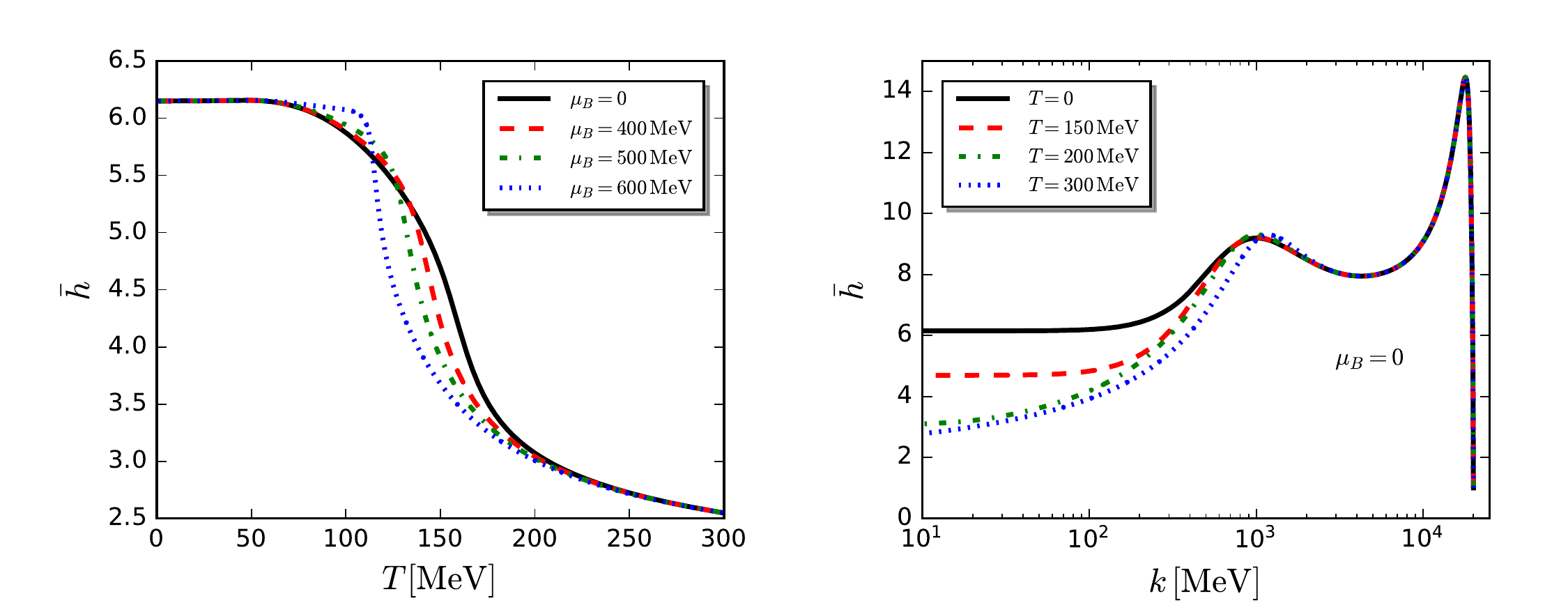}
	\caption{Left panel: Renormalized Yukawa coupling of $N_f=2+1$ flavor QCD at vanishing RG scale $k=0$ as a function of the temperature with several different values of $\mu_B$ obtained in fRG. Right panel: Renormalized Yukawa coupling of $N_f=2+1$ flavor QCD as a function of the RG scale with several different values of $T$ and $\mu_B=0$. The plots are adopted from \cite{Fu:2019hdw}.}\label{fig:h}
\end{figure*}
%

%
\begin{figure}[t]
	\includegraphics[width=0.98\columnwidth]{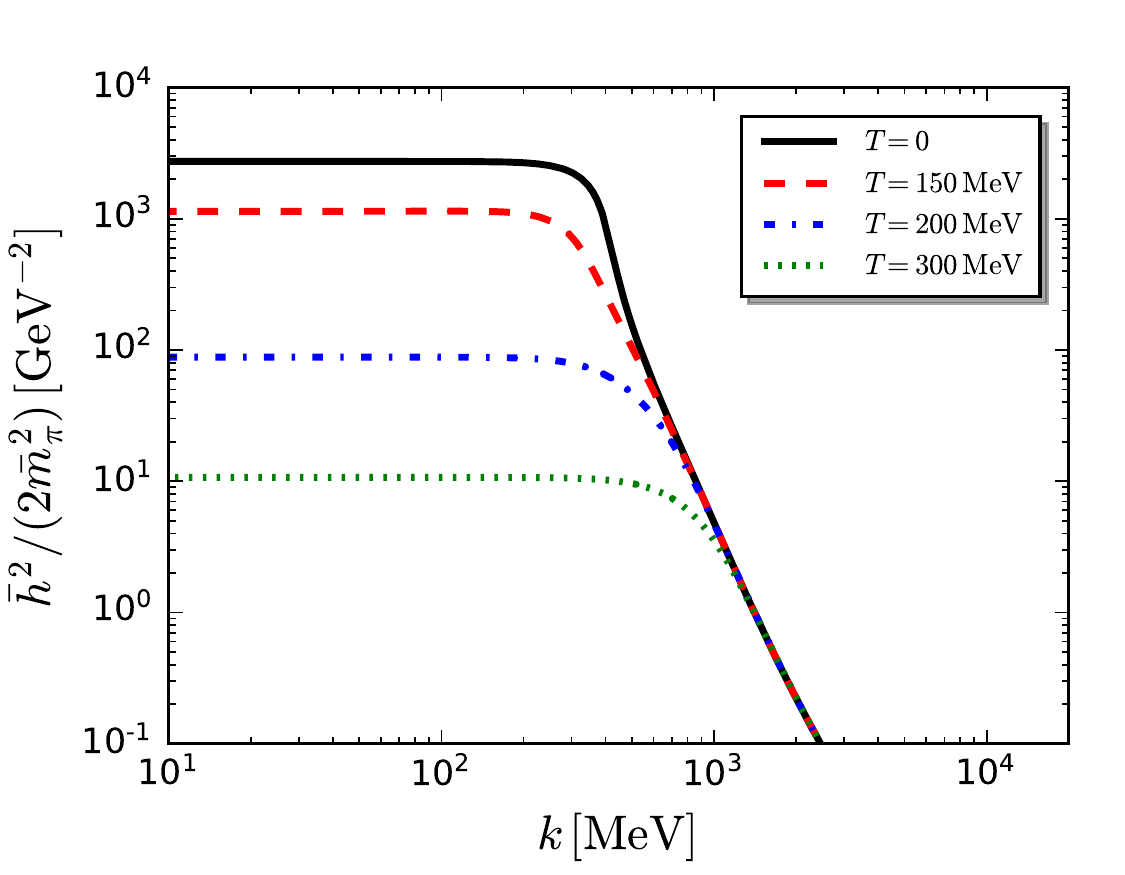}
	\caption{Effective four-quark coupling of $N_f=2+1$ flavor QCD in the pseudoscalar channel $\bar h_k^2/(2\bar m_{\pi,k}^2)$ as a function of the RG scale with several different values of temperature and $\mu_B=0$. The plot is adopted from \cite{Fu:2019hdw}.}\label{fig:lambda}
\end{figure}
%

%
\begin{figure}[t]
	\includegraphics[width=0.98\columnwidth]{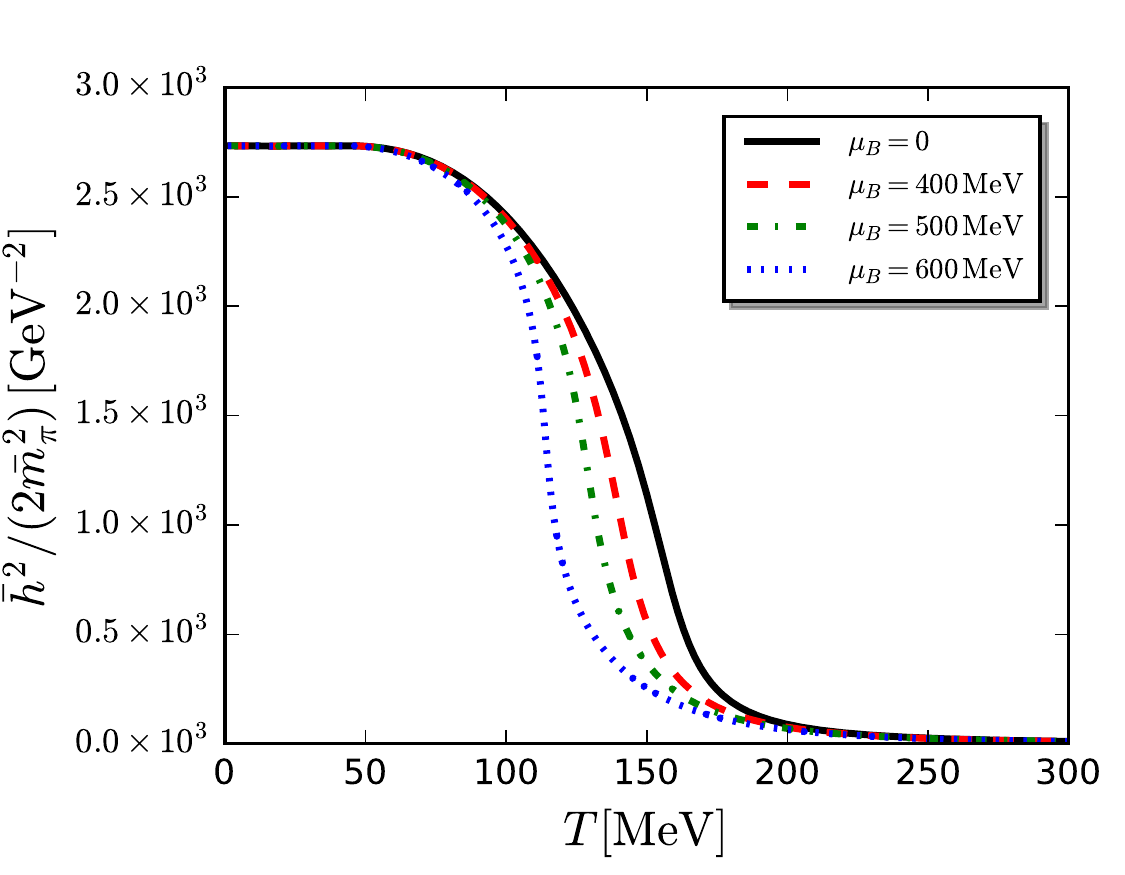}
	\caption{Effective four-quark coupling of $N_f=2+1$ flavor QCD in the pseudoscalar channel at vanishing RG scale $\bar h_{k=0}^2/(2\bar m_{\pi,k=0}^2)$ as a function of the temperature with several different values of baryon chemical potential. The plot is adopted from \cite{Fu:2019hdw}.}\label{fig:lambdamuB}
\end{figure}
%

Following \sec{subsec:dynhadron}, one performs the dynamical hadronization for the $\sigma-\pi$ channel, and use
\begin{align}
  \langle   \partial_t \hat \phi_k\rangle =&\dot{A}_k \,\bar{q}\tau q\,, \label{eq:HadAB0}
\end{align}
with $\tau=(T^0, i \gamma_5\bm{T})$ as shown in \Eq{eq:QCDaction}, where $\dot{A}_k$ is called as the hadronization function. It is more convenient to adopt the dimensionless, renormalized hadronization function and four-quark coupling \begin{align}
  \dot{\tilde A}=& \frac{Z_{\phi,k}^{1/2}}{Z_{q,k}}k^2 \dot{A}_k \,,\qquad \tilde\lambda_{q,k}= \frac{k^2\lambda_{q,k}}{Z_{q,k}^2}\,, \label{eq:RenCoup}
\end{align}
and the renormalized Yukawa coupling,
\begin{align}
  \bar h_k=&  \frac{h_k}{Z_{\phi,k}^{1/2} Z_{q,k}}\,. \label{}
\end{align}

Inserting the effective action in \Eq{eq:QCDaction} into \Eq{eq:FlowQCD} and performing a projection on the four-quark interaction in the $\sigma-\pi$ channel, one is left with
\begin{align}
  \partial_{t} \tilde\lambda_{q,k}=&2(1+\eta_{q,k})\tilde\lambda_{q,k}+\overline{\mathrm{Flow}}^{(4)}_{(\bar q\tau q)^2}+\dot{\tilde A}\, \bar h_k\,, \label{eq:flow4q}
\end{align}
where 
\begin{align}
  \overline{\mathrm{Flow}}^{(4)}_{(\bar q\tau q)^2}\equiv &-\frac{k^2}{Z_{q,k}^2} \left(\partial_t \Gamma_{k,(\bar q\tau q)^2}^{(4)}\right)_{\dot{A}_k=0}\,, \label{eq:Flowbqqbqq}
\end{align}
is the flow of the four-quark coupling in the $\sigma-\pi$ channel, whose contributions have been depicted in \Fig{fig:v4quark-equ}. One can see that the contributed diagrams can be classified into three sets, which correspond to the three lines on the r.h.s. of the flow equation in \Fig{fig:v4quark-equ}. The first line is comprised of diagrams with two gluon exchanges, and the second line with two meson exchanges. The last line denotes the contributions from the mixed diagrams with one gluon exchange and one meson exchange. The mixed diagrams are negligible, since the dynamics of gluons and mesons dominate in different regimes of the RG scale, as shown in \Fig{fig:flow4fermi}. Therefore, the four-quark flow in \Eq{eq:Flowbqqbqq} can be further written as 
\begin{align}
  \overline{\mathrm{Flow}}^{(4)}_{(\bar q\tau q)^2}=&\overline{\mathrm{Flow}}^{(4),A}_{(\bar q\tau q)^2}+\overline{\mathrm{Flow}}^{(4),\phi}_{(\bar q\tau q)^2}\,, \label{eq:Flowbqqbqq2}
\end{align}
where the two terms on the r.h.s. correspond to the contributions from the two gluon exchanges and two meson exchanges, respectivley. Their explicit expressions are presented in \Eq{eq:dtlambdaA} and \Eq{eq:dtlambdaAphi}. In \Fig{fig:flow4fermi} the two terms on the r.h.s. of \Eq{eq:Flowbqqbqq2} and their ratios with respect to the total flow of the four-quark coupling are depicted as functions of the RG scale with $T=0$ and $\mu_B=0$. It is observed that the flow resulting from the gluon exchange is dominant in the region of high energy, while that from the meson exchange plays a significant role in lower energy.

The dynamical hadronization is done by demanding 
\begin{align}
  \tilde\lambda_{q,k}=&0\,,\label{}
\end{align}
for every value of the RG scale $k$, which is equivalent to the fact that the Hubbard-Stratonovich transformation is performed for every value of $k$. Thus from \Eq{eq:flow4q} one arrives at the hadronization function as follows
\begin{align}
  \dot{\tilde A}=&-\frac{1}{\bar h_k}\overline{\mathrm{Flow}}^{(4)}_{(\bar q\tau q)^2} \,. \label{eq:dottildeA}
\end{align}

Inserting the effective action in \Eq{eq:QCDaction} into \Eq{eq:FlowQCD} and performing a projection on the Yukawa interactions between quarks and $\sigma$-$\pi$ mesons, i.e., $\bar{q}\tau\cdot\phi q$, one is led to
\begin{align}
  \partial_{t} \bar h_k=&\left(\frac{1}{2}\eta_{\phi,k}+\eta_{q,k}\right)\bar h_k+\overline{\mathrm{Flow}}^{(3)}_{(\bar q \bm\tau q) \bm\pi}-\tilde m^2_{\pi,k}\dot{\tilde A}\,, \label{eq:hpiflow}
\end{align}
with the dimensionless and renormalized pion mass as follows
\begin{align}
  \tilde m^2_{\pi,k}=&\frac{m^2_{\pi,k}}{Z_{\phi,k} k^2}\,. \label{}
\end{align}
Here in \Eq{eq:hpiflow}
\begin{align}
  \overline{\mathrm{Flow}}^{(3)}_{(\bar q \bm\tau q) \bm\pi}\equiv &\frac{1}{Z_{\phi,k}^{1/2} Z_{q,k}} \left(\partial_t \Gamma_{k,(\bar q \bm\tau q) \bm\pi}^{(3)}\right)_{\dot{A}_k=0}\,, \label{eq:Flowbqqpion}
\end{align}
is the flow of the Yukawa coupling between the pion and quarks, which is shown in the third line of \Fig{fig:vertex3point-equ}. Its explicit expression is presented in \Eq{eq:hexp}. Therefore, substituting the hadronization function in \Eq{eq:dottildeA} into \Eq{eq:hpiflow}, one obtains the total flow of the Yukawa coupling finally. Evidently, the dynamics of resonance of quarks are stored in the interactions between quarks and mesons through the dynamic hadronization.

In the left panel of \Fig{fig:h} the renormalized Yukawa coupling with $k=0$ is shown as a function of the temperature with several different values of baryon chemical potential. It is observed that with the increase of $T$ or $\mu_B$, the Yukawa coupling decays rapidly due to the restoration of the chiral symmetry. In the right panel of \Fig{fig:h} the dependence of the renormalized Yukawa coupling on the RG scale for different values of temperature is investigated. One can see that the Yukawa coupling is stable in the region of $k\gtrsim 1$ GeV, and the effects of temperature play a role approximately in $k\lesssim 2\pi T$. The effective four-quark coupling in the $\sigma$-$\pi$ channel can be described by the ratio $\bar h_k^2/(2\bar m_{\pi,k}^2)$ \cite{Braun:2014ata, Fu:2019hdw}, which is shown as a function of the RG scale with several different values of temperature and $\mu_B=0$ in \Fig{fig:lambda}. It is observed that with the decrease of $k$ and entering the regime of the chiral symmetry breaking, a resonance occurs in the scalar-pseudoscalar channel, which results in a rapid increase of the effective four-quark coupling. Moreover, the dependence of the effective four-quark coupling on the baryon chemical potential is shown in \Fig{fig:lambdamuB}. One finds that the effective four-quark coupling decrease with the increasing temperature or baryon chemical potential.

\subsection{Natural emergence of LEFTs from QCD}
\label{subsec:LEFTemergQCD}

%
\begin{figure}[t]
	\includegraphics[width=0.98\columnwidth]{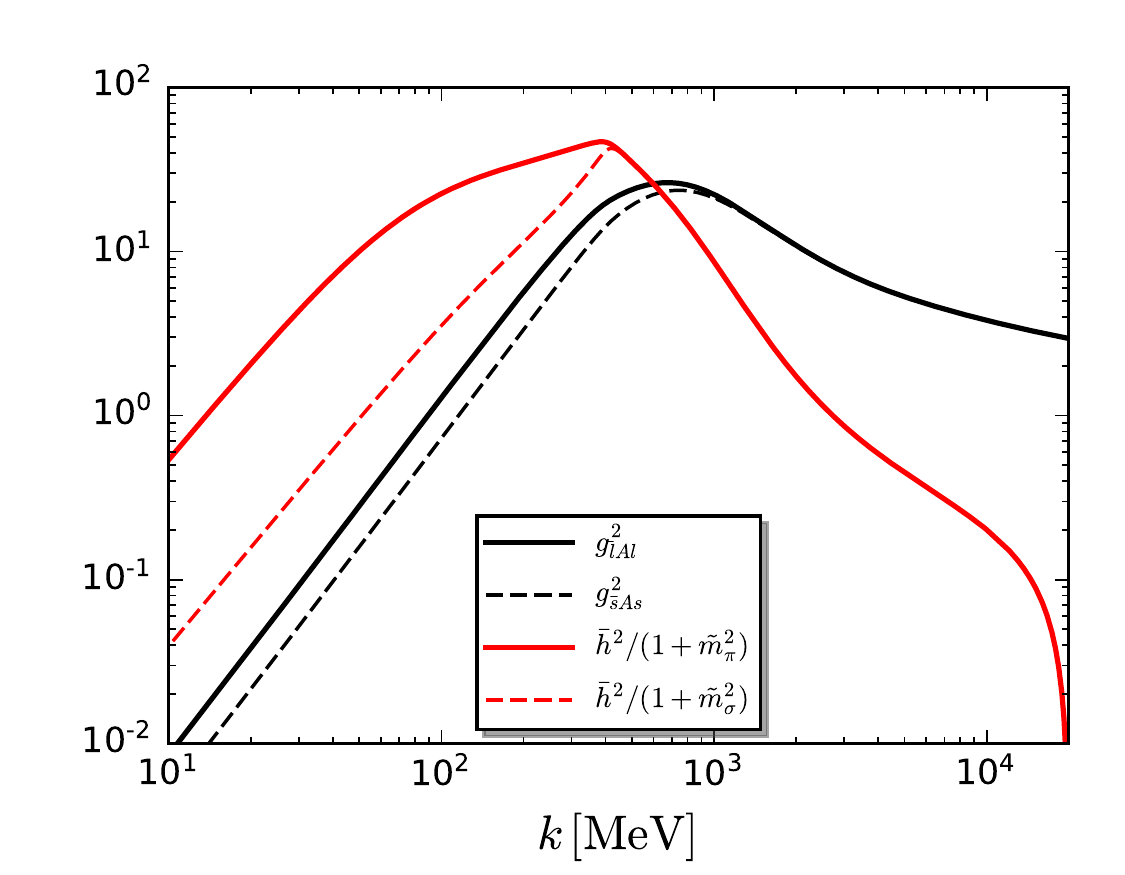}
	\caption{Four-quark single gluon exchange coupling for light quarks $\bar g^2_{\bar l l A,k}$ and strange quarks $\bar g^2_{\bar s s A,k}$, dimensionless four-quark single meson exchange coupling $\bar h_k^2 /(1+\tilde m_{\pi,k}^2)$ and $\bar h_k^2 /(1+\tilde m_{\sigma,k}^2)$  as functions of the RG scale, obtained in fRG for the $N_f=2+1$ flavor QCD in the vacuum. The plot is adopted from \cite{Fu:2019hdw}.}\label{fig:ExchangeCouplings}
\end{figure} 
%

%
\begin{figure}[t]
	\includegraphics[width=0.98\columnwidth]{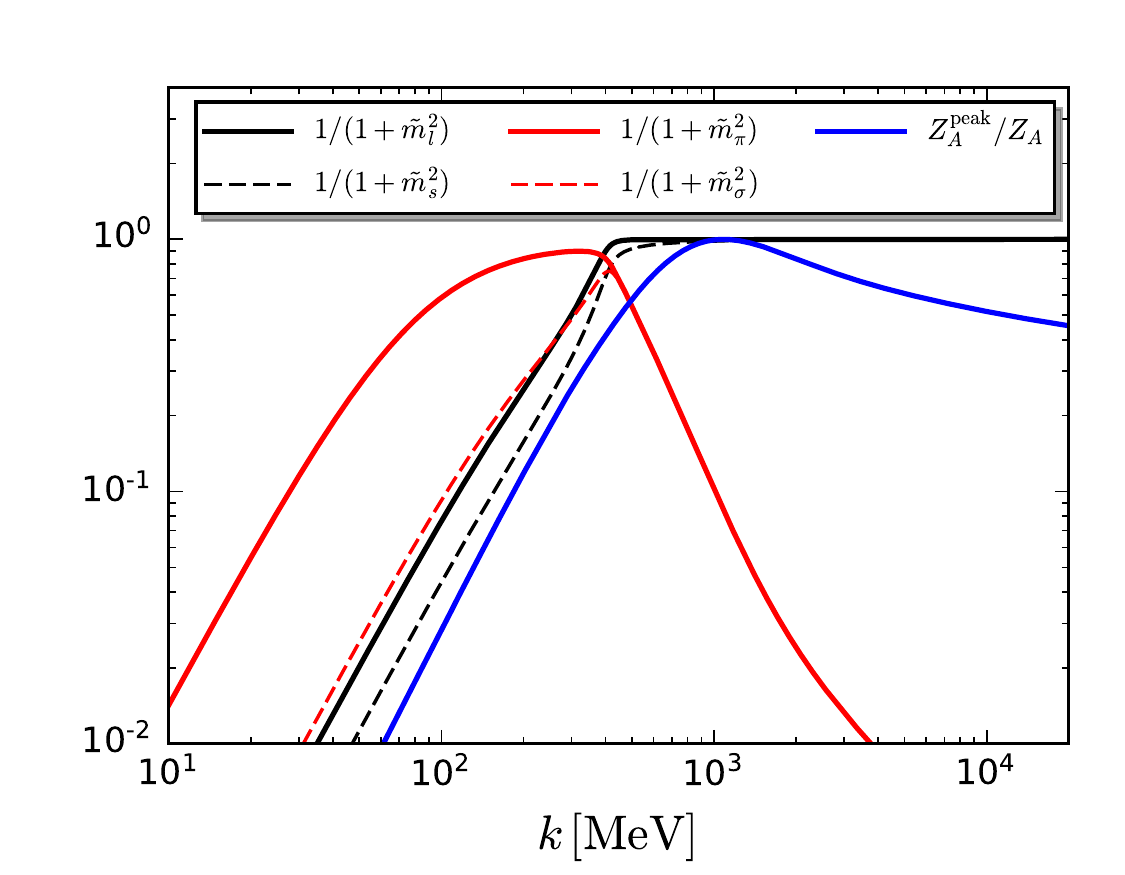}
	\caption{Dimensionless propagator gappings $1/(1+\tilde m^2_{\Phi_i,k})$ for $\Phi_i=l,s,\sigma, \pi$ as functions of the RG scale. The gluon dressing function $1/Z_{A,k}$ is also shown for comparison, which is normalized by its peak value $1/Z_A^{\mathrm{peak}}$ at $k=k_\mathrm{peak}$. The plot is adopted from \cite{Fu:2019hdw}.}\label{fig:PropagatorGapping}
\end{figure} 
%

The fRG approach to QCD with the dynamical hadronization discussed above, provide us with a method to study the transition of degrees of freedom from fundamental to composite ones. One is also able to observe a natural emergence of LEFTs from original QCD. To that end, in \Fig{fig:ExchangeCouplings} one shows the four-quark single gluon exchange coupling for light quarks $\bar g^2_{\bar l l A,k}$ and strange quarks $\bar g^2_{\bar s s A,k}$, dimensionless four-quark single meson exchange coupling $\bar h_k^2 /(1+\tilde m_{\pi,k}^2)$ and $\bar h_k^2 /(1+\tilde m_{\sigma,k}^2)$ as functions of the RG scale in the vacuum. Evidently, it is found that the gluonic exchange couplings are dominant in the perturbative regime of $k\gtrsim 1$ GeV. However, with the decrease of the RG scale the active dynamic is taken over gradually by the mesonic degrees of freedom, and one can see that the gluonic couplings and the Yukawa couplings are comparable to each other at $k\approx 600$. In \Fig{fig:PropagatorGapping} dimensionless propagator gappings $1/(1+\tilde m^2_{\Phi_i,k})$ for $\Phi_i=l,s,\sigma, \pi$ are shown as functions of the RG scale. The gluon dressing function is also show there for comparison. In \Fig{fig:PropagatorGapping} one observes the same information on decouplings as in \Fig{fig:ExchangeCouplings}, that is, as the RG scale evolves from the UV towards IR, the gluons decouple from the matter at first, then the quarks, and the mesons finally. For more related discussions see \cite{Fu:2019hdw} .

\subsection{Chiral condensate}
\label{subsec:condensateQCD}

%
\begin{figure*}[t]
	\includegraphics[width=0.98\textwidth]{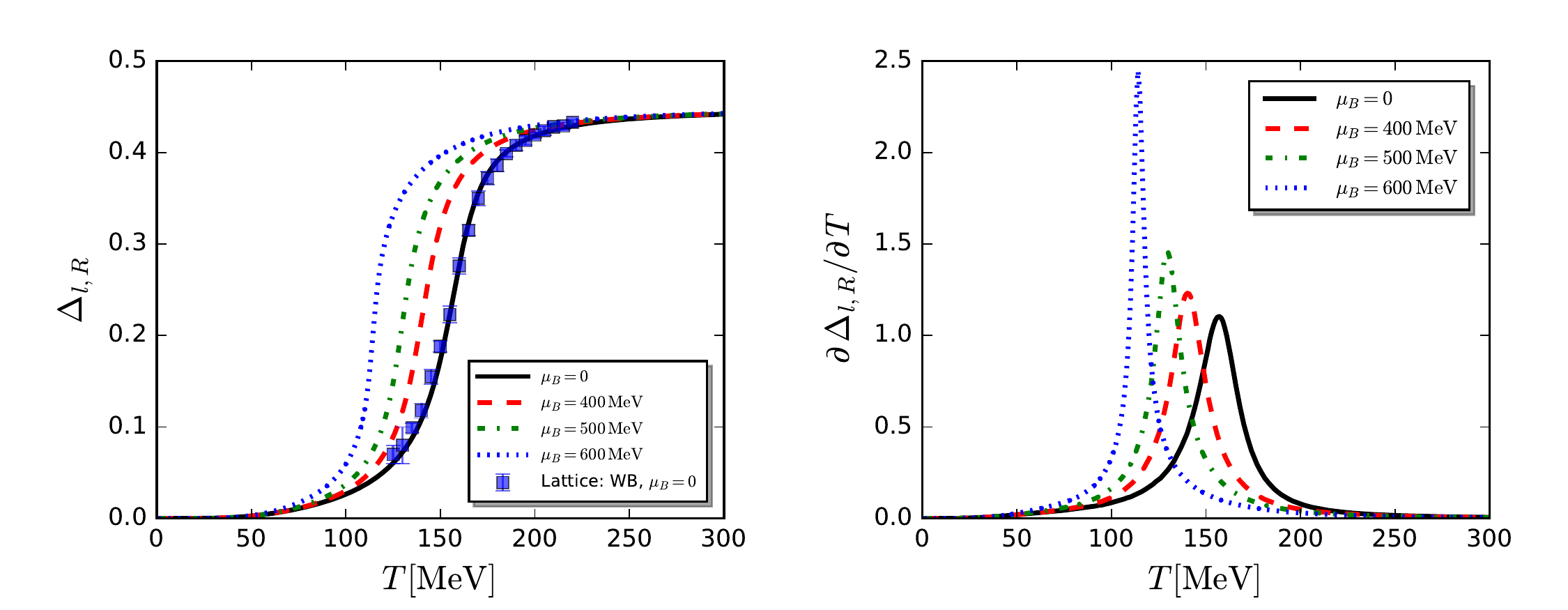}
	\caption{Left panel: Renormalized light quark chiral condensate $\Delta_{l,R}$ of $N_f=2+1$ flavor QCD as a function of the temperature with several different values of baryon chemical potential obtained in fRG, in comparison to the lattice results at vanishing $\mu_B$ \cite{Borsanyi:2010bp}. Right panel: Derivative of $\Delta_{l,R}$ with respect to the temperature, i.e., the thermal susceptibility of the renormalized light quark condensate, as a function of the temperature with several different values of baryon chemical potential. The plots are adopted from \cite{Fu:2019hdw}.}\label{fig:DeltalR}
\end{figure*}
%

%
\begin{figure}[t]
\includegraphics[width=0.95\columnwidth]{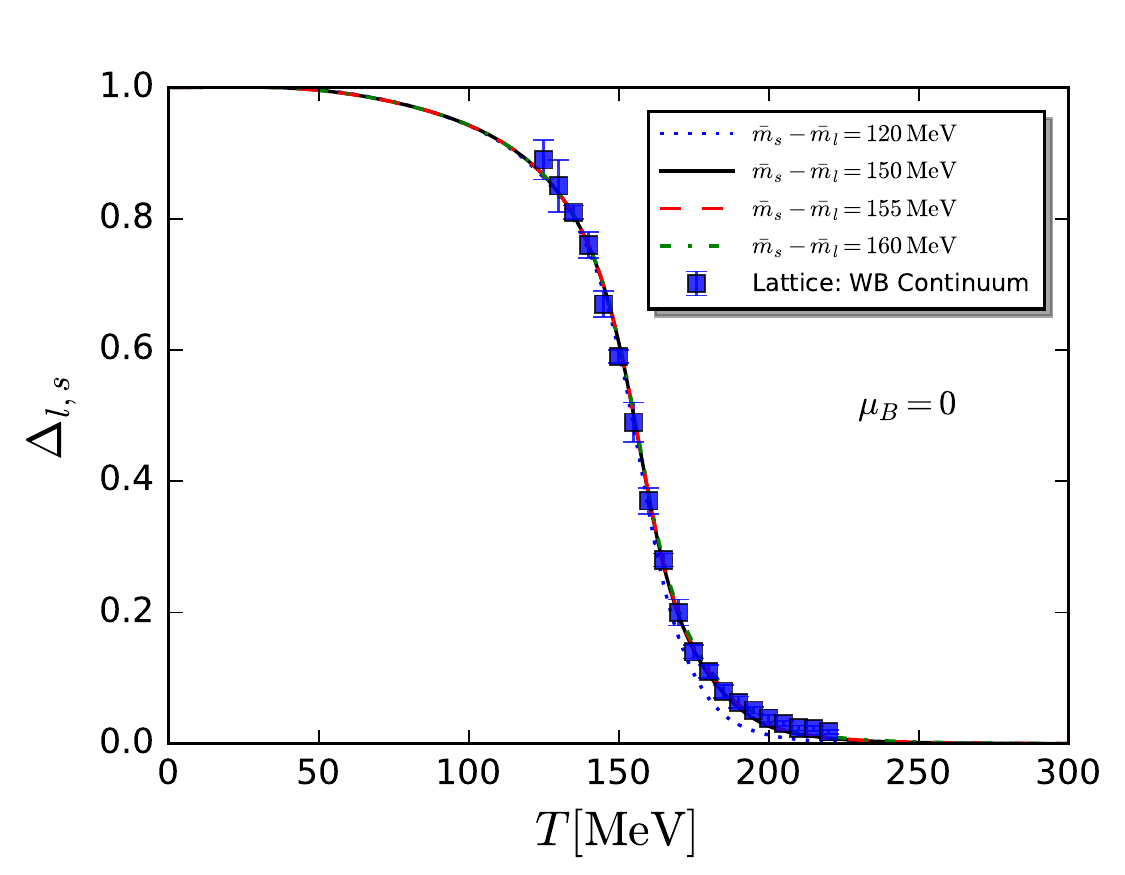}
\caption{Reduced condensate of $N_f=2+1$ flavor QCD as a function of the temperature with $\mu_B=0$ obtained in fRG, where the constituent quark mass differences are $\Delta\bar m_{sl}=\bar m_s-\bar m_l=120,150,155,160$\,MeV, corresponding to the current quark mass ratios of $m_s^0/m_l^0=c_{\sigma_s}/c_\sigma \approx 14,27,30,34$. The fRG results are compared with the lattice ones in \cite{Borsanyi:2010bp}. The plot is adopted from \cite{Fu:2019hdw}.}
\label{fig:Deltals}
\end{figure}
%

The chiral condensate of quark $q_i=u,d,s$ reads
\begin{align}
  \Delta_{q_i}= &  - m_{q_i}^0 T\sum_{n\in\mathbb{Z}} \int \frac{d^3 q}{(2 \pi)^3} \tr \,G_{q_i\bar q_i} (q)\,,\label{eq:chiralcondG}
\end{align}
up to a renormalization term, where no summation for the index $i$ is assumed, and $m_{q_i}^0$ is the current quark mass. Obviously, the light quark condensate is given by $\Delta_{l}=\Delta_{u}=\Delta_{d}$ with $m_{l}^0=m_{u}^0=m_{d}^0$. The renormalized condensate as follows
\begin{align}
  \Delta_{q_i,R} =& \frac{1}{{\cal N}_R}\left[\Delta_{q_i}(T,\mu_q)-\Delta_{q_i}(0,0)\right]\,, \label{eq:chiralcondren}
\end{align}
renders the vacuum part of the chiral condensate in \Eq{eq:chiralcondG} to be subtracted, where the dimensionless $\Delta_{q_i,R}$ is normalized by a constant ${\cal N}_R$, e.g., ${\cal N}_R\sim f_\pi^4$. In the present fRG approach to QCD, the light quark condensate is given by
\begin{align}
  \Delta_l =&\frac12 m_l^0\frac{\partial \Omega[\Phi_{\mathrm{EoM}};T,\mu_q]}{\partial {m_l^0}}
  = \frac12 c_\sigma \frac{\partial \Omega[\Phi_{\mathrm{EoM}};T,\mu_q]}{ \partial {c_\sigma}}\,,\label{eq:mDer-cDer}
\end{align}
where $\Omega$ is the thermodynamic potential with the field $\Phi_{\mathrm{EoM}}$ being on is equation of motion. From \Eq{eq:QCDaction}, one is led to
\begin{align}
  \Delta_{l}(T,\mu_q) =& -\frac{1}{2} c_\sigma \,\sigma_\mathrm{EoM}(T,\mu_q) \,, \label{eq:chiralcondSigma}
\end{align}
and 
\begin{align}
  \Delta_{l,R}(T,\mu_q) =& -\frac{c_\sigma}{{2 \,\cal N}_{R}}
  \Big[ \sigma_\mathrm{EoM}(T,\mu_q) -
  \sigma_\mathrm{EoM}(0,0)\Big]\,. \label{eq:RenCondSigma}
\end{align}

The reduced condensate $\Delta_{l,s}$ is defined as the weighted difference between the light and strange quark condensates as follows,
\begin{align}
  \Delta_{l,s}(T,\mu_q) =&\frac{1}{{\cal N}_{l,s}}\left[ \Delta_{l}(T,\mu_q) -\left(\frac{m_l^0}{m_s^0}\right)^2\Delta_s(T,\mu_q)\right]\,,\label{eq:Deltals}
\end{align}
which is usually normalized with its value in the vacuum, i.e., 
\begin{align}
  \Delta_{l,s}(T,\mu_q) =& \frac{ \Delta_{l}(T,\mu_q) -\left(\frac{m_l^0}{m_s^0}\right)^2\Delta_s(T,\mu_q)}{
  \Delta_{l}(0,0) - \left(\frac{m_l^0}{m_s^0}\right)^2\Delta_s(0,0)}\,.\label{eq:chiralcondred}
\end{align}
Similar with \Eq{eq:mDer-cDer}, one arrives at
\begin{align}
  \Delta_s =&m_s^0\frac{\partial \Omega[\Phi_{\mathrm{EoM}};T,\mu_q]}{\partial {m_s^0}}
  = c_{\sigma_s}\frac{\partial \Omega[\Phi_{\mathrm{EoM}};T, \mu_q]}{ \partial {c_{\sigma_s}}}\,,\label{eq:mDer-cDer-s}
\end{align}
and thus
\begin{align}
  \Delta_{s}(T,\mu_q) =& -\frac{1}{\sqrt{2} }c_{\sigma_s}\,\sigma_{s,\textrm{EoM}}(T,\mu_q)\,. \label{eq:CondSigma_s}
\end{align}
Finally, one is led to
\begin{align}
  \Delta_{l,s}(T,\mu_q) =& \frac{\left(\sigma -\sqrt{2}\frac{c_\sigma}{c_{\sigma_s}}\sigma_s\right)_{T,\mu_q}}{
  \left(\sigma - \sqrt{2}\frac{c_\sigma}{c_{\sigma_s}}\sigma_s\right)_{0,0}}\,,  \label{eq:RedCondSigma}
\end{align}
where one has used 
\begin{align}
  \frac{m_l^0}{m^0_s}=& \frac{c_\sigma}{c_{\sigma_s}} \,, \label{eq:ctomApp}
\end{align}

In the left panel \Fig{fig:DeltalR}, the renormalized light quark condensate is shown as a function of the temperature with several different values of baryon chemical potential. In the case of vanishing baryon chemical potential, the fRG result is compared with the continuum extrapolated result of lattice QCD \cite{Borsanyi:2010bp}, and excellent agreement is found. In the right panel of \Fig{fig:DeltalR}, the derivative of $\Delta_{l,R}$ with respect to the temperature, i.e., the thermal susceptibility of the renormalized light quark condensate, are shown.  The peak position of the thermal susceptibility can be used to define the pseudocritcal temperature, which is found to be $T_c=156$\,MeV for $\mu_B=0$, in good agreement with the lattice result. In \Fig{fig:Deltals} one shows the reduced condensate as a function of the temperature with $\mu_B=0$, in comparison to the lattice simulations. One finds that for the physical current quark mass ratio, i.e., $m_s^0/m_l^0=c_{\sigma_s}/c_\sigma \approx 27$ \cite{Aoki:2013ldr},  the fRG results are in quantitative agreement with the lattice ones.

\subsection{Phase structure}
\label{subsec:phaseStruQCD}

%
\begin{table*}[t]
  \begin{center}
  \begin{tabular}{cccccccc}
    \hline\hline & & & & & & &  \\[-2ex]   
     &  fRG:  \cite{Fu:2019hdw} & DSE:  \cite{Gao:2020fbl} & DSE:  \cite{Gunkel:2021oya} & Lattice (HotQCD):  \cite{Bazavov:2018mes} & Lattice (WB):  \cite{Bellwied:2015rza}  & Lattice (WB): \cite{Borsanyi:2020fev} & Lattice:  \cite{Bonati:2018nut} \\[1ex]
    \hline & & & & & & & \\[-2ex]
    $\kappa$ & 0.0142(2) &0.0147(5) & 0.0173 & 0.015(4)& 0.0149(21)  & 0.0153(18) & 0.0144(26)  \\[1ex]
    \hline & & & & & & & \\[-2ex]
    $ (T,{\mu_B})_{_{\tiny{\mathrm{CEP}}}}$ [MeV] & (107, 635) &(109, 610)& (112, 636) & &   &  & \\[1ex]
    \hline\hline
  \end{tabular}
  \caption{Curvature $\kappa$ of the phase boundary (second line) and location of CEP (third line) for $N_f=2+1$ flavor QCD, obtained from different approaches. fRG: \cite{Fu:2019hdw} (Fu {\it et al.}); DSE: \cite{Gao:2020fbl}  (Gao {\it et al.}), \cite{Gunkel:2021oya} (Gunkel {\it et al.}); Lattice QCD: \cite{Bazavov:2018mes} (HotQCD), \cite{Bellwied:2015rza, Borsanyi:2020fev} (WB), \cite{Bonati:2018nut} (Bonati {\it et al.}); .} \label{tab:kappas}
  \end{center}\vspace{-0.5cm}
\end{table*}
%

%
\begin{figure}[t]
  \includegraphics[width=.98\columnwidth]{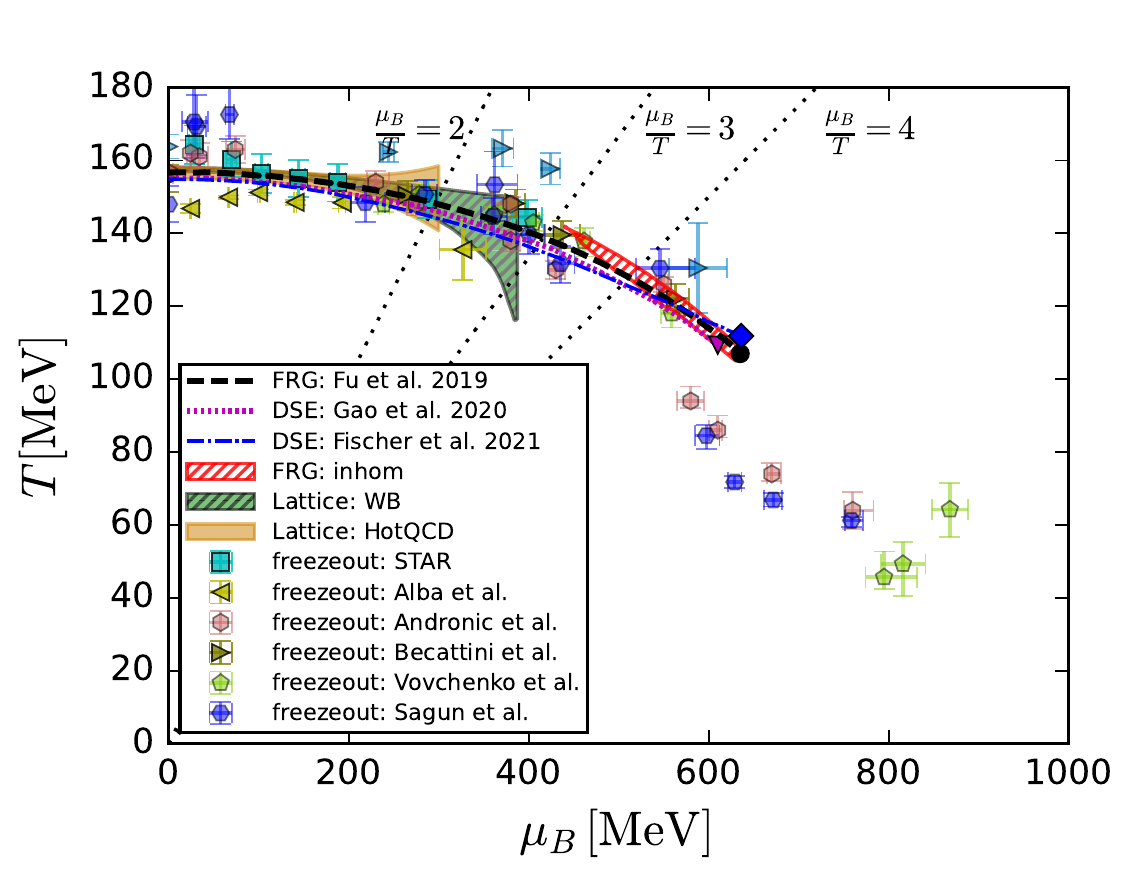}
  \caption{Phase diagram of $N_f=2+1$ flavor QCD in the plane of the temperature and the baryon chemical potential. The fRG results \cite{Fu:2019hdw} are compared with those from Dyson-Schwinger Equations \cite{Gao:2020fbl, Gunkel:2021oya}, lattice QCD \cite{Bellwied:2015rza, Bazavov:2018mes}. The hatched red area denotes the region of inhomogeneous instability for the chiral condensate found in the calculations of fRG. Some freeze-out data are also shown in the phase diagram: \cite{Adamczyk:2017iwn} (STAR), \cite{Alba:2014eba} (Alba {\it et al.}), \cite{Andronic:2017pug} (Andronic {\it et al.}), \cite{Becattini:2016xct} (Becattini {\it et al.}), \cite{Vovchenko:2015idt} (Vovchenko {\it et al.}), and \cite{Sagun:2017eye} (Sagun {\it et al.}).}\label{fig:phasediagramIIA}
\end{figure}
%

The QCD phase diagram in the plane of $T$ and $\mu_B$ is shown in \Fig{fig:phasediagramIIA}. The first-principle fRG results in \cite{Fu:2019hdw} are in comparison to state-of-the-art calculations of other functional approach, e.g., the Dyson-Schwinger Equations (DSE) \cite{Gao:2020fbl, Gunkel:2021oya}, and lattice QCD \cite{Bellwied:2015rza, Bazavov:2018mes}. The phase boundary in the regime of continuous crossover at small or medium $\mu_B$ shows the dependence of the pseudocritical temperature on the value of the baryon chemical potential. In the calculations of fRG, this pseudocritical temperature is determined by the thermal susceptibility of the renormalized light quark chiral condensate, $\partial \Delta_{l,R}/\partial T$, as discussed in \sec{subsec:condensateQCD}. Expanding the pseudocritical temperature around $\mu_B=0$, one arrives at
\begin{align}
  \frac{T_c(\mu_B)}{T_c}&=1-\kappa \left(\frac{\mu_B}{T_c}\right)^2+\lambda \left(\frac{\mu_B}{T_c}\right)^4+\cdots\,,
 \label{eq:curv}
\end{align}
with $T_c=T_c(\mu_B=0)$, where the quadratic expansion coefficient $\kappa$ is usually called as the curvature of the phase boundary. It is a sensitive measure for the QCD dynamics at finite temperature and densities. Therefore, it provides a benchmark test for functional approaches to make a comparison of the curvature at small baryon chemical potential with lattice simulations. The curvature values of the phase boundary obtained from fRG, DSE, and lattice QCD are summarized in the second line of \tab{tab:kappas}. It is found recent results of the curvature from state-of-the-art functional approaches, e.g., fRG \cite{Fu:2019hdw}, DSE \cite{Gao:2020fbl, Gunkel:2021oya}, have already been comparable to the lattice results. By contrast, those obtained from relatively former calculations of functional methods, e.g., \cite{Fischer:2014ata, Gao:2015kea, Fischer:2018sdj}, are significantly larger than the values of the curvature obtained from lattice simulations.

Besides the curvature of the phase boundary, another key ingredient of the QCD phase structure is the critical end point (CEP) of the first-order phase transition line at large baryon chemical potential, which is currently searched for with lots of efforts in experiments \cite{Adamczyk:2013dal, Adamczyk:2014fia, Luo:2015ewa, Luo:2017faz, Adamczyk:2017wsl, Adam:2019xmk, Adam:2020unf, STAR:2021rls}. Lattice simulations, however, are restricted in the region of $\mu_B/T\lesssim 2 \sim 3$ because of the sign problem at finite chemical potentials, and in this region no signal of CEP is observed \cite{Karsch:2019mbv}. Passing lattice benchmark tests at small baryon chemical potentials, functional approaches are able to provide relatively reliable estimates for the location of the CEP. The recent results of CEP after benchmark testing are presented in the third line of \tab{tab:kappas}, and also depicted in the phase diagram in \Fig{fig:phasediagramIIA}. Remarkably, estimates of CEP from fRG and DSE converge in a rather small region at baryon chemical potentials of about 600 MeV. Note that results of \cite{Gunkel:2021oya} in \tab{tab:kappas} are obtained without the dynamics of sigma and pion, and the relevant results are $\kappa=0.0167$ and $ (T,{\mu_B})_{_{\tiny{\mathrm{CEP}}}}=(117, 600)$ MeV when they are included, and see also, e.g., \cite{Fischer:2012vc,Fischer:2014ata,Fischer:2018sdj} for related discussions. It should be reminded that errors of functional approaches increase significantly when $\mu_B/T\gtrsim 4$, for a detailed discussion see \cite{Fu:2019hdw, Gao:2020qsj}, and thus one arrives at a more reasonable estimation for the location of CEP as $450\, \mathrm{MeV}\lesssim{\mu_B}_{\tiny{\mathrm{CEP}}}\lesssim 650\, \mathrm{MeV}$.

\subsubsection{	Region of inhomogeneous instability at large baryon chemical potential}
\label{subsubsec:inhomoRegionQCD}

%
\begin{figure*}[t]
	\includegraphics[width=0.98\textwidth]{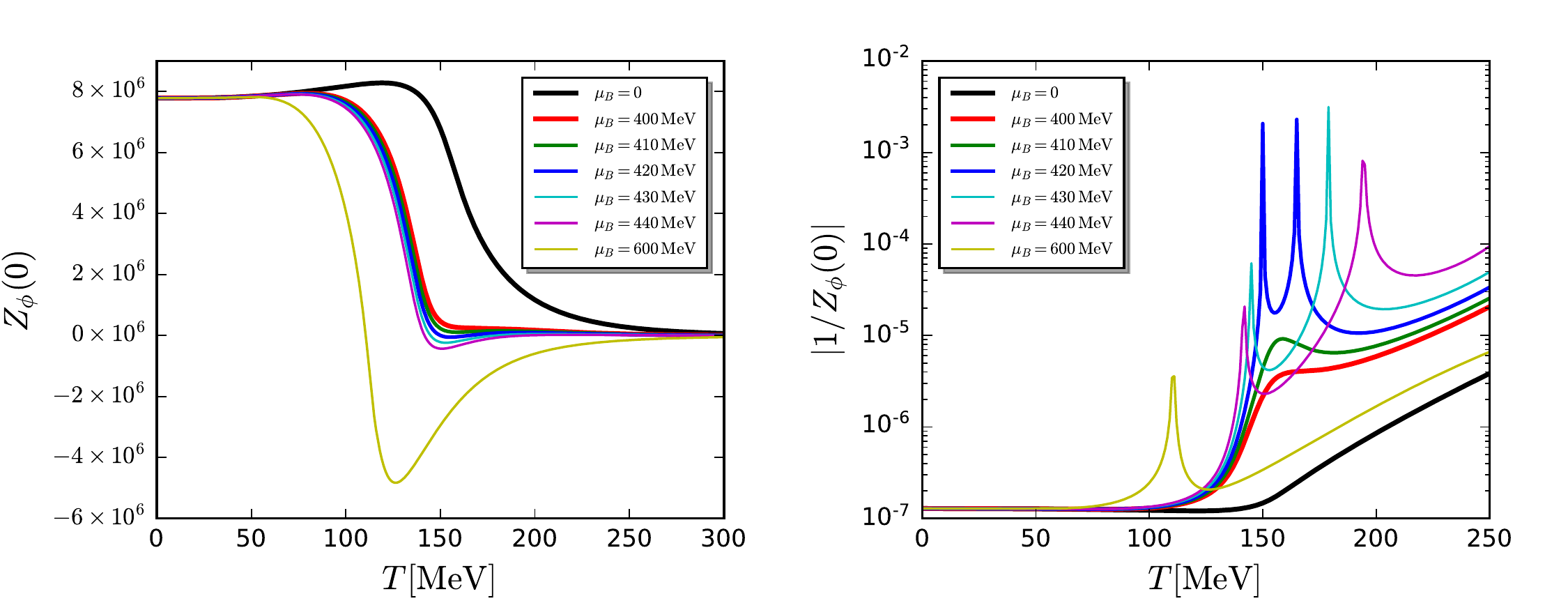}
	\caption{Left panel: Mesonic wave function renormalization at vanishing external momenta $Z_{\phi, k=0}(0)$ as a function of the temperature with several different values of the baryon chemical potential, obtained in the fRG approach to $N_f=2+1$ flavor QCD. Right panel: $\big|1/Z_{\phi, k=0}(0)\big|$ for the data in the left panel. The spikes are used to locate the region of negative $Z_{\phi, k=0}(0)$. The plots are adopted from \cite{Fu:2019hdw}.}\label{fig:Zphip0}
\end{figure*}
%

%
\begin{figure}[t]
  \includegraphics[width=1\columnwidth]{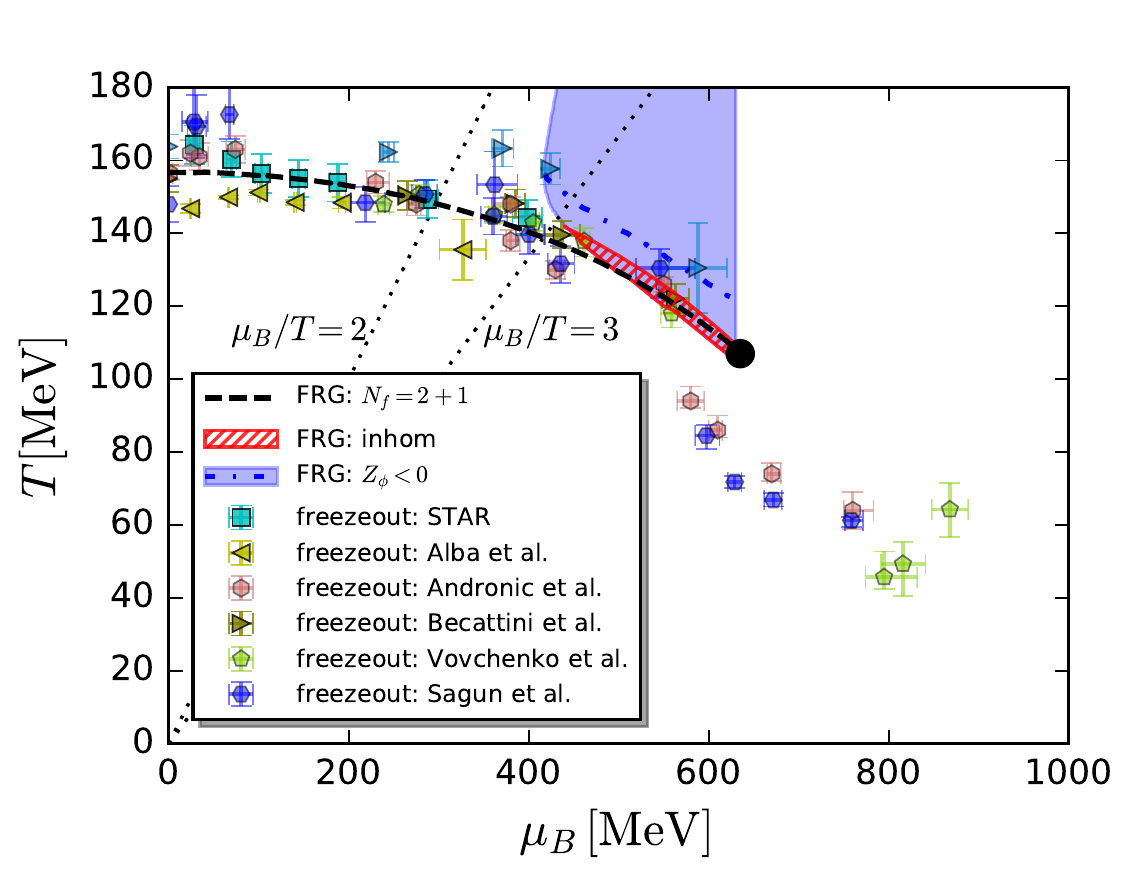}
  \caption{Phase diagram of $N_f=2+1$ flavor QCD obtained in the fRG approach to QCD in comparison to freeze-out data. See also the caption of \Fig{fig:phasediagramIIA}. The blue area denotes the region of negative mesonic wave function renormalization at vanishing external momenta $Z_{\phi, k=0}(0)$, which is an indicator for the inhomogeneous instability. The red hatched area stands for the regime with negative $Z_{\phi, k=0}(0)$ and also sizable chiral condensate. The plot is adopted from \cite{Fu:2019hdw}.}\label{fig:PhasediagramInhom}
\end{figure}
%

At large baryon chemical potentials, it is found within the fRG approach to QCD, that the mesonic wave function renormalization in \Eq{eq:Gam2pi2} and \Eq{eq:Gam2sigma2} develops negative values at small momenta \cite{Fu:2019hdw}. As shown in \Fig{fig:Zphip0} the mesonic wave function renormalization at vanishing external momenta $Z_{\phi, k=0}(0)$ is depicted as a function of the temperature with several different values of $\mu_B$. One can see a negative $Z_{\phi, k=0}(0)$ begins to appear for a temperature region when $\mu_B\gtrsim 420$ MeV. The negative regime is clearly shown in the right plot of $\big|1/Z_{\phi, k=0}(0)\big|$ between the two spikes, and it widens with the increase of the baryon chemical potential. From the two-point correlation functions of mesons as shown in Eqs. (\ref{eq:Gam2pi2}) and (\ref{eq:Gam2sigma2}), the negative $Z_{\phi, k=0}(0)$ implies that, for the dispersion relations of mesons, there is a minimum at a finite $\bm p^2 \neq 0$. This nontrivial behavior that the minimum of dispersion is located at a finite momentum is indicative of an inhomogeneous instability, for instance, the formation of a spatially modulated chiral condensate. However, it should be reminded that a negative $Z_{\phi, k=0}(0)$ is not bound to the formation of an inhomogeneous phase, it can also serve as an precursor for the inhomogeneous phase. See, e.g., \cite{Schon:2000qy, Thies:2006ti, Anglani:2013gfu, Carignano:2014jla, Buballa:2014tba, Braun:2015fva, Roscher:2015xha, Yokota:2017uzu, Tripolt:2017zgc, Pisarski:2019cvo, Pisarski:2021qof, Rennecke:2021ovl} for more related discussions. Most notably, very recently consequence of the inhomogeneous instability indicated by the negative wave function renormalization on the phenomenology of heavy-ion collisions has been studied. It is found that this inhomogeneous instability would result in a moat regime in both the particle $p_T$ spectrum and the two-particle correlation, where the peaks are located at nonzero momenta \cite{Pisarski:2021qof, Rennecke:2021ovl}. The region of negative $Z_{\phi, k=0}(0)$ is shown in \Fig{fig:PhasediagramInhom} by the blue area, while the red hatched area shows where this region overlaps with a sizable chiral condensate.

\subsection{Magnetic equation of state}
\label{subsec:magneticEosQCD}

%
\begin{figure}[t]
  \includegraphics[width=0.9\columnwidth]{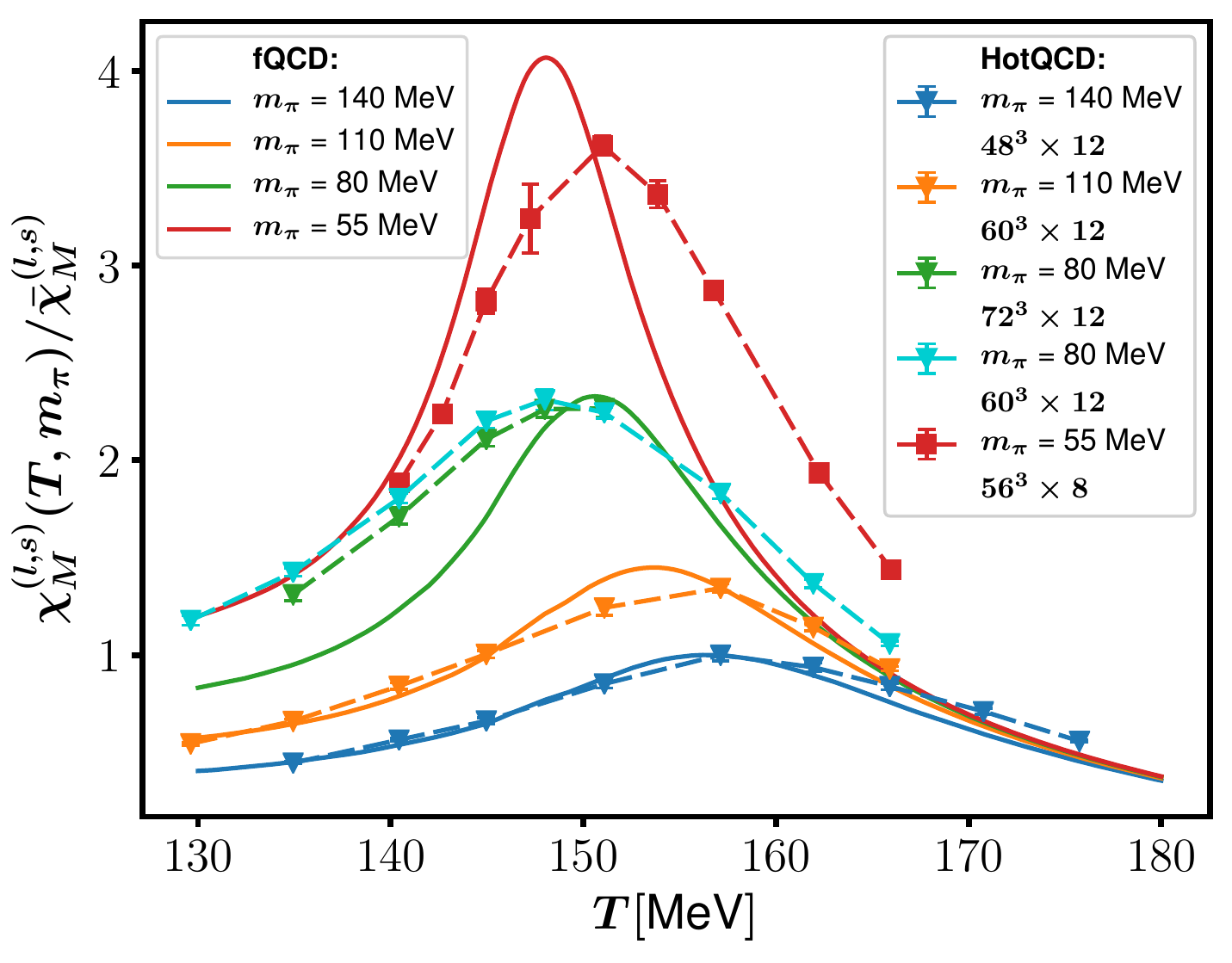}
  \caption{Susceptibility for the reduced condensate obtained in the fRG approach to $N_f=2+1$ flavor QCD (fQCD) as a function of the temperature, in comparison to the lattice results in \cite{Ding:2019prx,Ding:2019fzc}. The plot is adopted from \cite{Braun:2020ada}.}\label{fig:chiMls}
\end{figure}
%

%
\begin{figure}[t]
  \includegraphics[width=0.9\columnwidth]{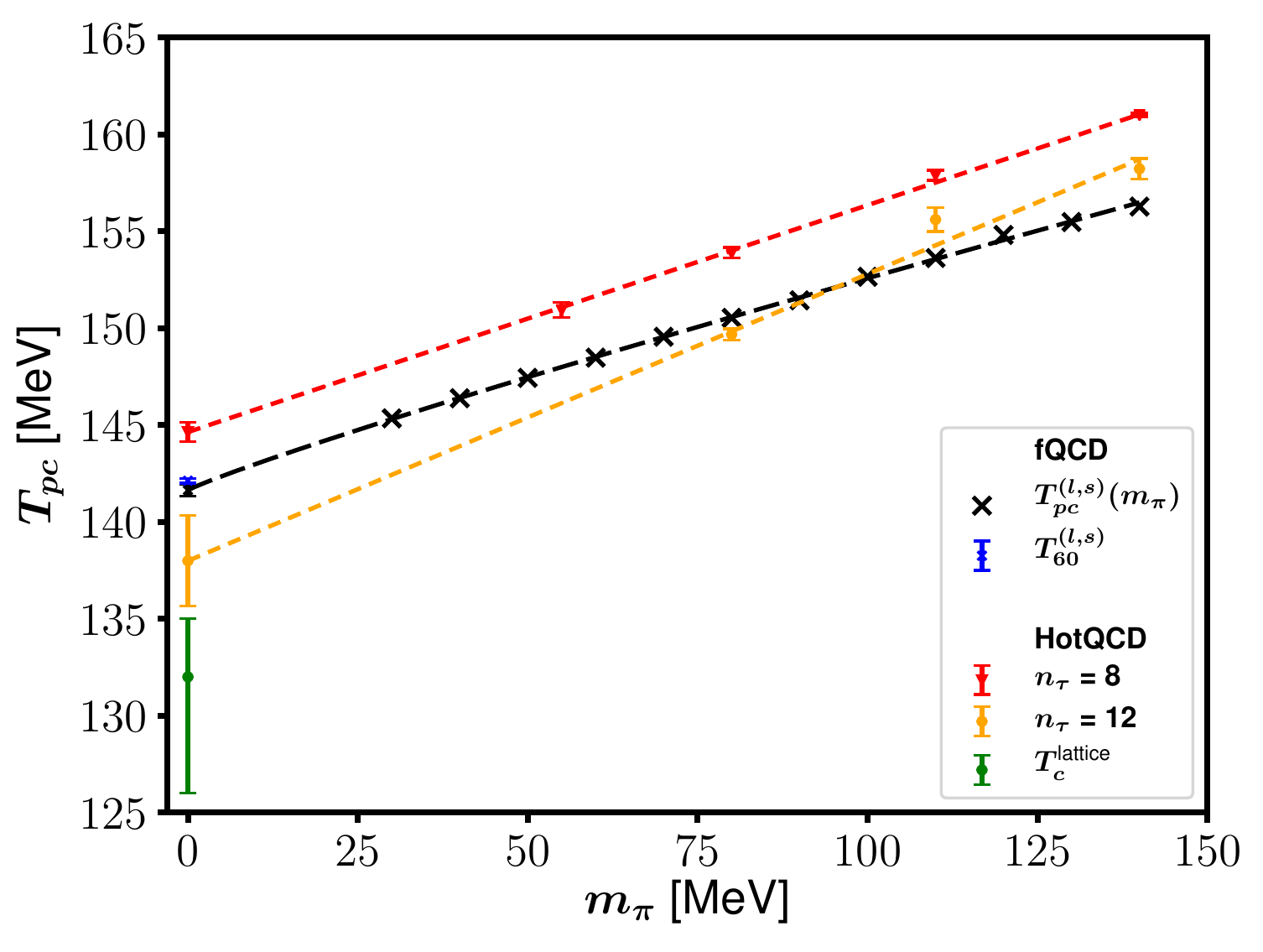}
  \caption{Pseudocritical temperature as a function of the pion mass obtained in fRG (fQCD) in comparison to the lattice results \cite{Ding:2019prx}. The critical temperature in the chiral limit $T_{\mathrm{c}}$ has been obtained from an extrapolation of the pseudocritical temperature $T_{\mathrm{pc}}$ to $m_{\pi}\to 0$. The plot is adopted from \cite{Braun:2020ada}.}\label{fig:tpc_mpi}
\end{figure}
%

From \Eq{eq:mDer-cDer}, \Eq{eq:RenCondSigma}, and \Eq{eq:Deltals}, it is convenient to define the corresponding susceptibilities for the light quark condensate, the renormalized light quark condensate, and the reduced condensate, i.e.,
\begin{align}
\chi_{M}^{(i)}(T) = -\frac{\partial}{\partial m_l^{0}}\left(\frac{\Delta_{i}(T)}{m_l^0}\right)\,, \label{eq:MagSus}
\end{align}
with $(i)=(l),\,(l,R),\,(l,s)$ \cite{Braun:2020ada}. Similar with the difference between the light quark condensate and the renormalized light quark condensate, the susceptibilities for them differ by the vacuum contribution. Among these three susceptibilities, the reduce susceptibility for the reduced condensate defined in fRG and lattice QCD, e.g., \cite{Ding:2019prx}, matches better.

In \Fig{fig:chiMls} the susceptibility obtained from the reduced condensate as a function of the temperature for different pion masses in the fRG approach \cite{Braun:2020ada}, is compared with the lattice results \cite{Ding:2019prx}. One can see that the fRG results are in good agreement with the lattice ones for pion masses $m_{\pi}\gtrsim 100$ MeV. There are some slight deviations from the two approaches for small pion masses. From the dependence of the reduced susceptibility on the temperature for a fixed pion mass in \Fig{fig:chiMls}, one is able to define the pseudocritical temperature for the chiral crossover at such value of pion mass, via the peak position of the curves, denoted by $T_{\mathrm{pc}}$. 

In \Fig{fig:tpc_mpi} one shows the dependence of the pseudocritical temperature on the pion mass, obtained both from the fRG approach and the lattice QCD. The values of pseudocritical temperature at finite pion masses are also extrapolated to the chiral limit, i.e., $m_{\pi}\to 0$, which leaves us with the critical temperature $T_{\mathrm{c}}$ in the chiral limit. One obtains $T_c^{\mathrm{fRG}}\approx142$ MeV from the fRG approach \cite{Braun:2020ada}, and $T_c^{\mathrm{lattice}}=132_{-6}^{+3}$ MeV from the lattice QCD \cite{Ding:2019prx}, that is, the critical temperature obtained in fRG is a bit larger than the lattice result. Recently, the critical temperature in the chiral limit from DSE is found to be $T_c^{\mathrm{DSE}}\approx 141$ MeV \cite{Gao:2021vsf}.

From the general scaling hypothesis, see e.g., \cite{Ma:2020a}, in the critical regime the pseudocritical temperature scales with the pion mass as
\begin{align}
T_{\mathrm{pc}}(m_{\pi})\approx T_{\mathrm{c}} + c\,m_{\pi}^{p}\,,\label{eq:TpcRfct}
\end{align}
where $c$ is a non-universal coefficient, while the exponent $p$ is related to the critical exponents $\beta$ and $\delta$ through $p= 2/(\beta\delta)$ \cite{Braun:2020ada}. Inputting the values of $\beta$ and $\delta$ for the 3-$d$ $O(4)$ universality class \cite{Kanaya:1994qe, Guida:1998bx, Hasenbusch:2000ph}, one arrives at $p\approx 1.08$. On the contrary, fitting of the fRG data in \Fig{fig:tpc_mpi} leads to $p \approx 0.91_{-0.03}^{+0.03}$ \cite{Braun:2020ada}. This discrepancy indicates that the pion masses under investigation in \cite{Braun:2020ada}, i.e., $m_{\pi} \gtrsim 30$ MeV are beyond the critical regime. In fact, the critical regime is found to be extremely small, $m_{\pi} \lesssim 1$ MeV, in fRG studies of LEFTs, and see, e.g., \cite{Schaefer:2006ds, Braun:2007td, Braun:2010vd, Chen:2021iuo} for a more detailed discussion.

	
\section{Real-time fRG}
\label{sec:RealtimefRG}

In this section we would like to give a brief introduction about the real-time fRG, based on a combination of the fRG approach and the formalism of Schwinger-Keldysh path integral, where the flow equations are formulated on the closed time path. The Schwinger-Keldysh path integral is devised to study real-time quantum dynamics \cite{Schwinger:1960qe, Keldysh:1964ud}, which has been proved be very powerful to cope with problems of both equilibrium and nonequilibrium many-body systems, see, e.g., \cite{Chou:1984es, Blaizot:2001nr, Berges:2004yj, Sieberer:2015svu} for relevant reviews.

Within the fRG approach on the closed time path, nonthermal fixed points of the $O(N)$ scalar theory are investigated \cite{Berges:2008sr, Berges:2012ty}. The transition from unitary to dissipative dynamics is studied \cite{Mesterhazy:2015uja}. Spectral functions in a scalar field theory with $d$=0+1 dimensions are computed within the real-time fRG approach \cite{Huelsmann:2020xcy}. Moreover, it has also been employed to study the nonequilibrium transport, dynamical critical behavior, etc. in open quantum systems \cite{Jakobs:2007, Sieberer:2013}, see also \cite{Sieberer:2015svu, Dupuis:2020fhh} for related reviews. Recently, the real-time fRG is compared with other real-time methods \cite{Roth:2021nrd}.

Furthermore, it should be mentioned that another conceptually new fRG with the Keldysh functional integral is put forward in \cite{Gasenzer:2007za, Gasenzer:2010rq, Corell:2019jxh, Heller:2021wan}, and the regulation of the RG scale there is replaced with that of the time, which is also called as the temporal renormalization group. Noteworthily, recently the K\"{a}ll\'{e}n-Lehmann spectral representation of correlation functions has been used in the approach of DSE, to study the spectral functions in the $\phi^4$-theory and Yang-Mills theory \cite{Horak:2020eng, Horak:2021pfr, Horak:2022myj}. The spectral functions provide important informations on the time-like properties of correlation functions, see, e.g., \cite{Haas:2013hpa, Christiansen:2014ypa, Cyrol:2018xeq, Binosi:2019ecz}. Besides the real-time fRG, another commonly adopted approach is the analytically continued functional renormalization group, where the Euclidean flow equations are analytically continued into the Minkowski spacetime on the level of analytic expressions with some specific truncations, and see, e.g., \cite{Tripolt:2013jra, Tripolt:2014wra, Pawlowski:2015mia, Jung:2016yxl, Jung:2021ipc} for more details.

In this section we would like to discuss the fRG formulated on the Keldysh path at the example of the $O(N)$ scalar theory \cite{Tan:2021zid}.

\subsection{fRG with the Keldysh functional integral}
\label{subsec:fRGKeldysh}

On the closed time path the classical action in \Eq{eq:Zk} for a closed system reads
\begin{align}
  S[\hat\Phi]&=\int_x \Big(\mathcal{L}[\hat\Phi_+]-\mathcal{L}[\hat\Phi_-]\Big)\,,\label{}
\end{align}
with $\int_x\equiv \int_{-\infty}^{\infty}d t \int d^3 x$, where the subscripts $\pm$ denote variables on the forward and backward time branches, respectively \cite{Chou:1984es}. Now the Keldysh generating functional reads
\begin{align}
  &Z[J_+,J_-]\nonumber \\[2ex]
  =&\int \big(\mathcal{D} \hat\Phi_+\mathcal{D} \hat\Phi_-\big) \exp\Big\{i \Big(S[\hat\Phi]+\big(J_+^i\hat\Phi_{i,+}-J_-^i\hat\Phi_{i,-}\big)\Big)\Big\}\,.\label{eq:ZJpm}
\end{align}
It is more convenient to adopt the physical representation in terms of the ``classical'' and ``quantum'' variables, denoted by subscripts $c$ and $q$, respectively, which are related to variables on the two time branches by a Keldysh rotation, that is,
\begin{align}
  &\left \{\begin{array}{l}
 \hat\Phi_{i,+}=\frac{1}{\sqrt{2}}(\hat\Phi_{i,c}+\hat\Phi_{i,q}),\\[3ex]
 \hat\Phi_{i,-}=\frac{1}{\sqrt{2}}(\hat\Phi_{i,c}-\hat\Phi_{i,q}),
\end{array} \right.  \label{eq:phipmTophy}
\end{align}
and 
\begin{align}
  &\left \{\begin{array}{l}
 J_+^i=\frac{1}{\sqrt{2}}(J_c^i+J_q^i),\\[3ex]
 J_-^i=\frac{1}{\sqrt{2}}(J_c^i-J_q^i),
\end{array} \right.  \label{}
\end{align}
Then, \Eq{eq:ZJpm} reads
\begin{align}
  &Z[J_c,J_q]\nonumber \\[2ex]
  =&\int \big(\mathcal{D} \hat\Phi_c \mathcal{D} \hat\Phi_q\big)\exp\Big\{i \Big(S[\hat\Phi]+\big(J_q^i\hat\Phi_{i,c}+J_c^i\hat\Phi_{i,q}\big)\Big)\Big\}\,.\label{eq:ZJcq}
\end{align}
The bilinear regulator term in \Eq{eq:DeltaS} can be chosen as
\begin{align}
  \Delta S_k[\hat\Phi]=&\frac{1}{2}(\hat\Phi_{i,c},\hat\Phi_{i,q})\begin{pmatrix} 0 &R_k^{ij}\\ (R_k^{ij})^* & 0 \end{pmatrix}\begin{pmatrix} \hat\Phi_{j,c} \\ \hat\Phi_{j,q} \end{pmatrix}\nonumber \\[2ex]
  =&\frac{1}{2}\Big(\hat\Phi_{i,c}R_k^{ij}\hat\Phi_{j,q}+\hat\Phi_{i,q} (R_k^{ij})^*\hat\Phi_{j,c}\Big)\,.  \label{eq:DeltaSk}
\end{align}
Then the RG scale $k$-dependent generating function \Eq{eq:Zk} with the Keldysh functional integral reads
\begin{align}
  Z_k[J_c,J_q]
  =&\int \big(\mathcal{D} \hat\Phi_c \mathcal{D} \hat\Phi_q\big)\exp\Big\{i \Big(S[\hat\Phi]+\Delta S_k[\hat\Phi]\nonumber \\[2ex]
  &+(J_q^i\hat\Phi_{i,c}+J_c^i\hat\Phi_{i,q})\Big)\Big\}\,.\label{}
\end{align}
The Schwinger functional in \Eq{eq:Wk} reads
\begin{align}
  W_k[J_c,J_q]&=-i\ln Z_k[J_c,J_q]\,.\label{eq:WkJcJq} 
\end{align}
Combining the indices $c,\, q$ and $i$ for other degrees of freedom into one single label, say $a$, i.e.,
\begin{align}
   \{\hat\Phi_a\}&= \big\{ \{\hat\Phi_{i,c}\}, \{\hat\Phi_{i,q}\} \big\}\,, \\[2ex]
   \{J^a\}&=\big\{ \{J_q^i\}, \{J_c^i\} \big\}\,.\label{}
\end{align}
one is able to simplify notations significantly, for instance, the regulator term in \Eq{eq:DeltaSk} now reading
\begin{align}
  \Delta S_k[\varphi]=&\frac{1}{2}\hat\Phi_{a}R_k^{ab}\hat\Phi_{b}\,,  \label{eq:DeltaSk2}
\end{align}
with
\begin{align}
  R_k^{ab}\equiv&\begin{pmatrix} 0 &R_k^{ij}\\ (R_k^{ij})^* & 0 \end{pmatrix}\,.  \label{eq:Rab}
\end{align}
In the same way, the expectation value of the field reads
\begin{align}
   \Phi_a&\equiv\langle\hat\Phi_a\rangle=\frac{\delta W_k[J]}{\delta J^a}\,,\label{eq:dWkdJ}
\end{align}
and the propagator is given by
\begin{align}
   G_{k,ab}\equiv& -i\langle \hat\Phi_a \hat\Phi_b \rangle_c=-i\big[\langle \hat\Phi_a \hat\Phi_b \rangle-\langle \hat\Phi_a\rangle\langle \hat\Phi_b\rangle\big]\nonumber\\[2ex]
  =&-\frac{\delta^2 W_k[J]}{\delta J^a J^b}\,.\label{eq:d2Wkd2J}
\end{align}

The effective action is obtained from the Schwinger functional upon a Legendre transformation, viz.,
\begin{align}
  \Gamma_k[\Phi]&=W_k[J]-\Delta S_k[\Phi]-J^a\Phi_a\,.  \label{eq:Gammak}
\end{align}
Similar with \Eq{eq:dGamdPhi} and \Eq{eq:invgam2}, one has
\begin{align}
  \frac{\delta(\Gamma_{k}[\Phi]+\Delta S_k[\Phi])}{\delta \Phi_a}&=-\gamma^a_{\hspace{0.15cm}b} J^b\,,  \label{}
\end{align}
and
\begin{align}
  G_{k,ab}&=\gamma^c_{\hspace{0.15cm}a}\Big[\big(\Gamma_{k}^{(2)}[\Phi]+\Delta S_k^{(2)}[\Phi]\big)^{-1}\Big]_{cb}\,.  \label{eq:invgam2rt}
\end{align}
with
\begin{align}
  \Big(\Gamma_{k}^{(2)}[\Phi]+\Delta S_k^{(2)}[\Phi]\Big)^{ab}&\equiv\frac{\delta^2(\Gamma_{k}[\Phi]+\Delta S_k[\Phi])}{\delta \Phi_a \delta \Phi_b}\,.  \label{eq:d2GadPhi2rt}
\end{align}
Then it is left to specify the flow equation of Schwinger functional with the Keldysh functional integral, that reads
\begin{align}
  \partial_\tau W_k[J]=&\frac{1}{2}iG_{k,ab} \partial_\tau R^{ab}_k+\frac{1}{2}\Phi_{a}\big(\partial_\tau R^{ab}_k\big)\Phi_{b}\nonumber \\[2ex]
  =&\frac{i}{2}\mathrm{STr}\Big[\big(\partial_\tau R^{\mathrm{T}}_k\big) G_{k}\Big]+\frac{1}{2}\Phi_{a}\big(\partial_\tau R^{ab}_k\big)\Phi_{b}\,, \label{eq:dtWkrt}
\end{align}
where the RG time is denoted by $\tau\equiv\ln (k/\Lambda)$, and $R^{\mathrm{T}}_k$ stands for the transpose of the regulator only in the $c$-$q$ space as shown in \Eq{eq:Rab}. Finally, the flow equation of the effective action reads
\begin{align}
  \partial_\tau \Gamma_k[\Phi]&=\partial_\tau W_k[J]-\partial_\tau\Delta S_k[\Phi]\nonumber \\[2ex]
  &=\frac{i}{2}\mathrm{STr}\Big[\big(\partial_\tau R^{\mathrm{T}}_k\big) G_{k}\Big]\,.  \label{eq:actionflow}
\end{align}

\subsection{Real-time $O(N)$ scalar theory}
\label{subsec:ONmodel}

%
\begin{figure}[t]
\includegraphics[width=0.98\columnwidth]{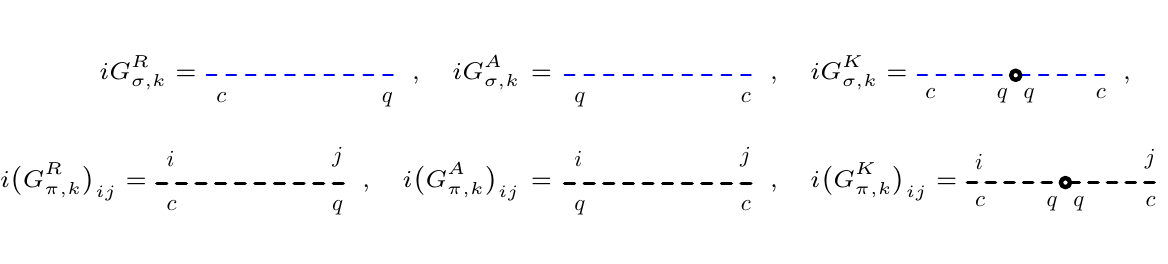}
\caption{Diagrammatic representation of the three different propagators for the sigma and pion mesons. A line associated with two end points labelled with ``$c,q$'' denotes the retarded propagator, while that with 
``$q,c$''  denotes the advanced propagator. A line with an empty circle inserted in the middle is used to denote the Keldysh propagator, which results from \Eq{eq:GK}. The plot is adopted from \cite{Tan:2021zid}.}\label{fig:FeynRule-meson-prop}
\end{figure}
%

The Keldysh effective action in \Eq{eq:Gammak} for the $O(N)$ scalar theory reads
\begin{align}
  \Gamma_k[\phi_c,\phi_q]=&\int_{x}\Big[Z_{\phi,k} (\partial_\mu \phi_{q})\cdot(\partial^\mu \phi_{c})-U_k(\phi_c,\phi_q)\Big]\,, \label{eq:GammaON}
\end{align}
where the potential is given by
\begin{align}
  U_k(\phi_c,\phi_q)=&V_k(\rho_+)- V_k(\rho_-)\,, \label{eq:Uk}
\end{align}
with $\rho_\pm=\phi_\pm^2/2$. Here $\phi_{i,\pm}$ and $\phi_{i,c/q}$ ($i=0,1,...N-1$) denote the fields of $N$ components on the closed time branches or in the physical representation, respectively, and their relations are given in \Eq{eq:phipmTophy}. Note that in \Eq{eq:GammaON} a local potential approximation with a $k$-dependent wave function renormalization $Z_{\phi,k}$ has been adopted, which is usually called the LPA' approximation.

When the $O(N)$ symmetry is broken into the $O(N-1)$ one in the direction of component $i=0$, the expectation values of the fields read
\begin{align}
   \bar \phi_{i,c}=&\bar \phi_{c}\delta_{i0}\,,\qquad \bar \phi_{i,q}=0\,.\label{eq:barphic}
\end{align}
Then, the sigma and pion fields are give by
\begin{align}
   \sigma_c=&\phi_{0,c}-\bar \phi_{c}\,,\qquad \sigma_q=\phi_{0,q}\,,\label{eq:sigmaField}
\end{align}
and 
\begin{align}
   \pi_{i,c}=&\phi_{i,c}\,,\qquad \pi_{i,q}=\phi_{i,q}\,, \qquad (i\ne 0)\,.\label{eq:piField}
\end{align}
The sigma and pion masses read
\begin{align}
   m_{\sigma,k}^2\equiv&V'_k(\bar \rho_{c})+2\bar \rho_{c}V^{(2)}_k(\bar \rho_{c})\,,\\[2ex]
   m_{\pi,k}^2\equiv&V'_k(\bar \rho_{c})\,,\label{eq:masspi2}
\end{align}
with $\bar \rho_{c}\equiv{\bar \phi_{c}}^2/4$. 

The regulator in \Eq{eq:Rab} in the case of the $O(N)$ scalar theory reads
\begin{align}
  R_k(q)=&\begin{pmatrix} 0 &R_k^{A}(q)\\[2ex] R_k^{R}(q) & 0 \end{pmatrix}\,,  \label{eq:regulator2}
\end{align}
with
\begin{align}
  R_k^{R}(q)&=R_k^{A}(q)=\begin{pmatrix} R_{\sigma,k}(q) &0\\[2ex] 0 & R_{\pi,k}(q) \end{pmatrix}\,.  \label{eq:regulator}
\end{align}
where one has
\begin{align}
  R_{\sigma,k}(q)=&R_{\phi,k}(q)\,,\qquad   \big(R_{\pi,k}\big)_{ij}(q)=R_{\phi,k}(q)\delta_{ij}\,, \label{eq:Rsigma}
\end{align}
with
\begin{align}
  R_{\phi,k}(q)=&Z_{\phi,k}\Big(-{\bm{q}}^2r_{B}\big(\frac{{\bm{q}}^2}{k^2}\big)\Big)\,.  \label{}
\end{align}
Here, the $3d$ flat regulator is used. 

The flow equation of effective action in \Eq{eq:actionflow} with the regulator in \Eq{eq:regulator2} can be reformulated as
\begin{align}
  \partial_\tau \Gamma_k[\Phi]&=\frac{i}{2}\mathrm{STr}\Big[\tilde\partial_\tau \ln \big(\Gamma_{k}^{(2)}[\Phi]+R_{k}\big)\Big]\,,  \label{eq:actionflow2}
\end{align}
with
\begin{align}
    \Big(\Gamma_{k}^{(2)}[\Phi]\Big)_{ab}&\equiv \frac{\delta^2\Gamma_{k}[\Phi]}{\delta \Phi_a \delta \Phi_b}\,,\label{}
\end{align}
where the fluctuation matrix can be decomposed into a sum of the inverse propagator $\mathcal{P}_{k}$ and the interaction $\mathcal{F}_{k}$ as shown in \Eq{eq:flucmatrdecom}. In thermal equilibrium at a temperature $T$, the $\mathcal{P}_{k}$ matrix reads
\begin{align}
  \mathcal{P}_{k}=&\begin{pmatrix} 0 &\mathcal{P}^{A}_{k}\\[1ex] \mathcal{P}^{R}_{k} & \mathcal{P}^{K}_{k} \end{pmatrix}\,,\label{eq:invprop}
\end{align}
where the inverse retarded propagator is given by
\begin{align}
  \mathcal{P}^{R}_{k}=&\begin{pmatrix} \mathcal{P}^{R}_{\sigma,k} &0\\[2ex] 0 & \mathcal{P}^{R}_{\pi,k} \end{pmatrix}\,,  \label{}
\end{align}
with
\begin{align}
  \mathcal{P}^{R}_{\sigma,k}=&Z_{\phi,k}\bigg[q_0^2-{\bm{q}}^2\Big(1+r_{B}\big(\frac{{\bm{q}}^2}{k^2}\big)\Big)\bigg]-m_{\sigma,k}^2\nonumber\\[2ex]
&+\mathrm{sgn}(q_0)i\epsilon\,, \label{eq:PRsigma}\\[2ex]
  \big(\mathcal{P}^{R}_{\pi,k}\big)_{ij}=&\Bigg\{Z_{\phi,k}\bigg[q_0^2-{\bm{q}}^2\Big(1+r_{B}\big(\frac{{\bm{q}}^2}{k^2}\big)\Big)\bigg]-m_{\pi,k}^2\nonumber\\[2ex]
&+\mathrm{sgn}(q_0)i\epsilon\Bigg\}\delta_{ij}\,.\label{eq:PRpi}
\end{align}
The inverse advanced propagator is related to the retarded one through a complex conjugate, to wit,
\begin{align}
  \mathcal{P}^{A}_{k}=&(\mathcal{P}^{R}_{k})^*\,.  \label{}
\end{align}
The $qq$ component of the inverse propagator in \Eq{eq:invprop}, i.e., $\mathcal{P}^{K}_{k}$, reads
\begin{align}
  \mathcal{P}^{K}_{k}=&\begin{pmatrix} \mathcal{P}^{K}_{\sigma,k} &0\\[2ex] 0 & \mathcal{P}^{K}_{\pi,k} \end{pmatrix}\,,  \label{}
\end{align}
with
\begin{align}
  \mathcal{P}^{K}_{\sigma,k}=&2i\epsilon \,\mathrm{sgn}(q_0)\coth\Big(\frac{q_0}{2T}\Big)\,,\\[2ex]
  \big(\mathcal{P}^{K}_{\pi,k}\big)_{ij}=&\bigg[2i\epsilon \,\mathrm{sgn}(q_0)\coth\Big(\frac{q_0}{2T}\Big)\bigg]\delta_{ij}\,.\label{}
\end{align}
Then, one can obtain the propagator matrix as follows
\begin{align}
  G_{k}=&\big(\mathcal{P}_{k}\big)^{-1}=\begin{pmatrix} G^{K}_{k} &G^{R}_{k}\\[1ex] G^{A}_{k} & 0 \end{pmatrix}\,,\label{eq:prop}
\end{align}
with the retarded and advanced propagator being
\begin{align}
  G^{R}_{k}=&(\mathcal{P}^{R}_{k})^{-1}\,,\qquad G^{A}_{k}=(\mathcal{P}^{A}_{k})^{-1}\,,\label{eq:GRGA}
\end{align}
and the correlation function or Keldysh propagator being
\begin{align}
  G^{K}_{k}=&-(\mathcal{P}^{R}_{k})^{-1}\mathcal{P}^{K}_{k}(\mathcal{P}^{A}_{k})^{-1}=-G^{R}_{k}\mathcal{P}^{K}_{k}G^{A}_{k}\,.\label{eq:GK}
\end{align}
Finally, one can verify the fluctuation-dissipation relation in thermal equilibrium, as follows
\begin{align}
  G^{K}_{k}=&\big(G^{R}_{k}-G^{A}_{k}\big)\coth\Big(\frac{q_0}{2T}\Big)\,.\label{eq:FDR}
\end{align}

From \Eq{eq:prop} one arrives at the three different propagators which read
\begin{align}
  i G^{R}_{\phi,k}=&\langle T_p\phi_c(x)\phi_q(y)\rangle\,,\quad i G^{A}_{\phi,k}=\langle T_p\phi_q(x)\phi_c(y)\rangle\,,\\[2ex]
  i G^{K}_{\phi,k}=&\langle T_p\phi_c(x)\phi_c(y)\rangle=\big(iG^{R}_{\phi, k}\big)\big(i\mathcal{P}^{K}_{\phi, k}\big)\big(iG^{A}_{\phi, k}\big)\,,\label{eq:GKphi}
\end{align}
where $T_p$ denotes the time ordering from the positive to negative time branch in the Keldysh closed time path, and \Eq{eq:GKphi} follows directly from \Eq{eq:GK}. Diagrammatic representation of these propagators is shown in \Fig{fig:FeynRule-meson-prop}.

%
\begin{figure}[t]
\includegraphics[width=0.48\textwidth]{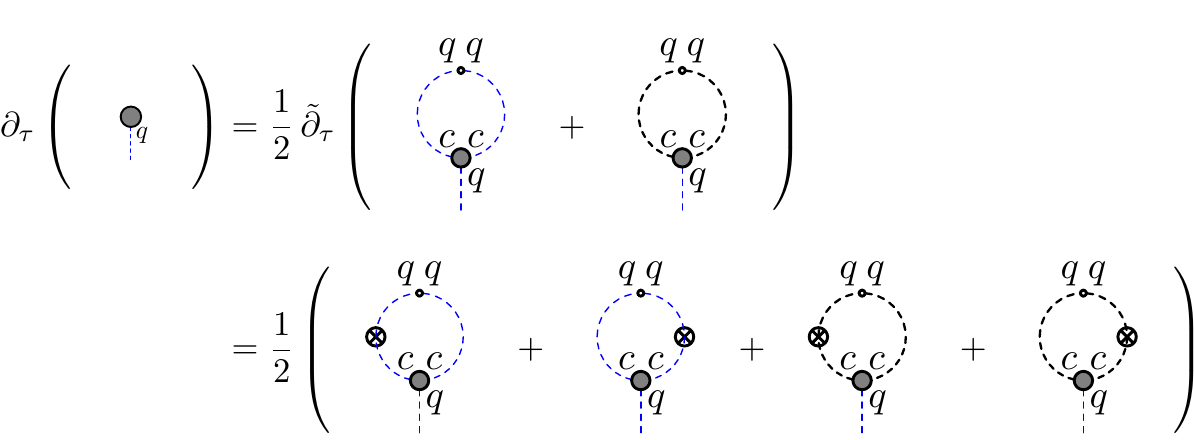}
\caption{Diagrammatic representation of the flow equation of the effective potential in the real-time $O(N)$ scalar theory. The vertices are denoted by gray blobs, and their legs are distinguished for the ``$q$'' and ``$c$'' fields. the crossed circles stand for the regulator insertion. The plot is adopted from \cite{Tan:2021zid}.}\label{fig:Gam1-sigq-equ}
\end{figure}
%

Projecting the flow equation of the effective action in \Eq{eq:actionflow2} onto a derivative with respect to the quantum sigma field $\sigma_{q}$, i.e.,
\begin{align}
    \left.\partial_\tau \left(\frac{i\delta \Gamma_{k}[\Phi]}{\delta \sigma_{q}}\right)\right |_{\Phi=0} \,,\label{}
\end{align}
with $\Phi=(\sigma_{c},\{\pi_{i,c}\},\sigma_{q},\{\pi_{i,q}\})$, one is led to
\begin{align}
  &\partial_\tau V'_k(\bar \rho_{c})\nonumber\\[2ex]
=&\frac{i}{4}\int \frac{d^{4}q}{(2\pi)^{4}}\tilde\partial_\tau \Big[ \frac{\partial m_{\sigma,k}^2}{\partial \bar \rho_{c}} G^{K}_{\sigma,k}(q)+\frac{\partial m_{\pi,k}^2}{\partial \bar \rho_{c}}\big(G^{K}_{\pi,k}\big)_{ii}(q)\Big]\nonumber\\[2ex]
=&\frac{\partial}{\partial \bar \rho_{c}}\bigg\{-\frac{i}{4}\int \frac{d^{4}q}{(2\pi)^{4}}\big(\partial_\tau R_{\phi,k}(q)\big)\Big[G^{K}_{\sigma,k}(q)\nonumber\\[2ex]
&+\big(G^{K}_{\pi,k}\big)_{ii}(q)\Big]\bigg\}\,,\label{eq:flowV1}
\end{align}
which is diagrammatically shown in \Fig{fig:Gam1-sigq-equ}. Integrating both sides of \Eq{eq:flowV1} over $\bar \rho_{c}$, one arrives at 
\begin{align}
  &\partial_\tau V_k(\bar \rho_{c})\nonumber\\[2ex]
=&-\frac{i}{4}\int \frac{d^{4}q}{(2\pi)^{4}}\big(\partial_\tau R_{\phi,k}(q)\big)\Big[G^{K}_{\sigma,k}(q)+\big(G^{K}_{\pi,k}\big)_{ii}(q)\Big]\,,\label{}
\end{align}
up to a term independent of $\bar \rho_{c}$. Inserting \Eq{eq:PRsigma} and \Eq{eq:PRpi}, one has
\begin{align}
  \partial_\tau V_k(\bar \rho_{c})=&\frac{k^4}{4\pi^2} \bigg [l^{(B,4)}_{0}(\tilde{m}^{2}_{\sigma,k},\eta_{\phi,k};T) \nonumber\\[2ex]
&+\big(N-1\big) l^{(B,4)}_{0}(\tilde{m}^{2}_{\pi,k},\eta_{\phi,k};T)\bigg]\,, \label{eq:flowV2}
\end{align}
with the threshold function given by
\begin{align}
  &l^{(B,4)}_{0}(\tilde{m}^{2}_{\phi,k},\eta_{\phi,k};T)\nonumber\\[2ex]
  =&\frac{2}{3}\left(1-\frac{\eta_{\phi,k}}{5}\right)\frac{1}{\sqrt{1+\tilde{m}^{2}_{\phi,k}}}\left(\frac{1}{2}+n_B(\tilde{m}^{2}_{\phi,k};T)\right)\,, \label{}
\end{align}
and the renormalized dimensionless meson masses reading
\begin{align}
  \tilde{m}^{2}_{\sigma,k}=&\frac{m^{2}_{\sigma,k}}{k^2 Z_{\phi,k}}\,,\qquad \tilde{m}^{2}_{\pi,k}=\frac{m^{2}_{\pi,k}}{k^2 Z_{\phi,k}}\,,\label{}
\end{align}
where the bosonic distribution function reads
\begin{align}
  n_B(\tilde{m}^{2}_{\phi,k};T)&=\frac{1}{\exp\bigg\{\frac{k}{T}\sqrt{1+\tilde{m}^{2}_{\phi,k}}\bigg\}-1}\,.
\end{align} 
Note that \Eq{eq:flowV2} is just the flow equation of the effective potential in the LPA' approximation obtained in the conventional Euclidean formalism of fRG.

\subsection{Flows of the two- and four-point correlation functions}
\label{subsec:propandver}

%
\begin{figure}[t]
\includegraphics[width=0.5\textwidth]{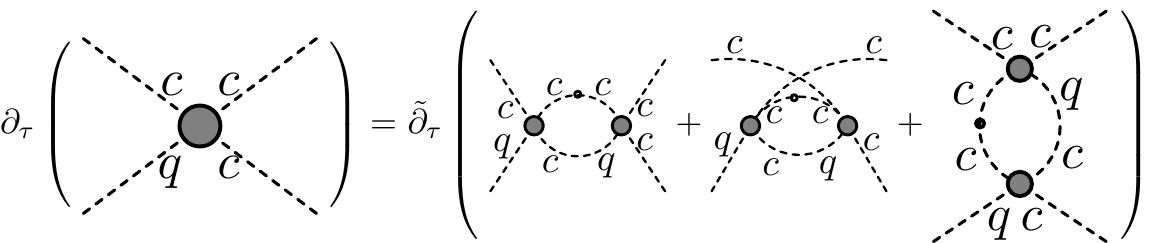}
\caption{Diagrammatic representation of the flow equation of the four-point vertex in the symmetric phase. The plot is adopted from \cite{Tan:2021zid}.} \label{fig:Gam4-equ-phi4}
\end{figure}
%

The 1PI $n$-point correlation function or vertex in the Euclidean field theory is given in \Eq{eq:1PIVert}, and the counterpart in the Keldysh field theory reads
\begin{align}
  i \Gamma_{k,\Phi_{a_1}\cdots\Phi_{a_n}}^{(n)}&=\left. \left(\frac{i\delta^{n} \Gamma_{k}[\Phi]}{\delta \Phi_{a_1}\cdots\delta \Phi_{a_n}}\right)\right |_{\Phi=\langle \Phi\rangle}\,.\label{eq:1PIVertKeldysh}
\end{align}
Note that a label ``$c$'' or ``$q$'' is associated with an external leg of the vertex to distinguish the classical or quantum field, as shown in \Fig{fig:Gam1-sigq-equ}. In the following we consider the symmetric phase with $\bar \phi_{c}=0$ in \Eq{eq:barphic}. In this case the pion and sigma fields are degenerate, and they are collectively denoted by $\phi_i$ ($i=0,1,...N-1$). The diagrammatic representation of the four-point vertex $i \Gamma_{\phi_{q}\phi_{c}\phi_{c}\phi_{c}}^{(4)}$ reads
\begin{align}
  i \Gamma_{k,\phi_{i,q}\phi_{j,c}\phi_{k,c}\phi_{l,c}}^{(4)}(p_i,p_j,p_k,p_l) &\equiv \parbox[c]{0.15\textwidth}{\includegraphics[width=0.12\textwidth]{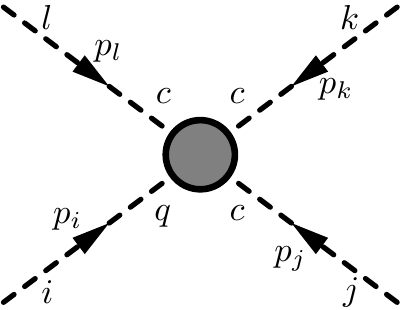}}\,,\label{}
\end{align}
whose flow equation is given in \Fig{fig:Gam4-equ-phi4}. According to the generic interchange symmetry of the external legs, the four-point vertex can be parameterized as
\begin{align}
  &i \Gamma_{k,\phi_{i,q}\phi_{j,c}\phi_{k,c}\phi_{l,c}}^{(4)}(p_i,p_j,p_k,p_l)\nonumber\\[2ex]
=&-\frac{i}{3}\Big[\lambda_{4\pi,k}^{\mathrm{eff}}(p_i,p_j,p_k,p_l)\delta_{il}\delta_{jk}+\lambda_{4\pi,k}^{\mathrm{eff}}(p_i,p_k,p_l,p_j)\delta_{ij}\delta_{kl}\nonumber\\[2ex]
&+\lambda_{4\pi,k}^{\mathrm{eff}}(p_i,p_l,p_j,p_k)\delta_{ik}\delta_{jl}\Big]\,,\label{eq:Gamma4}
\end{align}
where an effective four-point coupling $\lambda_{4\pi,k}^{\mathrm{eff}}$ is introduced. In the following, one adopts the truncation that the momentum dependence of the four-point vertex on the r.h.s. of the flow equation in \Fig{fig:Gam4-equ-phi4} is neglected, and the four-point vertex there is identified as
\begin{align}
  \lambda_{4\pi,k}&=\lambda_{4\pi,k}^{\mathrm{eff}}(0)\,.\label{}
\end{align}
Then, from the flow of the four-point vertex in \Fig{fig:Gam4-equ-phi4}, one arrives at 
\begin{align}
  &\partial_\tau \lambda_{4\pi,k}^{\mathrm{eff}}(p_i,p_j,p_k,p_l)\nonumber\\[2ex]
=& \frac{\lambda_{4\pi,k}^2}{3}\Big[(N+4) \tilde\partial_\tau I_k(-p_i-p_l)+2\,\tilde\partial_\tau I_k(-p_i-p_k)\nonumber\\[2ex]
&+2\,\tilde\partial_\tau I_k(-p_i-p_j)\Big]\,,  \label{eq:dtlameff}
\end{align}
with
\begin{align}
  I_k(p)&\equiv i\int \frac{d^{4}q}{(2\pi)^{4}} G^{K}_{\pi,k}(q)G^{A}_{\pi,k}(q-p)\,.  \label{eq:Ifun}
\end{align}
Note that the wave function renormalization $Z_{\phi,k}=1$ is adopted to simplify the calculation of $\tilde\partial_\tau I_k(p)$ in \Eq{eq:dtlameff}. The explicit expression of $I_k(p)$ can be found in \cite{Tan:2021zid}.

With the four-point vertex in \Eq{eq:Gamma4} one is able to obtain the self-energy as follows
\begin{align}
  -i \Sigma_{k,ij}(p) &\equiv \frac{1}{2}\parbox[c]{0.15\textwidth}{\includegraphics[width=0.15\textwidth]{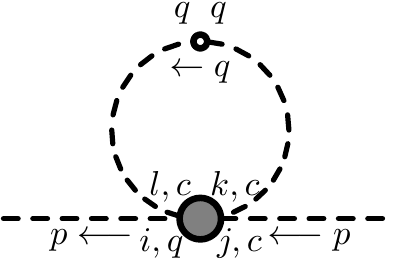}}\,,\label{eq:iSigma}
\end{align}
which reads
\begin{align}
  -i \Sigma_{k,ij}(p)&=\delta_{ij}(-\frac{i}{6})(N+2)\nonumber\\[2ex]
&\times\int \frac{d^{4}q}{(2\pi)^{4}}i G^{K}_{\pi,k}(q)\bar\lambda_{4\pi,k}^{\mathrm{eff}}(p_0,|\bm{p}|,q_0,|\bm{q}|,\cos \theta)\,. \label{}
\end{align}
Here, the function $\bar\lambda_{4\pi,k}^{\mathrm{eff}}$ is given by
\begin{align}
  &\bar\lambda_{4\pi,k}^{\mathrm{eff}}(p_0,|\bm{p}|,q_0,|\bm{q}|,\cos \theta)\nonumber\\[2ex]
=&\frac{1}{N+2}\Big[N \lambda_{4\pi,k}^{\mathrm{eff}}(-p,-q,q,p)+\lambda_{4\pi,k}^{\mathrm{eff}}(-p,p,-q,q)\nonumber\\[2ex]
&+\lambda_{4\pi,k}^{\mathrm{eff}}(-p,q,p,-q)\Big]\,, \label{eq:barlam4pi}
\end{align}
with $\theta$ being the angle between the two momenta $\bm{p}$ and $\bm{q}$. The inverse retarded propagator, i.e., the two-point correlation function with one $q$ field and one $c$ field, is given by
\begin{align}
  i \Gamma_{k,\phi_{i,q}\phi_{j,c}}^{(2)}(p)\equiv& i\frac{\delta^2 \Gamma_k[\phi]}{\delta \phi_{i,q}(p)\delta \phi_{j,c}(-p)}\nonumber\\[2ex]
=&i\delta_{ij} \Big(Z_{\phi,k}(p^2)p^2-m_{\pi,k}^2\Big)\,, \label{eq:iGamma2}
\end{align}
and its flow equation reads
\begin{align}
  &\partial_\tau\Gamma_{k,\phi_{q}\phi_{c}}^{(2)}(p)=-\tilde\partial_\tau \Sigma_{k}(p) \nonumber\\[2ex]
=&(-\frac{i}{6})(N+2)\int \frac{d^{4}q}{(2\pi)^{4}}\tilde\partial_\tau\Big(G^{K}_{\pi,k}(q)\Big)\nonumber\\[2ex]
&\times\bar\lambda_{4\pi,k}^{\mathrm{eff}}(p_0,|\bm{p}|,q_0,|\bm{q}|,\cos \theta)\,. \label{eq:dtGamma2}
\end{align}
Substituting the Keldysh propagator in \Eq{eq:GK}, one arrives at
\begin{align}
  &\partial_\tau\Gamma_{k,\phi_{q}\phi_{c}}^{(2)}(p_0,|\bm{p}|)\nonumber\\[2ex]
=&\partial_\tau\Gamma_{k,\phi_{q}\phi_{c}}^{(2)\mathrm{I}}(p_0,|\bm{p}|)+\partial_\tau\Gamma_{k,\phi_{q}\phi_{c}}^{(2)\mathrm{II}}(p_0,|\bm{p}|)\,.\label{eq:dtGamma2b}
\end{align}
The first part on the r.h.s. of \Eq{eq:dtGamma2b} reads
\begin{align}
  &\partial_\tau\Gamma_{k,\phi_{q}\phi_{c}}^{(2)\mathrm{I}}(p_0,|\bm{p}|))\nonumber\\[2ex]
=&-\frac{1}{24}\frac{(N+2)}{(2\pi)^2}\bigg[-\frac{\coth\Big(\frac{E_{\pi,k}(k)}{2T}\Big)}{\big(E_{\pi,k}(k)\big)^3}-\frac{\mathrm{csch}^2\Big(\frac{E_{\pi,k}(k)}{2T}\Big)}{2T\big(E_{\pi,k}(k)\big)^2}\bigg] \nonumber\\[2ex]
&\times (2k^2)\int_0^k d |\bm{q}| |\bm{q}|^2  \int_{-1}^1 d\cos\theta  \Big[\bar\lambda_{4\pi,k}^{\mathrm{eff}}\big|_{q_0=E_{\pi,k}(k)}\nonumber\\[2ex]
&+\bar\lambda_{4\pi,k}^{\mathrm{eff}}\big|_{q_0=-E_{\pi,k}(k)}\Big]\,,\label{eq:dtGamma2I}
\end{align}
and the second part is given by
\begin{align}
  &\partial_\tau\Gamma_{k,\phi_{q}\phi_{c}}^{(2)\mathrm{II}}(p_0,|\bm{p}|)\nonumber\\[2ex]
=&-\frac{1}{24}\frac{(N+2)}{(2\pi)^2}\frac{\coth\Big(\frac{E_{\pi,k}(k)}{2T}\Big)}{\big(E_{\pi,k}(k)\big)^2}(2k^2)\int_0^k d |\bm{q}| |\bm{q}|^2  \nonumber\\[2ex]
&\times\int_{-1}^1 d\cos\theta \Big[\frac{\partial}{\partial q_0}\bar\lambda_{4\pi,k}^{\mathrm{eff}}\big|_{q_0=E_{\pi,k}(k)}\nonumber\\[2ex]
&-\frac{\partial}{\partial q_0}\bar\lambda_{4\pi,k}^{\mathrm{eff}}\big|_{q_0=-E_{\pi,k}(k)}\Big]\,.\label{eq:dtGamma2II}
\end{align}
with
\begin{align}
  E_{\pi,k}(k)=&\Big(k^2+m^{2}_{\pi,k}\Big)^{1/2}\,,\label{eq:Epik}
\end{align}
where $m_{\pi,k}^2$ can be extracted from the two-point correlation function at vanishing momentum, viz.,
\begin{align}
  m_{\pi,k}^2=&-\Gamma_{k, \phi_{q} \phi_{c}}^{(2)}(0)\,.\label{}
\end{align}
Note that the second part in \Eq{eq:dtGamma2II} is negligible if the momentum dependence of the vertex is mild.

\subsection{Spectral functions and dynamical critical exponent}
\label{subsec:spec-fun}

%
\begin{figure*}[t]
\includegraphics[width=0.9\textwidth]{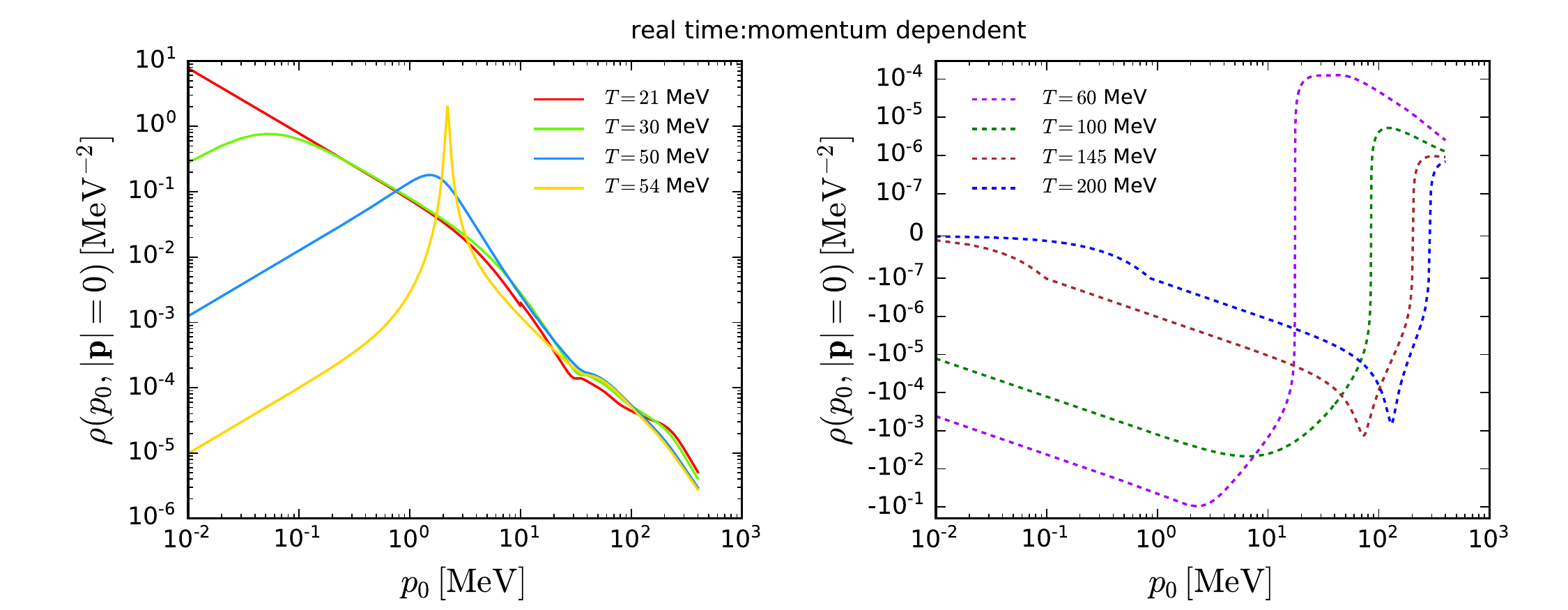}
\caption{Spectral function $\rho(p_0,|\bm{p}|=0)$ as a function of $p_0$ with several small (left panel, close to the critical temperature $T_c=20.4$ MeV) and large (right panel) values of temperature obtained in the real-time $O(4)$ scalar theory within the fRG approach. The plots are adopted from \cite{Tan:2021zid}.}\label{fig:spec_T}
\end{figure*}
%

%
\begin{figure*}[t]
\includegraphics[width=0.9\textwidth]{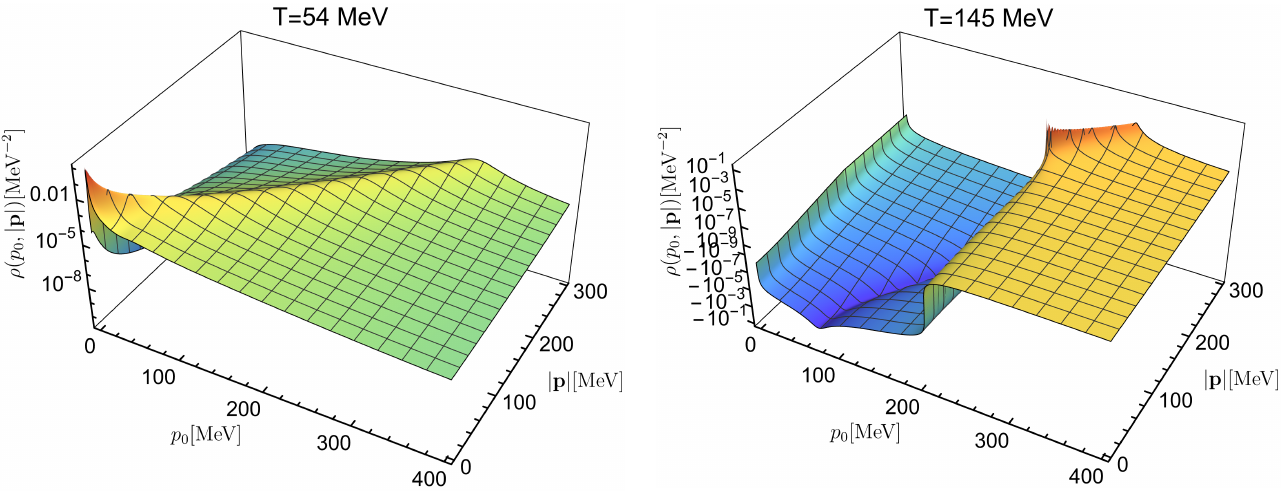}
\caption{3D plots of the spectral function as a function of $p_0$ and $|\bm{p}|$ at temperature $T=54$ MeV (left panel) and $145$ MeV (right panel), obtained in the real-time $O(4)$ scalar theory within the fRG approach. The plots are adopted from \cite{Tan:2021zid}.}\label{fig:spec3D}
\end{figure*}
%

%
\begin{figure}[t]
\includegraphics[width=0.48\textwidth]{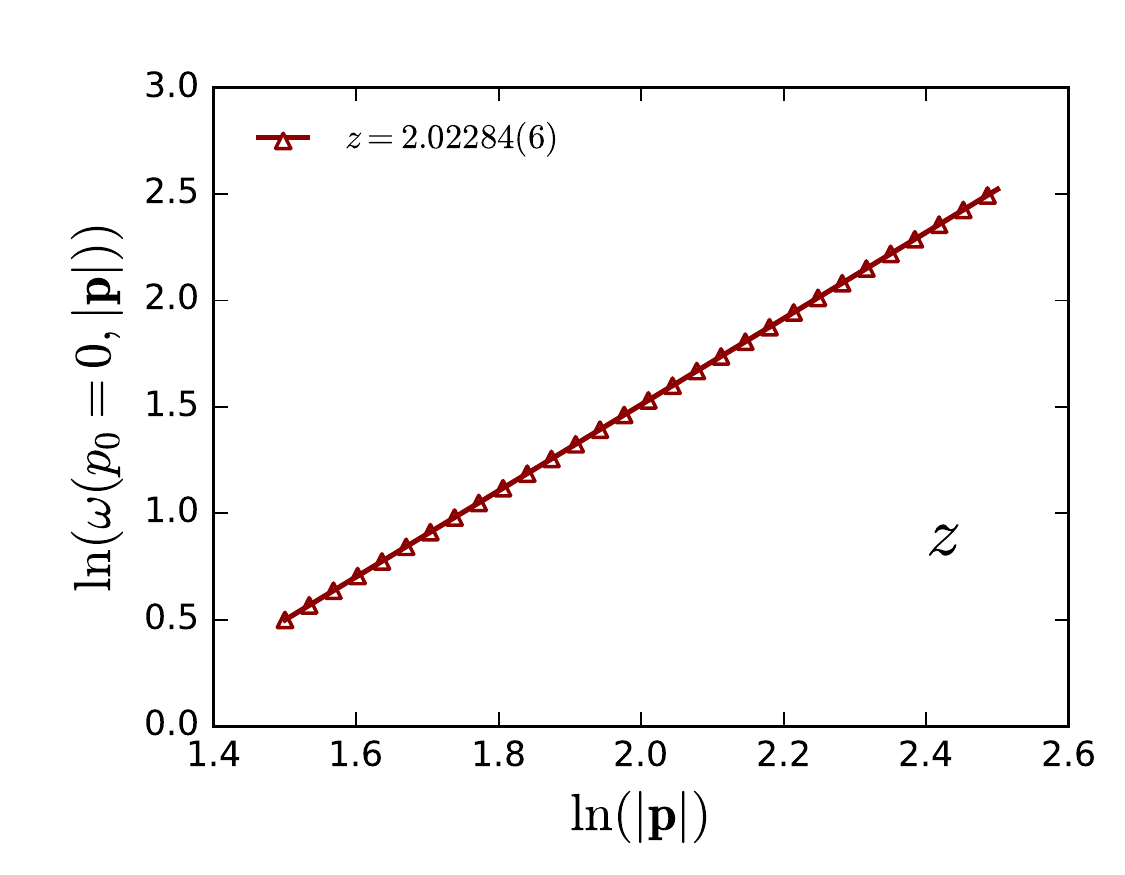}
\caption{Double logarithm plot of the relaxation rate $\omega(|\bm{p}|)$ in \Eq{eq:relax-rate} as a function of the spacial momentum with $p_0=0$ at the critical temperature $T_c=20.4$ MeV, obtained in the real-time 3$d$ $O(4)$ scalar theory within the fRG approach. The plot is adopted from \cite{Tan:2021zid}.}\label{fig:dynamicalCE}
\end{figure}
%

The retarded propagator in the K\"{a}ll\'{e}n-Lehmann spectral representation is related to the spectral function $\rho$ via a relation as follows
\begin{align}
  G_R(p_0,|\bm{p}|)=&-\int_{-\infty}^{\infty}\frac{d p_0^{\prime}}{2\pi}\frac{\rho(p_0^{\prime},|\bm{p}|)}{p_0^{\prime}-(p_0+i\epsilon)}\,. \label{}
\end{align}
Thus, the spectral function is proportional to the imaginary part of the retarded propagator, i.e.,
\begin{align}
  \rho(p_0,|\bm{p}|)=&-2\Im\, G_R(p_0,|\bm{p}|)\,. \label{}
\end{align}
Notice that here the IR limit $k\to 0$ is tacitly assumed. The retarded propagator is just the inverse of the two-point correlation function, to wit,
\begin{align}
 G_R(p_0,|\bm{p}|)=&\left[\Gamma_{\phi_{q}\phi_{c}}^{(2)}(p_0,|\bm{p}|)\right]^{-1}\,. \label{}
\end{align}
Consequently, the spectral function can be expressed in terms of the real and imaginary parts of the two-point correlation function, that is,
\begin{align}
  \rho(p_0,|\bm{p}|)=&\frac{2\Im\,\Gamma_{\phi_{q}\phi_{c}}^{(2)}(p_0,|\bm{p}|)}{\left[\Re\,\Gamma_{\phi_{q}\phi_{c}}^{(2)}(p_0,|\bm{p}|)\right]^2+\left[\Im\,\Gamma_{\phi_{q}\phi_{c}}^{(2)}(p_0,|\bm{p}|)\right]^2}\,. \label{}
\end{align}
Evidently, one has
\begin{align}
  \rho(-p_0,|\bm{p}|)=&-\rho(p_0,|\bm{p}|)\,. \label{}
\end{align}

In \Fig{fig:spec_T}, the spectral function $\rho(p_0,|\bm{p}|=0)$ is shown as a function of $p_0$ with different values of the temperature. In the left panel, the temperatures are above but close to the critical temperature $T_c=20.4$ MeV for the phase transition, where the symmetry is broken from $O(N)$ to $O(N-1)$ when $T < T_c$, while in the right panel, the temperatures are far larger than $T_c$. One can see that when temperature is large, one has a negative spectral function in the regime of small $p_0$, and there is a minus peak structure around the pole mass. It is found that the negative spectral function results from the contributions of the Landau damping, and see \cite{Tan:2021zid} for a more detailed discussion. Nonetheless, when the temperature is decreased below about 60 MeV, the Landau damping is dominated by the process of creation and annihilation of particles, and the spectral function is positive. Moreover, when the temperature is more and more close to the critical temperature, the peak structure on the spectral function becomes more and more wider, and finally disappears at the critical temperature. The 3D plots of the spectral function as a function of $p_0$ and $|\bm{p}|$ are shown in \Fig{fig:spec3D} with two different values of the temperature.

We proceed to the discussion of the dynamical critical exponent, and begin with the kinetic coefficient $\Gamma(|\bm{p}|)$ that is defined as
\begin{align}
  \frac{1}{\Gamma(|\bm{p}|)}=&-i\frac{\partial \Gamma_{\phi_{q}\phi_{c}}^{(2)}(p_0,|\bm{p}|)}{\partial p_0}\bigg|_{p_0=0}\nonumber\\[2ex]
=&\frac{\partial \Im\,\Gamma_{\phi_{q}\phi_{c}}^{(2)}(p_0,|\bm{p}|)}{\partial p_0}\bigg|_{p_0=0}\,, \label{eq:kin-coeff}
\end{align}
where the parity properties of the real and imaginary parts of the two-point correlation function have been used, i.e.,
\begin{align}
  \Re\,\Gamma_{\phi_{q}\phi_{c}}^{(2)}(-p_0,|\bm{p}|)=&\Re\,\Gamma_{\phi_{q}\phi_{c}}^{(2)}(p_0,|\bm{p}|)\,,\\[2ex]
  \Im\,\Gamma_{\phi_{q}\phi_{c}}^{(2)}(-p_0,|\bm{p}|)=&-\Im\,\Gamma_{\phi_{q}\phi_{c}}^{(2)}(p_0,|\bm{p}|)\,. \label{}
\end{align}
The relaxation rate, or dissipative characteristic frequency, is given by
\begin{align}
  \omega(|\bm{p}|)=&\Gamma(|\bm{p}|)\left(-\Gamma_{\phi_{q}\phi_{c}}^{(2)}(p_0=0,|\bm{p}|)\right)\nonumber\\[2ex]
=&-\Gamma(|\bm{p}|)\Re\,\Gamma_{\phi_{q}\phi_{c}}^{(2)}(p_0=0,|\bm{p}|)\,. \label{eq:relax-rate}
\end{align}
When the momentum is larger than the correlation length $|\bm{p}|>\xi^{-1}$, with $\xi\sim m_{\phi}^{-1}$, the relaxation rate scales as 
\begin{align}
  \omega(|\bm{p}|)\propto& |\bm{p}|^z\,, \label{eq:relax-rate}
\end{align}
through which one can extract the dynamical critical exponent $z$ \cite{Hohenberg:1977ym}. From \Fig{fig:dynamicalCE} the value of the dynamical critical exponent in the real-time 3$d$ $O(4)$ scalar theory is extracted, and one arrives at $z=2.02284(6)$, where the numerical error is shown in the bracket.

According to the standard classification for universalities of the critical dynamics \cite{Hohenberg:1977ym}, it is argued that the critical dynamics of the relativistic $O(4)$ scalar theory belongs to the universality of Model G \cite{Rajagopal:1992qz}, see also \cite{Schlichting:2019tbr}, which leaves us with $z=3/2$ in three dimensions. The dynamic critical exponent for the $O(4)$ model is also calculated in real-time classical-statistical lattice simulations, and it is found that $z$ is in favor of 2, but there is still a sizable numerical error \cite{Schlichting:2019tbr}. The dynamic critical exponent in a relativistic $O(N)$ vector model is also found to be close to 2 \cite{Mesterhazy:2015uja}. Similar result is found in a $O(3)$ model in \cite{Duclut:2016jct}. Moreover, $z=1.92(11)$ is found for Model A in three spatial dimensions from real-time classical-statistical lattice simulations \cite{Schweitzer:2020noq}. In short, the dynamic critical exponent is far from clear and conclusive in comparison to the static critical exponent, and more studies of the critical dynamics from different approaches, including the real-time fRG, are necessary and desirable in the future.

\section{Conclusions}
\label{sec:summary}

In this paper we present an overview on recent progress in studies of QCD at finite temperature and densities within the fRG approach. After a brief introduction of the formalism, the fRG approach is applied in low energy effective field theories (LEFTs) and the first-principle QCD. The mechanism of quark mass production and natural emergence of bound states is well illustrated within the fRG approach, and a set of self-consistent flow equations of different correlation functions provide the necessary resummations, and plays the same role as the quark gap equation and Bethe-Salpeter equation of bound states.

We present results for the QCD phase structure and the location of the critical end point (CEP), the QCD equation of state (EoS), the magnetic EoS, baryon number fluctuations confronted with recent experimental measurements, various critical exponents, etc. It is found that the non-monotonic dependence of the kurtosis of the net-baryon or net-proton (proxy for net-baryon in experiments) number distributions could arise from the increasingly sharp crossover with the decrease of the beam collision energy, which in turn indicates that the non-monotonicity observed in experiments is highly non-trivial. Furthermore, recent estimates of the location of the CEP from first-principle QCD calculations within fRG and Dyson-Schwinger Equations, which passes through lattice benchmark tests at small baryon chemical potentials, converge in a rather small region at baryon chemical potentials of about 600 MeV. But it should be reminded that errors of functional approaches increase significantly in the regime of $\mu_B/T\gtrsim 4$, and thus one arrives at a more reasonable estimation for the location of CEP as $450\, \mathrm{MeV}\lesssim{\mu_B}_{\tiny{\mathrm{CEP}}}\lesssim 650\, \mathrm{MeV}$. Moreover, a region of inhomogeneous instability indicated by a negative wave function renormalization is found with $\mu_B\gtrsim 420$ MeV, and its consequence on the phenomenology of heavy-ion collisions has been investigated very recently \cite{Pisarski:2021qof, Rennecke:2021ovl}. It is found that this inhomogeneous instability would result in a moat regime in both the particle $p_T$ spectrum and the two-particle correlation. By investigating the critical behavior and extracting critical exponents in the vicinity of the CEP, it is found that the size of the critical region is extremely small, quite smaller than $\sim 1$ MeV.

In this review we also discuss the real-time fRG, in which the flow equations are formulated on the Schwinger-Keldysh closed time path. By organizing the effective action and the flow equations in terms of the ``classical'' and ``quantum'' fields, one is able to give a concise diagrammatic representation for the flow equations of propagators and vertices in the real-time fRG. The spectral functions of the $O(N)$ scalar theory in the critical regime in the proximity of the critical temperature are obtained. The dynamical critical exponent in the $O(4)$ scalar theory in $3+1$ dimensions is found to be $z\simeq 2.023$.


\begin{acknowledgments}
I would like to thank Heng-Tong Ding, Xiaofeng Luo, Jan M. Pawlowski, Bernd-Jochen Schaefer, Yue-Liang Wu for reading this manuscript and providing helpful comments. I thank Jens Braun, Yong-rui Chen, Chuang Huang, Zi-Wei Lin, Yu-xin Liu, Xiaofeng Luo, Guo-Liang Ma, Jan M. Pawlowski, Fabian Rennecke, Bernd-Jochen Schaefer, Yang-yang Tan, Rui Wen, Yue-Liang Wu, Shi Yin for collaborating on related subjects. I am also grateful to Jens Braun, Yong-rui Chen, Heng-Tong Ding, Fei Gao, De-fu Hou, Chuang Huang, Mei Huang, Zi-Wei Lin, Yu-xin Liu, Xiaofeng Luo, Guo-Liang Ma, Jan M. Pawlowski, Fabian Rennecke, Bernd-Jochen Schaefer, Huichao Song, Yang-yang Tan, Rui Wen, Nicolas Wink, Yue-Liang Wu, Shi Yin, Yu Zhang, Peng-fei Zhuang for discussions on related subjects. I am very grateful to the authors of \cite{Otto:2022} for providing me with the unpublished results in \Fig{fig:phaseline-Otto2022}. I would like to thank the authors of \cite{Fu:2022a} for providing unpublished results in \Fig{fig:flowdiagram}, \Fig{fig:lam-k4d-10ch}, \Fig{fig:mq-k4d-10ch},  \Fig{fig:resonance} and \Fig{fig:invlam-p2-3d}. Moreover, I also would like to thank the members of the fQCD collaboration \cite{fQCD} for discussions and collaborations. This work is supported by the National Natural Science Foundation of China under Grant No. 12175030.
\end{acknowledgments}

	
\appendix
\section{Flow equations of the gluon and ghost self-energies in Yang-Mills theory at finite temperature}
\label{app:gluGhoSelfEner}

\subsection{Feynman rules}
\label{appsubsec:FeynRules}

First of all, we present the Feynman rules for relevant propagators and vertices at finite temperature and densities. 

\subsubsection{Gluon propagator}
\label{appsubsubsec:GluonProp}

The kinetic term for the gluon field in \Eq{eq:QCDaction} in momentum space is given by
\begin{align}
  \Gamma_{k,2A}\equiv &\frac{1}{2}\int \frac{d^4 q}{(2\pi)^4}A^a_{\mu}(-q)q^2\Big[\big(Z_{A,k}^{\mathrm{M}}(q)\Pi_{\mu\nu}^{\mathrm{M}}(q)\nonumber\\[2ex]
  &+Z_{A,k}^{\mathrm{E}}(q)\Pi_{\mu\nu}^{\mathrm{E}}(q)\big)+\frac{1}{\xi}\Pi_{\mu\nu}^{\parallel}(q)\Big]A^a_{\nu}(q)\,,\label{eq:ga2A}
\end{align}
where the magnetic projection operator reads
\begin{align}
  \Pi_{\mu\nu}^{\mathrm{M}}(q)=&(1-\delta_{\mu 0})(1-\delta_{\nu 0})\Big(\delta_{\mu\nu}-\frac{q_{\mu}q_{\nu}}{\bm{q}^2}\Big)\,,\label{eq:magproj}
\end{align}
and the electric projection operator
\begin{align}
  \Pi_{\mu\nu}^{\mathrm{E}}(q)=&\Pi_{\mu\nu}^{\perp}(q)-\Pi_{\mu\nu}^{\mathrm{M}}(q)\,,\label{eq:eleproj}
\end{align}
with the transverse and longitudinal tensors given by
\begin{align}
  \Pi_{\mu\nu}^{\perp}(q)=&\delta_{\mu\nu}-\frac{q_{\mu}q_{\nu}}{q^2},\quad \mathrm{and}\quad\Pi_{\mu\nu}^{\parallel}(q)=\frac{q_{\mu}q_{\nu}}{q^2}\,,\label{}
\end{align}
respectively. Note that in \Eq{eq:ga2A} different magnetic and electric gluonic dressing functions, i.e., $Z_{A,k}^{\mathrm{M}}(q)\ne Z_{A,k}^{\mathrm{E}}(q)$, are assumed at finite temperature. Differentiating \Eq{eq:ga2A} w.r.t. the gluon field twice, one obtains
\begin{align}
  &\big(\Gamma_{k,2A}^{(2)AA}\big)^{ab}_{\mu\nu}(q^{\prime},q)\equiv \frac{\delta^2 \Gamma_{k,2A}}{\delta A^a_{\mu}(q^{\prime})\delta A^b_{\nu}(q)}\nonumber \\[2ex]
 =&q^2\Big[\big(Z_{A,k}^{\mathrm{M}}(q)\Pi_{\mu\nu}^{\mathrm{M}}(q)+Z_{A,k}^{\mathrm{E}}(q)\Pi_{\mu\nu}^{\mathrm{E}}(q)\big)+\frac{1}{\xi}\Pi_{\mu\nu}^{\parallel}(q)\Big]\nonumber\\[2ex]
  &\times\delta^{ab}(2\pi)^4\delta^4(q^{\prime}+q)\,.\label{eq:Ga2AA}
\end{align}
Moreover, the regulator for the gluon reads
\begin{align}
  &\big(R_{k}^{AA}\big)^{ab}_{\mu\nu}(q^{\prime},q)\nonumber\\[2ex]
  =&\Big[\big(Z_{A,k}^{\mathrm{M}}q^2 r_{B}(q^2/k^2)\Pi_{\mu\nu}^{\mathrm{M}}(q)+Z_{A,k}^{\mathrm{E}}q^2 r_{B}(q^2/k^2)\Pi_{\mu\nu}^{\mathrm{E}}(q)\big)\nonumber \\[2ex]
 &+\frac{1}{\xi}q^2 r_{B}(q^2/k^2)\Pi_{\mu\nu}^{\parallel}(q)\Big]\delta^{ab}(2\pi)^4\delta^4(q^{\prime}+q)\,,\label{}
\end{align}
where $Z_{A,k}^{\mathrm{M}}$ and $Z_{A,k}^{\mathrm{E}}$ are independent of momentum $q$ by contrast to those in \Eq{eq:Ga2AA}. The threshold function $r_{B}(x)$ can be chosen to be that in \Eq{eq:regulatorExp2} or \Eq{eq:regulatorOpt2}. Therefore, the gluon propagator is readily obtained from \Eq{eq:invgam2}, as follows
\begin{align}
  \big(G_{k}^{AA}\big)^{ab}_{\mu\nu}(q,q^{\prime})&=(2\pi)^4\delta^4(q^{\prime}+q)\big(G_{k}^{AA}\big)^{ab}_{\mu\nu}(q)\,,\label{}
\end{align}
with
\begin{align}
  \big(G_{k}^{AA}\big)^{ab}_{\mu\nu}(q)&=\Big(G_{A,k}^{\mathrm{M}}(q)\Pi_{\mu\nu}^{\mathrm{M}}(q)+G_{A,k}^{\mathrm{E}}(q)\Pi_{\mu\nu}^{\mathrm{E}}(q)\Big)\delta^{ab}\,,\label{}
\end{align}
where one has
\begin{align}
  G_{A,k}^{\mathrm{M}}(q)&=\frac{1}{q^2\big[Z_{A,k}^{\mathrm{M}}(q)+Z_{A,k}^{\mathrm{M}}r_{B}(q^2/k^2)\big]}\,,\\[2ex]
  G_{A,k}^{\mathrm{E}}(q)&=\frac{1}{q^2\big[Z_{A,k}^{\mathrm{E}}(q)+Z_{A,k}^{\mathrm{E}}r_{B}(q^2/k^2)\big]}\,.\label{}
\end{align}
Here the gauge parameter is chosen to be $\xi=0$.

\subsubsection{Ghost propagator}
\label{appsubsubsec:GhostProp}

In the same way the kinetic term for the ghost field in \Eq{eq:QCDaction} reads
\begin{align}
  \Gamma_{k,2c}&\equiv \int \frac{d^4 q}{(2\pi)^4}Z_{c,k}(q)\bar{c}^a(-q)q^2 c^a(q)\,,\label{}
\end{align}
where we have used the convention of Fourier transformation for the Grassmann fields as follows
\begin{align}
   \bar{c}^a(x)=&\int\frac{d^{4}q}{(2\pi)^{4}} \bar{c}^a(q)e^{iqx}\,,\label{}\\[2ex]
   c^a(x)=&\int\frac{d^{4}q}{(2\pi)^{4}} c^a(q)e^{iqx}\,.\label{}
\end{align}
Then, it follows that 
\begin{align}
  &\big(\Gamma_{k,2c}^{(2)\bar{c}c}\big)^{ab}(q^{\prime},q)\equiv \frac{\overrightarrow{\delta}}{\delta\bar{c}^a(q^{\prime})}\Gamma_{k,2c}\frac{\overleftarrow{\delta}}{\delta c^b(q)}\nonumber\\[2ex]
  =&Z_{c,k}(q) q^2 \delta^{ab}(2\pi)^4\delta^4(q^{\prime}+q)\,,\label{}
\end{align}
and
\begin{align}
  &\big(\Gamma_{k,2c}^{(2)c\bar{c}}\big)^{ab}(q^{\prime},q)\equiv \frac{\overrightarrow{\delta}}{\delta c^a(q^{\prime})}\Gamma_{k,2c}\frac{\overleftarrow{\delta}}{\delta \bar{c}^b(q)}\nonumber\\[2ex]
=&-Z_{c,k}(q^{\prime}) {q^{\prime}}^2 \delta^{ab}(2\pi)^4\delta^4(q^{\prime}+q)\nonumber \\[2ex]
=&-\big(\Gamma_{k,2c}^{(2)\bar{c}c}\big)^{ba}(q,q^{\prime})=-\Big[\big(\Gamma_{k,2c}^{(2)\bar{c}c}\big)^T\Big]^{ab}(q^{\prime},q)\,,\label{}
\end{align}
Note that, for a Grassmann field, the relation such as
\begin{align}
  \Gamma_{k,2c}^{(2)c\bar{c}}&=-\big(\Gamma_{k,2c}^{(2)\bar{c}c}\big)^T\,,\label{}
\end{align}
always holds. The regulator for the ghost field reads
\begin{align}
  \big(R_{k}^{\bar{c}c}\big)^{ab}(q^{\prime},q)&=Z_{c,k}q^2 r_{B}(q^2/k^2)\delta^{ab}(2\pi)^4\delta^4(q^{\prime}+q)\,,\label{}
\end{align}
and $R_{k}^{c\bar{c}}= -\big(R_{k}^{\bar{c}c}\big)^T$. As a consequence, the $\mathcal{P}$ matrix in \Eq{eq:flucmatrdecom} for the ghost field reads
\begin{align}
  \mathcal{P}_c&=\begin{pmatrix} 0 & -(\Gamma_{k,2c}^{(2)\bar{c}c}+R_{k}^{\bar{c}c})^T \\ \Gamma_{k,2c}^{(2)\bar{c}c}+R_{k}^{\bar{c}c} & 0 \end{pmatrix}\,, \label{}
\end{align}
whose inverse matrix is just the ghost propagator which is given by
\begin{align}
  \left(\frac{1}{\mathcal{P}}\right)_c&=\begin{pmatrix} 0 & G_{k}^{c\bar{c}} \\ -(G_{k}^{c\bar{c}})^T & 0 \end{pmatrix}\,, \label{}
\end{align}
with 
\begin{align}
  G_{k}^{c\bar{c}}&=\frac{1}{\Gamma_{k,2c}^{(2)\bar{c}c}+R_{k}^{\bar{c}c}}\,.\label{}
\end{align}
Recovering the indices, one obtains the ghost propagator as follows
\begin{align}
  \big(G_{k}^{c\bar{c}}\big)^{ab}(q,q^{\prime})&=\big(G_{k}^{c\bar{c}}\big)^{ab}(q)(2\pi)^4\delta^4(q+q^{\prime})\nonumber \\[2ex]
 &=G_{k}^{c}(q)\delta^{ab}(2\pi)^4\delta^4(q+q^{\prime})\,,\label{}
\end{align}
with
\begin{align}
  G_{k}^{c}(q)&=\frac{1}{q^2 \big[Z_{c,k}(q)+Z_{c,k}r_{B}(q^2/k^2)\big]}\,.\label{}
\end{align}

\subsubsection{Quark propagator}
\label{appsubsubsec:QuarkProp}

The kinetic term for the quark field in \Eq{eq:QCDaction} reads
\begin{align}
  \Gamma_{k,2q}&\equiv \int \frac{d^4 q}{(2\pi)^4}\bar{q}(-q)\Big[Z_{q,k}(q)i \gamma\cdot q+m_{q,k}(q)\Big]q(q)\,.\label{}
\end{align}
It follows that
\begin{align}
  &\big(\Gamma_{k,2q}^{(2)\bar{q}q}\big)_{ij}(q^{\prime},q)=\frac{\overrightarrow{\delta}}{\delta\bar{q}_i(q^{\prime})}\Gamma_{k,2q}\frac{\overleftarrow{\delta}}{\delta q_j(q)}\nonumber \\[2ex]
 =&\Big[Z_{q,k}(q) i q_{\mu}(\gamma_{\mu})_{ij}+m_{q,k}(q)\delta_{ij}\Big](2\pi)^4\delta^4(q^{\prime}+q)\,,\label{eq:Gamma2bqq2}
\end{align}
and
\begin{align}
  &\big(\Gamma_{k,2q}^{(2)q\bar{q}}\big)_{ij}(q^{\prime},q)=\frac{\overrightarrow{\delta}}{\delta q_i(q^{\prime})}\Gamma_{k,2q}\frac{\overleftarrow{\delta}}{\delta \bar{q}_j(q)}\nonumber \\[2ex]
 =&-\Big[Z_{q,k}(q^{\prime}) i q^{\prime}_{\mu}(\gamma_{\mu})_{ji}+m_{q,k}(q^{\prime})\delta_{ji}\Big](2\pi)^4\delta^4(q+q^{\prime})\nonumber \\[2ex]
 &=-\Big[\big(Z_{q,k}(q^{\prime}) i q^{\prime}_{\mu}\gamma_{\mu}+m_{q,k}(q^{\prime})\big)^T\Big]_{ij}(2\pi)^4\delta^4(q+q^{\prime})\nonumber \\[2ex]
 &=-\Big[\big(\Gamma_{k,2q}^{(2)\bar{q}q}\big)^T\Big]_{ij}(q^{\prime},q)\,.\label{}
\end{align}
The regulator reads
\begin{align}
  \big(R_{k}^{\bar{q}q}\big)_{ij}(q^{\prime},q)&=Z_{q,k} i q \cdot \gamma_{ij}r_{F}(q^2/k^2)(2\pi)^4\delta^4(q^{\prime}+q)\,.\label{}
\end{align}
The $\mathcal{P}$ matrix for the quark field reads
\begin{align}
  \mathcal{P}_q&=\begin{pmatrix} 0 & -(\Gamma_{k,2q}^{(2)\bar{q}q}+R_{k}^{\bar{q}q})^T \\ \Gamma_{k,2q}^{(2)\bar{q}q}+R_{k}^{\bar{q}q} & 0 \end{pmatrix}\,, \label{}
\end{align}
and the relevant propagator is given by
\begin{align}
  \left(\frac{1}{\mathcal{P}}\right)_q&=\begin{pmatrix} 0 & G_{k}^{q\bar{q}} \\ -(G_{k}^{q\bar{q}})^T & 0 \end{pmatrix}\,, \label{}
\end{align}
with
\begin{align}
  G_{k}^{q\bar{q}}&=\frac{1}{\Gamma_{k,2q}^{(2)\bar{q}q}+R_{k}^{\bar{q}q}}\,.\label{}
\end{align}
Displaying the momentum explicitly, one arrives at the quark propagator as follows
\begin{align}
  G_{k}^{q\bar{q}}(q,q^{\prime})&=G_{k}^{q\bar{q}}(q)(2\pi)^4\delta^4(q+q^{\prime})\nonumber \\[2ex]
 &=G_{k}^{q}(q)(2\pi)^4\delta^4(q+q^{\prime})\,,\label{}
\end{align}
with
\begin{align}
  G^{q}_k(q)&=\frac{1}{Z_{q,k}(q)i \gamma\cdot q+Z_{q,k}r_F(q^2/k^2)i \gamma\cdot q+m_{q,k}(q)}\,.\label{}
\end{align}

\subsubsection{Ghost-gluon vertex}
\label{appsubsubsec:GhostGluV}

First of all, we consider the case in the vacuum. The action relevant to the ghost-gluon interaction in \Eq{eq:QCDaction} reads
\begin{align}
  &\Gamma_{k,2cA}\equiv \int_x (-g) f^{abc} \partial_{\mu}\bar{c}^a(x) c^b(x) A^c_{\mu}(x)\nonumber \\[2ex]
 =&\int \frac{d^4 p}{(2\pi)^4} \frac{d^4 q}{(2\pi)^4}(-g) f^{abc} i p_{\mu} \bar{c}^a(p) c^b(q) A^c_{\mu}(-p-q)\,,\label{}
\end{align}
which yields
\begin{align}
  &\big(\Gamma_{k,2cA}^{(3)\bar{c}cA}\big)^{abc}_{\mu}(p,q,k)\equiv \frac{\delta}{\delta A^c_{\mu}(k)}\frac{\overrightarrow{\delta}}{\delta\bar{c}^a(p)}\Gamma_{k,2cA}\frac{\overleftarrow{\delta}}{\delta c^b(q)}\nonumber \\[2ex]
 =&-g f^{abc} i p_{\mu} (2\pi)^4\delta^4(p+q+k)\,.\label{}
\end{align}
Thus, the Feynman rule for the ghost-gluon vertex in the vacuum is given as follows
\begin{align}
  &\parbox[c]{0.12\textwidth}{\includegraphics[width=0.12\textwidth]{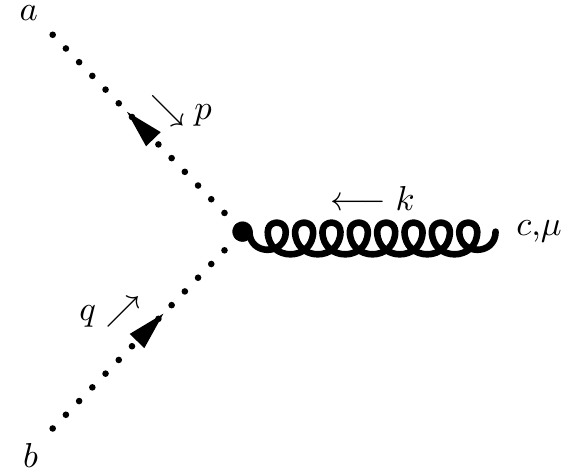}} \equiv -\big(\Gamma_{k,2cA}^{(3)\bar{c}cA}\big)^{abc}_{\mu}(p,q,k)\nonumber \\[2ex]
 =&-g\big(S^{(3)}_{\bar{c}cA}\big)^{abc}_{\mu}(p,q,k)\,.\label{}
\end{align}
with the classical tensor of the ghost-gluon vertex defined by
\begin{align}
  \big(S^{(3)}_{\bar{c}cA}\big)^{abc}_{\mu}(p,q,k)&\equiv -i p_{\mu}f^{abc} \,,\label{}
\end{align}
At finite temperature, the external leg of gluon would be split into the magnetic and electric sectors, and thus the general form for the ghost-gluon vertex at finite temperature reads
\begin{align}
  \parbox[c]{0.12\textwidth}{\includegraphics[width=0.12\textwidth]{ghost-gluon-vert}}&=-\big(S^{(3)}_{\bar{c}cA}\big)^{abc}_{\mu'}(p,q,k)\big(\lambda_{\bar{c}cA}\big)_{\mu'\mu}(k)\,.\label{eq:VertexbccA}
\end{align}
with 
\begin{align}
  \big(\lambda_{\bar{c}cA}\big)_{\mu'\mu}(k)&\equiv \lambda^{\mathrm{M}}_{\bar{c}cA}\Pi_{\mu'\mu}^{\mathrm{M}}(k)+\lambda^{\mathrm{E}}_{\bar{c}cA}\Pi_{\mu'\mu}^{\mathrm{E}}(k)\,,\label{}
\end{align}
where $\lambda^{\mathrm{M}}_{\bar{c}cA}$ and $\lambda^{\mathrm{E}}_{\bar{c}cA}$ are the dressing couplings of the ghost-gluon vertex for the magnetic and electric components, respectively. Note that the couplings are dependent on the RG scale, and their subscript $k$ has been suppressed.

\subsubsection{Three- and four-gluon vertices}
\label{appsubsubsec:GluonV}

From the effective action in \Eq{eq:QCDaction}, one is able to obtain the three-gluon vertex in the vacuum, as follows
\begin{align}
  &\big(\Gamma_{k}^{(3)A^3}\big)^{a_1a_2a_3}_{\mu_1\mu_2\mu_3}(q_1,q_2,q_3)\nonumber \\[2ex]
 \equiv &\frac{\delta^3 \Gamma_{k}}{\delta A^{a_1}_{\mu_1}(q_1)\delta A^{a_2}_{\mu_2}(q_2)\delta A^{a_3}_{\mu_3}(q_3)}\bigg\vert_{A=0}\nonumber \\[2ex]
 =&(2\pi)^4\delta^4(q_1+q_2+q_3)g\big(S^{(3)}_{A^3}\big)^{a_1a_2a_3}_{\mu_1\mu_2\mu_3}(q_1,q_2,q_3)\,,\label{}
\end{align}
with the classical three-gluon tensor
\begin{align}
  &\big(S^{(3)}_{A^3}\big)^{a_1a_2a_3}_{\mu_1\mu_2\mu_3}(q_1,q_2,q_3)\nonumber \\[2ex]
 \equiv &-i f^{a_1a_2a_3}\Big[\delta_{\mu_1\mu_2}(q_1-q_2)_{\mu_3}+\delta_{\mu_2\mu_3}(q_2-q_3)_{\mu_1}\nonumber \\[2ex]
 &+\delta_{\mu_3\mu_1}(q_3-q_1)_{\mu_2}\Big] \,.\label{}
\end{align}
In the same way, the four-gluon vertex reads
\begin{align}
  &\big(\Gamma_{k}^{(4)A^4}\big)^{a_1a_2a_3a_4}_{\mu_1\mu_2\mu_3\mu_4}(q_1,q_2,q_3,q_4)\nonumber \\[2ex]
 \equiv &\frac{\delta^4 \Gamma_{k,A}}{\delta A^{a_1}_{\mu_1}(q_1)\delta A^{a_2}_{\mu_2}(q_2)\delta A^{a_3}_{\mu_3}(q_3)\delta A^{a_4}_{\mu_4}(q_4)}\bigg\vert_{A=0}\nonumber \\[2ex]
 =&(2\pi)^4\delta^4(q_1+q_2+q_3+q_4)g^2\big(S^{(4)}_{A^4}\big)^{a_1a_2a_3a_4}_{\mu_1\mu_2\mu_3\mu_4}\,,\label{}
\end{align}
with the classical four-gluon tensor
\begin{align}
  \big(S^{(4)}_{A^4}\big)^{a_1a_2a_3a_4}_{\mu_1\mu_2\mu_3\mu_4}&\equiv f^{ea_1a_2}f^{ea_3a_4}(\delta_{\mu_1\mu_3}\delta_{\mu_2\mu_4}-\delta_{\mu_1\mu_4}\delta_{\mu_2\mu_3})\nonumber \\[2ex]
  &+f^{ea_1a_3}f^{ea_2a_4}(\delta_{\mu_1\mu_2}\delta_{\mu_3\mu_4}-\delta_{\mu_1\mu_4}\delta_{\mu_2\mu_3})\nonumber \\[2ex]
  &+f^{ea_1a_4}f^{ea_2a_3}(\delta_{\mu_1\mu_2}\delta_{\mu_4\mu_3}-\delta_{\mu_1\mu_3}\delta_{\mu_4\mu_2})\,.\label{}
\end{align}

Similar with the case of ghost-gluon vertex, each external leg of the three- and four-gluon vertices should be split into a sum of the magnetic and electric components at finite temperature, and see, e.g., \cite{Cyrol:2017qkl} for more details. Consequently, the three-gluon vertex at finite temperature reads
\begin{align}
  &\parbox[c]{0.12\textwidth}{\includegraphics[width=0.12\textwidth]{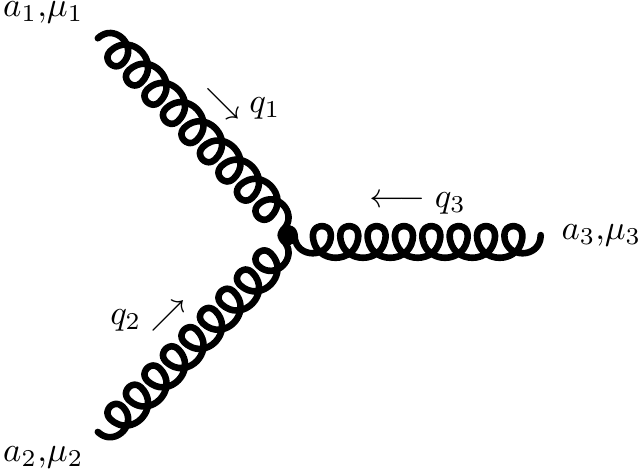}} \equiv -\big(\Gamma_{k}^{(3)A^3}\big)^{a_1a_2a_3}_{\mu_1\mu_2\mu_3}(q_1,q_2,q_3)\nonumber \\[2ex]
 &=-\big(S^{(3)}_{A^3}\big)^{a_1a_2a_3}_{\mu_1'\mu_2'\mu_3'}(q_1,q_2,q_3)\big(\lambda_{A^3}\big)^{\mu_1'\mu_2'\mu_3'}_{\mu_1\mu_2\mu_3}(q_1,q_2,q_3)\,,\label{eq:VertexA3}
\end{align}
with 
\begin{align}
  &\big(\lambda_{A^3}\big)^{\mu_1'\mu_2'\mu_3'}_{\mu_1\mu_2\mu_3}(q_1,q_2,q_3) \nonumber \\[2ex]
 \equiv & \lambda^{\mathrm{MMM}}_{A^3}\big(\Pi^{\mathrm{MMM}}\big)^{\mu_1'\mu_2'\mu_3'}_{\mu_1\mu_2\mu_3}(q_1,q_2,q_3)\nonumber \\[2ex]
 &+\lambda^{\mathrm{EMM}}_{A^3}\big(\Pi^{\mathrm{EMM}}\big)^{\mu_1'\mu_2'\mu_3'}_{\mu_1\mu_2\mu_3}(q_1,q_2,q_3)\nonumber \\[2ex]
 &+\lambda^{\mathrm{EEM}}_{A^3}\big(\Pi^{\mathrm{EEM}}\big)^{\mu_1'\mu_2'\mu_3'}_{\mu_1\mu_2\mu_3}(q_1,q_2,q_3)\nonumber \\[2ex]
 &+\lambda^{\mathrm{EEE}}_{A^3}\big(\Pi^{\mathrm{EEE}}\big)^{\mu_1'\mu_2'\mu_3'}_{\mu_1\mu_2\mu_3}(q_1,q_2,q_3) \,,\label{}
\end{align}
where $\lambda^{\mathrm{MMM}}_{A^3}$, $\lambda^{\mathrm{EMM}}_{A^3}$, $\lambda^{\mathrm{EEM}}_{A^3}$, $\lambda^{\mathrm{EEE}}_{A^3}$ are the dressing three-gluon couplings for different components, and the relevant projectors reads
\begin{align}
  &\big(\Pi^{\mathrm{MMM}}\big)^{\mu_1'\mu_2'\mu_3'}_{\mu_1\mu_2\mu_3}(q_1,q_2,q_3)
 \equiv \Pi_{\mu_1'\mu_1}^{\mathrm{M}}(q_1)\Pi_{\mu_2'\mu_2}^{\mathrm{M}}(q_2)\Pi_{\mu_3'\mu_3}^{\mathrm{M}}(q_3)\,,\\[2ex]
   &\big(\Pi^{\mathrm{EEE}}\big)^{\mu_1'\mu_2'\mu_3'}_{\mu_1\mu_2\mu_3}(q_1,q_2,q_3)
 \equiv \Pi_{\mu_1'\mu_1}^{\mathrm{E}}(q_1)\Pi_{\mu_2'\mu_2}^{\mathrm{E}}(q_2)\Pi_{\mu_3'\mu_3}^{\mathrm{E}}(q_3)\,,\label{}
\end{align}
the projector with one electric gluon and two magnetic gluons
\begin{align}
  &\big(\Pi^{\mathrm{EMM}}\big)^{\mu_1'\mu_2'\mu_3'}_{\mu_1\mu_2\mu_3}(q_1,q_2,q_3)\nonumber \\[2ex]
 \equiv &\Pi_{\mu_1'\mu_1}^{\mathrm{E}}(q_1)\Pi_{\mu_2'\mu_2}^{\mathrm{M}}(q_2)\Pi_{\mu_3'\mu_3}^{\mathrm{M}}(q_3)+\Pi_{\mu_1'\mu_1}^{\mathrm{M}}(q_1)\Pi_{\mu_2'\mu_2}^{\mathrm{E}}(q_2)\nonumber \\[2ex]
 &\times\Pi_{\mu_3'\mu_3}^{\mathrm{M}}(q_3)+\Pi_{\mu_1'\mu_1}^{\mathrm{M}}(q_1)\Pi_{\mu_2'\mu_2}^{\mathrm{M}}(q_2)\Pi_{\mu_3'\mu_3}^{\mathrm{E}}(q_3)\,,\label{}
\end{align}
the projector with two electric gluons and one magnetic gluon
\begin{align}
  &\big(\Pi^{\mathrm{EEM}}\big)^{\mu_1'\mu_2'\mu_3'}_{\mu_1\mu_2\mu_3}(q_1,q_2,q_3)\nonumber \\[2ex]
 \equiv &\Pi_{\mu_1'\mu_1}^{\mathrm{E}}(q_1)\Pi_{\mu_2'\mu_2}^{\mathrm{E}}(q_2)\Pi_{\mu_3'\mu_3}^{\mathrm{M}}(q_3)+\Pi_{\mu_1'\mu_1}^{\mathrm{E}}(q_1)\Pi_{\mu_2'\mu_2}^{\mathrm{M}}(q_2)\nonumber \\[2ex]
 &\times\Pi_{\mu_3'\mu_3}^{\mathrm{E}}(q_3)+\Pi_{\mu_1'\mu_1}^{\mathrm{M}}(q_1)\Pi_{\mu_2'\mu_2}^{\mathrm{E}}(q_2)\Pi_{\mu_3'\mu_3}^{\mathrm{E}}(q_3)\,.\label{}
\end{align}

The four-gluon vertex at finite temperature reads
\begin{align}
  &\parbox[c]{0.12\textwidth}{\includegraphics[width=0.12\textwidth]{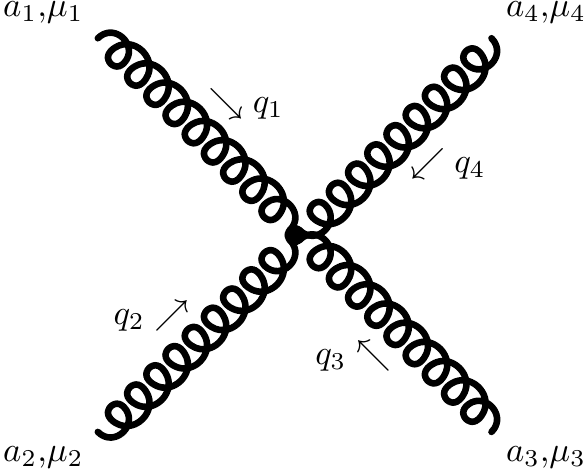}} \equiv -\big(\Gamma_{k}^{(4)A^4}\big)^{a_1a_2a_3a_4}_{\mu_1\mu_2\mu_3\mu_4}(q_1,q_2,q_3,q_4)\nonumber \\[2ex]
 &=-\big(S^{(4)}_{A^4}\big)^{a_1a_2a_3a_4}_{\mu_1'\mu_2'\mu_3'\mu_4'}\big(\lambda_{A^4}\big)^{\mu_1'\mu_2'\mu_3'\mu_4'}_{\mu_1\mu_2\mu_3\mu_4}(q_1,q_2,q_3,q_4)\,,\label{eq:VertexA4}
\end{align}
with 
\begin{align}
  &\big(\lambda_{A^4}\big)^{\mu_1'\mu_2'\mu_3'\mu_4'}_{\mu_1\mu_2\mu_3\mu_4}(q_1,q_2,q_3,q_4)\nonumber \\[2ex]
 \equiv & \lambda^{\mathrm{MMMM}}_{A^4}\big(\Pi^{\mathrm{MMMM}}\big)^{\mu_1'\mu_2'\mu_3'\mu_4'}_{\mu_1\mu_2\mu_3\mu_4}(q_1,q_2,q_3,q_4)\nonumber \\[2ex]
 &+\lambda^{\mathrm{EMMM}}_{A^4}\big(\Pi^{\mathrm{EMMM}}\big)^{\mu_1'\mu_2'\mu_3'\mu_4'}_{\mu_1\mu_2\mu_3\mu_4}(q_1,q_2,q_3,q_4)\nonumber \\[2ex]
 &+\lambda^{\mathrm{EEMM}}_{A^4}\big(\Pi^{\mathrm{EEMM}}\big)^{\mu_1'\mu_2'\mu_3'\mu_4'}_{\mu_1\mu_2\mu_3\mu_4}(q_1,q_2,q_3,q_4)\nonumber \\[2ex]
 &+\lambda^{\mathrm{EEEM}}_{A^4}\big(\Pi^{\mathrm{EEEM}}\big)^{\mu_1'\mu_2'\mu_3'\mu_4'}_{\mu_1\mu_2\mu_3\mu_4}(q_1,q_2,q_3,q_4) \nonumber \\[2ex]
 &+\lambda^{\mathrm{EEEE}}_{A^4}\big(\Pi^{\mathrm{EEEE}}\big)^{\mu_1'\mu_2'\mu_3'\mu_4'}_{\mu_1\mu_2\mu_3\mu_4}(q_1,q_2,q_3,q_4)\,,\label{}
\end{align}
where $\lambda^{\mathrm{MMMM}}_{A^4}$, $\lambda^{\mathrm{EMMM}}_{A^4}$, $\lambda^{\mathrm{EEMM}}_{A^4}$, $\lambda^{\mathrm{EEEM}}_{A^4}$, and $\lambda^{\mathrm{EEEE}}_{A^4}$ are the dressing four-gluon couplings for different components, and the relevant projectors reads
\begin{align}
  &\big(\Pi^{\mathrm{MMMM}}\big)^{\mu_1'\mu_2'\mu_3'\mu_4'}_{\mu_1\mu_2\mu_3\mu_4}(q_1,q_2,q_3,q_4)\nonumber \\[2ex]
 \equiv& \Pi_{\mu_1'\mu_1}^{\mathrm{M}}(q_1)\Pi_{\mu_2'\mu_2}^{\mathrm{M}}(q_2)\Pi_{\mu_3'\mu_3}^{\mathrm{M}}(q_3)\Pi_{\mu_4'\mu_4}^{\mathrm{M}}(q_4)\,,\label{}
\end{align}
and
\begin{align}
   &\big(\Pi^{\mathrm{EEEE}}\big)^{\mu_1'\mu_2'\mu_3'\mu_4'}_{\mu_1\mu_2\mu_3\mu_4}(q_1,q_2,q_3,q_4)\nonumber \\[2ex]
 \equiv& \Pi_{\mu_1'\mu_1}^{\mathrm{E}}(q_1)\Pi_{\mu_2'\mu_2}^{\mathrm{E}}(q_2)\Pi_{\mu_3'\mu_3}^{\mathrm{E}}(q_3)\Pi_{\mu_4'\mu_4}^{\mathrm{E}}(q_4)\,,\label{}
\end{align}
the projector with one electric gluon and three magnetic gluons
\begin{align}
  &\big(\Pi^{\mathrm{EMMM}}\big)^{\mu_1'\mu_2'\mu_3'\mu_4'}_{\mu_1\mu_2\mu_3\mu_4}(q_1,q_2,q_3,q_4)\nonumber \\[2ex]
 \equiv &\Pi_{\mu_1'\mu_1}^{\mathrm{E}}(q_1)\Pi_{\mu_2'\mu_2}^{\mathrm{M}}(q_2)\Pi_{\mu_3'\mu_3}^{\mathrm{M}}(q_3)\Pi_{\mu_4'\mu_4}^{\mathrm{M}}(q_4)\nonumber \\[2ex]
 &+\Pi_{\mu_1'\mu_1}^{\mathrm{M}}(q_1)\Pi_{\mu_2'\mu_2}^{\mathrm{E}}(q_2)\Pi_{\mu_3'\mu_3}^{\mathrm{M}}(q_3)\Pi_{\mu_4'\mu_4}^{\mathrm{M}}(q_4)\nonumber \\[2ex]
 &+\Pi_{\mu_1'\mu_1}^{\mathrm{M}}(q_1)\Pi_{\mu_2'\mu_2}^{\mathrm{M}}(q_2)\Pi_{\mu_3'\mu_3}^{\mathrm{E}}(q_3)\Pi_{\mu_4'\mu_4}^{\mathrm{M}}(q_4)\nonumber \\[2ex]
 &+\Pi_{\mu_1'\mu_1}^{\mathrm{M}}(q_1)\Pi_{\mu_2'\mu_2}^{\mathrm{M}}(q_2)\Pi_{\mu_3'\mu_3}^{\mathrm{M}}(q_3)\Pi_{\mu_4'\mu_4}^{\mathrm{E}}(q_4)\,,\label{}
\end{align}
the projector with three electric gluon and one magnetic gluons
\begin{align}
  &\big(\Pi^{\mathrm{EEEM}}\big)^{\mu_1'\mu_2'\mu_3'\mu_4'}_{\mu_1\mu_2\mu_3\mu_4}(q_1,q_2,q_3,q_4)\nonumber \\[2ex]
 \equiv &\Pi_{\mu_1'\mu_1}^{\mathrm{E}}(q_1)\Pi_{\mu_2'\mu_2}^{\mathrm{E}}(q_2)\Pi_{\mu_3'\mu_3}^{\mathrm{E}}(q_3)\Pi_{\mu_4'\mu_4}^{\mathrm{M}}(q_4)\nonumber \\[2ex]
 &+\Pi_{\mu_1'\mu_1}^{\mathrm{E}}(q_1)\Pi_{\mu_2'\mu_2}^{\mathrm{E}}(q_2)\Pi_{\mu_3'\mu_3}^{\mathrm{M}}(q_3)\Pi_{\mu_4'\mu_4}^{\mathrm{E}}(q_4)\nonumber \\[2ex]
 &+\Pi_{\mu_1'\mu_1}^{\mathrm{E}}(q_1)\Pi_{\mu_2'\mu_2}^{\mathrm{M}}(q_2)\Pi_{\mu_3'\mu_3}^{\mathrm{E}}(q_3)\Pi_{\mu_4'\mu_4}^{\mathrm{E}}(q_4)\nonumber \\[2ex]
 &+\Pi_{\mu_1'\mu_1}^{\mathrm{M}}(q_1)\Pi_{\mu_2'\mu_2}^{\mathrm{E}}(q_2)\Pi_{\mu_3'\mu_3}^{\mathrm{E}}(q_3)\Pi_{\mu_4'\mu_4}^{\mathrm{E}}(q_4)\,,\label{}
\end{align}
the projector with two electric gluons and two magnetic gluons
\begin{align}
  &\big(\Pi^{\mathrm{EEMM}}\big)^{\mu_1'\mu_2'\mu_3'\mu_4'}_{\mu_1\mu_2\mu_3\mu_4}(q_1,q_2,q_3,q_4)\nonumber \\[2ex]
 \equiv &\Pi_{\mu_1'\mu_1}^{\mathrm{E}}(q_1)\Pi_{\mu_2'\mu_2}^{\mathrm{E}}(q_2)\Pi_{\mu_3'\mu_3}^{\mathrm{M}}(q_3)\Pi_{\mu_4'\mu_4}^{\mathrm{M}}(q_4)\nonumber \\[2ex]
 &+\Pi_{\mu_1'\mu_1}^{\mathrm{E}}(q_1)\Pi_{\mu_2'\mu_2}^{\mathrm{M}}(q_2)\Pi_{\mu_3'\mu_3}^{\mathrm{E}}(q_3)\Pi_{\mu_4'\mu_4}^{\mathrm{M}}(q_4)\nonumber \\[2ex]
 &+\Pi_{\mu_1'\mu_1}^{\mathrm{E}}(q_1)\Pi_{\mu_2'\mu_2}^{\mathrm{M}}(q_2)\Pi_{\mu_3'\mu_3}^{\mathrm{M}}(q_3)\Pi_{\mu_4'\mu_4}^{\mathrm{E}}(q_4)\nonumber \\[2ex]
 &+\Pi_{\mu_1'\mu_1}^{\mathrm{M}}(q_1)\Pi_{\mu_2'\mu_2}^{\mathrm{E}}(q_2)\Pi_{\mu_3'\mu_3}^{\mathrm{E}}(q_3)\Pi_{\mu_4'\mu_4}^{\mathrm{M}}(q_4)\nonumber \\[2ex]
 &+\Pi_{\mu_1'\mu_1}^{\mathrm{M}}(q_1)\Pi_{\mu_2'\mu_2}^{\mathrm{E}}(q_2)\Pi_{\mu_3'\mu_3}^{\mathrm{M}}(q_3)\Pi_{\mu_4'\mu_4}^{\mathrm{E}}(q_4)\nonumber \\[2ex]
 &+\Pi_{\mu_1'\mu_1}^{\mathrm{M}}(q_1)\Pi_{\mu_2'\mu_2}^{\mathrm{M}}(q_2)\Pi_{\mu_3'\mu_3}^{\mathrm{E}}(q_3)\Pi_{\mu_4'\mu_4}^{\mathrm{E}}(q_4)\,.\label{}
\end{align}

\subsection{Gluon self-energy}
\label{appsubsec:GluSelf-ener}

With the Feynman rules discussed in \sec{appsubsec:FeynRules}, it is straightforward to write down expressions for the loop diagrams of the gluon self-energy as shown on the r.h.s. of flow equation in \Fig{fig:A2-equ}. The first gluon loop reads
\begin{align}
  \parbox[c]{0.12\textwidth}{\includegraphics[width=0.12\textwidth]{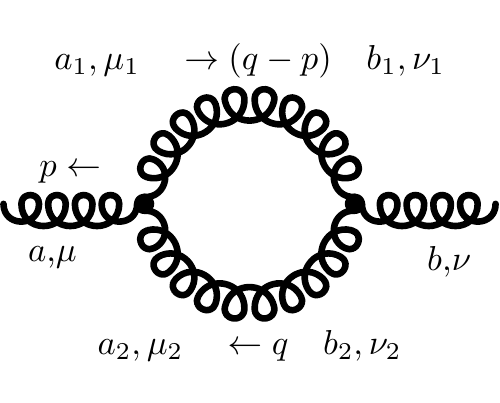}}&\equiv \big(\Sigma^{AA}_a\big)_{\mu \nu}^{ab}(p)\,.\label{}
\end{align}
with 
\begin{align}
  &\big(\Sigma^{AA}_a\big)_{\mu \nu}^{ab}(p)\nonumber \\[2ex]
 =&\int \frac{d^4 q}{(2\pi)^4} (-)\big(S^{(3)}_{A^3}\big)^{aa_1a_2}_{\mu'\mu_1'\mu_2'}(-p,-q+p,q)\nonumber \\[2ex]
 &\times\big(\lambda_{A^3}\big)^{\mu'\mu_1'\mu_2'}_{\mu\mu_1\mu_2}(-p,-q+p,q)\Big[G_{A}^{\mathrm{M}}(q)\Pi_{\mu_2\nu_2}^{\mathrm{M}}(q)\nonumber \\[2ex]
 &+G_{A}^{\mathrm{E}}(q)\Pi_{\mu_2\nu_2}^{\mathrm{E}}(q)\Big]\delta^{a_2b_2} (-)\big(S^{(3)}_{A^3}\big)^{bb_1b_2}_{\nu'\nu_1'\nu_2'}(p,q-p,-q)\nonumber \\[2ex]
 &\times\big(\lambda_{A^3}\big)^{\nu'\nu_1'\nu_2'}_{\nu\nu_1\nu_2}(p,q-p,-q) \Big[G_{A}^{\mathrm{M}}(q-p)\Pi_{\nu_1\mu_1}^{\mathrm{M}}(q-p)\nonumber \\[2ex]
 &+G_{A}^{\mathrm{E}}(q-p)\Pi_{\nu_1\mu_1}^{\mathrm{E}}(q-p)\Big]\delta^{b_1a_1}\,.\label{eq:SigAAa}
\end{align}
Projecting \Eq{eq:SigAAa} onto the magnetic component, one arrives at 
\begin{align}
  &\big(\Sigma^{AA}_a\big)_{\mu \nu}^{ab}(p)\big(\delta^{ab}\Pi_{\mu\nu}^{\mathrm{M}}(p)\big)\nonumber \\[2ex]
 =&N_c(N_c^2-1)\int \frac{d^4 q}{(2\pi)^4}\bigg[{\lambda^{\mathrm{MMM}}_{A^3}}^2 G_{A}^{\mathrm{M}}(q) G_{A}^{\mathrm{M}}(q-p)C^{M}_{MM}\nonumber \\[2ex]
 &+{\lambda^{\mathrm{EEM}}_{A^3}}^2 G_{A}^{\mathrm{E}}(q) G_{A}^{\mathrm{E}}(q-p)C^{M}_{EE}+{\lambda^{\mathrm{EMM}}_{A^3}}^2 \nonumber \\[2ex]
 &\times\Big(G_{A}^{\mathrm{M}}(q) G_{A}^{\mathrm{E}}(q-p)C^{M}_{ME}+G_{A}^{\mathrm{E}}(q)G_{A}^{\mathrm{M}}(q-p)C^{M}_{EM}\Big)\bigg]\,.\label{eq:SigAAaM}
\end{align}
Here we have defined several coefficients, which reads
\begin{align}
  C^{M}_{MM}=&\frac{2 \sin^2\theta}{p_s^2+q_s^2-2 p_s q_s\cos\theta}\bigg\{11 p_s^2 q_s^2+4 p_s^4+4 q_s^4\nonumber \\[2ex]
 &+p_s q_s \left[p_s q_s\cos2 \theta-8\left(p_s^2+q_s^2\right)\cos\theta\right]\bigg\}\,,\label{}
\end{align}
where the two 4-momenta are $p=(p_0, \bm{p})$ and $q=(q_0, \bm{q})$, with $p_s=|\bm{p}|$, $q_s=|\bm{q}|$ and $\cos\theta=\bm{p}\cdot\bm{q}/(p_s q_s)$. In the same way, we have
\begin{align}
  C^{M}_{ME}=&\frac{2 (\cos2 \theta+3) }{p_s^2+q_s^2-2 p_s q_s\cos\theta}\nonumber \\[2ex]
 &\times\frac{\Big[q_0 p_s^2+p_0 q_s^2- p_s q_s\left(p_0+q_0\right)\cos\theta\Big]^2}{\left(p_0-q_0\right)^2+p_s^2+q_s^2-2 p_s q_s\cos\theta}\,.\label{}
\end{align}
The coefficient $C^{M}_{EM}$ could be deduced from $C^{M}_{ME}$ through the replacement as follows
\begin{align}
  &C^{M}_{EM}=C^{M}_{ME}\Big|_{q\rightarrow -q+p}\nonumber \\[2ex]
 =&\frac{2 \left(p_s q_0\cos\theta -p_0 q_s\right)^2 }{\left(q_0^2+q_s^2\right) \left(p_s^2-2 p_s q_s\cos\theta+q_s^2\right)}\nonumber \\[2ex]
 &\times\bigg[4 p_s^2-8 p_s
   q_s\cos\theta +(3+\cos 2\theta) q_s^2\bigg]\,.\label{}
\end{align}
Finally, the last one is given by
\begin{align}
  C^{M}_{EE}=&\frac{4 \sin ^2\theta}{\left(q_0^2+q_s^2\right) \left(p_s^2-2 p_s q_s\cos\theta +q_s^2\right)}\nonumber \\[2ex]
 &\times\frac{1}{p_s^2+\left(p_0-q_0\right)^2-2 p_s q_s\cos\theta+q_s^2}\nonumber\\[2ex]
 &\times\Big\{q_s^4+p_s^2 q_0^2+\left(p_0^2+p_s^2-p_0 q_0+q_0^2\right) q_s^2\nonumber \\[2ex]
 &-p_s q_s \left[q_0 \left(p_0+q_0\right)+2
   q_s^2\right]\cos\theta\Big\}^2\,.\label{}
\end{align}

Projection of \Eq{eq:SigAAa} onto the electric component yields 
\begin{align}
  &\big(\Sigma^{AA}_a\big)_{\mu \nu}^{ab}(p)\left(\delta^{ab}\Pi_{\mu\nu}^{\mathrm{E}}(p)\right)\nonumber \\[2ex]
=&N_c(N_c^2-1)\int \frac{d^4 q}{(2\pi)^4}\bigg[{\lambda^{\mathrm{EEE}}_{A^3}}^2 G_{A}^{\mathrm{E}}(q) G_{A}^{\mathrm{E}}(q-p)C^{E}_{EE}\nonumber\\[2ex]
 &+{\lambda^{\mathrm{EMM}}_{A^3}}^2 G_{A}^{\mathrm{M}}(q) G_{A}^{\mathrm{M}}(q-p)C^{E}_{MM}+{\lambda^{\mathrm{EEM}}_{A^3}}^2 \nonumber\\[2ex]
 &\times \Big(G_{A}^{\mathrm{M}}(q) G_{A}^{\mathrm{E}}(q-p)C^{E}_{ME}+G_{A}^{\mathrm{E}}(q) G_{A}^{\mathrm{M}}(q-p) C^{E}_{EM}\Big)\bigg]\,.\label{eq:SigAAaE}
\end{align}
There are four coefficients as well, and their explicit expressions are given as follows,
\begin{align}
C^{E}_{EE}=&\frac{1}{\left(p_0^2+p_s^2\right) \left(q_0^2+q_s^2\right) }\frac{1}{p_s^2-2 p_s q_s\cos\theta +q_s^2}\nonumber\\[2ex]
 &\times\frac{1}{p_s^2+\left(p_0-q_0\right)^2-2p_s q_s\cos\theta+q_s^2}\nonumber\\[2ex]
 &\times\Bigg\{p_s q_s \bigg[\Big(2 p_s^2 q_0+p_0 q_0 \left(p_0+q_0\right)+2 p_0 q_s^2\Big)\cos 2\theta\nonumber \\[2ex]
 &+\left(p_0+q_0\right) \Big(2 p_0^2-3 p_0 q_0+2 \left(p_s^2+q_0^2+q_s^2\right)\Big)\bigg]\nonumber\\[2ex]
 &-2 \bigg[p_s^2 q_0 \Big(p_0^2+p_s^2-p_0 q_0+q_0^2\Big)+\Big(p_0^3-p_0^2 q_0\nonumber \\[2ex]
 &+2 p_s^2 q_0+p_0 \left(2 p_s^2+q_0^2\right)\Big) q_s^2+p_0 q_s^4\bigg]\cos\theta
\Bigg\}^2
\,,\label{}
\end{align}
coefficient $C^{E}_{ME}$
\begin{align}
C^{E}_{ME}=&\frac{4 \sin ^2\theta}{\left(p_0^2+p_s^2\right) \left(p_s^2-2 p_s q_s\cos\theta +q_s^2\right)}\nonumber\\[2ex]
 &\times\frac{1}{p_s^2+\left(p_0-q_0\right)^2-2  p_s q_s\cos\theta +q_s^2}\nonumber\\[2ex]
 &\times\bigg[p_s^2 \left(p_0^2+p_s^2-p_0 q_0+q_0^2\right)\nonumber\\[2ex]
 &-p_s q_s\Big(2 p_s^2+p_0 \left(p_0+q_0\right)\Big)\cos\theta +\left(p_0^2+p_s^2\right) q_s^2\bigg]^2\,,\label{}
\end{align}
coefficient $C^{E}_{EM}$
\begin{align}
  &C^{E}_{EM}=C^{E}_{ME}\Big|_{q\rightarrow -q+p}\nonumber \\[2ex]
 =&\frac{4 \sin ^2\theta}{\left(p_0^2+p_s^2\right) \left(q_0^2+q_s^2\right)}\frac{1}{p_s^2-2 p_s q_s\cos\theta +q_s^2}\nonumber\\[2ex]
 &\times\bigg[p_s^2\left(q_0^2+q_s^2\right)+p_0^2 q_s^2-p_0 p_s q_0 q_s\cos\theta\bigg]^2\,,\label{}
\end{align}
and coefficient $C^{E}_{MM}$
\begin{align}
  C^{E}_{MM}=&\frac{2 \left(p_s q_0-p_0 q_s\cos\theta \right)^2}{\left(p_0^2+p_s^2\right) \left(p_s^2-2 p_s q_s\cos\theta +q_s^2\right)}\nonumber\\[2ex]
 &\times\left(3 p_s^2+p_s^2\cos2 \theta -8 p_s q_s\cos\theta +4 q_s^2\right)\,.\label{}
\end{align}

The ghost loop of the gluon self-energy, i.e., the second diagram on the r.h.s. of flow equation in \Fig{fig:A2-equ}, reads
\begin{align}
  \parbox[c]{0.12\textwidth}{\includegraphics[width=0.12\textwidth]{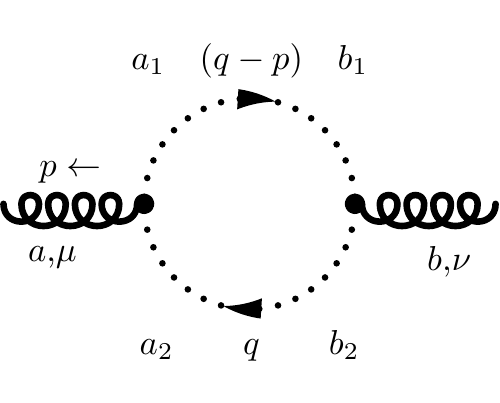}}&\equiv \big(\Sigma^{AA}_b\big)_{\mu \nu}^{ab}(p)\,.\label{}
\end{align}
with
\begin{align}
  &\big(\Sigma^{AA}_b\big)_{\mu \nu}^{ab}(p)\nonumber \\[2ex]
 =&\int \frac{d^4 q}{(2\pi)^4} (-)\big(S^{(3)}_{\bar{c}cA}\big)^{a_1a_2a}_{\mu'}(-q+p,q,-p)\big(\lambda_{\bar{c}cA}\big)_{\mu'\mu}(-p)\nonumber \\[2ex]
 &\times G_{c}(q)\delta^{a_2b_2} (-)\big(S^{(3)}_{\bar{c}cA}\big)^{b_2b_1b}_{\nu'}(-q,q-p,p)\big(\lambda_{\bar{c}cA}\big)_{\nu'\nu}(p)\nonumber \\[2ex]
 &\times G_{c}(q-p)\delta^{b_1a_1}\,.\label{eq:SigAAb}
\end{align}
Projecting \Eq{eq:SigAAb} onto the magnetic component leads us to
\begin{align}
  &\big(\Sigma^{AA}_b\big)_{\mu \nu}^{ab}(p)\left(\delta^{ab}\Pi_{\mu\nu}^{\mathrm{M}}(p)\right)\nonumber \\[2ex]
=&\int \frac{d^4 q}{(2\pi)^4} N_c(N_c^2-1)(\lambda^{\mathrm{M}}_{\bar{c}cA})^2G_{c}(q)G_{c}(q-p)C^{g}_{M}\,,\label{eq:SigAAbM}
\end{align}
with
\begin{align}
  C^{g}_{M}&=q_s^2\sin^2\theta\,.\label{}
\end{align}
And the electric component is given by
\begin{align}
  &\big(\Sigma^{AA}_b\big)_{\mu \nu}^{ab}(p)\left(\delta^{ab}\Pi_{\mu\nu}^{\mathrm{E}}(p)\right)\nonumber \\[2ex]
=&\int \frac{d^4 q}{(2\pi)^4} N_c(N_c^2-1)(\lambda^{\mathrm{E}}_{\bar{c}cA})^2G_{c}(q)G_{c}(q-p)C^{g}_{E}\,,\label{eq:SigAAbE}
\end{align}
with
\begin{align}
  C^{g}_{E}&=\frac{\big(p_sq_0-p_0q_s\cos\theta\big)^2}{p_0^2+p_s^2}\,.\label{}
\end{align}

The tadpole diagram of the gluon self-energy due to the four-gluon vertex, i.e., the third diagram on the r.h.s. of flow equation in \Fig{fig:A2-equ}, reads
\begin{align}
  \parbox[c]{0.08\textwidth}{\includegraphics[width=0.08\textwidth]{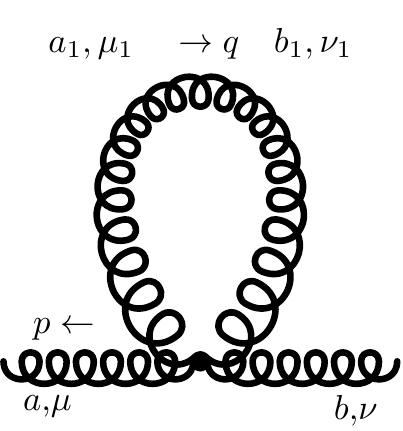}}&\equiv \big(\Sigma^{AA}_c\big)_{\mu \nu}^{ab}(p)\,.\label{}
\end{align}
with
\begin{align}
  &\big(\Sigma^{AA}_c\big)_{\mu \nu}^{ab}(p)\nonumber \\[2ex]
 =&\int \frac{d^4 q}{(2\pi)^4} (-)\big(S^{(4)}_{A^4}\big)^{aa_1b_1b}_{\mu'\mu_1'\nu_1'\nu'}\big(\lambda_{A^4}\big)^{\mu'\mu_1'\nu_1'\nu'}_{\mu\mu_1\nu_1\nu}(-p,-q,q,p)\nonumber \\[2ex]
 &\times\Big(G_{A}^{\mathrm{M}}(q)\Pi_{\mu_1\nu_1}^{\mathrm{M}}(q)+G_{A}^{\mathrm{E}}(q)\Pi_{\mu_1\nu_1}^{\mathrm{E}}(q)\Big)\delta^{a_1b_1}\,.\label{eq:SigAAc}
\end{align}
Projecting \Eq{eq:SigAAc} onto the magnetic component leaves us with
\begin{align}
  &\big(\Sigma^{AA}_c\big)_{\mu \nu}^{ab}(p)\left(\delta^{ab}\Pi_{\mu\nu}^{\mathrm{M}}(p)\right)\nonumber \\[2ex]
=&\int \frac{d^4 q}{(2\pi)^4} N_c(N_c^2-1)\Big[\lambda^{\mathrm{MMMM}}_{A^4}G_{A}^{\mathrm{M}}(q)C^{t}_{MM}\nonumber \\[2ex]
 &+\lambda^{\mathrm{EEMM}}_{A^4}G_{A}^{\mathrm{E}}(q)C^{t}_{ME}\Big]\,,\label{eq:SigAAcM}
\end{align}
with
\begin{align}
  C^{t}_{MM}&=\cos(2\theta) -5\,,\\[2ex]
  C^{t}_{ME}&=-\frac{2 \big(q_0^2+q_0^2\cos ^2\theta +2q_s^2\big)}{q_0^2+q_s^2}\,.\label{}
\end{align}
The electric component reads
\begin{align}
  &\big(\Sigma^{AA}_c\big)_{\mu \nu}^{ab}(p)\left(\delta^{ab}\Pi_{\mu\nu}^{\mathrm{E}}(p)\right)\nonumber \\[2ex]
=&\int \frac{d^4 q}{(2\pi)^4} N_c(N_c^2-1)\Big[\lambda^{\mathrm{EEMM}}_{A^4}G_{A}^{\mathrm{M}}(q)C^{t}_{EM}\nonumber \\[2ex]
 &+\lambda^{\mathrm{EEEE}}_{A^4}G_{A}^{\mathrm{E}}(q)C^{t}_{EE}\Big]\,,\label{eq:SigAAcE}
\end{align}
with
\begin{align}
  C^{t}_{EM}&=-\frac{2 \big(p_0^2+p_0^2\cos ^2\theta +2p_s^2\big)}{p_0^2+p_s^2}\,,\\[2ex]
  C^{t}_{EE}&=-\frac{2 \left[p_s^2 q_0^2-2 p_0 p_s q_0q_s\cos\theta + p_0^2 \left(q_0^2\sin ^2\theta +q_s^2\right)\right]}{\left(p_0^2+p_s^2\right)\left(q_0^2+q_s^2\right)}\,,\label{}
\end{align}

Substituting Eqs. (\ref{eq:SigAAaM}), (\ref{eq:SigAAaE}), (\ref{eq:SigAAbM}), (\ref{eq:SigAAbE}), (\ref{eq:SigAAcM}), (\ref{eq:SigAAcE}) into the flow equation in \Fig{fig:A2-equ}, one obtains the flow equation of the magnetic gluon dressing function 
\begin{align}
  &\partial_t Z_{A,k}^{\mathrm{M}}(p)\nonumber \\[2ex]
  =&-\frac{N_c}{2p^2} \int \frac{d^4 q}{(2\pi)^4} \Bigg\{\bigg[{\lambda^{\mathrm{MMM}}_{A^3}}^2 (\tilde{\partial}_t G_{A}^{\mathrm{M}}(q)) G_{A}^{\mathrm{M}}(q-p)C^{M}_{MM}\nonumber \\[2ex]
 &+{\lambda^{\mathrm{EEM}}_{A^3}}^2 (\tilde{\partial}_t G_{A}^{\mathrm{E}}(q)) G_{A}^{\mathrm{E}}(q-p)C^{M}_{EE}+{\lambda^{\mathrm{EMM}}_{A^3}}^2 \Big((\tilde{\partial}_t G_{A}^{\mathrm{M}}(q))\nonumber \\[2ex]
 &\times G_{A}^{\mathrm{E}}(q-p)C^{M}_{ME}+(\tilde{\partial}_t G_{A}^{\mathrm{E}}(q))G_{A}^{\mathrm{M}}(q-p)C^{M}_{EM}\Big)\bigg]\nonumber \\[2ex]
 &-2(\lambda^{\mathrm{M}}_{\bar{c}cA})^2(\tilde{\partial}_t G_{c}(q))G_{c}(q-p)C^{g}_{M}+\frac{1}{2}\Big[\lambda^{\mathrm{MMMM}}_{A^4}\nonumber \\[2ex]
 &\times (\tilde{\partial}_t G_{A}^{\mathrm{M}}(q))C^{t}_{MM}+\lambda^{\mathrm{EEMM}}_{A^4}(\tilde{\partial}_t G_{A}^{\mathrm{E}}(q))C^{t}_{ME}\Big]\Bigg\}\,,\label{}
\end{align}
and the flow equation of the electric gluon dressing function 
\begin{align}
  &\partial_t Z_{A,k}^{\mathrm{E}}(p)\nonumber \\[2ex]
  =&-\frac{N_c}{p^2} \int \frac{d^4 q}{(2\pi)^4}\Bigg\{\bigg[{\lambda^{\mathrm{EEE}}_{A^3}}^2 (\tilde{\partial}_t G_{A}^{\mathrm{E}}(q)) G_{A}^{\mathrm{E}}(q-p)C^{E}_{EE}\nonumber \\[2ex]
 &+{\lambda^{\mathrm{EMM}}_{A^3}}^2 (\tilde{\partial}_t G_{A}^{\mathrm{M}}(q)) G_{A}^{\mathrm{M}}(q-p)C^{E}_{MM}+{\lambda^{\mathrm{EEM}}_{A^3}}^2 \Big((\tilde{\partial}_t G_{A}^{\mathrm{M}}(q))\nonumber \\[2ex]
 &\times G_{A}^{\mathrm{E}}(q-p)C^{E}_{ME}+(\tilde{\partial}_t G_{A}^{\mathrm{E}}(q)) G_{A}^{\mathrm{M}}(q-p) C^{E}_{EM}\Big)\bigg]\nonumber \\[2ex]
 &-2(\lambda^{\mathrm{E}}_{\bar{c}cA})^2(\tilde{\partial}_t G_{c}(q))G_{c}(q-p)C^{g}_{E}+\frac{1}{2}\Big[\lambda^{\mathrm{EEMM}}_{A^4}\nonumber \\[2ex]
 &\times (\tilde{\partial}_t G_{A}^{\mathrm{M}}(q))C^{t}_{EM}+\lambda^{\mathrm{EEEE}}_{A^4}(\tilde{\partial}_t G_{A}^{\mathrm{E}}(q))C^{t}_{EE}\Big]\Bigg\}\,,\label{}
\end{align}

\subsection{Ghost self-energy}
\label{appsubsec:GhoSelf-ener}

%
\begin{figure}[t]
\includegraphics[width=0.35\textwidth]{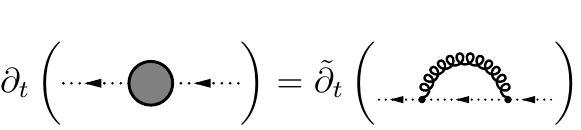}
\caption{Diagrammatic representation of the flow equation for the ghost self-energy in Yang-Mills theory.}\label{fig:cbc-equ}
\end{figure}
%

The flow equation of the ghost self-energy in Yang-Mills theory is depicted in \Fig{fig:cbc-equ}. The one-loop diagram on the r.h.s. reads
\begin{align}
  \parbox[c]{0.15\textwidth}{\includegraphics[width=0.15\textwidth]{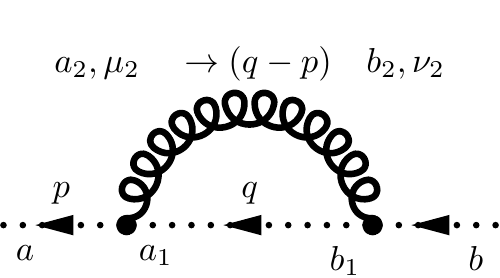}}&\equiv \big(\Sigma^{\bar c c}\big)^{ab}(p)\,.\label{}
\end{align}
with
\begin{align}
  &\big(\Sigma^{\bar c c}\big)^{ab}(p)\nonumber \\[2ex]
 =&\int \frac{d^4 q}{(2\pi)^4} (-)\big(S^{(3)}_{\bar{c}cA}\big)^{aa_1a_2}_{\mu_2'}(-p,q,-q+p)\nonumber \\[2ex]
 &\times \big(\lambda_{\bar{c}cA}\big)_{\mu_2'\mu_2}(-q+p)G_{c}(q)\delta^{a_1b_1} \nonumber \\[2ex]
 &\times (-)\big(S^{(3)}_{\bar{c}cA}\big)^{b_1bb_2}_{\nu_2'}(-q,p,q-p)\nonumber \\[2ex]
 &\times \big(\lambda_{\bar{c}cA}\big)_{\nu_2'\nu_2}(q-p) \Big[G_{A}^{\mathrm{M}}(q-p)\Pi_{\mu_2\nu_2}^{\mathrm{M}}(q-p)\nonumber \\[2ex]
 &+G_{A}^{\mathrm{E}}(q-p)\Pi_{\mu_2\nu_2}^{\mathrm{E}}(q-p)\Big]\delta^{a_2b_2}\,.\label{eq:Sigcbc}
\end{align}
Tracing the color indices, one arrives at
\begin{align}
  &\big(\Sigma^{\bar c c}\big)^{ab}(p)\delta^{ab}\nonumber \\[2ex]
=&\int \frac{d^4 q}{(2\pi)^4} N_c(N_c^2-1)\Big[(\lambda^{\mathrm{M}}_{\bar{c}cA})^2G_{A}^{\mathrm{M}}(q-p)G_{c}(q)C^{\bar c c}_{M}\nonumber \\[2ex]
 &+(\lambda^{\mathrm{E}}_{\bar{c}cA})^2G_{A}^{\mathrm{E}}(q-p)G_{c}(q)C^{\bar c c}_{E}\Big]\,,\label{eq:SigcbcTra}
\end{align}
with
\begin{align}
  C^{\bar c c}_{M}=&\frac{p_s^2 q_s^2\sin ^2\theta }{p_s^2-2 p_s q_s\cos \theta +q_s^2}\,,\\[2ex]
  C^{\bar c c}_{E}=&\frac{1}{\Big(p_s^2-2 p_s q_s\cos \theta +q_s^2\Big) }\nonumber \\[2ex]
 &\times \frac{\Big[p_s^2 q_0-p_s q_s \left(p_0+q_0\right)\cos \theta +p_0 q_s^2\Big]^2}{p_s^2+\left(p_0-q_0\right)^2-2 p_s q_s\cos\theta +q_s^2}\,.\label{}
\end{align}
In the following we also need the expression in \Eq{eq:SigcbcTra} with the replacement of the internal momentum $q\rightarrow-q+p$, which reads
\begin{align}
  &\big(\Sigma^{\bar c c}\big)^{ab}(p)\delta^{ab}\nonumber \\[2ex]
=&\int \frac{d^4 q}{(2\pi)^4} N_c(N_c^2-1)\Big[(\lambda^{\mathrm{M}}_{\bar{c}cA})^2G_{A}^{\mathrm{M}}(q)G_{c}(q-p)C'^{\bar c c}_{M}\nonumber \\[2ex]
 &+(\lambda^{\mathrm{E}}_{\bar{c}cA})^2G_{A}^{\mathrm{E}}(q)G_{c}(q-p)C'^{\bar c c}_{E}\Big]\,,\label{eq:SigcbcTra2}
\end{align}
with
\begin{align}
  C'^{\bar c c}_{M}&=C^{\bar c c}_{M}\Big|_{q\rightarrow-q+p}\,,\quad C'^{\bar c c}_{E}&=C^{\bar c c}_{E}\Big|_{q\rightarrow-q+p}\,.\label{}
\end{align}
Finally, inserting Eqs. (\ref{eq:SigcbcTra}) and (\ref{eq:SigcbcTra2}) into the flow equation in \Fig{fig:cbc-equ}, one is led to the flow equation of the ghost dressing function, as follows
\begin{align}
  &\partial_t Z_{c,k}(p)\nonumber \\[2ex]
 =&-\frac{N_c}{p^2} \int \frac{d^4 q}{(2\pi)^4}\Bigg\{\Big[(\lambda^{\mathrm{M}}_{\bar{c}cA})^2G_{A}^{\mathrm{M}}(q-p)(\tilde{\partial}_t G_{c}(q))C^{\bar c c}_{M}\nonumber \\[2ex]
 &+(\lambda^{\mathrm{E}}_{\bar{c}cA})^2G_{A}^{\mathrm{E}}(q-p)(\tilde{\partial}_t G_{c}(q))C^{\bar c c}_{E}\Big]\nonumber \\[2ex]
 &+\Big[(\lambda^{\mathrm{M}}_{\bar{c}cA})^2(\tilde{\partial}_t G_{A}^{\mathrm{M}}(q))G_{c}(q-p)C'^{\bar c c}_{M}\nonumber \\[2ex]
 &+(\lambda^{\mathrm{E}}_{\bar{c}cA})^2(\tilde{\partial}_t G_{A}^{\mathrm{E}}(q))G_{c}(q-p)C'^{\bar c c}_{E}\Big]\Bigg\}\,.\label{}
\end{align}

\section{Fierz-complete basis of four-quark interactions of $N_f=2$ flavors}
\label{app:Fierzbasis}

Here we list the Fierz-complete basis of four-quark interactions of $N_f=2$ flavors \Eq{eq:NJLaction}, see also e.g., \cite{Mitter:2014wpa, Cyrol:2017ewj, Fu:2022a}. The ten different channels can be classified into four different subsets according to their invariance or not, under the global transformations of the groups $SU_{\mathrm{V}}(N_f)$, $U_{\mathrm{V}}(1)$, $SU_{\mathrm{A}}(N_f)$, and $U_{\mathrm{A}}(1)$. Of the ten channels, several are invariant under all the transformations mentioned above, which read
\begin{align}
 \mathcal{O}^{(V-A)}_{ijlm}{\bar{q}}_iq_j{\bar{q}}_lq_m=&(\bar{q}\gamma_\mu T^0 q)^2-(\bar{q}i\gamma_\mu\gamma_5 T^0 q)^2\,, \label{eq:VmA}\\[2ex]
 \mathcal{O}^{(V+A)}_{ijlm}{\bar{q}}_iq_j{\bar{q}}_lq_m=&(\bar{q}\gamma_\mu T^0 q)^2+(\bar{q}i\gamma_\mu\gamma_5 T^0 q)^2\,, \\[2ex]
 \mathcal{O}^{(S-P)_{+}}_{ijlm}{\bar{q}}_iq_j{\bar{q}}_lq_m=&(\bar{q}\,T^0 q)^2-(\bar{q}\,\gamma_5 T^0 q)^2\nonumber\\[2ex]
&+(\bar{q}\,T^a q)^2-(\bar{q}\,\gamma_5 T^a q)^2\,,\label{eq:SmPp}\\[2ex]
 \mathcal{O}^{(V-A)^{\mathrm{adj}}}_{ijlm}{\bar{q}}_iq_j{\bar{q}}_lq_m=&(\bar{q}\gamma_\mu T^0 t^a q)^2-(\bar{q}i\gamma_\mu\gamma_5 T^0 t^a q)^2\,,\label{eq:VmAadjCh}
\end{align}
where generators of the flavor $SU(N_f)$ group and the color $SU(N_c)$ group are denoted by $T^a$'s and $t^a$'s, respectively, and summation for the indices is assumed. Furthermore, one has $T^0=1/\sqrt{2 N_f} \mathbb{1}_{N_f\times N_f}$. Another two channels, given by
\begin{align}
 \mathcal{O}^{(S+P)_{-}}_{ijlm}{\bar{q}}_iq_j{\bar{q}}_lq_m=&(\bar{q}\,T^0 q)^2+(\bar{q}\,\gamma_5 T^0 q)^2\nonumber\\[2ex]
&-(\bar{q}\,T^a q)^2-(\bar{q}\,\gamma_5 T^a q)^2\,,\label{eq:SpPm}\\[2ex]
 \mathcal{O}^{(S+P)_{-}^{\mathrm{adj}}}_{ijlm}{\bar{q}}_iq_j{\bar{q}}_lq_m=&(\bar{q}\,T^0t^a q)^2+(\bar{q}\,\gamma_5 T^0t^a q)^2\nonumber\\[2ex]
&-(\bar{q}\,T^a t^bq)^2-(\bar{q}\,\gamma_5 T^a t^b q)^2\,,\label{eq:SpPmadjCh}
\end{align}
break the symmetry of $U_\mathrm{A}(1)$ while preserve $SU_\mathrm{V}(N_f)\otimes U_\mathrm{V}(1) \otimes SU_\mathrm{A}(N_f)$. The two channels that read
\begin{align}
 \mathcal{O}^{(S-P)_{-}}_{ijlm}{\bar{q}}_iq_j{\bar{q}}_lq_m=&(\bar{q}\,T^0 q)^2-(\bar{q}\,\gamma_5 T^0 q)^2\nonumber\\[2ex]
&-(\bar{q}\,T^a q)^2+(\bar{q}\,\gamma_5 T^a q)^2\,,\label{eq:SmPm}\\[2ex]
 \mathcal{O}^{(S-P)_{-}^{\mathrm{adj}}}_{ijlm}{\bar{q}}_iq_j{\bar{q}}_lq_m=&(\bar{q}\,T^0t^a q)^2-(\bar{q}\,\gamma_5 T^0t^a q)^2\nonumber\\[2ex]
&-(\bar{q}\,T^a t^b q)^2+(\bar{q}\,\gamma_5 T^a t^b q)^2\,,\label{}
\end{align}
break $SU_\mathrm{A}(N_f)$ while preserve $SU_\mathrm{V}(N_f)\otimes U_\mathrm{V}(1) \otimes U_\mathrm{A}(1)$. The last two independent channels, viz.,
\begin{align}
 \mathcal{O}^{(S+P)_{+}}_{ijlm}{\bar{q}}_iq_j{\bar{q}}_lq_m=&(\bar{q}\,T^0 q)^2+(\bar{q}\,\gamma_5 T^0 q)^2\nonumber\\[2ex]
&+(\bar{q}\,T^a q)^2+(\bar{q}\,\gamma_5 T^a q)^2\,,\label{eq:SpPp}\\[2ex]
 \mathcal{O}^{(S+P)_{+}^{\mathrm{adj}}}_{ijlm}{\bar{q}}_iq_j{\bar{q}}_lq_m=&(\bar{q}\,T^0t^a q)^2+(\bar{q}\,\gamma_5 T^0t^a q)^2\nonumber\\[2ex]
&+(\bar{q}\,T^a t^b q)^2+(\bar{q}\,\gamma_5 T^a t^b q)^2\,.\label{eq:SpPpadj}
\end{align}
break both $U_\mathrm{A}(1)$ and $SU_\mathrm{A}(N_f)$, while preserve $SU_\mathrm{V}(N_f)\otimes U_\mathrm{V}(1)$.

Moreover, it is also useful to combine linearly several different channels of the four-quark couplings to form new independent elements of basis. For instance, the four scalar-pseudoscalar channels as follows
\begin{align}
 \mathcal{O}^{\sigma}_{ijlm}{\bar{q}}_iq_j{\bar{q}}_lq_m=&(\bar{q}\,T^0 q)^2\,,\label{eq:sigmaCh}\\[2ex]
 \mathcal{O}^{\pi}_{ijlm}{\bar{q}}_iq_j{\bar{q}}_lq_m=&-(\bar{q}\,\gamma_5 T^a q)^2\,,\\[2ex]
 \mathcal{O}^{a}_{ijlm}{\bar{q}}_iq_j{\bar{q}}_lq_m=&(\bar{q}\,T^a q)^2\,,\\[2ex]
 \mathcal{O}^{\eta}_{ijlm}{\bar{q}}_iq_j{\bar{q}}_lq_m=&-(\bar{q}\,\gamma_5 T^0 q)^2\,,\label{eq:etaCh}
\end{align}
are obtained from linear combinations of $\mathcal{O}^{(S-P)_{+}}$ in \Eq{eq:SmPp}, $\mathcal{O}^{(S+P)_{-}}$ in \Eq{eq:SpPm}, $\mathcal{O}^{(S-P)_{-}}$ in \Eq{eq:SmPm}, and $\mathcal{O}^{(S+P)_{+}}$ in \Eq{eq:SpPp}.

\section{Some flow functions}
\label{app:flowFunc}

In this appendix we present explicit expressions for some flow equations. The threshold functions involved in this appendix can be found in, e.g., \cite{Fu:2019hdw, Yin:2019ebz}.

The anomalous dimension of quarks $\eta_{q,k}$ in \eq{eq:etapsi} reads
\begin{align}
  \eta_{q,k}=
  &\frac{1}{24\pi^2N_{f}}(4-\eta_{\phi,k})\bar{h}_{k}^{2}\nonumber \\[1ex] 
  &\times\bigg\{(N_{f}^{2}-1)
    \mathcal{FB}_{(1,2)}(\tilde{m}_{q,k}^{2},\tilde{m}_{\pi,k}^{2};
    T,\mu_q,p_{0,\text{\tiny{ex}}})\nonumber \\[1ex] 
  &+\mathcal{FB}_{(1,2)}(\tilde{m}_{q,k}^{2},\tilde{m}_{\sigma,k}^{2};
    T,\mu_q,p_{0,\text{\tiny{ex}}})
    \bigg\}\nonumber \\[1ex] 
  &+\frac{1}{24\pi^2}\frac{N_c^2-1}{2N_c} g_{\bar q A q, k}^{2}\nonumber \\[1ex] 
  &\times\bigg\{2(4-\eta_{A,k}) \mathcal{FB}_{(1,2)}(\tilde{m}_{q,k}^{2},0;
    T,\mu_q,p_{0,\text{\tiny{ex}}})\nonumber \\[1ex] 
  &+3(3-\eta_{q,k})\Big( \mathcal{FB}_{(1,1)}(\tilde{m}_{q,k}^{2},0;
    T,\mu_q,p_{0,\text{\tiny{ex}}})\nonumber \\[1ex] 
  &-2 \mathcal{FB}_{(2,1)}(\tilde{m}_{q,k}^{2},0;T,\mu_q,p_{0,\text{\tiny{ex}}})
    \Big)\bigg\},\label{eq:etapsiexp}
\end{align}
with $N_{f}=2$ and $N_c=3$. The anomalous dimension of mesons at $p=0$ in \Eq{eq:etaphi} reads
\begin{align}\nonumber 
  \eta_{\phi,k}(0) =
  &\, \frac{\bar Z_{\phi,k}}{Z_{\phi,k}(0)} \\[1ex]
  &\,\times \frac{1}{6\pi^2}\bigg
    \{\frac{4}{k^{2}}\bar{\kappa}_{k}(\bar{V}^{''}_{k}(
    \bar{\kappa}_{k}))^{2}\mathcal{BB}_{(2,2)}(\tilde{m}_{\pi,k}^{2},
    \tilde{m}_{\sigma,k}^{2};T)\nonumber \\[1ex]
  &\,+N_{c}\bar{h}_k^{2}\Big[(2\eta_{q,k}-3)\mathcal{F}_{(2)}(
    \tilde{m}_{q,k}^{2};T,\mu_q)\nonumber\\[1ex] 
  &\,-4(\eta_{q,k}-2)\mathcal{F}_{(3)}(\tilde{m}_{q,k}^{2};T,\mu_q)\Big]\bigg\}\,,
\label{eq:etaphiexp}
\end{align}
and that at $p_0=0$ and $\bm{p}^2=k^2$ in \Eq{eq:etaphipk} reads
\begin{align}
  &\eta_{\phi,k}(0,k)\nonumber \\[1ex]
  =&\frac{2}{3\pi^2}\frac{1}{k^{2}}\bar{\kappa}_{k}(\bar{V}^{''}_{k}(
     \bar{\kappa}_{k}))^{2}\mathcal{BB}_{(2,2)}(\tilde{m}_{\pi,k}^{2},
     \tilde{m}_{\sigma,k}^{2};T)\nonumber \\[1ex]
  &-\frac{N_{c}}{\pi^2}\bar h_k^2\int_0^1 d x\bigg[(1-\eta_{q,k}) \sqrt x
    +\eta_{q,k} x\bigg]\nonumber\\[1ex]
  &\times \int_{-1}^1 d \cos \theta\Bigg\{\bigg[
    \Big(\mathcal{FF}_{(1,1)}(\tilde{m}_{q,k}^{2},
    \tilde{m}_{q,k}^{2})-\mathcal{F}_{(2)}(
    \tilde{m}_{q,k}^{2}) \Big)\nonumber \\[1ex]
  &-\Big(\mathcal{FF}_{(2,1)}(\tilde{m}_{q,k}^{2},
    \tilde{m}_{q,k}^{2})-\mathcal{F}_{(3)}(\tilde{m}_{q,k}^{2})
    \Big)\bigg] \nonumber \\[1ex]
  &+\bigg[\Big(\sqrt x -\cos \theta\Big)
    \Big(1+r_F(x')\Big)\mathcal{FF}_{(2,1)}(
    \tilde{m}_{q,k}^{2},\tilde{m}_{q,k}^{2})\nonumber \\[1ex]
  &-\mathcal{F}_{(3)}(\tilde{m}_{q,k}^{2})\bigg]
    -\frac{1}{2}\bigg[\Big(\sqrt x -\cos \theta\Big)
    \Big(1+r_F(x')\Big)\nonumber \\[1ex]
  &\times\mathcal{FF}_{(1,1)}(\tilde{m}_{q,k}^{2},
    \tilde{m}_{q,k}^{2})-\mathcal{F}_{(2)}(
    \tilde{m}_{q,k}^{2})\bigg]\Bigg\}\,,
\label{eq:etaphipkexp}
\end{align}
with $x=\bm{q}^2/k^2$ and $x'=(\bm{q}-\bm{p})^2/k^2$, where $\bm{q}$ and $\bm{p}$ stand for the loop and external 3-momenta, respectively, and $\theta$ is the angle between them. The external momentum is chosen to be $|\bm{p}|=k$.

The contribution to the gluon anomalous dimension from the quark loop reads
\begin{align}
  \eta_{A}^{q}
  &=-\frac{N_f}{\pi^2} g_{\bar q A q,k}^{2}\int_0^1 d x
    \bigg[(1-\eta_{q,k}) \sqrt x+\eta_{q,k} x\bigg]\nonumber\\[1ex]
  &\times \int_{-1}^1 d \cos \theta\bigg[
    \Big(\mathcal{FF}_{(1,1)}(\tilde{m}_{q,k}^{2},\tilde{m}_{q,k}^{2})
    \nonumber\\[1ex]
  &-\mathcal{FF}_{(2,1)}(\tilde{m}_{q,k}^{2},\tilde{m}_{q,k}^{2})\Big)
    +\Big(\sqrt x \cos^2\!\theta-\cos \theta\Big)\nonumber\\[1ex]
  &\times\Big(1+r_F(x')\Big)\Big(\mathcal{FF}_{(2,1)}(\tilde{m}_{q,k}^{2},
    \tilde{m}_{q,k}^{2})\nonumber\\[1ex]
  &-\frac{1}{2}\mathcal{FF}_{(1,1)}(\tilde{m}_{q,k}^{2},\tilde{m}_{q,k}^{2})
    \Big)\bigg]\,.\label{eq:DeltaAqexpl}
\end{align}
As same as in \Eq{eq:etaphipkexp}, the external momentum in \Eq{eq:DeltaAqexpl} is chosen to be $|\bm{p}|=k$. The in-medium contribution from the light quarks, included in \Eq{eq:etaAT}, is given by
\begin{align} 
\Delta\eta_{A}^{q} &= \eta_{A}^{q} - \eta_{A}^{q}\big|_{T,\mu = 0}\,,\label{eq:DeltaAqexplMed}
\end{align}
with $\bar m_q = \bar m_l$, $N_f = 2$ and the light-quark--gluon coupling, $g_{\bar l A l}$ in \Eq{eq:DeltaAqexpl}. For the strange quark, since the vacuum contribution is also presented in \Eq{eq:etaAQCD2+1}, one needs
\begin{align} 
\eta_{A}^{s}&= \eta^s_{A,\mathrm{vac}}+\Delta\eta_{A}^{s}\,,\label{eq:etaAs}
\end{align}
which is obtained from \Eq{eq:DeltaAqexpl} with $\bar m_q = \bar m_s$, $N_f = 1$ and the strange-quark--gluon coupling $g_{\bar s A s}$.

The second term on the r.h.s. of \Eq{eq:etaAT}, $\Delta\eta_{A}^{\mathrm{glue}}$, denotes the contribution to the gluon anomalous dimension from the thermal part of glue sector, which is taken into account through the thermal screening mass of gluons. The modified gluon anomalous dimension reads
\begin{align}
  \bar \eta_{A}=&\eta_{A}+\frac{\Delta m_\textrm{scr}^2(k,T)}{\bar Z_{A}k^2}(2-\eta_{A})\nonumber\\[1ex]
  &-\frac{1}{\bar Z_{A}k^2}\partial_t \big(\Delta m_\textrm{scr}^2(k,T)\big)\,.\label{eq:barEtaA}
\end{align}
with $\bar \eta_{A}=-\partial_t \bar Z_{A}/ \bar Z_{A}$, where $\eta_{A}$ is the gluon anomalous dimension without the thermal screening mass, and see \cite{Fu:2019hdw} for a more detailed discussion. The screening mass reads
\begin{align}
  \Delta m_\textrm{scr}^2(k,T)&=(c\,T)^2 \exp\Big[-\Big(\frac{k}{\pi T}\Big)^n\Big]\,.\label{eq:screenmass}
\end{align}
where $c=2$ is adopted for $N_f=2+1$, which is consistent with the result in \cite{Cyrol:2017qkl}. Furthermore, $n=2$ is chosen in \eq{eq:screenmass}.

The flow of the quark-gluon couplings in \Eq{eq:dtgbllA} and \Eq{eq:dtgbssA} are given by
\begin{align}
  \overline{\textrm{Flow}}^{(3),A}_{(\bar q q A)}
  =&\frac{3}{8\pi^2N_c}
     \bar g_{\bar{q}qA,k}^{3}\tilde{m}_{q,k}^{2}\bigg
     \{\frac{2}{15}(5-\eta_{A,k})\mathcal{FB}_{(2,2)}(
     \tilde{m}_{q,k}^{2},0)\nonumber \\[1ex] 
   &+\frac{1}{3}(4-\eta_{q,k})\mathcal{FB}_{(3,1)}(
     \tilde{m}_{q,k}^{2},0)\bigg\}\nonumber \\[1ex] 
   &+\frac{3N_c}{8\pi^2}\bar g_{\bar{q}qA,k}^{2}\bar g_{A^3,k}
     \bigg\{\frac{1}{20} (5-\eta_{q,k})
     \mathcal{FB}_{(1,2)}(\tilde{m}_{q,k}^{2},0)\nonumber \\[1ex] 
   &-\frac{1}{6} (4-\eta_{q,k})\mathcal{FB}_{(2,1)}(\tilde{m}_{q,k}^{2},0)+
     \frac{1}{30} (5-2\eta_{q,k})\nonumber \\[1ex] 
   &\times\mathcal{FB}_{(2,2)}(\tilde{m}_{q,k}^{2},0)
     -\frac{4}{15}(5-\eta_{A,k})\nonumber \\[1ex] 
   &\times\mathcal{FB}_{(1,2)}(\tilde{m}_{q,k}^{2},0)
     +\frac{1}{30}(10-3\eta_{A,k})\nonumber \\[1ex] 
   &\times\mathcal{FB}_{(1,3)}(\tilde{m}_{q,k}^{2},0)\bigg\}\,. 
 \label{eq:dtgA}
\end{align}
and 
\begin{align}
  \overline{\textrm{Flow}}^{(3),\phi}_{{(\bar q  q A)}}
  =&-\frac{1}{8\pi^2N_f}\bar g_{\bar{q}qA,k}
     \bar{h}_{k}^{2}\bigg\{\frac{1}{6}(4-\eta_{q,k})\nonumber
  \\[1ex] 
   &\times\Big[\mathcal{FB}_{(2,1)}(\tilde{m}_{q,k}^{2},
     \tilde{m}_{\sigma,k}^{2})\nonumber \\[1ex] 
   &+2\tilde{m}_{q,k}^{2}\mathcal{FB}_{(3,1)}(\tilde{m}_{q,k}^{2},
     \tilde{m}_{\sigma,k}^{2})\Big]+\frac{2}{15}(5-\eta_{\phi,k})
     \nonumber \\[1ex] 
   &\times\Big[\mathcal{FB}_{(1,2)}(\tilde{m}_{q,k}^{2},\tilde{m}_{\sigma,k}^{2})
     \nonumber \\[1ex] 
   &+\tilde{m}_{q,k}^{2}\mathcal{FB}_{(2,2)}(\tilde{m}_{q,k}^{2},
     \tilde{m}_{\sigma,k}^{2})\Big]\bigg\}\nonumber \\[1ex] 
   &-\frac{N_f^2-1}{8\pi^2N_f}\bar g_{\bar{q}qA,k}\bar{h}_{k}^{2}
     \bigg\{\frac{1}{6}(4-\eta_{q,k})\nonumber \\[1ex] 
   &\times\Big[\mathcal{FB}_{(2,1)}(\tilde{m}_{q,k}^{2},
     \tilde{m}_{\pi,k}^{2})\nonumber \\[1ex] 
   &+2\tilde{m}_{q,k}^{2}\mathcal{FB}_{(3,1)}(\tilde{m}_{q,k}^{2},
     \tilde{m}_{\pi,k}^{2})\Big]+\frac{2}{15}(5-\eta_{\phi,k})
     \nonumber \\[1ex] 
   &\times\Big[\mathcal{FB}_{(1,2)}(\tilde{m}_{q,k}^{2},
     \tilde{m}_{\pi,k}^{2})\nonumber \\[1ex] 
   &+\tilde{m}_{q,k}^{2}\mathcal{FB}_{(2,2)}(\tilde{m}_{q,k}^{2},
     \tilde{m}_{\pi,k}^{2})\Big]\bigg\}\,.
 \label{eq:dtgphi}
\end{align}
Here for the light-quark--gluon coupling in \Eq{eq:dtgbllA} we use the light quark mass and $N_f = 2$, and for the strange-quark--gluon coupling in \Eq{eq:dtgbssA} we use the strange quark mass and $N_f = 1$.

The flow of the four-quark coupling in the $\sigma-\pi$ channel from two gluon exchanges in \Eq{eq:Flowbqqbqq2}, as diagrammatically shown in the first line of \Fig{fig:v4quark-equ}, reads
\begin{align}
  \overline{\mathrm{Flow}}^{(4),A}_{(\bar q\tau q)^2}
  &=-\frac{3}{2\pi^2}\frac{N_c^2-1}{2N_c}
    \Big(\frac{3}{4}-\frac{1}{N_c^2}\Big) \bar g_{\bar q A q,k}^{4}
    \nonumber \\[1ex] 
  &\quad\times\bigg\{\frac{2}{15}(5-\eta_{A,k})\Big[
    \mathcal{FB}_{(1,3)}(\tilde{m}_{q,k}^{2},0)
    \nonumber \\[1ex] 
  &\quad-\tilde{m}_{q,k}^{2}
    \mathcal{FB}_{(2,3)}(\tilde{m}_{q,k}^{2},0)
    \Big]\nonumber \\[1ex] 
  &\quad+\frac{1}{12}(4-\eta_{q,k})\Big[\mathcal{FB}_{(2,2)}(
    \tilde{m}_{q,k}^{2},0)\nonumber \\[1ex] 
  &\quad-2\tilde{m}_{q,k}^{2}\mathcal{FB}_{(3,2)}(
    \tilde{m}_{q,k}^{2},0)\Big]\bigg\}\,.
 \label{eq:dtlambdaA}
\end{align}
The contributions to the flow of four-quark coupling from two meson exchanges, as shown in the second line of \Fig{fig:v4quark-equ}, reads
\begin{align}
  \overline{\mathrm{Flow}}^{(4),\phi}_{(\bar q\tau q)^2}
  &=\frac{1}{32\pi^2}\frac{N_f^2-2}{N_fN_c}
    \bar{h}_{k}^{4}\bigg\{\frac{2}{15}(5-\eta_{\phi,k})
    \nonumber \\[1ex] 
  &\quad\times\Big[\Big(\mathcal{FBB}_{(1,1,2)}(
    \tilde{m}_{q,k}^{2},\tilde{m}_{\pi,k}^{2},
    \tilde{m}_{\sigma,k}^{2})\nonumber \\[1ex] 
  &\quad+\mathcal{FBB}_{(1,1,2)}(\tilde{m}_{q,k}^{2},
    \tilde{m}_{\sigma,k}^{2},
    \tilde{m}_{\pi,k}^{2})\nonumber \\[1ex] 
  &\quad-2\mathcal{FB}_{(1,3)}(\tilde{m}_{q,k}^{2},
    \tilde{m}_{\pi,k}^{2})
    \Big)\nonumber \\[1ex] 
  &\quad-\tilde{m}_{q,k}^{2}\Big(
    \mathcal{FBB}_{(2,1,2)}(\tilde{m}_{q,k}^{2},
    \tilde{m}_{\pi,k}^{2},\tilde{m}_{\sigma,k}^{2})
    \nonumber \\[1ex] 
  &\quad+\mathcal{FBB}_{(2,2,1)}(\tilde{m}_{q,k}^{2},\tilde{m}_{\pi,k}^{2},
    \tilde{m}_{\sigma,k}^{2})\nonumber \\[1ex] 
  &\quad-2\mathcal{FB}_{(2,3)}(\tilde{m}_{q,k}^{2},\tilde{m}_{\pi,k}^{2})\Big)\Big]
    +\frac{1}{6}(4-\eta_{q,k})\nonumber \\[1ex] 
  &\quad\times\Big[\Big(\mathcal{FBB}_{(2,1,1)}(\tilde{m}_{q,k}^{2},
    \tilde{m}_{\pi,k}^{2},\tilde{m}_{\sigma,k}^{2})\nonumber \\[1ex] 
  &\quad-\mathcal{FB}_{(2,2)}(\tilde{m}_{q,k}^{2},\tilde{m}_{\pi,k}^{2})
    \Big)\nonumber \\[1ex] 
  &\quad-2\tilde{m}_{q,k}^{2}\Big(\mathcal{FBB}_{(3,1,1)}(\tilde{m}_{q,k}^{2},
    \tilde{m}_{\pi,k}^{2},\tilde{m}_{\sigma,k}^{2})\nonumber \\[1ex] 
  &\quad-\mathcal{FB}_{(3,2)}(\tilde{m}_{q,k}^{2},\tilde{m}_{\pi,k}^{2})
    \Big)\Big]\bigg\}\,.
 \label{eq:dtlambdaAphi}
\end{align}

The flow of the Yukawa coupling in \Eq{eq:Flowbqqpion} reads
\begin{align}
  &\overline{\mathrm{Flow}}^{(3)}_{(\bar q \bm\tau q) \bm\pi}\nonumber\\[1ex]
  &\quad=\frac{1}{4\pi^2 N_f}\bar{h}_k^3
    \bigg[-(N_{f}^{2}-1)\nonumber \\[1ex] 
  &\qquad\times L_{(1,1)}^{(4)}(
    \tilde{m}_{q,k}^{2},\tilde{m}_{\pi,k}^{2},
    \eta_{q,k},\eta_{\phi,k};T,\mu_q,
    p_{0,\text{\tiny{ex}}})\nonumber\\[1ex]
  &\qquad+L_{(1,1)}^{(4)}(\tilde{m}_{q,k}^{2},\tilde{m}_{\sigma,k}^{2},
    \eta_{q,k},\eta_{\phi,k};T,\mu_q,p_{0,\text{\tiny{ex}}})
    \bigg]\nonumber\\[1ex]
  &\qquad-\frac{3}{2\pi^2}\frac{N_c^2-1}{2N_c}
    \bar g_{\bar q q A , k}^{2}\bar{h}_k
    \nonumber \\[1ex]  
  &\qquad\times L_{(1,1)}^{(4)}(\tilde{m}_{q,k}^{2},0,
    \eta_{q,k},\eta_{A,k};T,
    \mu_q,p_{0,\text{\tiny{ex}}})\,.\label{eq:hexp}
\end{align}

	
\bibliography{ref-lib}

\end{document}